\documentclass{ws-ijmpd}
\usepackage[super,compress]{cite}
\def\draftnote{\today\quad\currenttime\quad MG13\qquad\jobname}%
\usepackage{graphicx}

\def\PGRB{_{P\mbox{-}GRB}}

\begin{document}

\markboth{Remo Ruffini}
{Black Holes, Supernovae and Gamma Ray Bursts}

%
\catchline{}{}{}{}{}
%

\title{\uppercase{Black Holes, Supernovae and Gamma Ray Bursts}}

\author{\uppercase{Remo Ruffini}}

\address{Dip.\ di Fisica, Sapienza University of Rome and ICRA\\
Piazzale Aldo Moro 5, I--00185, Rome, Italy\\
ICRANet, Piazzale della Repubblica 10, I--65122 Pescara, Italy\\
Universit\'e de Nice Sophie Antipolis, Nice, CEDEX 2\\
Grand Ch\^ateau Parc Valrose\\
$^*$E-mail: ruffini@icra.it\\}

\maketitle

\begin{history}\received{?? ?? 2013}\revised{8 August 2013}\end{history}

\begin{abstract}
We review recent progress in our understanding of the nature of gamma ray bursts (GRBs) and in particular, 
of the relationship between short GRBs and long GRBs. 
The first example of a short GRB is described. 
The coincidental occurrence of a GRB with a supernova (SN) is explained within the induced gravitational collapse (IGC) paradigm, following the sequence: 
1) an initial binary system consists of a compact carbon-oxygen (CO) core star and a neutron star (NS); 
2) the CO core explodes as a SN, and part of the SN ejecta accretes onto the NS which reaches its critical mass and collapses to a black hole (BH) giving rise to a GRB; 
3) a new NS is generated by the SN as a remnant. 
The observational consequences of this scenario are outlined.


\end{abstract}



\section{Introduction}\label{aba:sec1}

While supernovae (SNe) have been known and studied for a long time, from $1054$ A.D. to the classic work of Baade and Zwicky in 1939, observations of GRBs only date from the detection by the Vela satellites in the early $1970$s, see e.g. Ref.~\refcite{1975ASSL...48.....G}. It has only  been after the observations by the Beppo-Sax satellite and the optical identification of GRBs that their enormous energetics, $10^3$--$10^4$ times larger than those of SNe, have been determined: energies of the order of $10^{54}$ erg, equivalent to the release of $\sim M_\odot c^2$ in few tens of seconds.
This situation has become even more interesting after the observation of a temporal coincidence between the emission of a GRB and a SN, see e.g. GRB 980425\cite{2000ApJ...536..778P} and SN 1998bw\cite{1998Natur.395..670G}.
The explanation of this coincidence has led to a many-cosmic-body-interaction and therefore to the introduction of a cosmic matrix: a C-matrix. 
This totally unprecedented situation has lead to the opening of a new understanding of a vast number of unknown domains of physics and astrophysics.

\subsection{CRAB --- pulsars and NS rotational energy}

Of all the objects in the sky none has been richer in results for physics, astronomy and astrophysics than the Crab Nebula. Although a result of the $1054$ A.D. supernova observed by Chinese, Japanese and Korean astronomers, the nebula itself was not identified till 1731, and not associated with that supernova until the last century, but it has
been of interest to astronomers, and later astrophysicists and theoretical physicists ever since,
even very recently, 
see e.g. the discovery by Agile of the giant flare discovered in September $2010$ \cite{2011Sci...331..736T}. It was only in 1968 that a pulsar was discovered at its center following the predicted existence of rapidly rotating NSs in that decade then soon after observed as pulsars.

However, there still remains to explain an outstanding physical process needed to model this object: the expulsion of the shell of the SN during the process of gravitational collapse to a NS. We are currently gaining some understanding of the physical processes governing NSs, motivated by the research on GRBs and BH formation which is being fully exploited to this end at the present time. Paradoxically the study of BHs was started by the discovery of the NS in the Crab Nebula. This study and the understanding of BH formation and consequently of the emission of GRBs is likely to lead, in this Faustian effort to learn the laws of nature, to the understanding of the process of NS formation and the expulsion of the remnant in the SN explosion.

\begin{figure}
\centering
\includegraphics[width=0.7\hsize,clip]{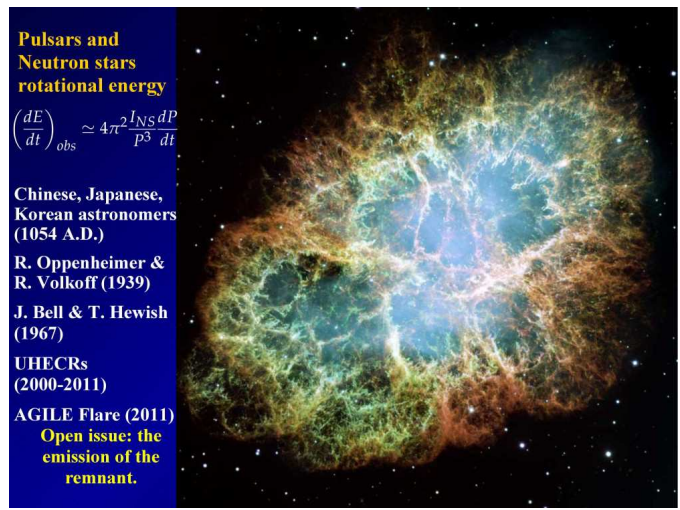}
\caption{Hubble Space Telescope photograph (2005) of the Crab Nebula.}
\label{Fig1}
\end{figure}

\begin{figure}
\centering
\includegraphics[width=0.7\hsize,clip]{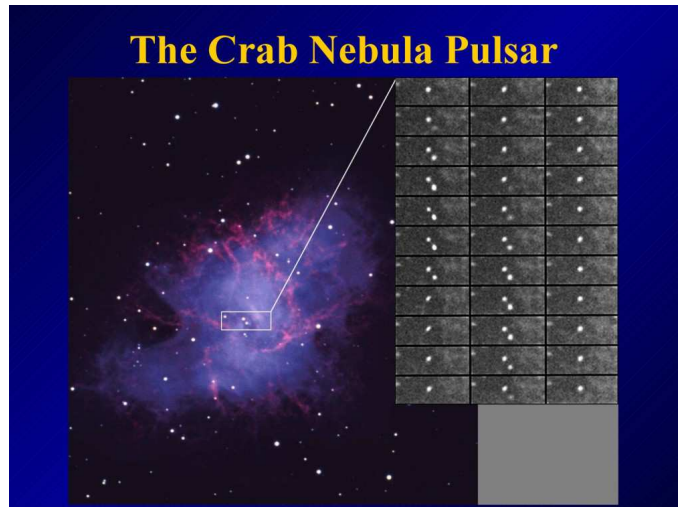}
\caption{The sequence of black and white images on the right is separated by one millisecond intervals, from which it is clear that the left star is a pulsar with a period of $P=33$ milliseconds. This period changes with a rate $dP/dt$ of $12.5$ microseconds per year. The fact that the loss of rotational energy of a neutron star with moment of inertia $I$ is given by $dE/dt\propto-I(1/P^3)dP/dt$ explains precisely the energetics of the pulsar and proves at once the existence of NSs \protect\cite{1968ApJ...153..865F}.}
\label{Fig2}
\end{figure}

That NSs exist in nature has been proven by the direct observation of pulsars. The year $1967$ marked the discovery of the first pulsar, observed at radio wavelengths in November $28$, $1967$ by Jocelyn Bell Burnell and Antony Hewish \cite{1968Natur.217..709H}. Just a few months later, the pulsar NP$0532$ was found in the center of the Crab Nebula (see Fig.~\ref{Fig1}) and observed first at radio wavelengths and soon after at optical wavelengths (see Fig.~\ref{Fig2}).

\begin{figure}
\centering
\includegraphics[width=0.7\hsize,clip]{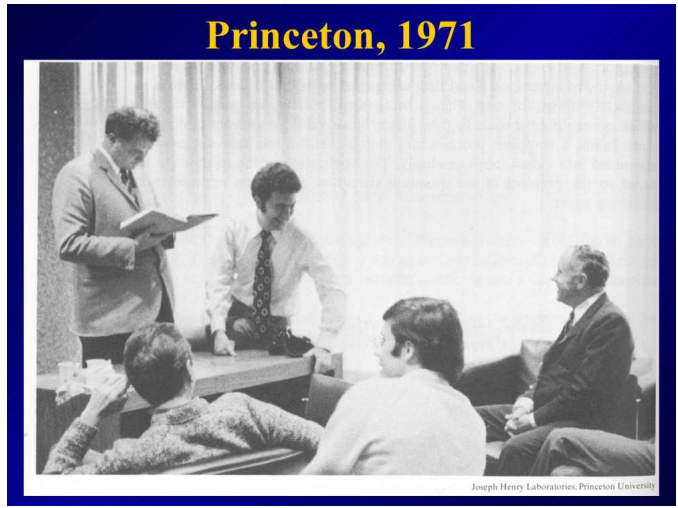}
\caption{Standing to the left Tullio Regge, sitting on the desk Remo Ruffini and sitting on the chair John Wheeler.}
\label{Fig3}
\end{figure}

The discovery of NSs led our small group working around John Wheeler in Princeton to direct our main attention to the study of continuous gravitational collapse introduced by Oppenheimer and his students (see Fig.~\ref{Fig3}). The work in Princeton addressed the topic of BHs, gravitational waves (GWs) and cosmology. A summary of that work can be found in Refs.~\refcite{LesHouches,1974bhgw.BOOK.....R}, where a vast number of topics of relativistic astrophysics was reconsidered, including the cross-sections of GW detectors, the possible sources of GWs and especially, an entirely new family of phenomena occurring around BHs.

\subsection{The BH mass-energy formula}

The most important result in understanding the physics and astrophysics of BHs has been the formulation of the BH mass-energy formula. From this formula, indeed, it became clear that up to $50 \%$ of the mass-energy of a BH could be extracted by using reversible transformations \cite{1971PhRvD...4.3552C}. It then followed that during the formation of a BH, some of the most energetic processes in the universe could exist, releasing an energy of the order of $\sim10^{54}$ erg for a $1 M_{\odot}$ BH.

\begin{figure}
\centering
\includegraphics[width=0.7\hsize,clip]{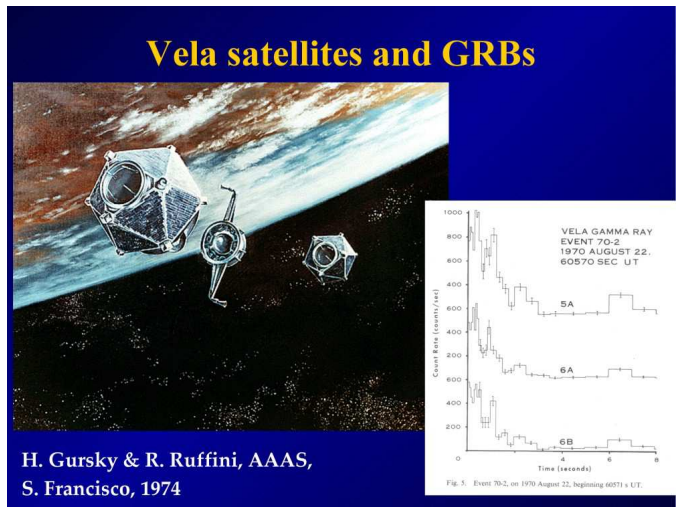}
\caption{The Vela satellites, see e.g. the Ian Strong chapter in Ref.~\protect\refcite{1975ASSL...48.....G}.}
\label{Fig5}
\end{figure}

\subsection{VELA satellites and GRBs}

\begin{figure}
\centering
\includegraphics[width=0.7\hsize,clip]{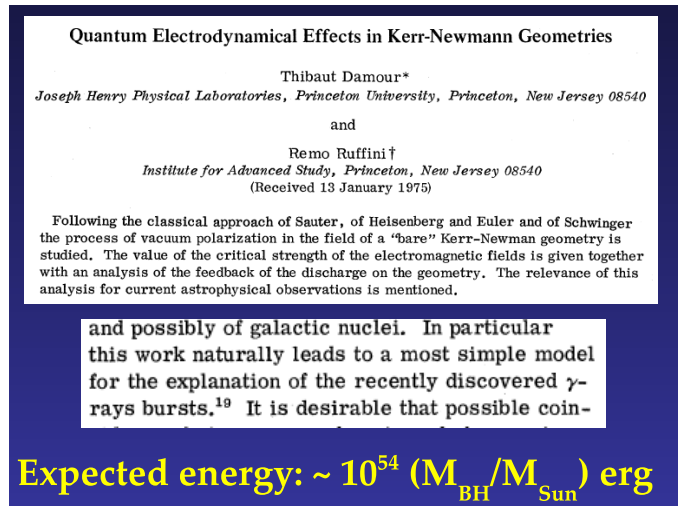}
\caption{The classic paper   Ref.~\protect\refcite{1975PhRvL..35..463D} by Damour and Ruffini on the extractable energy of a Kerr-Newman BH through vacuum polarization.}
\label{Fig6}
\end{figure}

In Ref.~\refcite{RRKerr} 
I described how the observations of the Vela satellites were fundamental in discovering GRBs, see Fig.~\ref{Fig5}. Initially it was difficult to model GRBs to understand their nature since their distance from the Earth was unknown, and thousands of models were presented \cite{RRKl} attempting to explain the mystery they presented. Just a few months after the public announcement of their discovery \cite{1975ASSL...48.....G}, with T. Damour, a collaborator at Princeton, I formulated a theoretical model based on the extractable energy of a Kerr-Newmann BH through a vacuum polarization process as the origin of GRBs, see Fig.~\ref{Fig6}. In our paper \cite{1975PhRvL..35..463D}, we pointed out that vacuum polarization occurring in the field of electromagnetic BHs could release a vast $e^+e^-$ plasma which self-accelerates and gives origin to the GRB phenomenon. Energetics for GRBs all the way up to $\sim10^{55}$ ergs was theoretically predicted for a $10$ $M_\odot$ BH. The dynamics of this $e^- e^+$ plasma was first studied by J.R. Wilson and myself with the collaboration of  S.-S Xue and J.D. Salmonson \cite{1999A&A...350..334R,2000A&A...359..855R}.

\subsection{The BATSE detectors and short and long GRBs}

\begin{figure}
\centering
\includegraphics[width=0.7\hsize,clip]{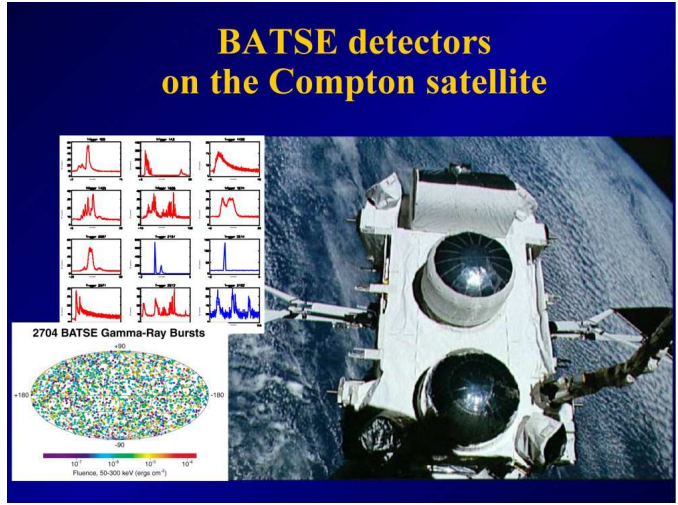}
\caption{The BATSE detectors on-board the Compton satellite (taken from the NASA website http://science.nasa.gov/science-at-nasa/1997/ast15jan97).}
\label{Fig8}
\end{figure}

\begin{figure}
\centering
\includegraphics[width=0.7\hsize,clip]{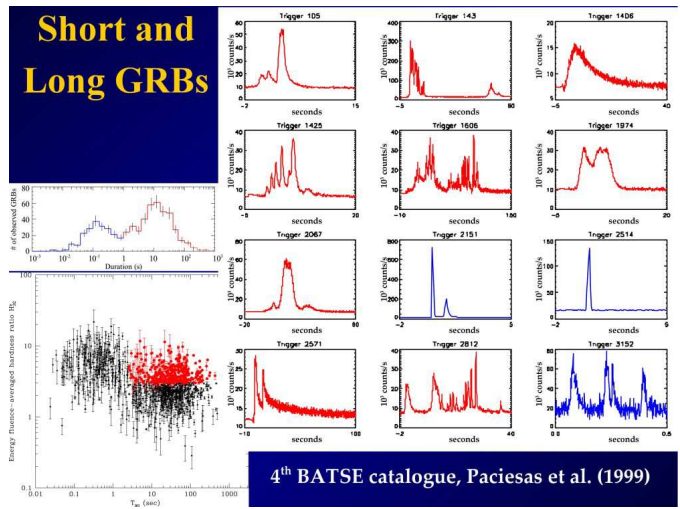}
\caption{Short and long GRB light curves and their temporal distribution from the 4$^{th}$ BATSE catalog, Ref.~\protect\refcite{1999ApJS..122..465P}}
\label{Fig9}
\end{figure}

The launching of the Compton satellite with the BATSE detectors on-board (see Fig.~\ref{Fig8}) led to the following important discoveries:
\begin{enumerate}
\item the homogeneus distribution of GRBs in the universe (see Fig.~\ref{Fig8});
\item the existence of short GRBs lasting less than 1 second (see Fig.~\ref{Fig9}); and
\item the existence of long GRBs, lasting more than 1 second (see Fig.~\ref{Fig9}). 
\end{enumerate}

The crucial contribution to interpreting GRBs came from the Beppo-Sax satellite which led to a much more precise definition of their position in the sky obtained using a wide field X-ray camera and narrow field instrumentation. This enabled the optical identification of GRBs and the determination of their cosmological redshifts, and consequently of their energetics, which turned out to be up to $\sim10^{55}$ erg, precisely the value predicted by Damour and myself in Ref.~\refcite{1975PhRvL..35..463D}. 
Since that time no fewer than ten different  X- and $\gamma$-ray observatory  missions and numerous observations at optical and radio wavelengths have allowed us to reach a deeper understanding of the nature of GRBs.

After reviewing in the next paragraphs some recent theoretical progress motivated by the study of GRBs, I will turn to the first example of a genuine short GRB 090227B \cite{2013ApJ...763..125M}. 
Then I will describe the analysis of the GRB 090618 in the fireshell scenario \cite{2012A&A...543A..10I} and illustrate the first application of the IGC paradigm to it \cite{2012A&A...548L...5I}. Finally I will indicate some recent results on a possible distance indicator inferred from a GRB-SN connection within the IGC paradigm \cite{2013A&A...552L...5P}, then giving some additional evidence coming from the identification of the NS created by the SN and its use as a cosmological candle.

\subsection{Some recent theoretical progress}

I would like just to present some key images and cite corresponding references to articles documenting some crucial progress we have made that is propedeutic for understanding the physics and astrophysics of GRBs.

\subsubsection{Mass, charge and angular momentum in a Kerr-Newman BH: the dyadotorus}

\begin{figure}
\centering
\includegraphics[width=0.7\hsize,clip]{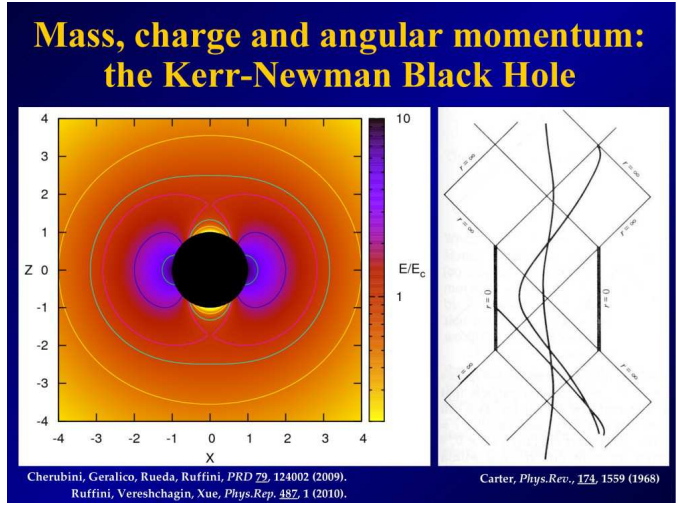}
\caption{On the left: the dyadotorus as introduced in Ref.~\protect\refcite{2009PhRvD..79l4002C}; on the right: the space-time diagram representing the region inside the horizon of a Kerr-Newman BH Ref.~\protect\refcite{1968PhRv..174.1559C}.}
\label{Fig10}
\end{figure}

Fig.~\ref{Fig10} summarizes the profound difference in analyzing the Kerr-Newman BH between the original paper of B. Carter \cite{1968PhRv..174.1559C} and our current approach to the physics of the dyadotorus. In Carter's approach attention was focused on geodesics crossing through the horizon of an eternally existing BH and reaching either the BH singularity or analytic extensions to other asymptotically flat space-times.
Instead our approach is directed to the fundamental physical processes occurring outside the horizon of a BH and to their possible detection in the dynamical phases of BH formation. Our major focus is to understand the quantum processes leading to vacuum polarization and pair creation and the resulting dynamical expansion to infinity. This mechanism is essential to extract energy from the BH, an amount which can be as high as $50\%$ of its total mass energy as already mentioned above.
To reach a theoretical understanding of this problem, it was necessary to introduce the dyadotorus, see Fig.~\ref{Fig10}.

\subsubsection{Thermalization of an electron-positron plasma}

A key result was obtained by analyzing the evolution of the $e^+e^-$ plasma created in the dyadotorus by vacuum polarization. Cavallo and Rees \cite{1978MNRAS.183..359C} envisaged that the sudden annihilation of the $e^+e^-$ pairs and the expansion of 
the thermal radiation
in the circumburst medium (CBM) would lead to an explosion very similar to an H-bomb, a scenario identified as the fireball model.

By considering the essential role of three-body interactions, we have proven that the $e^+e^-$ pairs do not annihilate all at once as claimed by Cavallo and Rees \cite{1978MNRAS.183..359C} but they thermalize with the photons \cite{2007PhRvL..99l5003A} and keep expanding in a shell until transparency of the $e^+e^-$ plasma is reached \cite{2001ApJ...555L.113R}, a new paradigm for GRBs called the fireshell model.

\begin{figure}
\centering
\includegraphics[width=0.7\hsize,clip]{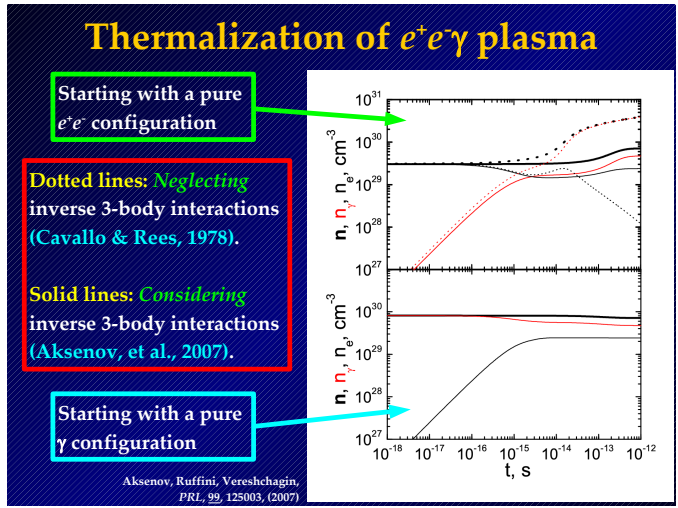}
\caption{The thermalization of a pure $e^+e^-\gamma$ plasma, taken from Ref.~\protect\refcite{2007PhRvL..99l5003A}.}
\label{Fig11}
\end{figure}

\subsubsection{The new approach to analyzing NS equilibrium configurations in an unified approach encompassing all fundamental interactions}

\begin{figure}
\centering
\includegraphics[width=0.8\hsize,clip]{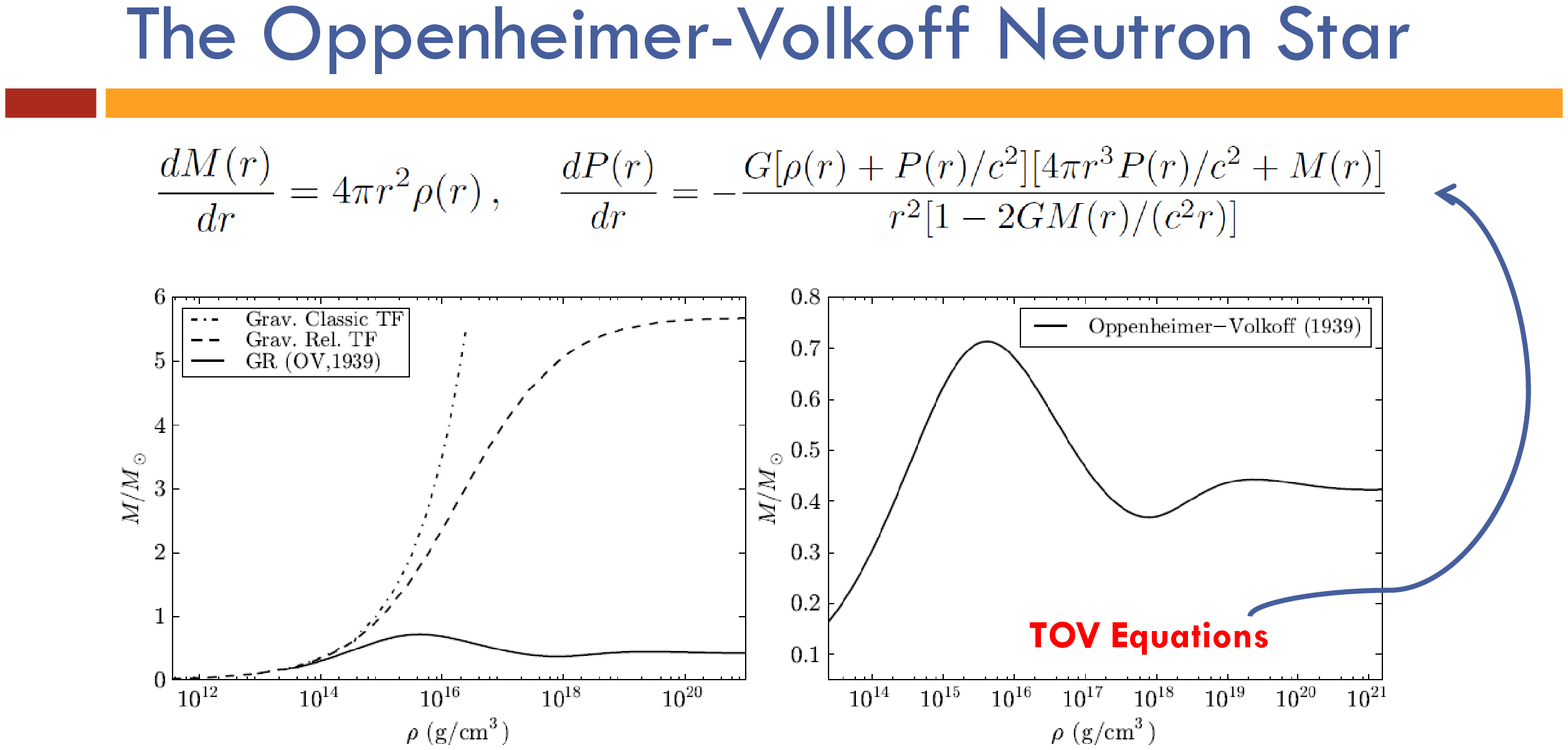}
\caption{The Oppenheimer-Volkoff NS, see Ref.~\protect\refcite{1939PhRv...55..374O}.}
\label{Fig12}
\end{figure}

\begin{figure}
\centering
\includegraphics[width=0.7\hsize,clip]{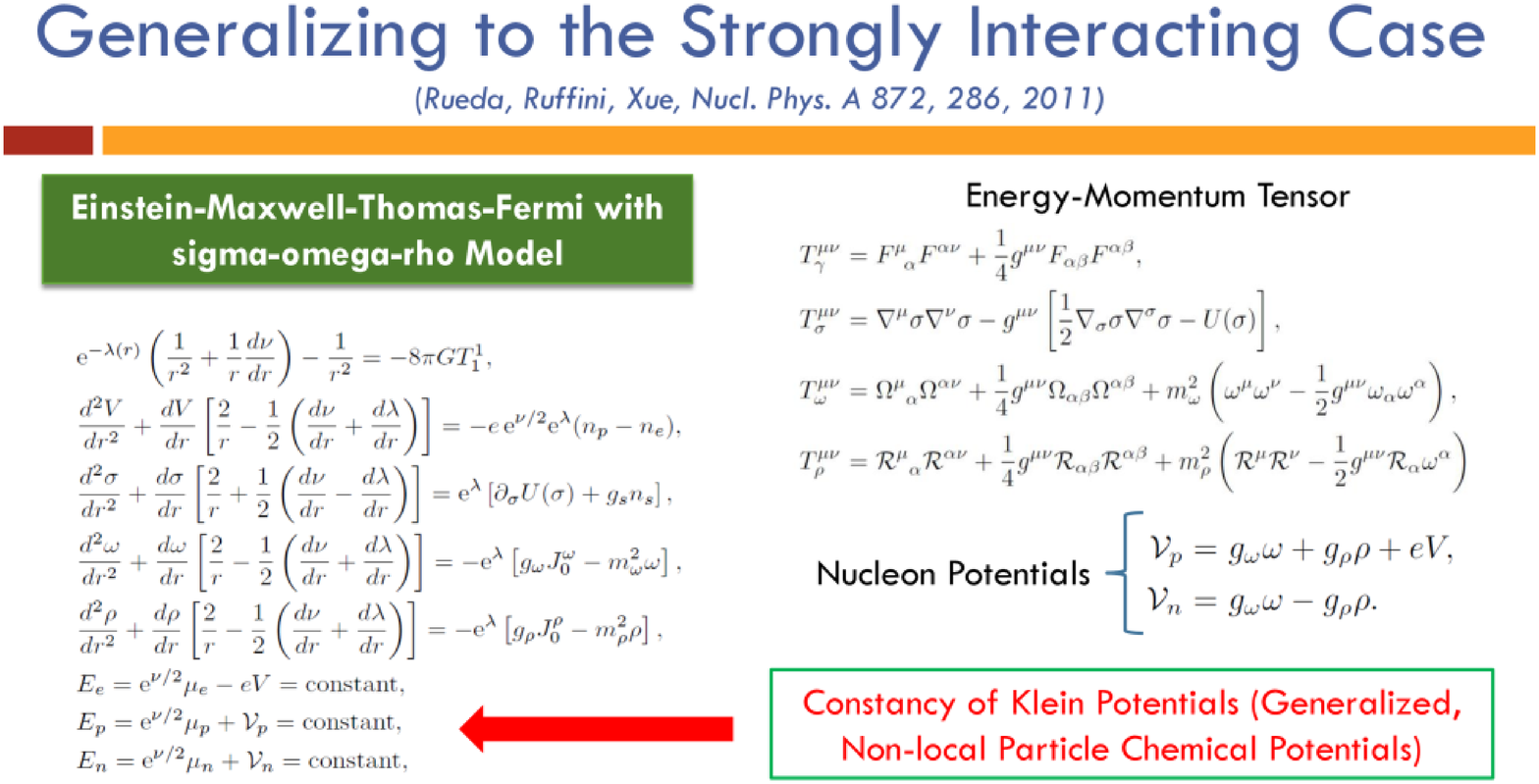}
\caption{From Ref.~\protect\refcite{2011NuPhA.872..286R}.}
\label{Fig13}
\end{figure}

\begin{figure}
\centering
\includegraphics[width=0.7\hsize,clip]{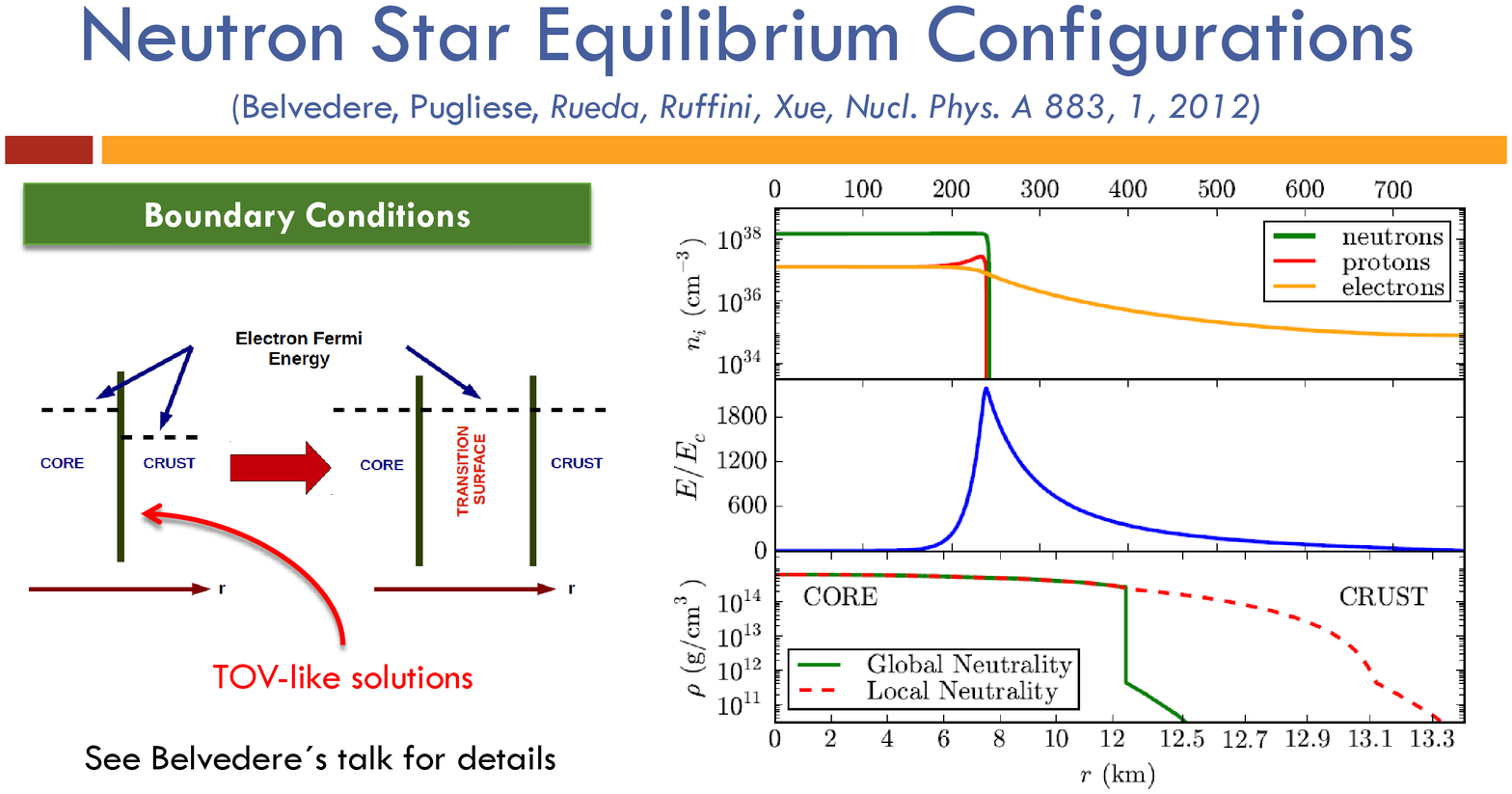}
\caption{From Ref.~\protect\refcite{2012NuPhA.883....1B}.}
\label{Fig14}
\end{figure}

A completely new approach to NS equilibrium configurations was advanced in recent years and has evolved into a much more complicated model, fulfilling the  criteria needed conceptually for the description of NSs \cite{2008pint.conf..207R,2012NuPhA.883....1B}. The first model for a NS was given by Gamow as a system entirely composed of neutrons governed by both Fermi statistics and Newtonian gravity. The extension of this model to general relativity was made by Oppenheimer and his students, leading to the classic Tolman-Oppenheimer-Volkoff (TOV) equilibrium equations \cite{1939PhRv...55..364T,1939PhRv...55..374O}. This was then extended to a system of three degenerate gases of neutrons, protons and electrons and solved by John Wheeler and his students and collaborators \cite{1966ARA&A...4..393W}. However, they assumed local charge neutrality for mathematical convenience. It was later realized that a more complete description was needed, since the previous analyses violated basic thermodynamic and general relativistic conditions required for conservation of the Klein potential \cite{1949RvMP...21..531K}. A new much more complete treatment appeared to be needed involving in a self-consistent way all the fundamental forces. A new model has since emerged, extending the general relativistic Thomas-Fermi equations to the strong and weak interactions throughout the entire NS \cite{2012NuPhA.883....1B} (see Fig.~\ref{Fig12}--\ref{Fig14}).
This complete model satisfies instead global charge neutrality of the entire configuration and not strict local charge neutrality, an erroneous assumption usually made in the existing literature on NS models.

With this short summary of the most relevant conceptual and theoretical issues, I now briefly summarize how some of them have allowed us to reach a new understanding of the short GRBs and the SN-GRB connection.

\section{GRB 090227B: The Missing Link between the Genuine Short and Long GRBs}\label{short_marco}

\subsection{Introduction}\label{short_marco_sec:0}

\begin{figure}
\centering
\includegraphics[width=0.7\hsize,clip]{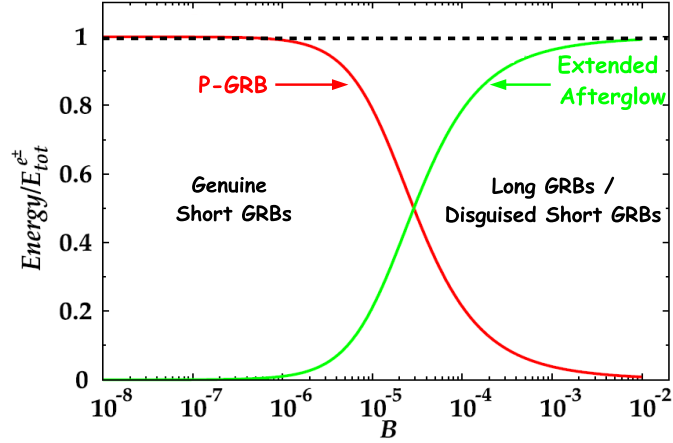}
\caption{The energy emitted in the extended afterglow (solid green curve) and in the P-GRB (solid red curve) in units of $E_{e^+e^-}^{tot} = 1.77 \times 10^{53}$ erg (dashed horizontal line), as functions of $B$. The crossing point, corresponding to the condition $E\PGRB \equiv 50\%E_{e^+e^-}^{tot}$, marks the division between the genuine short and the disguised short and long GRB regions.}
\label{short_marco_fig:1}
\end{figure}

Using the data obtained from the Fermi-GBM satellite \cite{2009ApJ...702..791M},  Ref.~\refcite{2013ApJ...763..125M} has proven the existence of yet another class of GRBs 
theoretically predicted by the fireshell model \cite{2001ApJ...555L.113R,2002ApJ...581L..19R}
which we define here as  the ``genuine short GRBs.''
This canonical class of GRBs is characterized by extremely small values of the Baryon Load $B \lesssim 10^{-5}$ (see Fig.~\ref{short_marco_fig:1}).
The energy emitted in the proper GRB (P-GRB) described below
is predominate 
with respect to the extended afterglow
and its characteristic duration \cite{2013ApJ...763..125M} is expected to be shorter than a fraction of a second (see Sec.~\ref{short_marco_sec:fireshell:genuine}).

A search has begun for these genuine short GRBs among the bursts detected by the Fermi-GBM instrument during the first three years of its mission. 
The initial list of short GRBs was reduced by requiring that no prominent X-ray or optical afterglow be observed. 
The GRB 090227B  has been identified among the remaining bursts.
A spectral analysis of its source has been performed
from its observed light curves, and  its cosmological redshift and all the basic parameters of the burst, as well as the isotropic energy, the Lorentz $\Gamma$ factor at transparency, and the intrinsic duration, have all been inferred from  theory.

In Sec.~\ref{short_marco_sec:fireshell} the relevant properties of the fireshell model are summarized.
In Sec.~\ref{short_marco_sec:analysis} the observations of GRB 090227B by various satellites and their data analysis are reviewed.
In Sec.~\ref{short_marco_sec:3} all the parameters characterizing this GRB within the fireshell scenario, including the redshift, are determined. 
In the conclusions we show that this GRB is the missing link between the genuine short and the long GRBs, with some common characteristics of both classes.
Further analysis of genuine short GRBs with a smaller value of $B$ should lead to a P-GRB with an even more pronounced thermal component.
The progenitor of GRB 090227B is identified as a symmetric binary system of two neutron stars, each of $\sim1.34M_\odot$, see e.g. Ref.~\refcite{2012arXiv1205.6915R}.

\subsection{The fireshell versus the fireball model}\label{short_marco_sec:fireshell}

\subsubsection{The GRB prompt emission in the fireball scenario}

A variety of models have been developed to theoretically explain the observational properties of GRBs, among which  the fireball model \cite{2004RvMP...76.1143P} is one of those most often used.
In Refs.~\refcite{1978MNRAS.183..359C,1986ApJ...308L..47G,1986ApJ...308L..43P} it was proposed that the sudden release of a large quantity of energy in a compact region can lead to an optically thick photon-lepton plasma and to the production of $e^+e^-$ pairs.
The sudden initial total annihilation of the $e^+e^-$ plasma was assumed by 
Cavallo and Rees \cite{1978MNRAS.183..359C}, leading to an enormous release of energy pushing on the  CBM: the ``fireball.'' 

An alternative approach, originating in the gravitational collapse to a BH, is the fireshell model, see e.g. Refs.~\refcite{PhysRep,2011IJMPD..20.1797R}.
Here the GRB originates from an optically thick $e^+e^-$ plasma in thermal equilibrium, with a total energy of $E_{tot}^{e^\pm}$.
This plasma is initially confined between the radius $r_h$ of a BH and the dyadosphere radius 
\begin{equation}\label{090618_eq:rh}
r_{ds}= r_h \left[2 \alpha \frac{E_{tot}^{e^+e^-}}{m_e c^2}\left(\frac{\hbar /m_e c}{r_h}\right)^3\right]^{1/4},
\end{equation}
where $\alpha$ is the usual fine structure constant, $\hbar$ the Planck constant, $c$ the speed of light, and $m_e$ the mass of the electron.
The lower limit of $E_{tot}^{e^\pm}$ is assumed to coincide with the
observed isotropic energy $E_{iso}$ emitted in X-rays and gamma rays alone in the GRB. 
The condition of thermal equilibrium assumed in this model \cite{2007PhRvL..99l5003A} distinguishes this approach from alternative ones, e.g. Ref.~\refcite{1978MNRAS.183..359C}.

In the fireball model, the prompt emission, including the sharp luminosity variations \cite{2000ApJ...539..712R}, are caused by the prolonged and variable activity of the ``inner engine'' \cite{1994ApJ...430L..93R,2004RvMP...76.1143P}.
The conversion of the fireball energy to radiation originates in shocks, either internal (when faster moving matter overtakes a slower moving shell, see Ref.~\refcite{1994ApJ...430L..93R}) or external (when the moving matter is slowed down by the external medium surrounding the burst, see Ref.~\refcite{1992MNRAS.258P..41R}).
Much attention has been given to synchrotron emission from relativistic electrons in the CBM, possibly accompanied by Self-Synchrotron Compton (SSC) emission, to explain the observed GRB spectrum. 
These processes were found to be consistent with the observational data of many GRBs \cite{1996ApJ...466..768T,2000ApJS..127...59F}. 
However, several limitations have been reported in relation with the low-energy spectral slopes of time-integrated spectra \cite{1997ApJ...479L..39C,2002ApJ...581.1248P,2002A&A...393..409G,2003A&A...406..879G} and the time-resolved
spectra \cite{2003A&A...406..879G}.
Additional limitations on SSC emission have also been pointed out in Refs.~\refcite{2008MNRAS.384...33K,2009MNRAS.393.1107P}.

The latest phases of the afterglow are described in the fireball model by assuming an equation of motion given by the  Blandford-McKee self-similar power-law solution \cite{1976PhFl...19.1130B}. 
The maximum Lorentz factor of the fireball is estimated from the temporal occurrence of the peak of the optical emission, which is identified with the peak of the forward external shock emission \cite{2007A&A...469L..13M,2009ApJ...702..489R} in the thin shell approximation \cite{1999ApJ...520..641S}.
Several partly alternative and/or complementary scenarios have been developed distinct from the fireball model, e.g. based on quasi-thermal Comptonization \cite{1999A&AS..138..527G}, Compton drag emission \cite{1991ApJ...366..343Z,1994MNRAS.269.1112S}, synchrotron emission from a decaying magnetic field \cite{2006ApJ...653..454P}, jitter radiation \cite{2000ApJ...540..704M}, Compton scattering of synchrotron self-absorbed photons \cite{2000ApJ...544L..17P,2004MNRAS.352L..35S}, and photospheric emission \cite{2000ApJ...529..146E,2000ApJ...530..292M,2002ARA&A..40..137M,2002MNRAS.336.1271D,2006A&A...457..763G,2009ApJ...702.1211R,2010ApJ...725.1137L}. 
In particular, it was pointed out in Ref.~\refcite{2009ApJ...702.1211R} that photospheric emission overcomes some of the difficulties of purely non-thermal emission models.

\subsubsection{The fireshell scenario}

\begin{figure}
\includegraphics[width=0.49\hsize,clip]{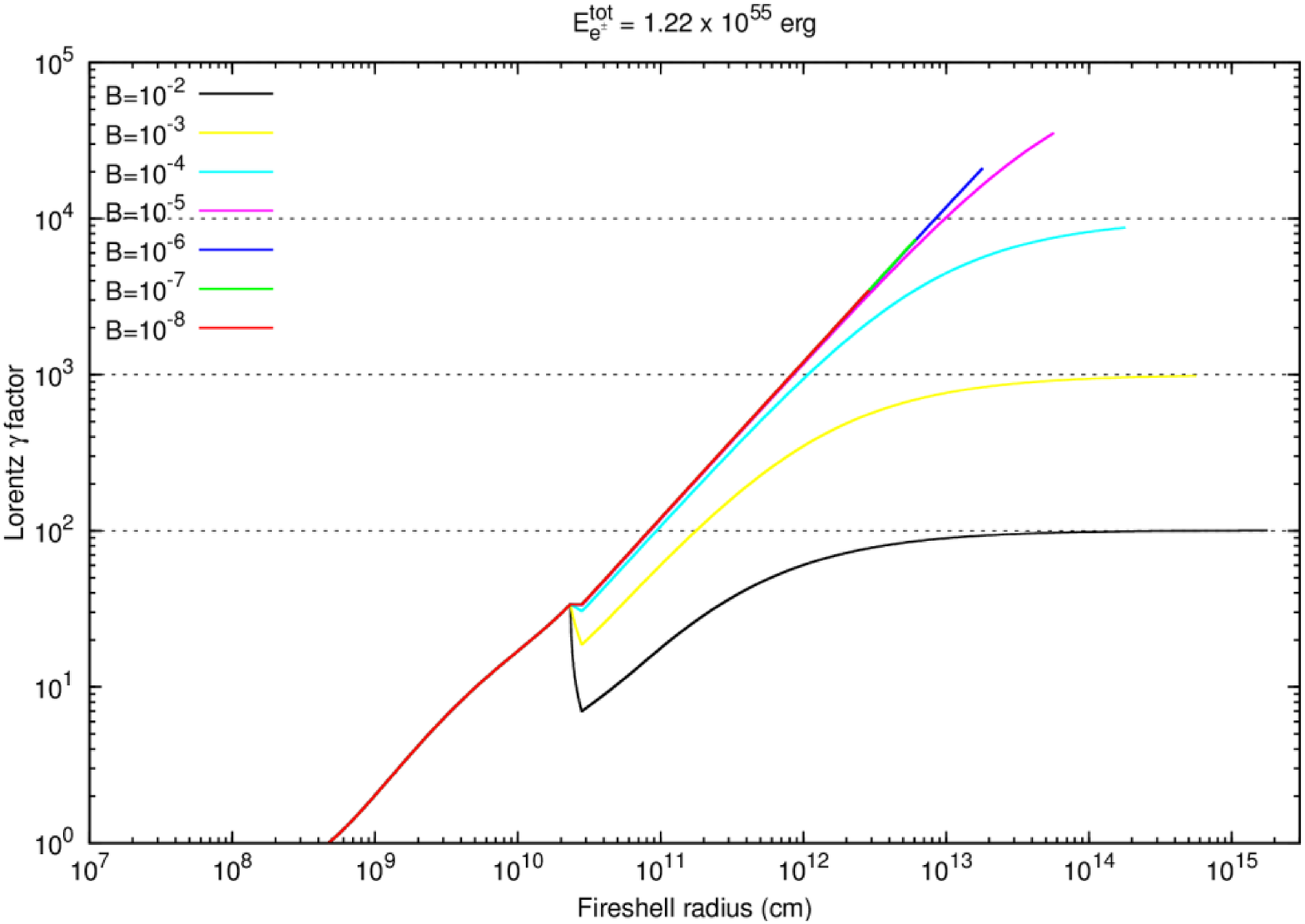}
\includegraphics[width=0.49\hsize,clip]{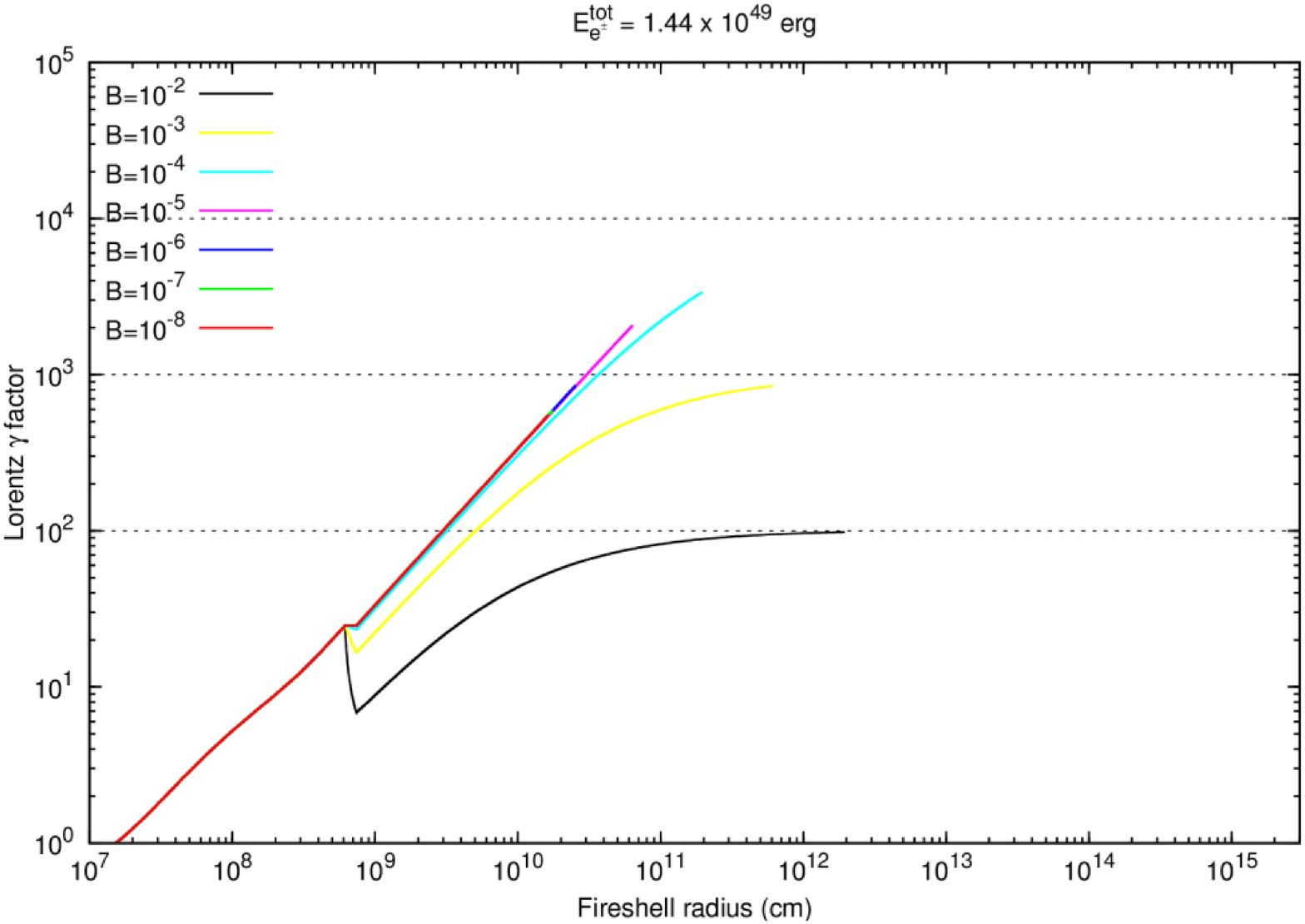}
\caption{Evolution of the Lorentz $\Gamma$ factor until the transparency emission for a GRB of a fixed $E_{tot}^{e^+e^-}$ = 1.22 $\times$ 10$^{55}$ (upper panel),and $E_{tot}^{e^+e^-}$ = 1.44 $\times$ 10$^{49}$, for different values of the baryon load $B$. This computation refers to a BH mass of 10 M$_{\odot}$ and the transparency condition $\tau \equiv \int_R dr (n_{e^{\pm}} + n_{e^-}^b) \sigma_T = 0.67$, where $\sigma_T$ is the Thomson cross-section and the integration is over the thickness of the fireshell \protect\cite{1999A&AS..138..513R}.
}
\label{090618_fig:no4g}
\end{figure}

In the fireshell model, the rate equation for the $e^+e^-$ pair plasma and its dynamics (the pair-electromagnetic pulse or PEM pulse for short) have been described in Ref.~\refcite{2000A&A...359..855R}. 
This plasma engulfs the baryonic material left over from the process of gravitational collapse having a mass $M_B$, still maintaining thermal equilibrium between electrons, positrons, and baryons. 
The baryon load is measured by the dimensionless parameter $B=M_B c^2/E_{tot}^{e^+e^-}$. 
Ref.~\refcite{1999A&AS..138..513R} showed that no relativistic expansion of the plasma exists for $B > 10^{-2}$. The fireshell is still optically thick and self-accelerates to ultrarelativistic velocities (the pair-electromagnetic-baryonic pulse or PEMB pulse for short \cite{1999A&AS..138..513R}). 
Then the fireshell becomes transparent and the P-GRB is emitted \cite{2001ApJ...555L.113R}. 
The final Lorentz gamma factor at transparency can vary over a wide range between $10^2$ and $10^4$ as a function of $E_{tot}^{e^+ e^-}$ and $B$, see Fig.~\ref{090618_fig:no4g}.
For its final determination it is necessary to explicitly integrate the rate equation for the $e^+ e^-$ annihilation process and evaluate, for a given BH mass and a given $e^+e^-$ plasma radius, at what point the transparency condition is reached \cite{2000A&A...359..855R} (see Fig.~\ref{090618_fig:no4}). 

The fireshell scenario does not require any prolonged activity of the inner engine.
After transparency, the remaining accelerated baryonic matter still expands ballistically and starts to slow down from collisions with the CBM of average density $n_{CBM}$. In the standard fireball scenario \cite{2006RPPh...69.2259M}, the spiky light curve is assumed to be caused by internal shocks.
In the fireshell model the entire extended-afterglow emission is assumed to originate from an expanding thin shell, which maintains energy and momentum conservation during its collision with the CBM. 
The condition of a fully radiative regime is assumed \cite{2001ApJ...555L.113R}.
This in turn allows one to estimate the characteristic inhomogeneities of the CBM, as well as its average value.

It is appropriate to point out another difference between our treatment and others in the current literature. 
The complete analytic solution of the equations of motion of the baryonic shell were developed in Refs.~\refcite{2004ApJ...605L...1B,2005ApJ...620L..23B}, while elsewhere the Blandford-McKee self-similar approximate solution is almost always adopted without justification  \cite{1993ApJ...415..181M,1997ApJ...489L..37S,1998ApJ...494L..49S,1997ApJ...491L..19W,1998ApJ...496L...1R,1999ApJ...513..679G,1998ApJ...493L..31P,1999ApJ...511..852G,2000ARA&A..38..379V,2002ARA&A..40..137M}. 
The analogies and differences between the two approaches have been explicitly explained in Ref.~\refcite{2005ApJ...633L..13B}.

In our general approach, a canonical GRB bolometric light curve is composed of two different parts: the P-GRB and the extended afterglow. 
The relative energetics of these two components and the observed temporal separation between the corresponding peaks is a function of the above three parameters $E_{tot}^{e^+e^-}$, $B$, and the average value of the $n_{CBM}$. The first two parameters are inherent to the accelerator characterizing the GRB, i.e., the optically thick phase, while the third one is inherent to the environment surrounding the GRB which gives rise to the extended-afterglow. 
For the observational properties of a relativistically expanding fireshell model, a crucial concept has been the introduction of the EQui-Temporal Surfaces (EQTS). Here too our model differs from those in the literature by having deriving an analytic expression of the EQTS obtained from the solutions to the equations of motion \cite{2005ApJ...633L..13B}.

\subsubsection{The emission of the P-GRB}\label{sec:pgrb}

\begin{figure}
\centering
\begin{tabular}{cc}
(a) & (b)\\
\includegraphics[width=0.50\hsize,clip]{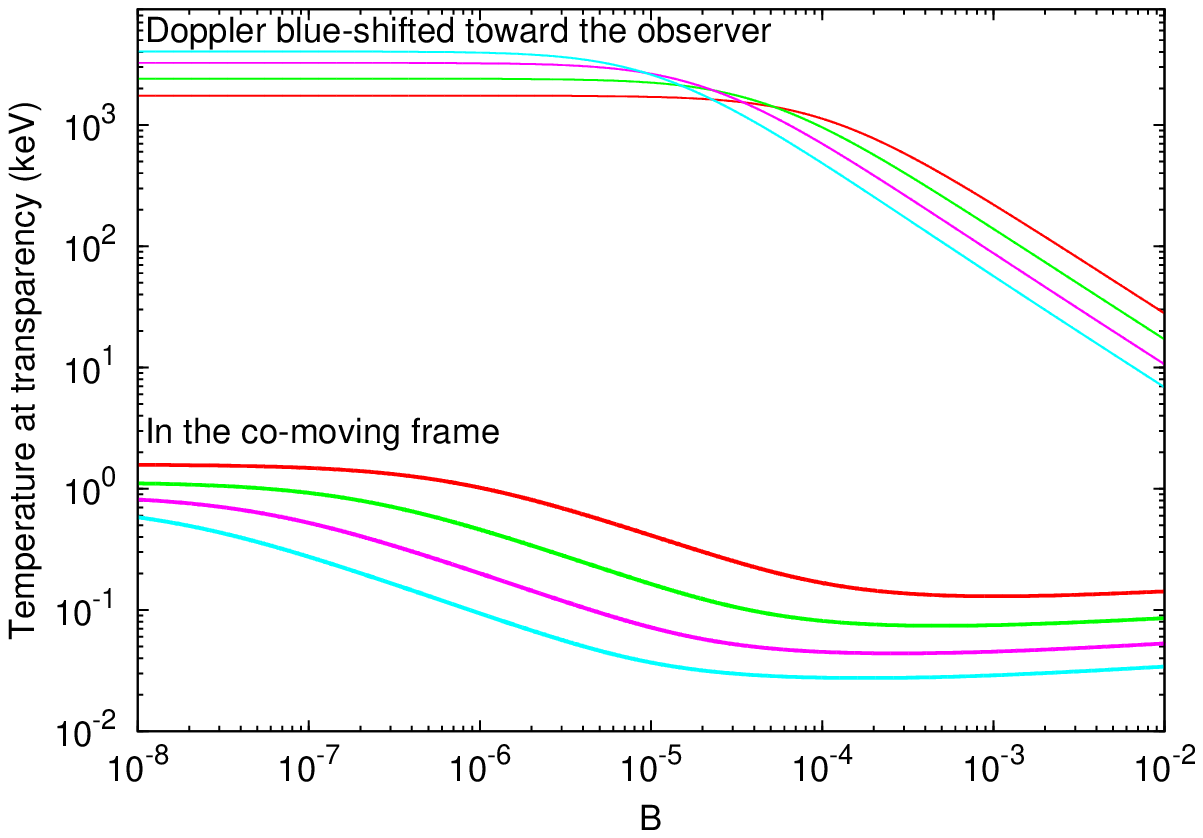}& \includegraphics[width=0.50\hsize,clip]{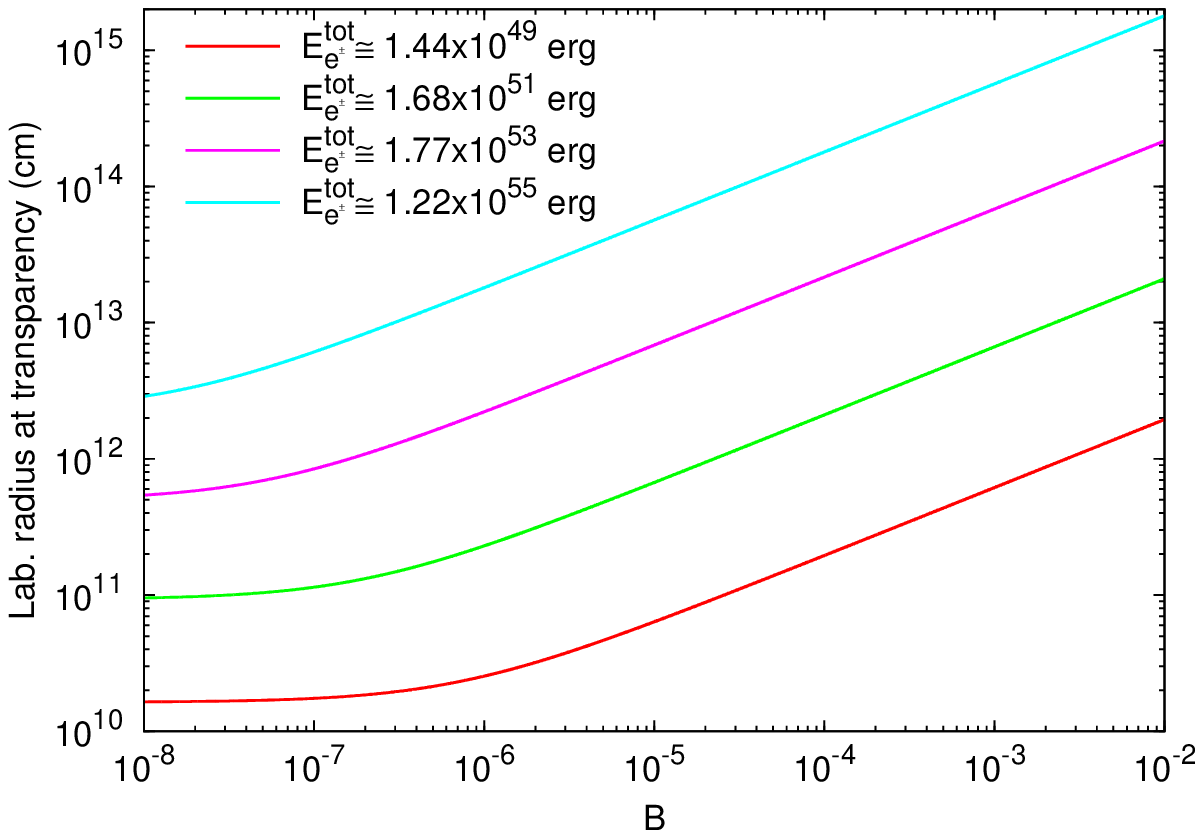}\\
(c) & (d)\\
\includegraphics[width=0.50\hsize,clip]{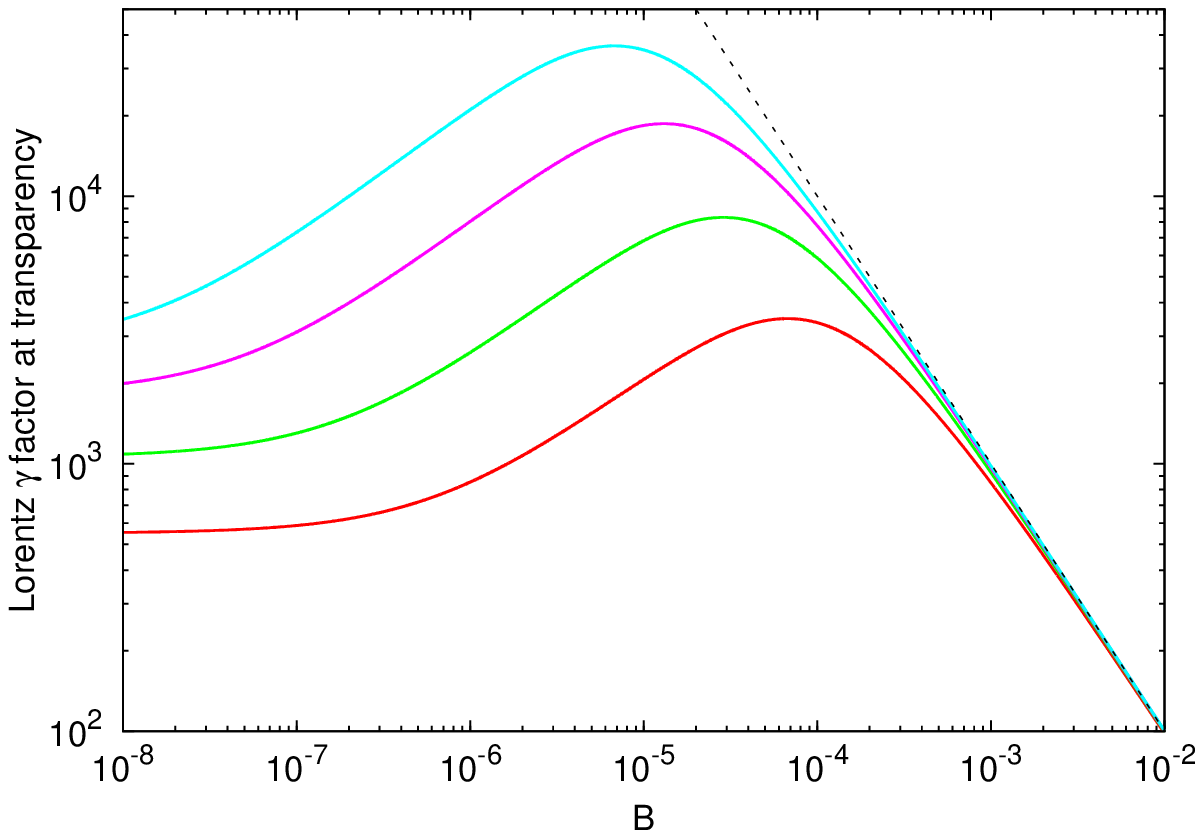}& \includegraphics[width=0.50\hsize,clip]{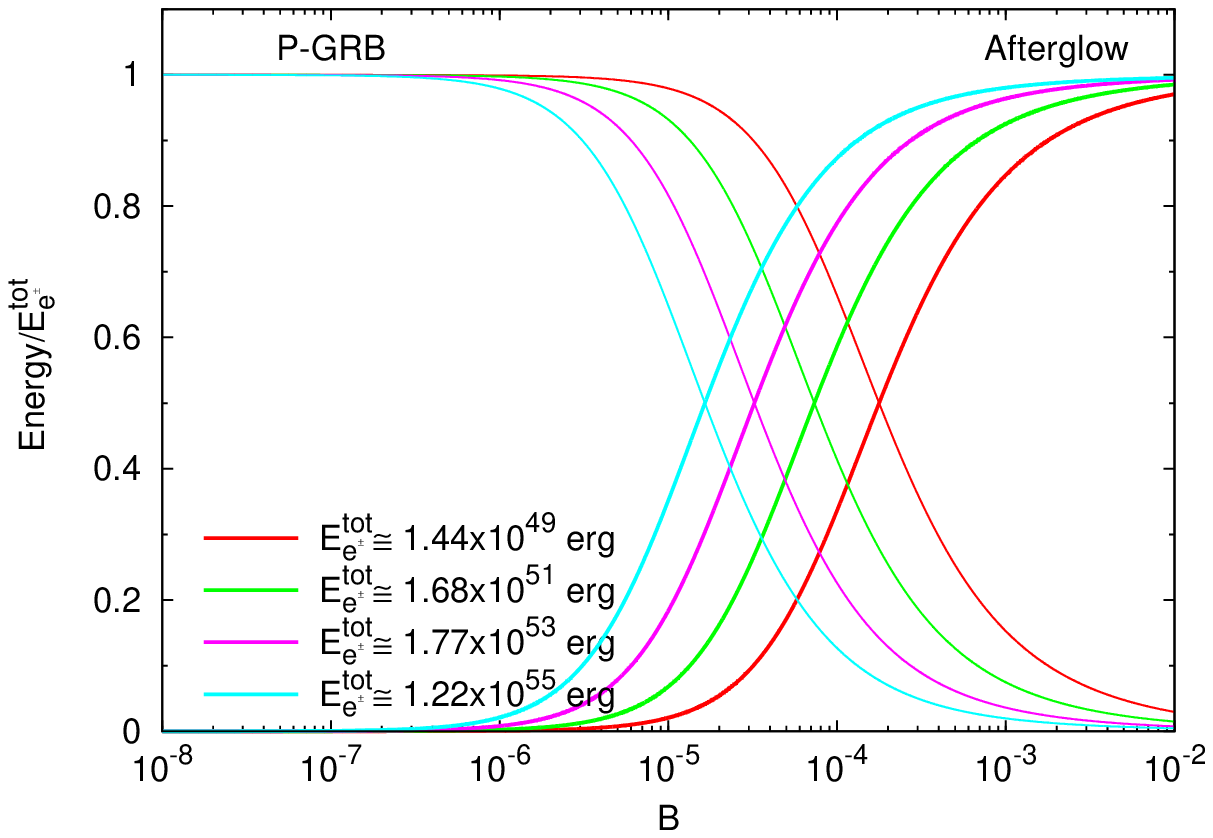}
\end{tabular}
\caption{fireshell temperature in the comoving and observer frame and the laboratory radius at the transparency emission (panels (a) and (b)), the Lorentz $\Gamma$ factor at the transparency (panel (c)) and the energy radiated in the P-GRB and in the afterglow in units of $E_{tot}^{e^+e^-}$ (panel (d)) as a function of the baryon load $B$ for four different values of $E_{tot}^{e^+e^-}$.}
\label{090618_fig:no4}
\end{figure}

The lower limit for $E_{tot}^{e^+e^-}$ is given by the observed isotropic energy $E_{iso}$ emitted in the GRB. 
The identification of the energy of the afterglow and of the P-GRB determines the baryon load $B$ and from these it is possible to determine the value of the Lorentz $\Gamma$ factor at transparency, the observed temperature as well as the temperature in the comoving frame and the laboratory radius at transparency, see Fig.~\ref{090618_fig:no4}.
We can indeed determine from the spectral analysis of the P-GRB candidate the temperature $kT_{obs}$ and the energy $E\PGRB $ emitted at the point of transparency.
The relation between these parameters cannot be expressed analytically, only through numerical integration of the entire set of fireshell equations of motion.
In practice we need to perform a trial-and-error procedure to find a set of values that fits the observations.

The direct measure of the temperature of the thermal component at transparency offers very important new information on the determination of the GRB parameters.
Two different phases are present in the emission of the P-GRB: one corresponding to the emission of the photons when transparency is reached and another corresponding to the early interaction of the ultra-relativistic protons and electrons with the CBM.
A spectral energy distribution with both a thermal and a non-thermal component should be expected to result from these two phases.

\subsubsection{The extended afterglow}\label{exaft}

The majority of articles in the current literature have analyzed the afterglow emission as the result of various combinations of synchrotron and inverse Compton processes \cite{2004RvMP...76.1143P}.
It appears, however, that this description is not completely satisfactory \cite{2003A&A...406..879G,2008MNRAS.384...33K,2009MNRAS.393.1107P}.

\begin{figure}
\centering
\includegraphics[width=0.49\hsize,clip]{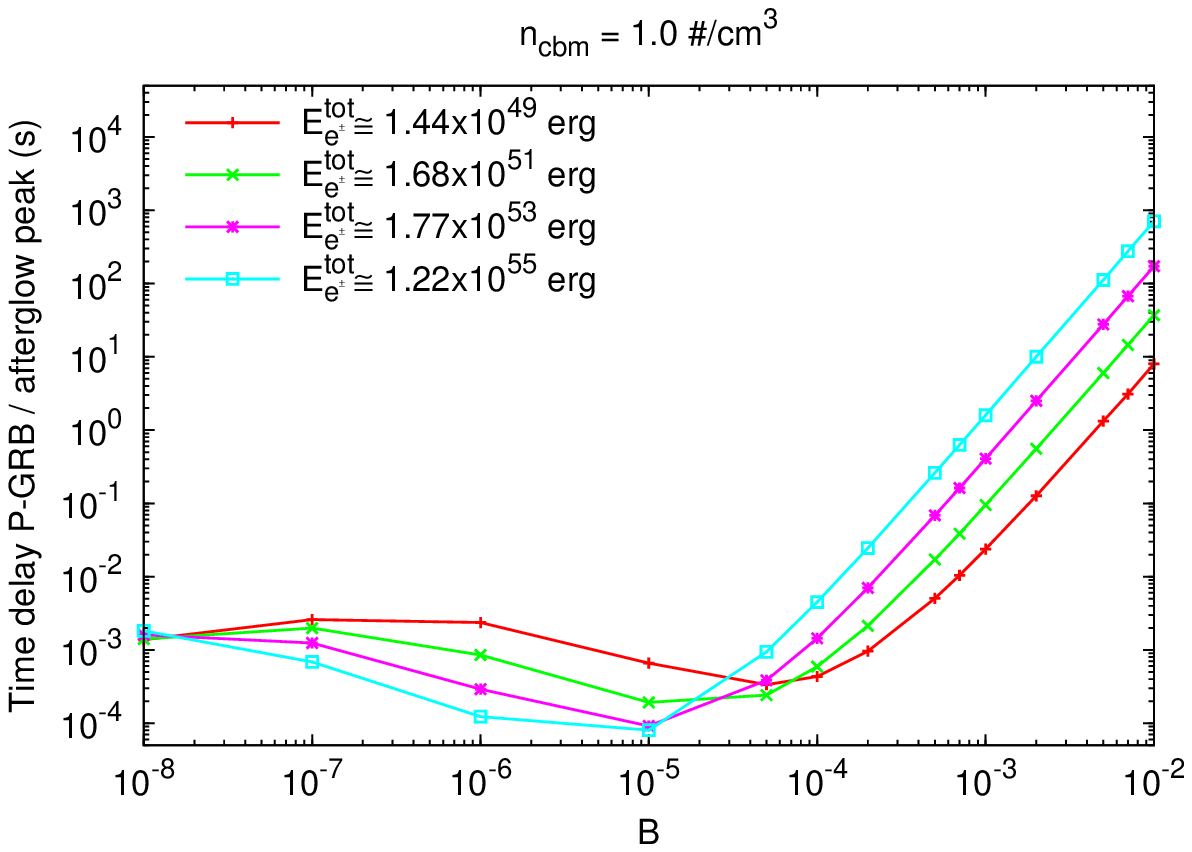}
\includegraphics[width=0.49\hsize,clip]{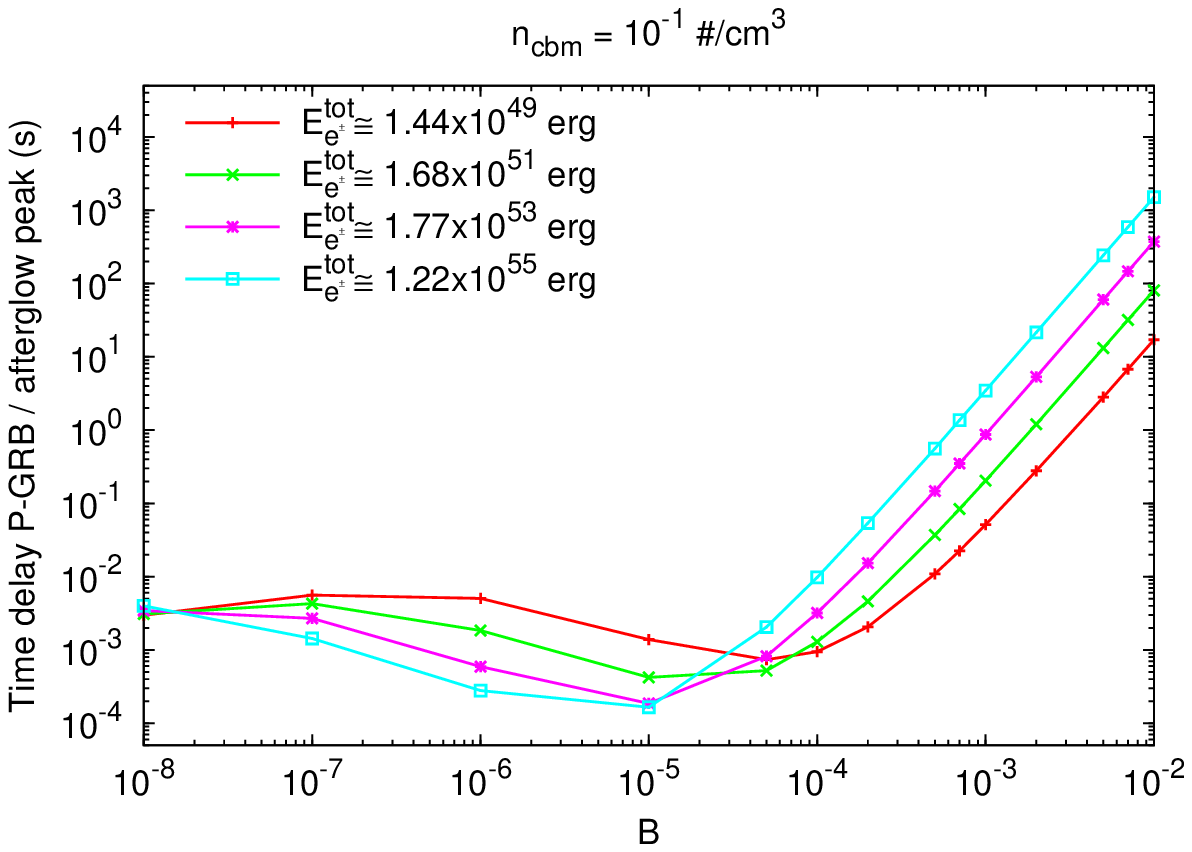}
\includegraphics[width=0.49\hsize,clip]{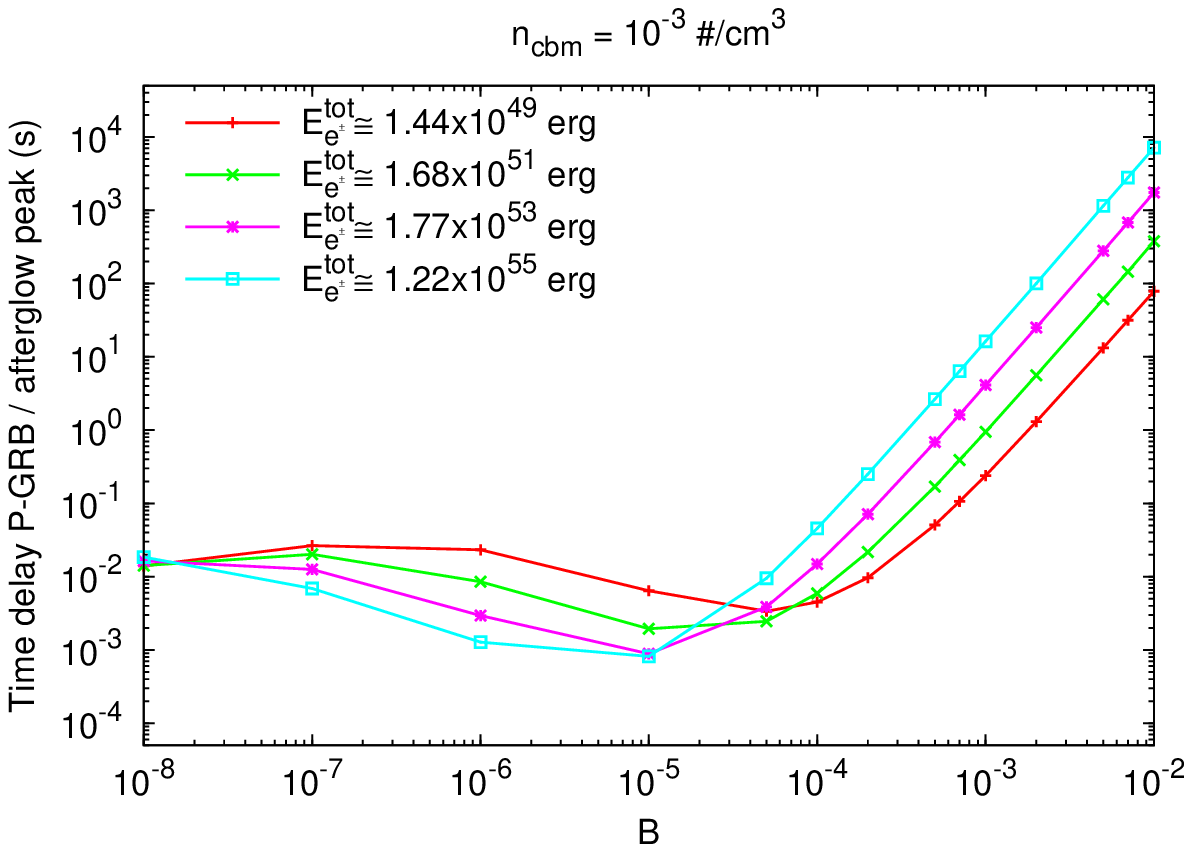}
\includegraphics[width=0.49\hsize,clip]{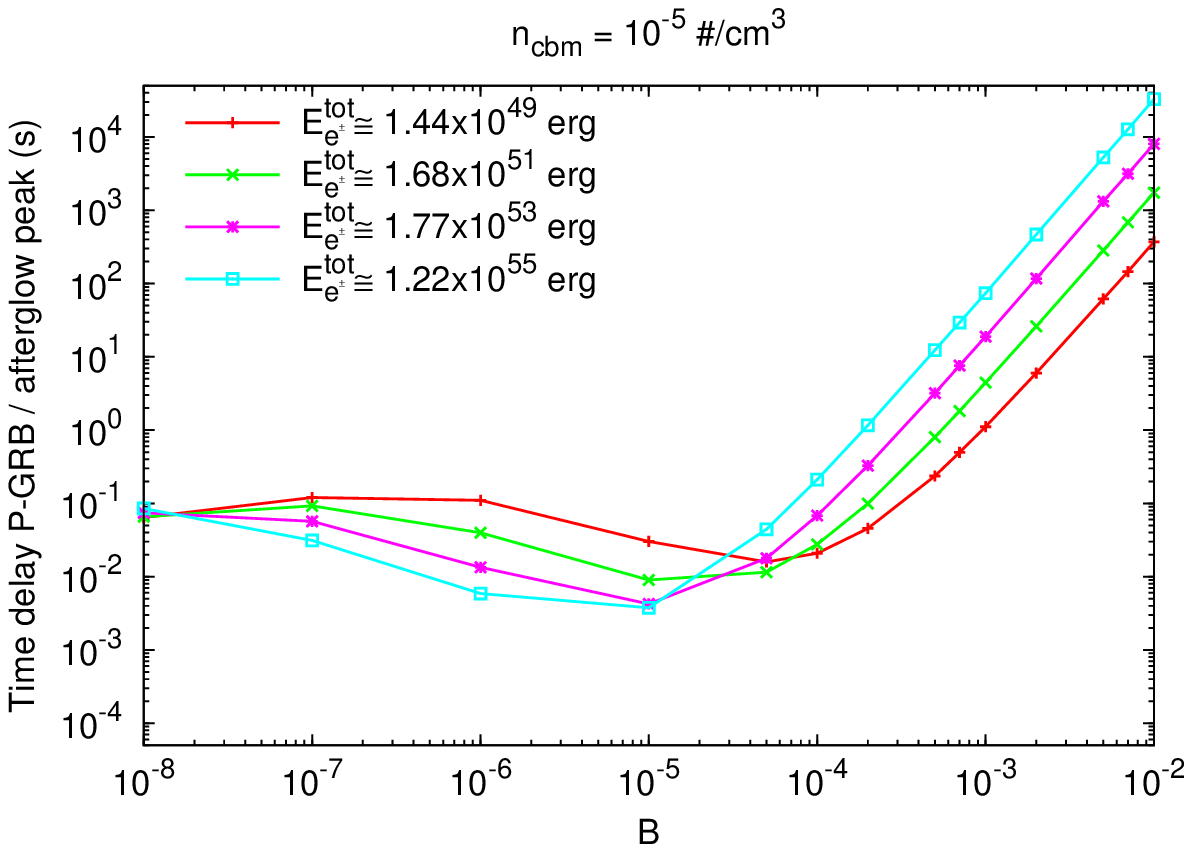}
\caption{Plots of the arrival time separation $\Delta t_a$ between the P-GRB and the peak of the extended afterglow as a function of $B$ for four different values of $E_{e^+e^-}^{tot}$, measured in the source cosmological rest frame. This computation has been performed assuming four values of the constant CBM density $n_{CBM} = 1.0,\,1.0\times 10^{-1},\,1.0\times 10^{-3},\,1.0\times 10^{-5}$ particles/cm$^3$.}
\label{short_marco_fig:2c}
\end{figure}

\begin{figure}
\centering
\includegraphics[width=0.7\hsize,clip]{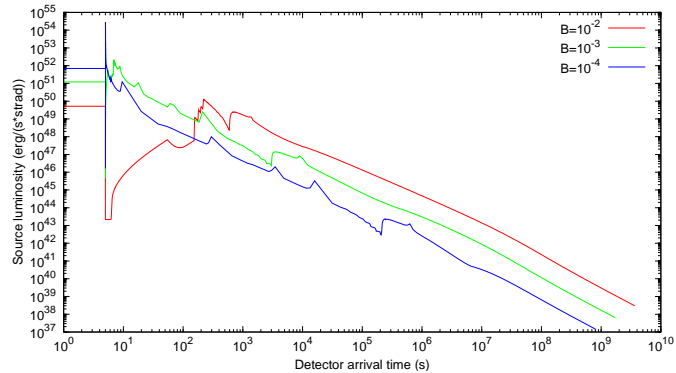} 
\caption{The dependence of the shape of the light curve on $B$. The computations have been performed assuming $E^{tot}_{e^+e^-} = 4.83\times10^{53}$ ergs, $\langle n_{CBM} \rangle = 1.0$ particles/cm$^3$, for three different values of the baryon load $B = 10^{-2},10^{-3},10^{-4}$ and the P-GRB duration fixed, i.e., $5$ s. For decreasing $B$, the extended afterglow light curve squeezes itself onto the P-GRB and the peak becomes sharper and higher.}
\label{short_marco_letizia}
\end{figure}

We adopted a pragmatic approach in our fireshell model by making full use of the knowledge of the equations of motion, of the EQTS formulations \cite{2005ApJ...620L..23B}, and of the correct relativistic transformations between the comoving frame of the fireshell and the observer frame.
These equations, which relate four distinct time variables, are necessary for interpreting the GRB data.
They are: a) the comoving time, b) the laboratory time, c) the arrival time, and d) the arrival time at the detector corrected for cosmological effects.
This is the content of the relative space-time transformation paradigm, essential for the interpretation of GRB data \cite{2001ApJ...555L.107R}.
This paradigm requires a global rather than a piecewise description in time of the GRB phenomenon \cite{2001ApJ...555L.107R} and has led to a new interpretation of the burst structure paradigm \cite{2001ApJ...555L.113R}. 
As mentioned in the introduction, a new conclusion arising from the burst structure paradigm has been that emission by the accelerated baryons interacting with the CBM is indeed occurring already in the prompt emission phase, just after the P-GRB emission.
This is the extended-afterglow emission, which exhibits in its ``light curve'' a rising part, a peak, and a decaying tail.
Following this paradigm, the prompt emission phase consists therefore of the P-GRB emission and the peak of the extended afterglow. 
Their relative energetics and observed time separation are functions of the energy $E_{e^+e^-}^{tot}$, of the baryon load $B$, and of the CBM density distribution $n_{CBM}$ (see Fig.~\ref{short_marco_fig:2c}). 
In particular, fordecreasing $B$ , the extended afterglow light curve ``squeezes'' itself on the P-GRB and the P-GRB peak luminosity increases (see Fig.~\ref{short_marco_letizia}).

To evaluate the extended-afterglow spectral properties, we adopted an ansatz for the spectral properties of the emission in the collisions between the baryons and the CBM in the comoving frame. We then evaluated all observational properties in the observer frame by integrating over the EQTS.
The initial ansatz of a thermal spectrum \cite{2001ApJ...555L.113R} has recently been modified to
\begin{equation}\label{090618_eq:no2}
\frac{d N_{\gamma}}{d V d \epsilon} = \left(\frac{8 \pi}{h^3 c^3}\right)\left(\frac{\epsilon}{k_B T}\right)^{\alpha}\frac{\epsilon^2}{exp\left(\frac{\epsilon}{k_B T}\right)-1},
\end{equation}
where $\alpha$ is a phenomenological parameter defined in the comoving frame of the fireshell \cite{2011IJMPD..20.1983P}, determined by the optimization of the simulation of the observed data.
It is well known that in the ultrarelativistic collision of protons and electrons with the CBM, collective processes of ultrarelativistic plasma physics are expected, which are not yet fully explored and understood (e.g. the Weibel instability, see Ref.~\refcite{1999ApJ...526..697M}).
Promising results along this line have already been obtained in Refs.~\refcite{2008ApJ...673L..39S,2009ApJ...700..956M} and may lead to the understanding of the physical origin of the $\alpha$ parameter in Eq.~\ref{090618_eq:no2}.

To take into due account the filamentary, clumpy and porous structure of the CBM, we introduced the additional parameter $\mathcal{R}$, which describes the fireshell surface-filling factor.
It is defined as the ratio between the effective emitting area of the fireshell $A_{eff}$ and its total visible area $A_{vis}$, see e.g. Refs.~\refcite{2002ApJ...581L..19R,2005AIPC..782...42R}.
 
One of the main features of the GRB afterglow has been the observation of hard-to-soft spectral variation, which is generally absent in the first spike-like emission, and which we have identified as the P-GRB \cite{2007A&A...474L..13B,2009A&A...498..501C,2010A&A...521A..80C,2011A&A...529A.130D}.
An explanation of the hard-to-soft spectral variation has been advanced on the grounds of two different contributions: the curvature effect and the intrinsic spectral evolution.
In particular, Ref.~\refcite{2011AN....332...92P} used the model developed in Ref.~\refcite{2002A&A...396..705Q} for the spectral lag analysis, taking into account an intrinsic band model for the GRBs and using a Gaussian profile for the GRB pulses to take into account angular effects, and they found that both provide a very good explanation for the observed time lags.
Within the fireshell model we can indeed explain a hard-to-soft spectral variation in the extended-afterglow emission very naturally.
Since the Lorentz $\Gamma$ factor decreases with time, the observed effective temperature of the fireshell will drop as the emission goes on, and consequently the peak of the emission will occur at lower energies. 
This effect is amplified by the curvature effect, which originates from the EQTS analysis.
Both these observed features are considered to be responsible for the time lag observed in GRBs.

\subsubsection{The simulation of a GRB light curve and spectra of the extended afterglow}

The simulation of a GRB light curve and the respective spectrum also requires the determination of the filling factor $\mathcal{R}$ and of the CBM density $n_{CBM}$.
These extra parameters are extrinsic and they are just functions of the radial coordinate from the source.
The parameter $\mathcal{R}$, in particular, determines the effective temperature in the comoving frame and the corresponding peak energy of the spectrum, while $n_{CBM}$ determines the temporal behavior of the light curve.
Particularly important is the determination of the average value of $n_{cbm}$. Values on the order of $0.1$-$10$ particles/cm$^3$ have been found for GRBs exploding inside star-forming region galaxies, while values on the order of $10^{-3}$ particles/cm$^3$ have been found for GRBs exploding in galactic halos \cite{2007A&A...474L..13B,2009A&A...498..501C,2011A&A...529A.130D}.
It is found that the CBM is typically formed of ``clumps''. This clumpy medium, already predicted in pioneering work by Fermi on the theoretical study of interstellar matter in our galaxy \cite{1949PhRv...75.1169F,1954ApJ...119....1F}, is by now well-established both from the GRB observations and by additional astrophysical observations, see e.g. the CBM observed in SNe \cite{1997AJ....114..258S}, or by theoretical considerations involving a super-giant massive star clumpy wind \cite{2009MNRAS.398.2152D}.

The determination of the parameter $\mathcal{R}$ and $n_{CBM}$ depends essentially on the reproduction of the shape of the extended-afterglow and of the respective spectral emission in a fixed energy range.
Clearly, the simulation of a source within the fireshell model is much more complicated than simply fitting the photon spectrum $N(E)$ of the burst (number of photons at a given energy) 
with analytic phenomenological formulas for a finite temporal range of the data.
It is a consistent picture, which has to find the best value for the parameters of the source, the P-GRB \cite{2001ApJ...555L.113R}, its spectrum, its temporal structure, as well as its energetics. 
For each spike in the light curve the parameters of the corresponding CBM clumps are computed, taking into account all the thousands of convolutions of comoving spectra over each EQTS that leads to the observed spectrum \cite{2005ApJ...620L..23B,2005ApJ...633L..13B}. 
It is clear that, since the EQTSs encompass emission processes occurring at different comoving times weighted by their Lorentz and Doppler factors, the ``fitting'' of a single spike is not only a function of the properties of the specific CBM clump but of the entire previous history of the source. 
Any mistake at any step of the simulation process affects the entire evolution that follows and conversely,
at any step a fit must be made consistently with the entire previous history: because of the nonlinearity of the system and the EQTSs, any change in the simulation produces observable effects up to a much later time. 
This leads to an extremely complicated trial and error procedure in the data simulation, in which the variation of the parameters defining the source are increasingly narrowed down, reaching uniqueness very quickly. 
Of course, we cannot expect the last parts of the simulation to be very accurate, since some of the basic hypotheses about the equations of motion and possible fragmentation of the shell can affect the procedure. 

In particular, the theoretical photon number spectrum to be compared with the observational data is obtained by an averaging procedure over instantaneous spectra. 
In turn, each instantaneous spectrum is linked to the simulation of the observed multiband light curves in the chosen time interval. 
Therefore, the simulation of the spectrum and of the observed multiband light curves have to be performed together and have optimized simultaneously.

\subsubsection{The canonical long GRBs}\label{short_marco_sec:fireshell:long}

According to the fireshell model theory, the canonical long GRBs are characterized by a baryon load varying in the range $3.0\times10^{-4} \lesssim B \leq 10^{-2}$ and they occur in a typical galactic CBM with an average density $\langle n_{CBM} \rangle \approx 1$ particle/cm$^3$.
As a result the extended afterglow is predominant with respect to the P-GRB (see Fig.~\ref{short_marco_fig:1}).

\subsubsection{The disguised short GRBs}\label{short_marco_sec:fireshell:disguised}

After the observations by Swift of GRB 050509B \cite{2005Natur.437..851G}, which was declared in the literature as the first short GRB with an extended emission ever observed, it has become clear that all such sources are actually disguised short GRBs \cite{2011A&A...529A.130D}. 
It is conceivable and probable that also a large fraction of the declared short duration GRBs in the BATSE catalog, observed before the discovery of the afterglow, are members of this class.
In the case of the disguised short GRBs the baryon load varies in the same range of the long bursts, while the CBM density is of the order of $10^{-3}$ particles/cm$^3$. 
As a consequence, the extended afterglow results in a ``deflated'' emission that can be exceeded in peak luminosity by the P-GRB \cite{2007A&A...474L..13B,2008AIPC..966....7B,2009A&A...498..501C,2010A&A...521A..80C,2011A&A...529A.130D}. 
Indeed the integrated emission in the extended afterglow is much larger than the one of the P-GRB (see Fig.~\ref{short_marco_fig:1}), as expected for long GRBs.
With these understandings long and disguised short GRBs are interpreted in terms of long GRBs exploding, respectively, in a typical galactic density or in a galactic halo density.
This interpretation has been supported by direct optical observations of GRBs located in the outskirts of the host galaxies \cite{1997Natur.387R.476S,1997Natur.386..686V,2006ApJ...638..354B,2008MNRAS.385L..10T,2010ApJ...708....9F,2011NewAR..55....1B,2012arXiv1203.1864K}.

\subsubsection{The class of genuine short GRBs}\label{short_marco_sec:fireshell:genuine}

The canonical genuine short GRBs occur in the limit of very low baryon load, e.g. $B \lesssim 10^{-5}$ with the P-GRB predominant with respect to the extended afterglow.
For such small values of $B$ the afterglow peak emission shrinks over the P-GRB and its flux is lower than that of the P-GRB (see Fig.~\ref{short_marco_letizia}).

Since the baryon load is small but not zero, in addition to the predominant role of the P-GRB, which has a thermal spectrum, a nonthermal component originating from the extended afterglow is expected. 

The best example of a genuine short GRB is GRB 090227B (see details in Ref.~\refcite{2013ApJ...763..125M}).

\subsection{Observations and data analysis of GRB 090227B}\label{short_marco_sec:analysis}

At 18:31:01.41 UT on February 27, 2009, the Fermi-GBM detector \cite{2009GCN..9568....1K} triggered and located the short and bright burst GRB 090227B (trigger 257452263/090227772). The on-ground calculated location, using the GBM trigger data, was (RA,\,Dec)(J2000)=(11$^h$48$^m$36$^s$,\,32$^{\rm{o}}10'12''$), with an uncertainty of 1.77$^{\rm{o}}$ (statistical only).  
The angle from the Fermi LAT boresight was 72$^{\rm{o}}$. 
The burst was also located by IPN \cite{2009GCN..8925....1G} and detected by Konus-Wind \cite{2009GCN..8926....1G}, showing a single pulse with duration $\sim 0.2$ s ($20$ keV -- $10$ MeV).
No X-rays or optical observations were reported on the GCN Circular Archive, so the redshift of the source is unknown. 

To obtain the Fermi-GBM light-curves and the spectrum in the energy range $8$ keV -- $40$ MeV, we made use of the \texttt{RMFIT} program.
For the spectral analysis, we have downloaded from the gsfc website \footnote{ftp://legacy.gsfc.nasa.gov/fermi/data/gbm/bursts} the \texttt{TTE} (Time-Tagged Events) files, suitable for short or highly structured events.
We used the light curves corresponding to the NaI-n2 ($8$ -- $900$ keV) and the BGO-b0 ($250$ keV -- $40$ MeV) detectors. 
The $64$ ms binned GBM light curves show one very bright spike with a short duration of $0.384$ s, in the energy range $8$ keV -- $40$ MeV, and a faint tail lasting up to $0.9$ s after the trigtime $T_0$ in the energy range $10$ keV -- $1$ MeV.
After the subtraction of the background, we have proceeded with the time-integrated and time-resolved spectral analyses. 

\subsubsection{Time-integrated spectral analysis}\label{short_marco_sec:timeint}

\begin{figure}
\centering 
\hfill
\includegraphics[width=0.49\hsize,clip]{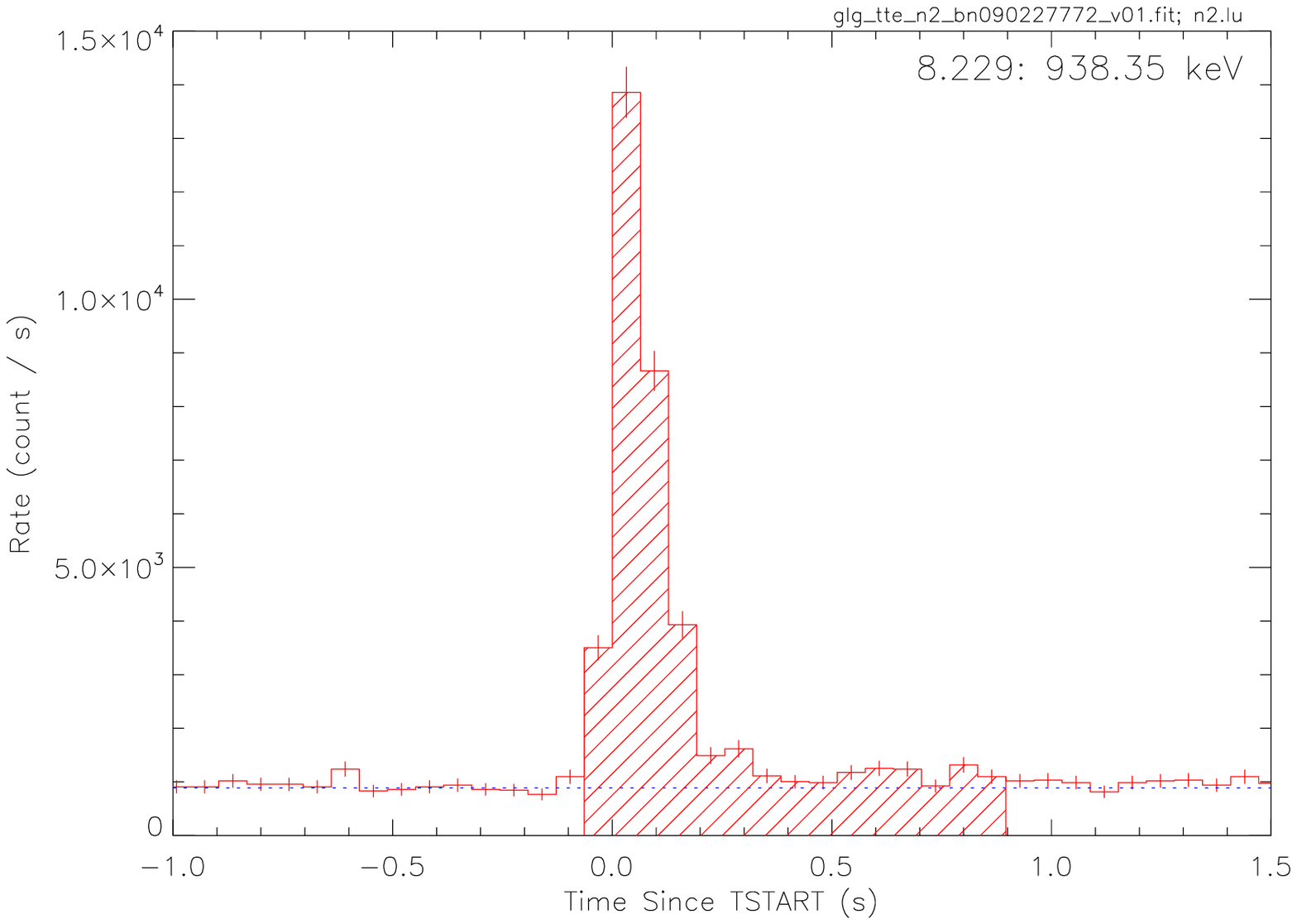}
\hfill
\includegraphics[width=0.49\hsize,clip]{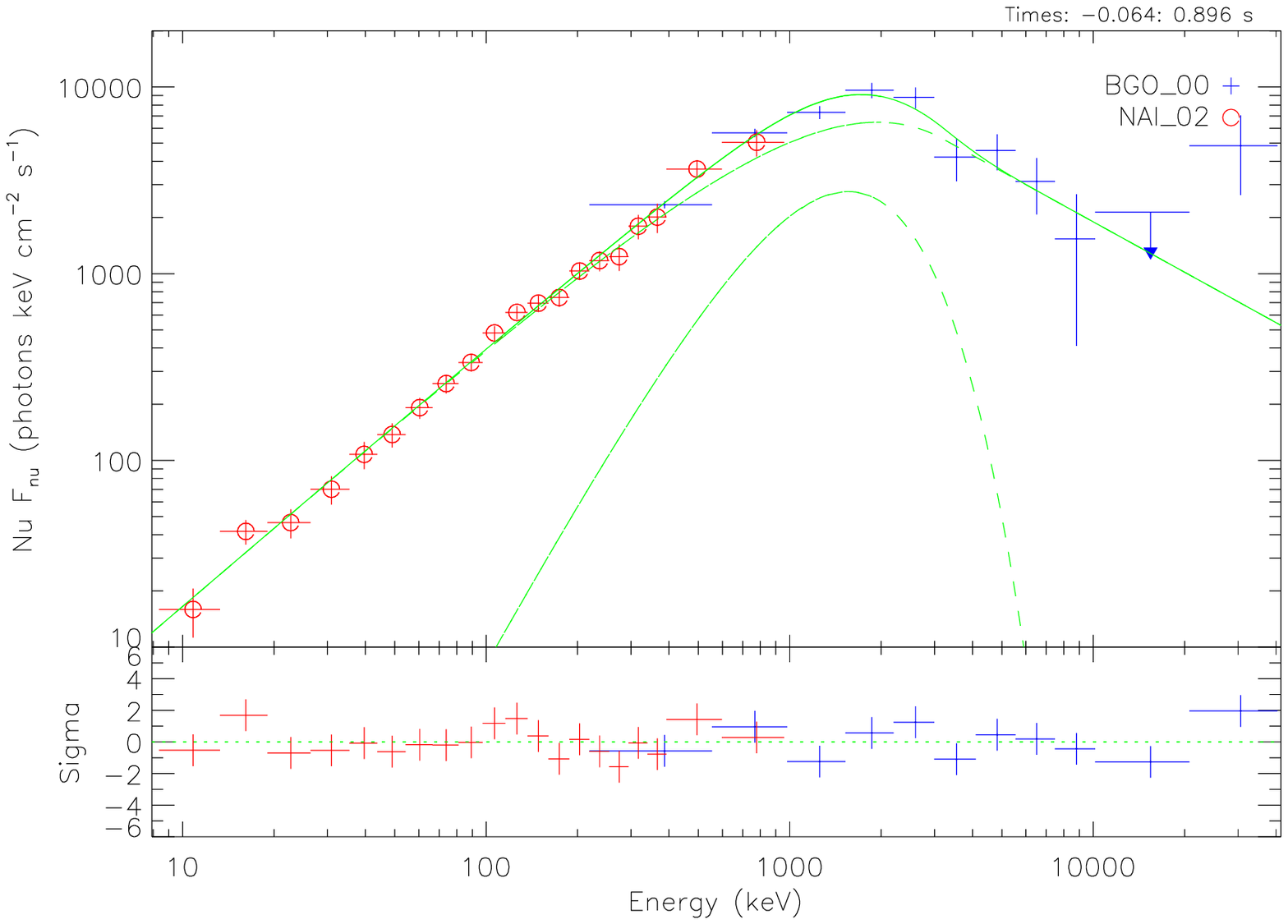} 
\hfill\null\\
\hfill
\includegraphics[width=0.49\hsize,clip]{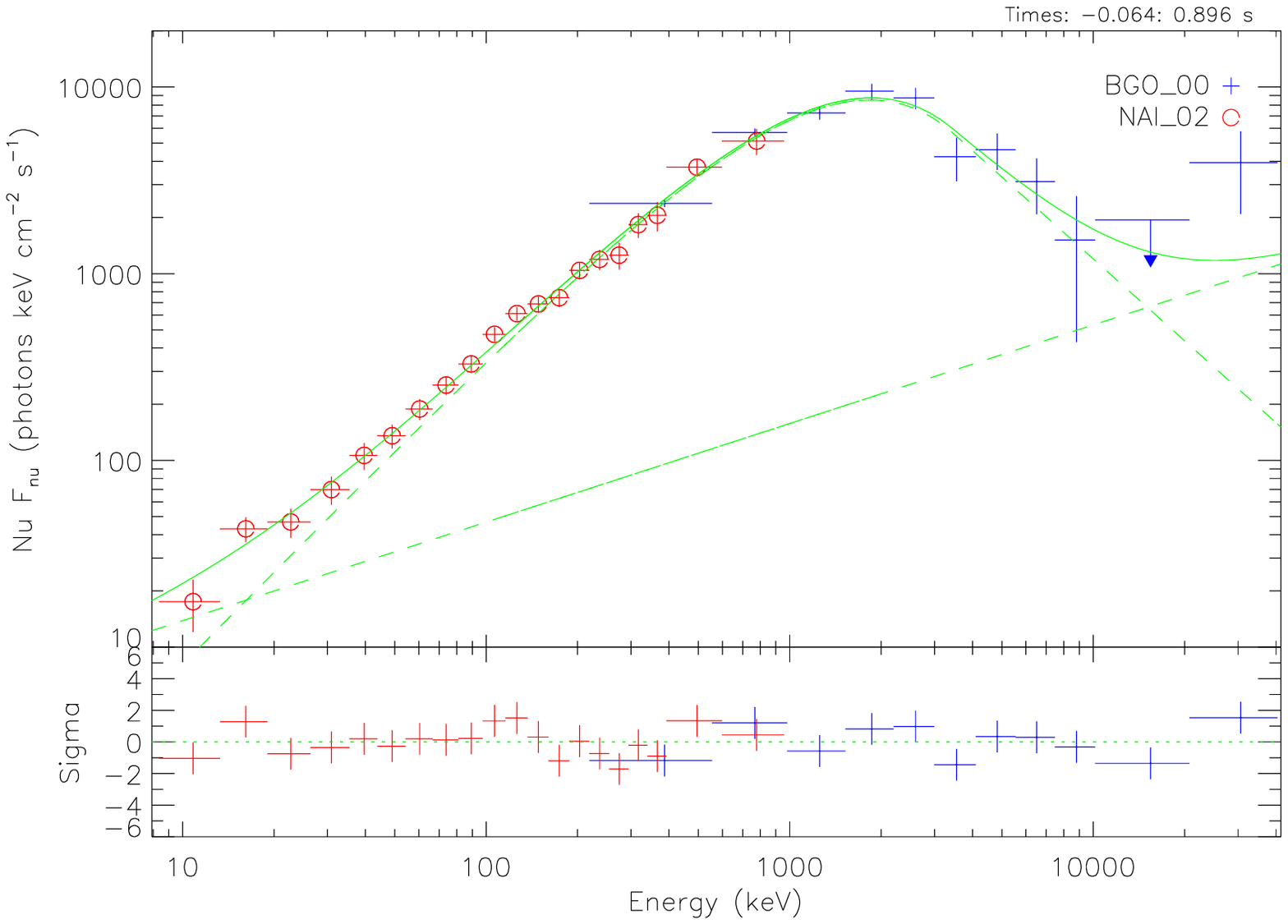}
\hfill
\includegraphics[width=0.49\hsize,clip]{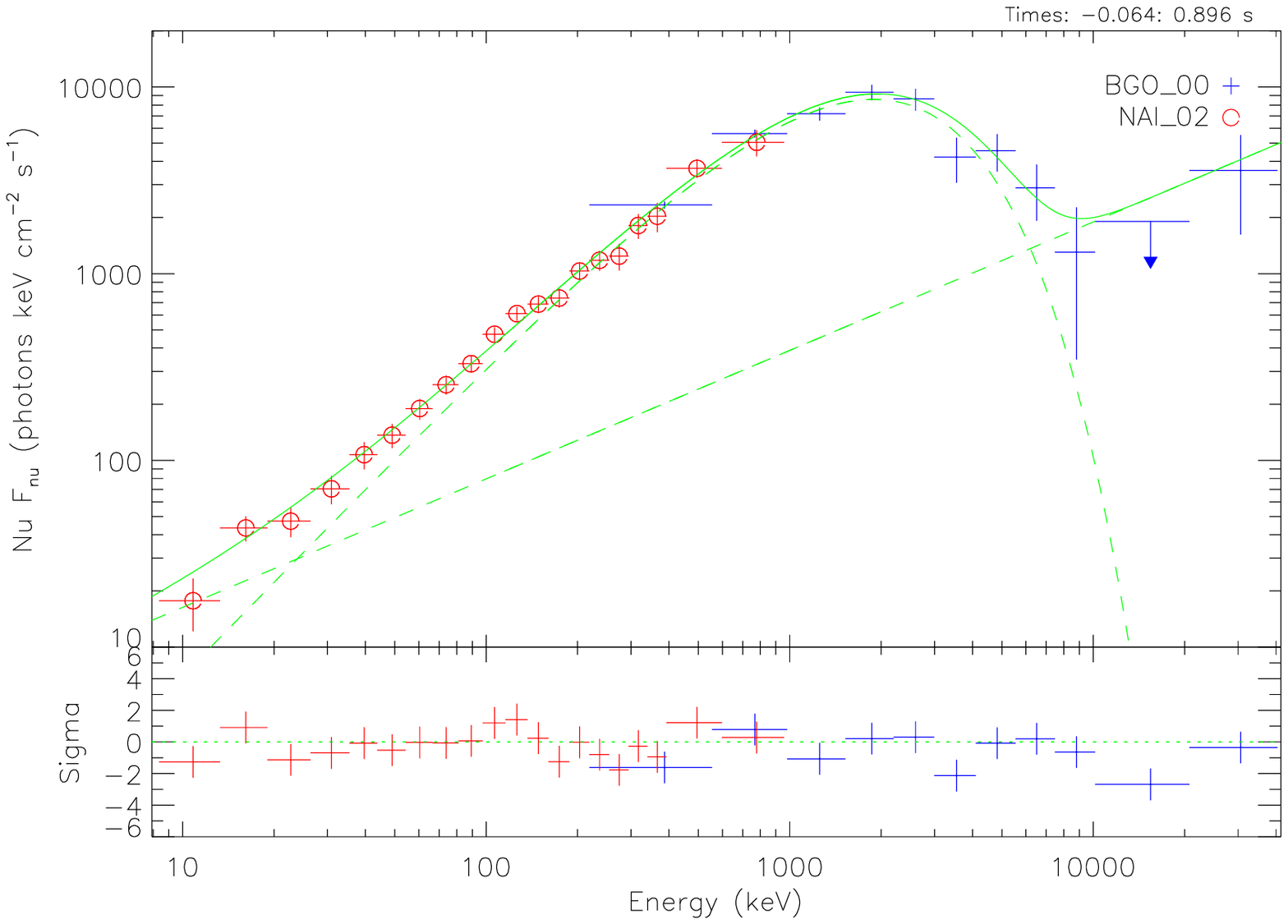} 
\hfill\null\\
\caption{The 64 ms time-binned NaI-n2 light curve (top left panel) and the NaI-n2+BGO-b0 $\nu F_\nu$ spectra (top right BB+Band, bottom left Band+PL, bottom right Compt+PL) of GRB 090227B in the $T_{90}$ time interval.}
\label{short_marco_fig:2a}
\end{figure}

We have performed a time-integrated spectral analysis in the time interval from $T_0-0.064$ s to $T_0+0.896$ s, which corresponds to the $T_{90}$ duration of the burst. 
We have fit the spectrum in this time interval considering the following models: comptonization (Compt) plus power-law (PL) and band \cite{1993ApJ...413..281B} plus PL, as outlined, e.g. in Ref.~\refcite{2010ApJ...725..225G}, as well as a combination of black body (BB) and band (see Fig.~\ref{short_marco_fig:2a}).
Within the $T_{90}$ time interval, the BB+Band model improves the fit with respect to the Compt+PL model at a confidence level of $5\%$. 
The comparison between Band+PL and Compt+PL models is outside such a confidence level (about $8\%$). 
The direct comparison between BB+Band and Band+PL models, which have the same number of degrees of freedom, 
provides almost the same C-STAT values for the BB+Band and Band+PL models ($\Delta\textnormal{C-STAT}\approx0.89$).
This means that all three models are viable. 
For the BB+Band model, the ratio between the fluxes of the thermal component and the non-thermal  (NT) component is $F_{BB}/F_{NT} \approx 0.22$.
The BB component is important for the determination of the peak of the $\nu F_\nu$ spectrum and has an observed temperature $kT = (397\pm70)$ keV.

\begin{figure}
\centering 
\hfill
\includegraphics[width=0.49\hsize,clip]{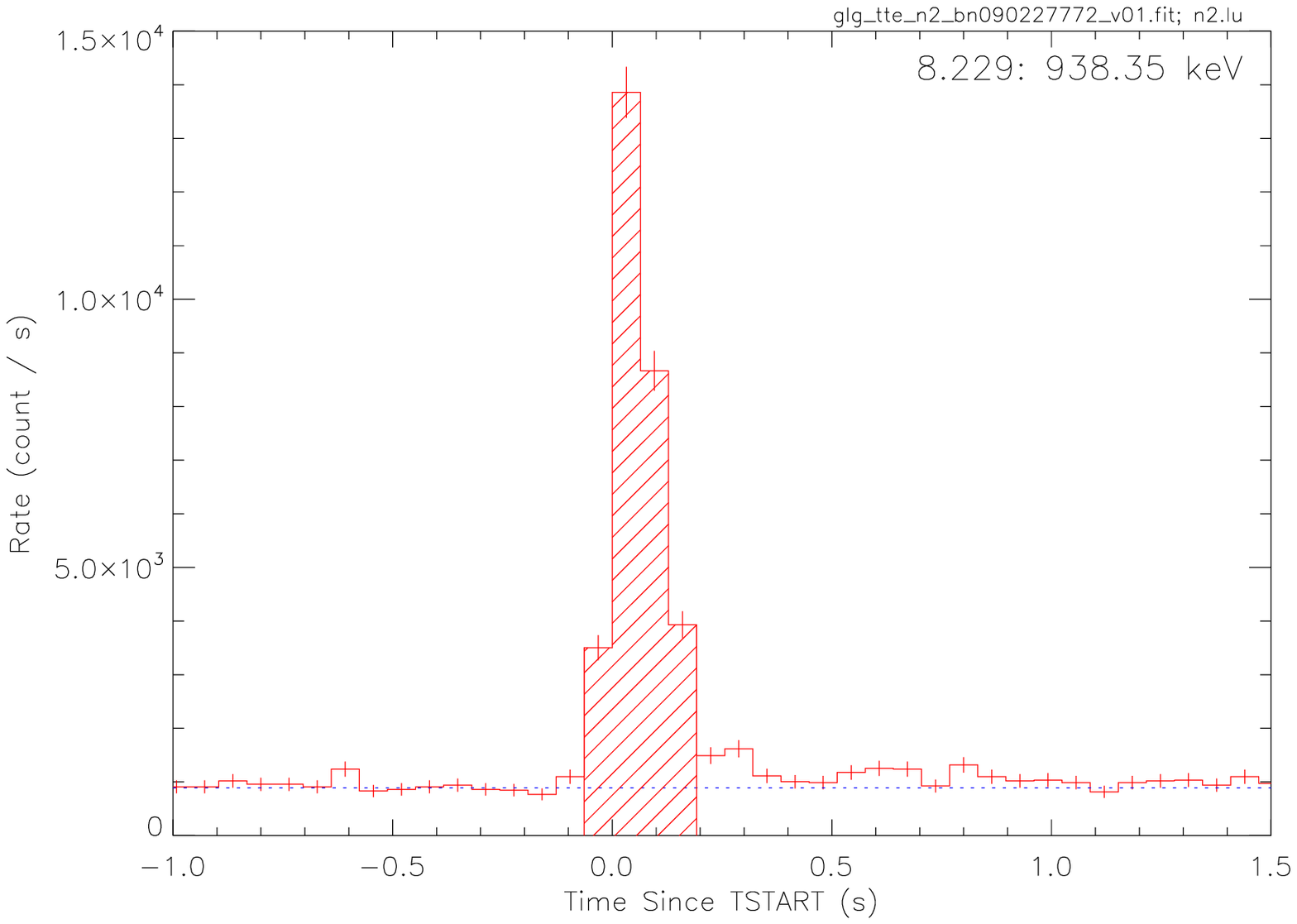}
\hfill
\includegraphics[width=0.49\hsize,clip]{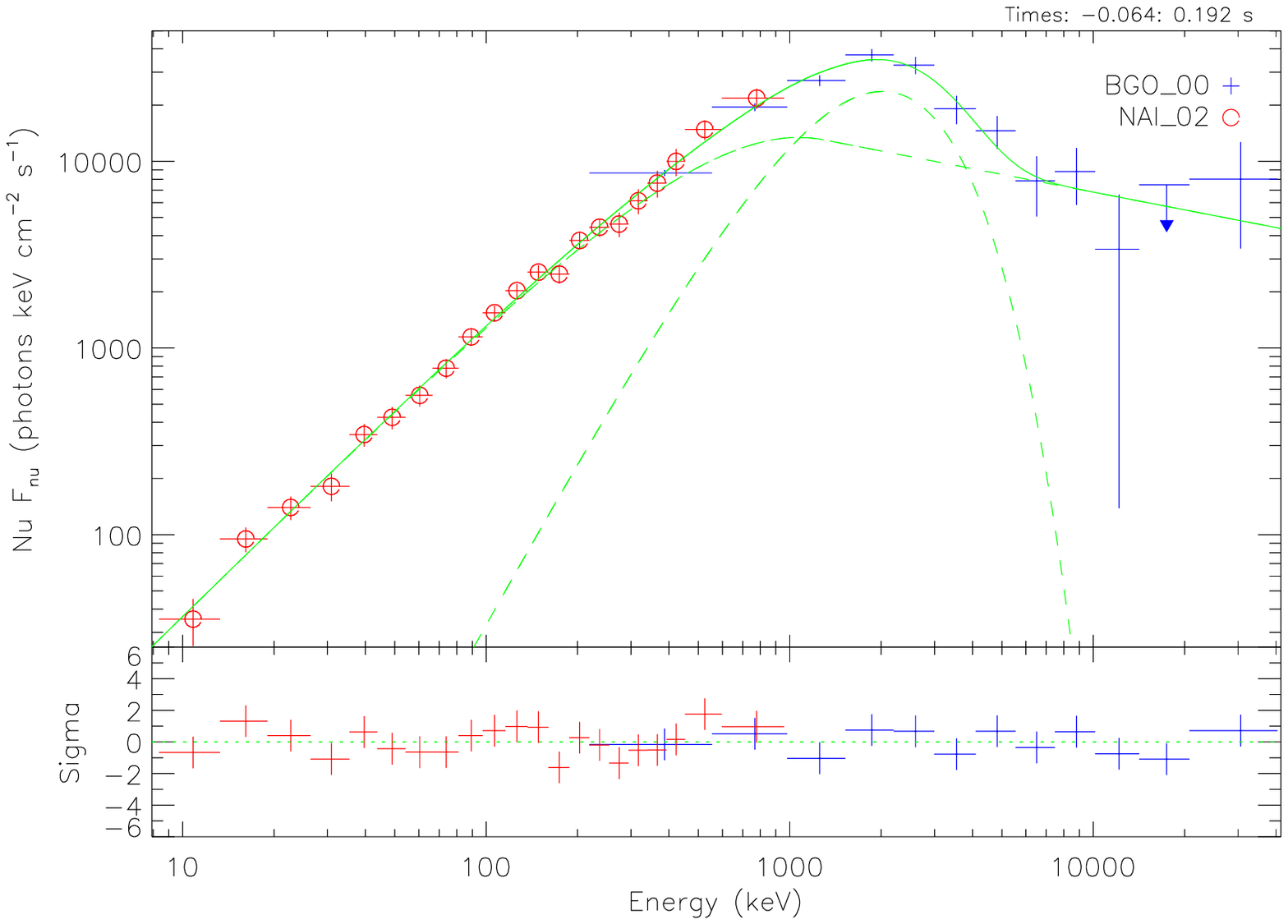} 
\hfill\null\\
\hfill
\includegraphics[width=0.49\hsize,clip]{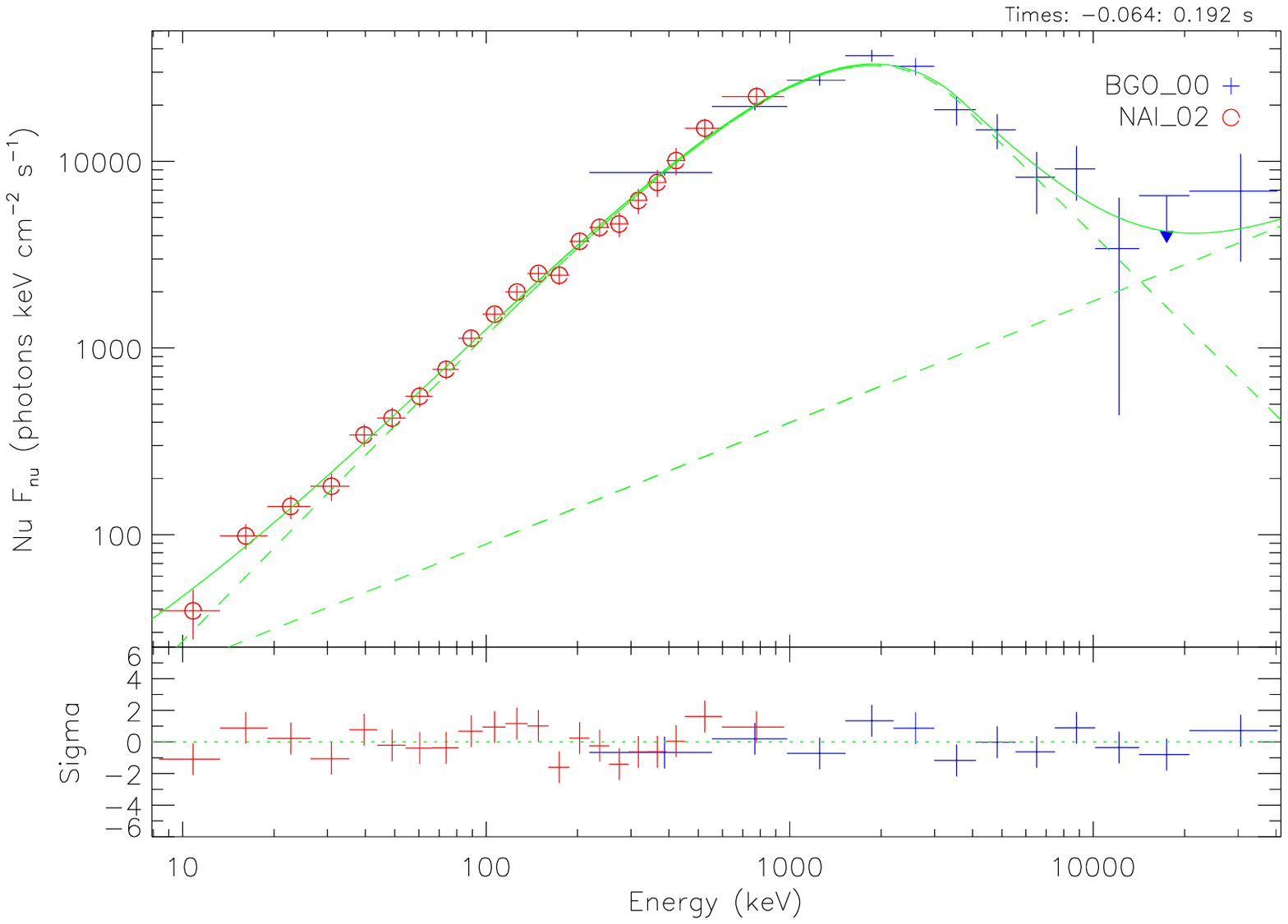}
\hfill
\includegraphics[width=0.49\hsize,clip]{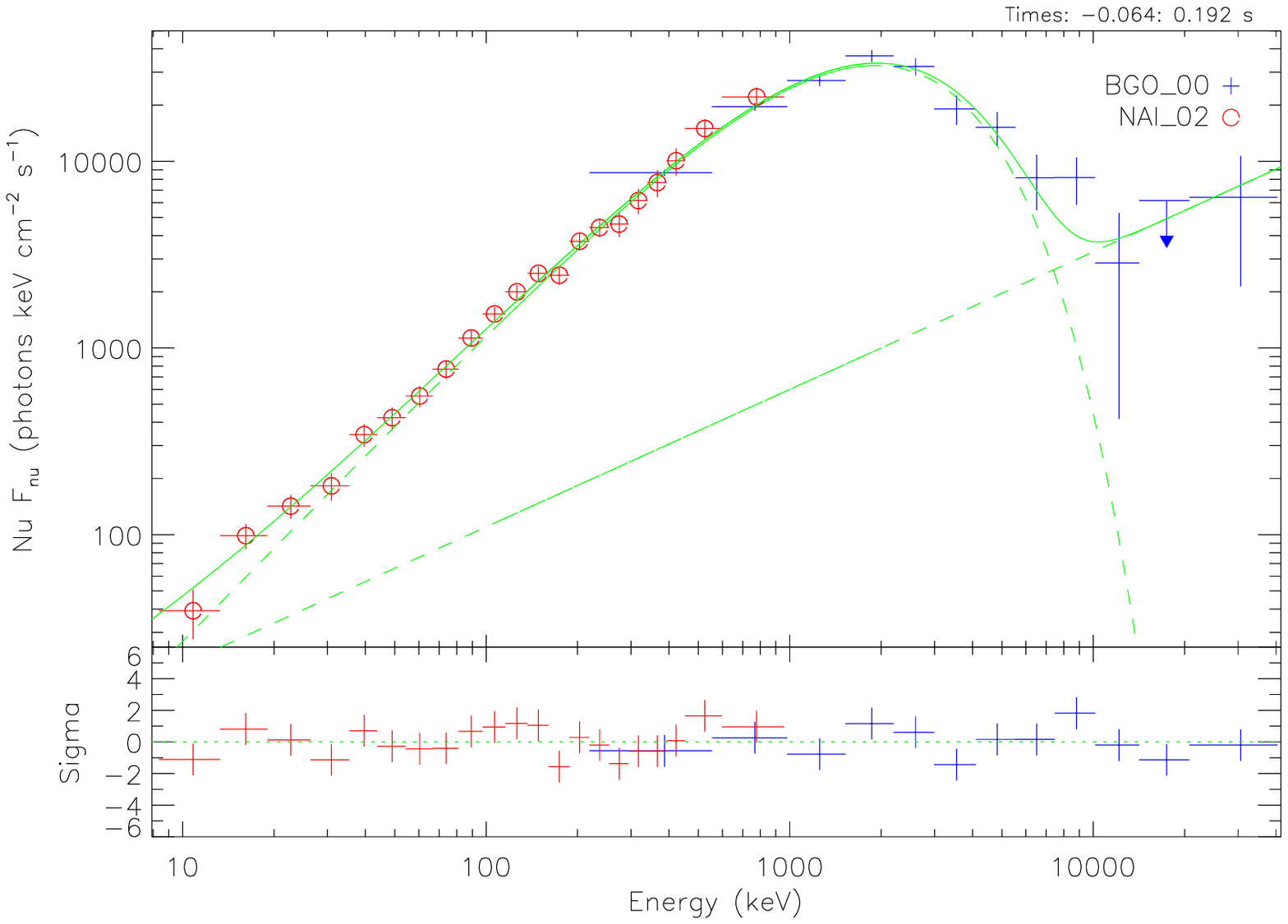}
\hfill\null\\
\caption{The same considerations as in Fig.~\ref{short_marco_fig:2a}, in the $T_{spike}$ time interval.}
\label{short_marco_fig:2aa}
\end{figure}

We have then focused our attention on the spike component, namely the time interval from $T_0-0.064$ s to $T_0+0.192$, which we indicate in the following as the $T_{spike}$.
We have repeated the time-integrated analysis considering the same spectral models of the previous interval (see Fig.~\ref{short_marco_fig:2aa}).
Within the $T_{spike}$ time interval, both the BB+Band and Band+PL models marginally improve the fit of the data with respect to the Compt+PL model within a confidence level of $5\%$. 
Again, the C-STAT values of the BB+Band and Band+PL models are almost the same ($\Delta\textnormal{C-STAT}\approx0.15$) and they are statically equivalent in the $T_{spike}$.
For the BB+Band model, the observed temperature of the thermal component is $kT = (515\pm28)$ keV and the flux ratio between the BB  and NT components increases up to $F_{BB}/F_{NT} \approx 0.69$.

\begin{figure}
\centering 
\hfill
\includegraphics[width=0.49\hsize,clip]{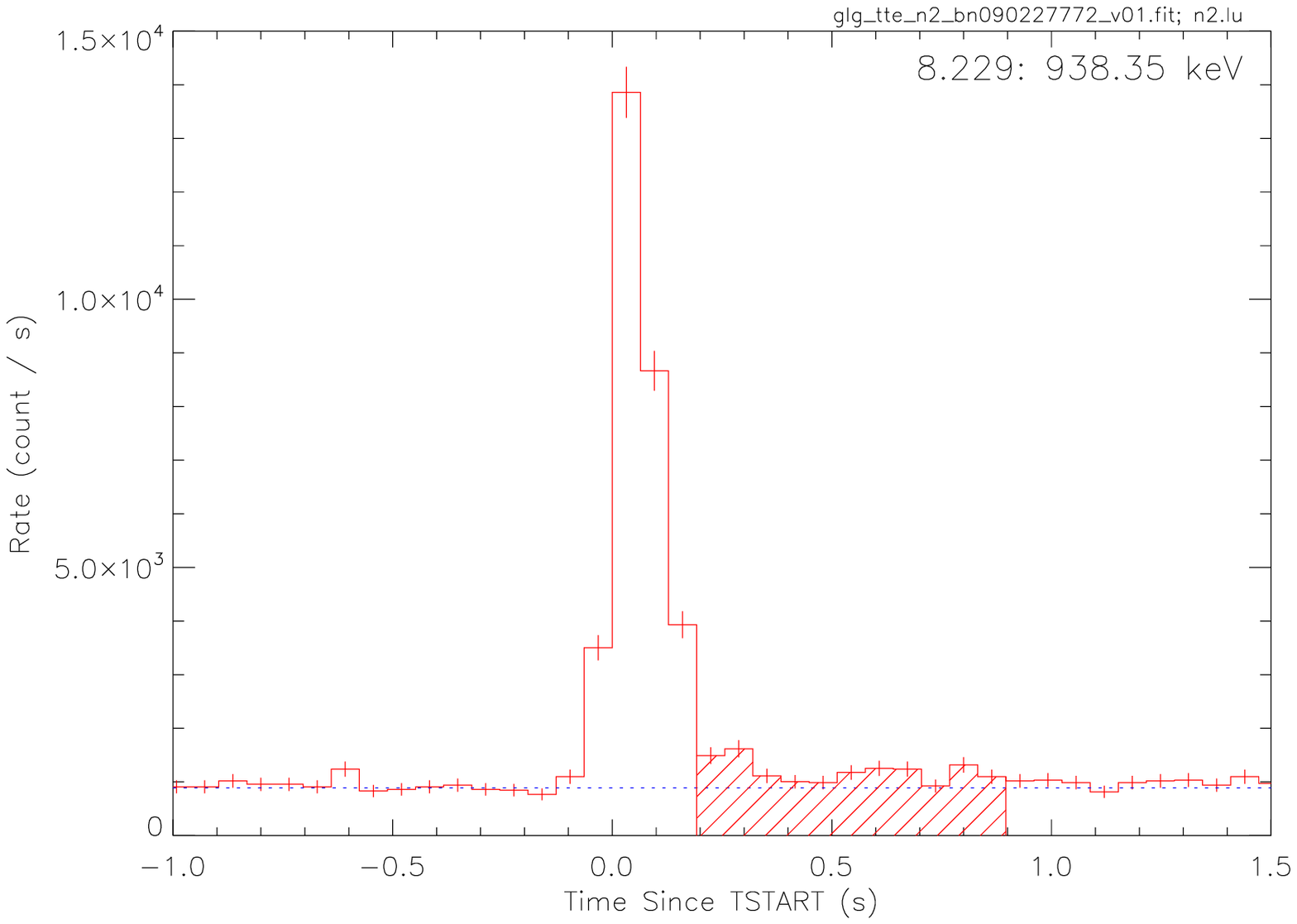}
\hfill
\includegraphics[width=0.49\hsize,clip]{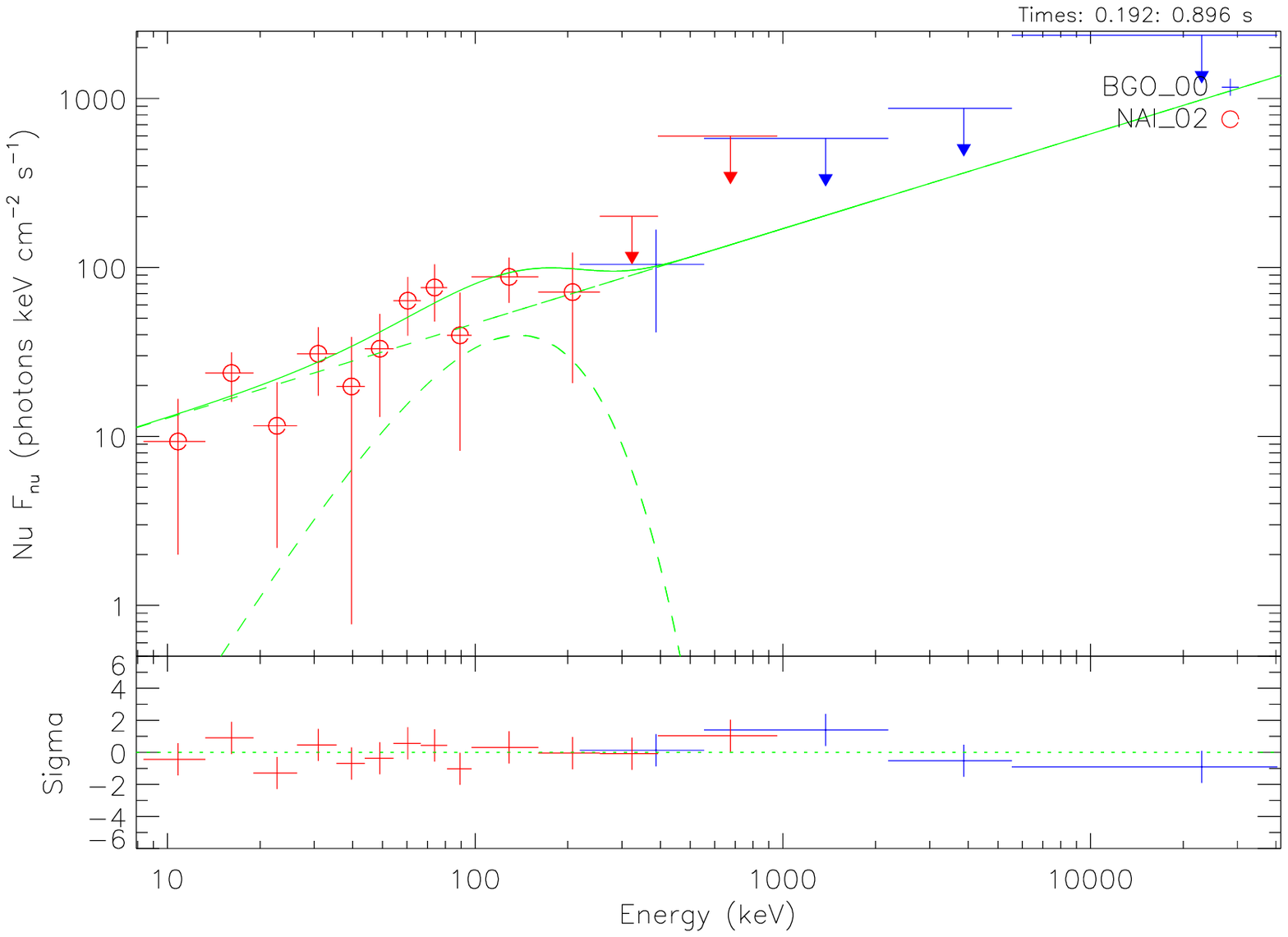} 
\hfill\null\\
\hfill
\includegraphics[width=0.49\hsize,clip]{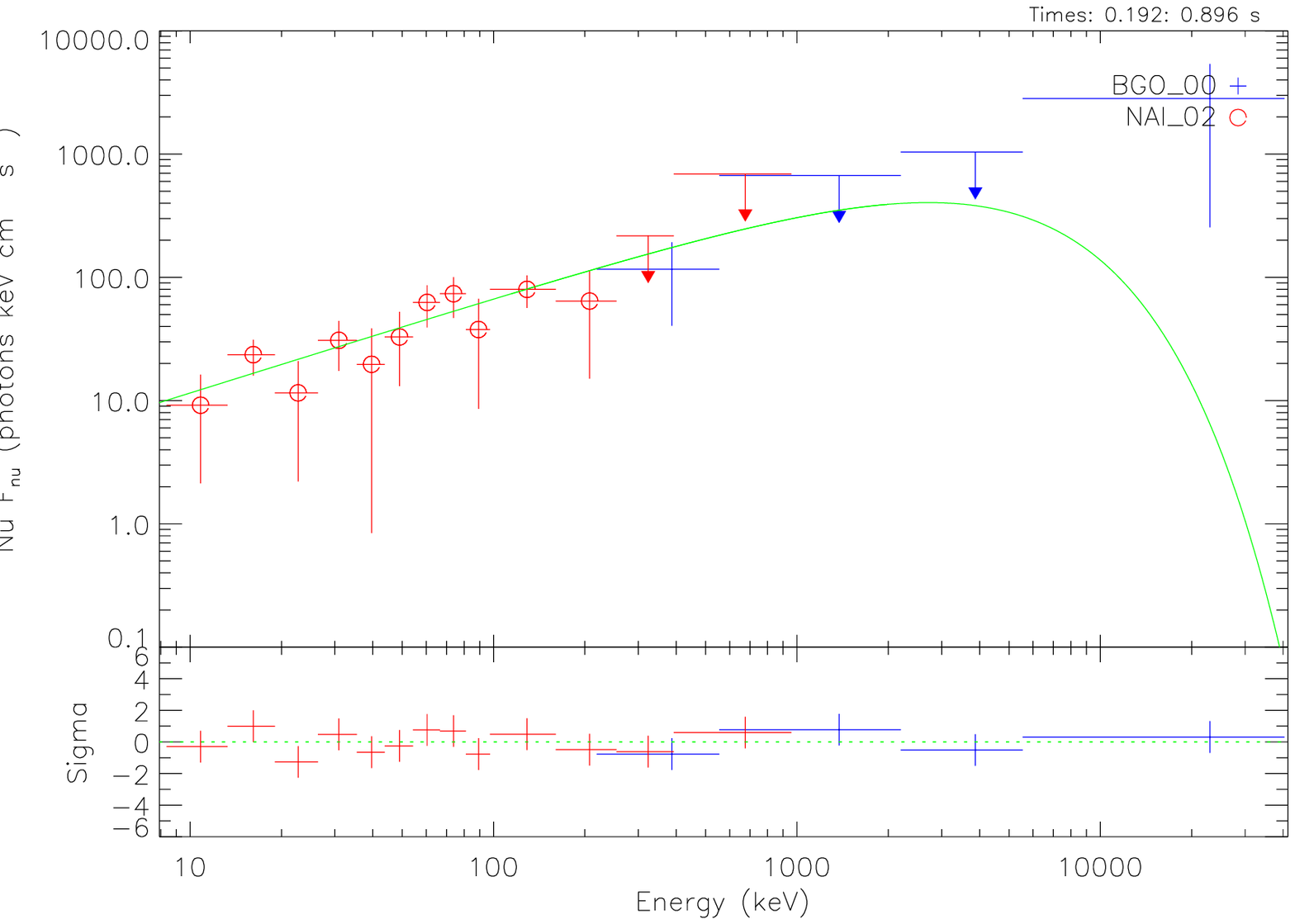}
\hfill
\includegraphics[width=0.49\hsize,clip]{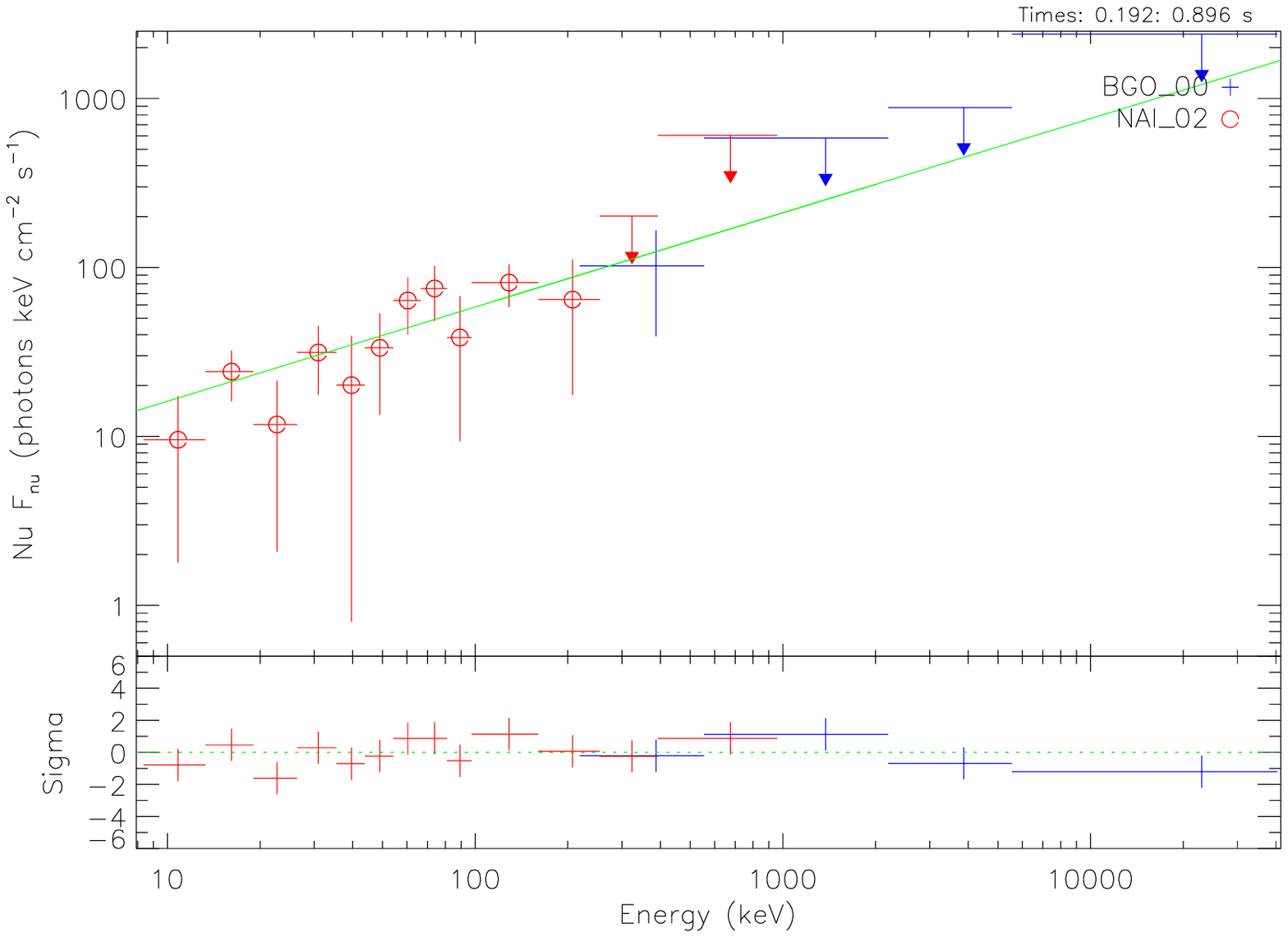} 
\hfill\null\\
\caption{The 64 ms time-binned NaI-n2 light curve (top left panel) and the NaI-n2+BGO-b0 $\nu F_\nu$ spectra (top right BB+PL, bottom left Compt, bottom right PL) of GRB 090227B in the $T_{tail}$ time interval.}
\label{short_marco_fig:2aaa}
\end{figure}

We have performed a further analysis in the time interval from $T_0+0.192$ s to $T_0+0.896$ s, which we indicate as $T_{tail}$, by considering the BB+PL, Compt and PL models (see Fig.~\ref{short_marco_fig:2aaa}). 
The statistical comparison shows that the best fit is the Compt model and a thermal component is ruled out.
For details, see Ref.~\refcite{2013ApJ...763..125M}.

In view of the above, we have focused our attention on the fit of the data of the BB+Band model within the fireshell scenario, being equally probable from a mere statistical point of view with the other two choices, namely the Band+PL and Compt+PL.
According to the fireshell scenario (see Sec.~\ref{sec:pgrb}), the emission within the $T_{spike}$ time interval is related to the P-GRB and is expected to be thermal.
In addition the transition between the transparency emission of the P-GRB and the extended afterglow is not sharp.
The time separation between the P-GRB and the peak of the extended afterglow depends on the energy of the $e^+e^-$ plasma $E_{e^+e^-}^{tot}$, the baryon load $B$ and the CBM density $n_{CBM}$ (see Fig.~\ref{short_marco_letizia}).
As shown in Figs.~\ref{short_marco_fig:2c} and \ref{short_marco_letizia}, for decreasing values of $B$ an early onset of the extended afterglow in the P-GRB spectrum occurs and thus an NT component in the $T_{spike}$ is expected.
As a further check, the theory of the fireshell model indeed predicts in the early part of the prompt emission of GRBs a thermal component due to the transparency of the $e^+e^-$ plasma (see Sec.~\ref{short_marco_sec:fireshell}), while in the extended afterglow no thermal component is expected (see Sec.~\ref{exaft}), as observed in the $T_{tail}$ time interval.

Our theoretical interpretation has shown to be consistent with the observational data and the statistical analysis. 
From an astrophysical point of view the BB+Band model is preferred over the other two models, statistically equivalent in view of the above theoretical considerations.

\subsubsection{Time-resolved spectral analysis}\label{short_marco_sec:timeres}

\begin{figure}
\centering
\includegraphics[width=0.7\hsize,clip]{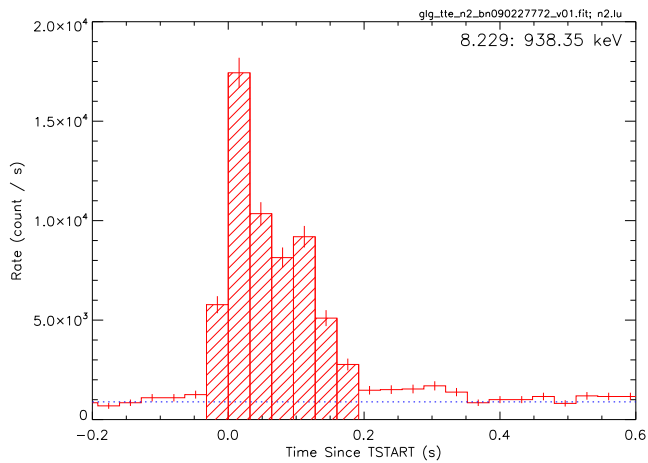}
\caption{The $32$ ms time-binned NaI-n2 light curve of GRB 090227B in the time interval from $T_0-0.032$ s to $T_0+0.192$ s; each time bin corresponds to the time-resolved interval considered in the Sec.~\ref{short_marco_sec:timeres}.}
\label{short_marco_fig:timeres}
\end{figure}

We have performed a time-resolved spectral analysis on selected shorter  time intervals of $32$ ms (see Fig.~\ref{short_marco_fig:timeres}) in order to correctly identify the P-GRB, namely finding out in which time interval the thermal component exceeds or at least has a comparable flux with respect to the NT one due to the onset of the extended afterglow. 
In this way we can single out the contribution of the NT component in the spectrum of the P-GRB.

Within the first time-resolved interval the BB+PL model has a thermal flux $(11.2\pm3.4)$ times bigger than the PL flux; the fit with the BB+Band provides $F_{BB}=(0.50\pm0.26)F_{NT}$, where the NT component is in this case the band model.
In the second and fourth intervals, the BB+Band model provides an improvement at a significance level of $5\%$ in the fitting procedure with respect to the simple band model. 
In the third time interval as well as in the remaining time intervals up to $T_0+0.192$ s the band spectral models provide better fits with respect to the BB+NT ones. 

This is exactly what we expect from our theoretical understanding: from $T_0-0.032$ s to $T_0+0.096$ s we have found the edge of the P-GRB emission, in which the thermal components have fluxes higher or comparable to the NT ones.
The third interval corresponds to the peak emission of the extended afterglow (see Fig.~\ref{short_marco_fig:2d1}).
The contribution of the extended afterglow in the remaining time intervals increases, while the thermal flux noticeably decreases.

\begin{figure}
\centering
\hfill
\includegraphics[width=0.49\hsize,clip]{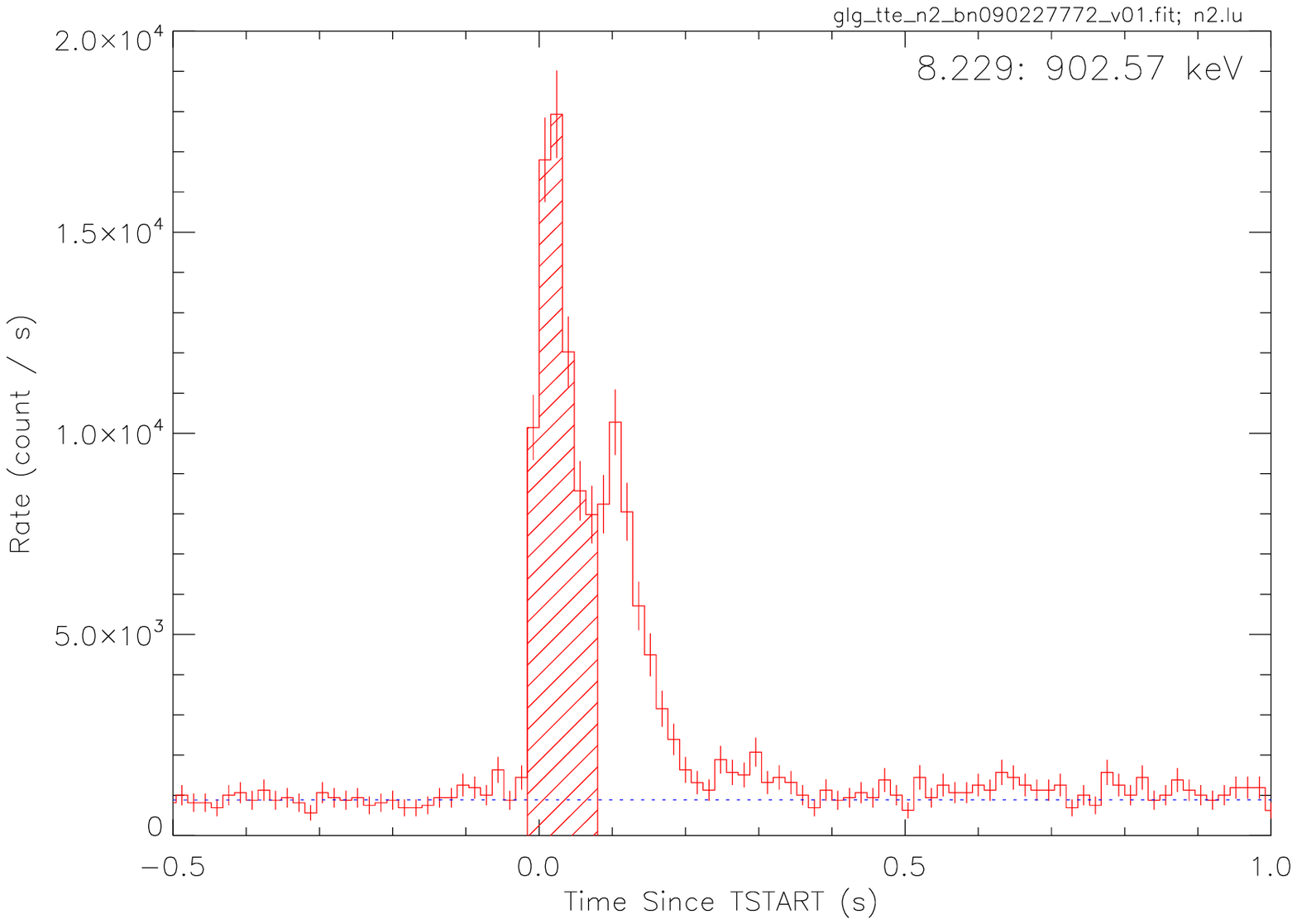}
\hfill
\includegraphics[width=0.49\hsize,clip]{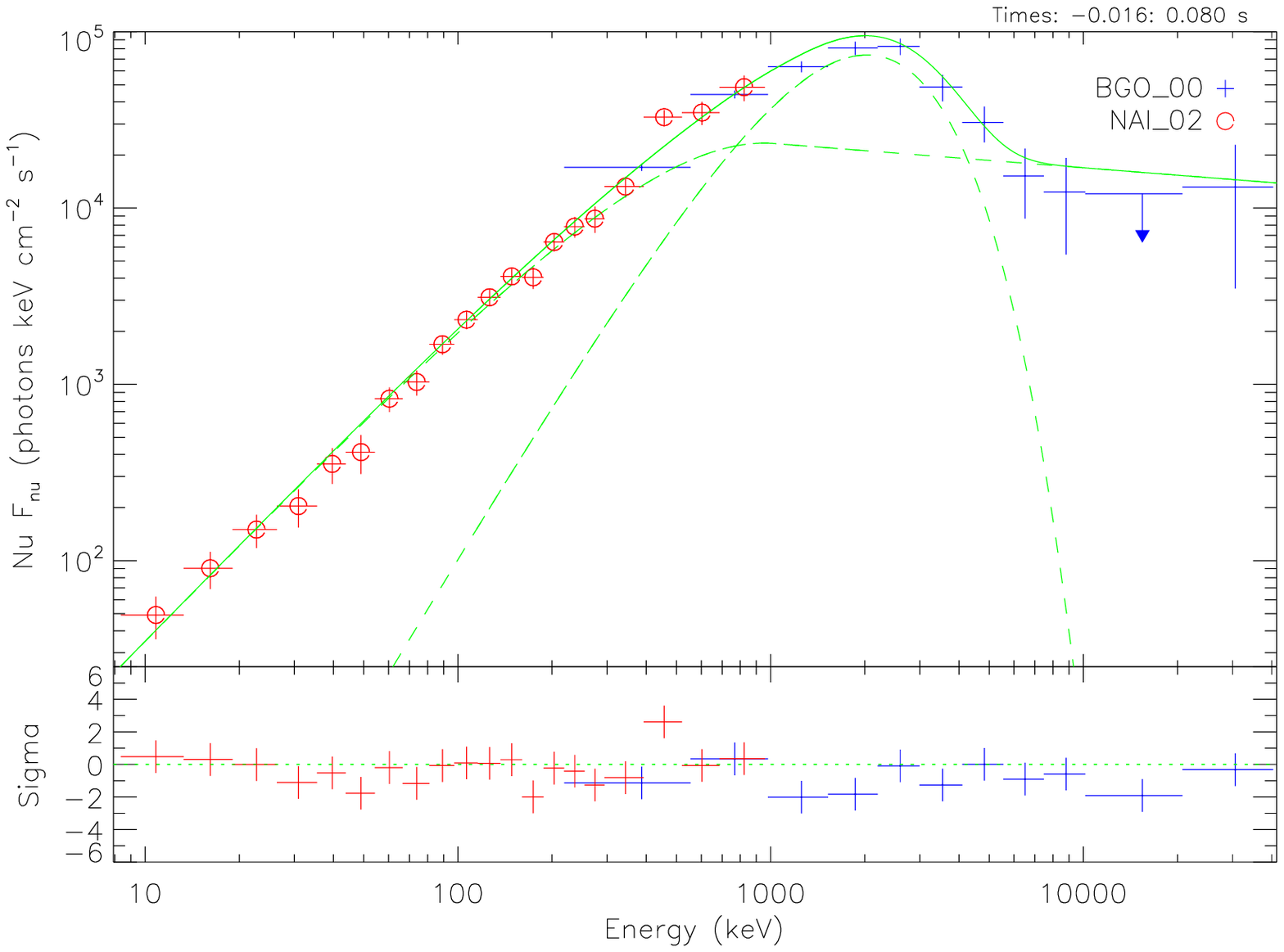} 
\hfill\null\\
\hfill
\includegraphics[width=0.49\hsize,clip]{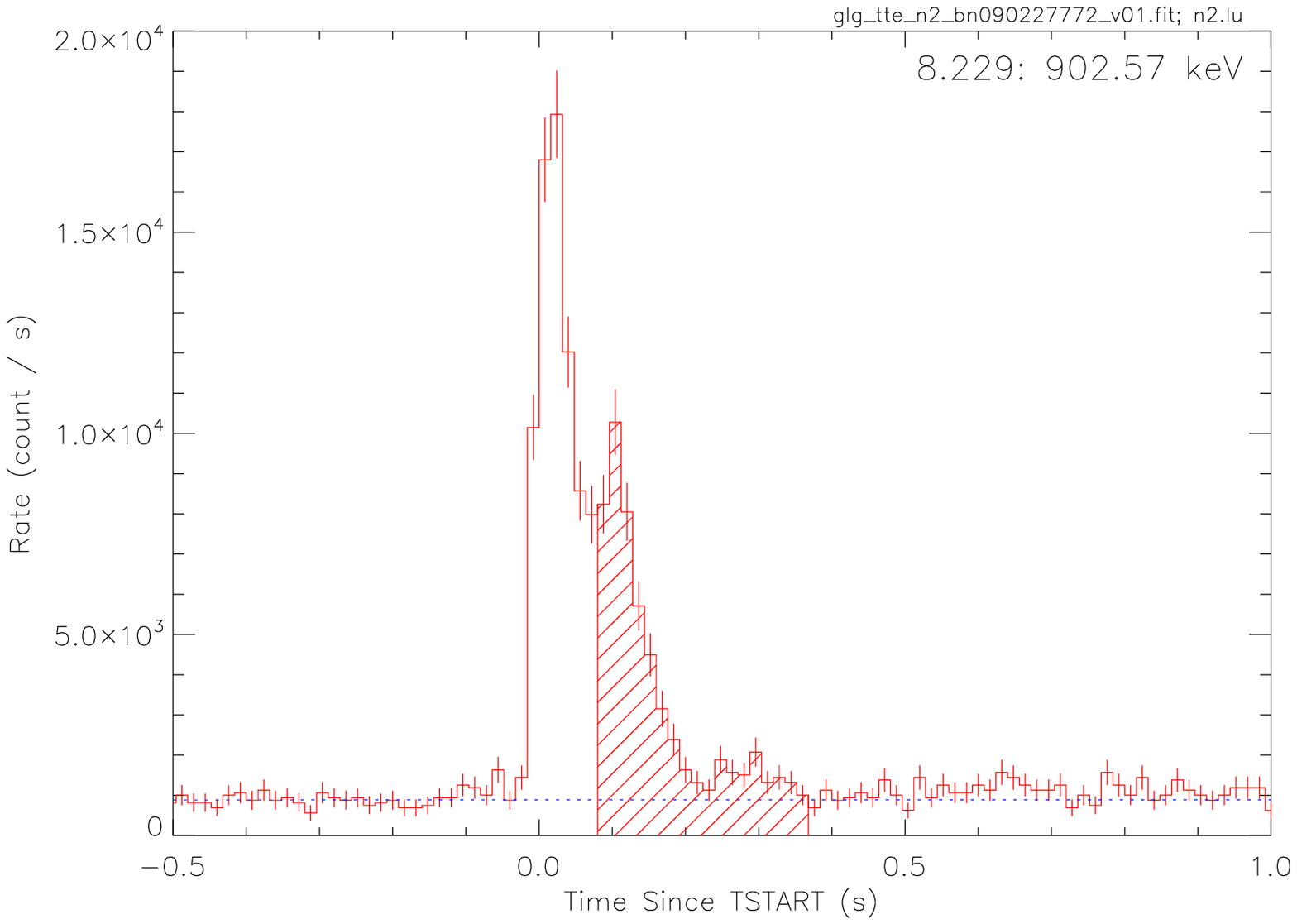}
\hfill
\includegraphics[width=0.49\hsize,clip]{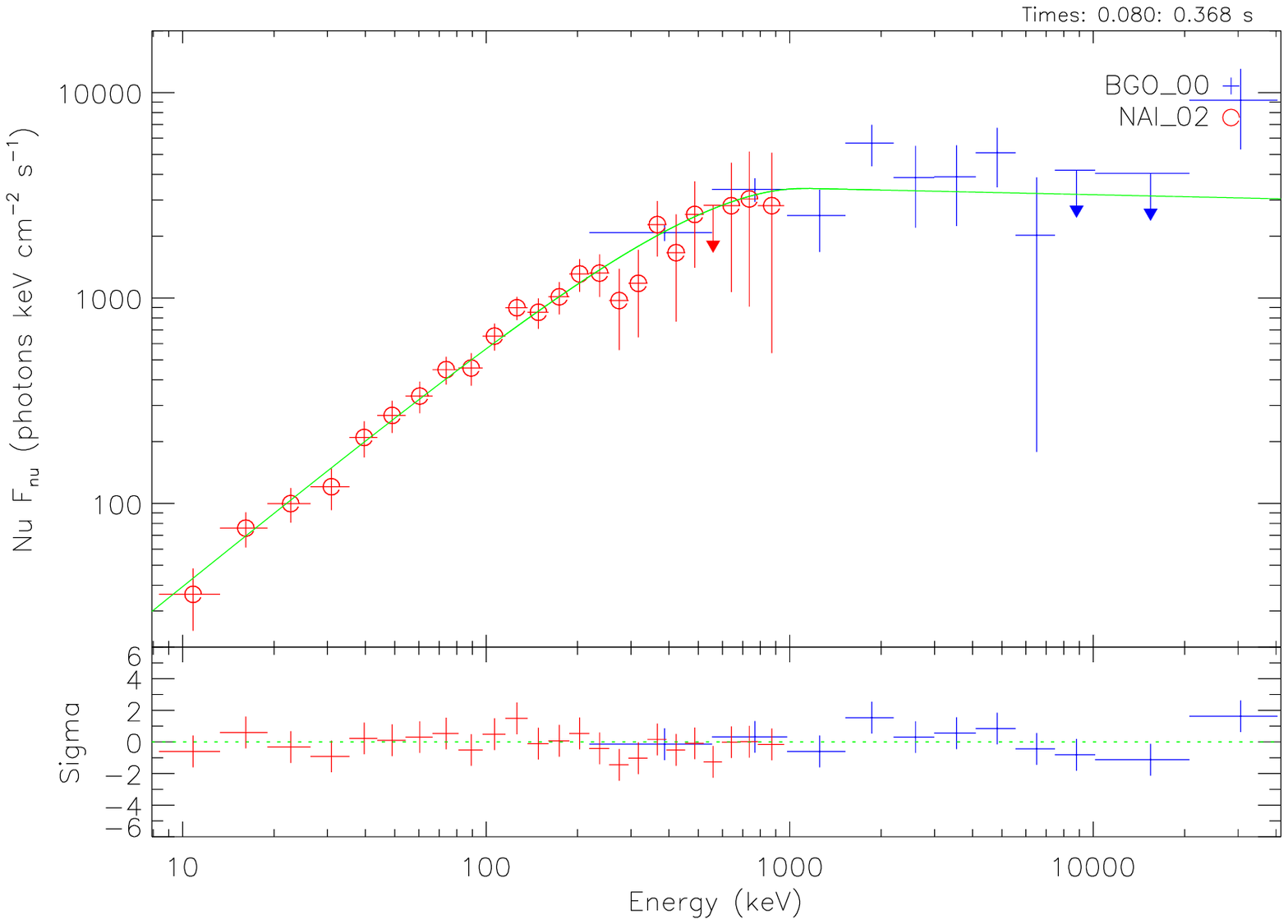} 
\hfill\null\\
\caption{The 16 ms time-binned NaI-n2 light curves of the P-GRB (left upper panel) and the extended afterglow (left lower panel) and their NaI-n2+BGO-b0 $\nu F_\nu$ spectra (on the right, the upper panel for the P-GRB and the lower one for the extended afterglow). The fit of the P-GRB is composed of a BB superimposed by a band spectrum; the extended afterglow is well fit by a simple band function.}
\label{short_marco_fig:2b}
\end{figure}

We have then explored the possibility of a further rebinning of the time interval $T_{spike}$, taking advantage of the large statistical content of each time bin. 
We have plotted the NaI-n2 light curve of GRB 090227B using time bins of 16 ms (see Fig.~\ref{short_marco_fig:2b}, left panels).
The re-binned light curves show two spike-like substructures. 
The duration of the first spike is $96$ ms and it is clearly distinct from the second spike.
In this time range the observed BB temperature is $kT = (517\pm 28)$ keV and the ratio between the fluxes of the thermal and  non-thermal  components is $F_{BB}/F_{NT} \approx 1.1$.
Consequently, we have interpreted the first spike as the P-GRB and the second spike as part of the extended afterglow.
Their spectra are shown in Fig.~\ref{short_marco_fig:2b}, right panels.

\subsection{Analysis of GRB 090227B in the fireshell model}\label{short_marco_sec:3}

The identification of the P-GRB is fundamental in order to determine the baryon load and the other physical quantities characterizing the plasma at the transparency point (see Fig.~\ref{090618_fig:no4}).
It is crucial to determine the cosmological redshift, which can be derived by combining the observed fluxes and the spectral properties of the P-GRB and of the extended afterglow with the equation of motion of our theory. 
From the cosmological redshift we derive $E_{e^+e^-}^{tot}$ and the relative energetics of the P-GRB and of the extended afterglow components (see Fig.~\ref{090618_fig:no4}).
Having so derived the baryon load $B$ and the energy $E_{e^+e^-}^{tot}$, we can constrain the total energy and simulate the canonical light curve of the GRBs with their characteristic pulses, modeled by a variable number density distribution of the CBM around the burst site.

\subsubsection{Estimation of the redshift of GRB 090227B}\label{short_marco_sec:z}

Having determined the redshift of the source, the analysis consists of equating $E_{e^+e^-}^{tot} \equiv E_{iso}$ (namely $E_{iso}$ is a lower limit on $E_{e^+e^-}^{tot}$) and inserting a value of the baryon load to complete the simulation.
The right set of $E_{e^+e^-}^{tot}$ and $B$ is determined when the theoretical energy and temperature of the P-GRB match the observed ones of the thermal emission [namely $E\PGRB  \equiv E_{BB}$ and $kT_{obs} = kT_{blue}/(1+z)$].

In the case of GRB 090227B we have estimated  (see Ref.~\refcite{2013ApJ...763..125M}) the ratio $E\PGRB /E_{e^+e^-}^{tot}$ from the observed fluences
\begin{equation}
\label{short_marco_fluences}
\frac{E\PGRB }{E_{e^+e^-}^{tot}} = \frac{4\pi d_l^2 F_{BB} \Delta t_{BB}/(1+z)}{4\pi d_l^2 F_{tot} \Delta t_{tot}/(1+z)} = \frac{S_{BB}}{S_{tot}}\ ,
\end{equation}
where $d_l$ is the luminosity distance of the source and $S = F \Delta t$ are the fluences.
The fluence of the BB component of the P-GRB is $S_{BB} = (1.54\pm0.45)\times10^{-5}$ erg/cm$^2$.
The total fluence of the burst is $S_{tot} = (3.79\pm0.20)\times10^{-5}$ erg/cm$^2$ and has been evaluated in the time interval from $T_0-0.016$ s to $T_0+0.896$ s.
This interval differs slightly from $T_{90}$ because of the new time boundaries defined after the rebinning of the light curve at a resolution of $16$ ms.
Therefore the observed energy ratio is $E\PGRB /E_{e^+e^-}^{tot} = (40.67\pm0.12)\%$.
As is clear from the bottom right diagram in Fig.~\ref{090618_fig:no4}, for each value of this ratio we have a range of possible parameters $B$ and $E^{tot}_{e^+e^-}$. 
In turn, for each of their values we can determine the theoretical blue-shifted toward the observer temperature $kT_{blue}$ (see the top right diagram in Fig.~\ref{090618_fig:no4}).
Correspondingly, for each pair of values of $B$ and $E^{tot}_{e^+e^-}$ we estimate the value of $z$ by the ratio between $kT_{blue}$ and the observed temperature of the P-GRB $kT_{obs}$,
\begin{equation}
\label{short_marco_zth} \frac{kT_{blue}}{kT_{obs}}=1+z\ .
\end{equation}
In order to remove the degeneracy $[E_{e^+e^-}^{tot}(z),B(z)]$, we have made use of the isotropic energy formula
\begin{equation}
\label{short_marco_correction}
E_{iso} = 4\pi d_l^2 \frac{S_{tot}}{(1+z)} \frac{\int^{E_{max}/(1+z)}_{E_{min}/(1+z)}{E\,N(E) dE}}{\int^{40000}_{8}{E\,N(E) dE}}\ ,
\end{equation}
in which $N(E)$ is the photon spectrum of the burst and the integrals are due to the bolometric correction on $S_{tot}$.
By imposing the condition $E_{iso} \equiv E_{e^+e^-}^{tot}$,  
we have found the values $z = 1.61\pm0.14$ for $B = (4.13\pm0.05)\times10^{-5}$ and $E_{e^+e^-}^{tot} = (2.83\pm0.15)\times10^{53}$ ergs.
The complete quantities determined in this way are summarized in Table~\ref{short_marco_par}.

\begin{table}
\tbl{The results of the simulation of GRB 090227B in the fireshell model.}
{\begin{tabular}{cc}
\hline\hline
\textbf{fireshell Parameter}                         & \textnormal{\textbf{Value}}     \\
\hline
$E^{tot}_{e^+e^-}$\,[erg]                            &  $(2.83 \pm 0.15)\times10^{53}$ \\
$B$                                                  &  $(4.13\pm0.05)\times10^{-5}$   \\
$\Gamma_{tr}$                                        &  $(1.44\pm0.01)\times10^4$      \\
$r_{tr}$\,[cm]                                       &  $(1.76\pm0.05)\times10^{13}$   \\
$kT_{blue}$\,[keV]                                   &  $(1.34\pm0.01)\times10^3$      \\
$z$                                                  &  $1.61 \pm 0.14$                \\
\hline
$\langle n \rangle$ [\textnormal{particles/cm}$^3$]  &  $(1.90\pm0.20)\times10^{-5}$   \\
$\langle \delta n/n \rangle$                         &  $0.82\pm0.11$                  \\
\hline
\end{tabular}
\label{short_marco_par}}
\end{table}

\subsubsection{The analysis of the extended afterglow and the observed spectrum of the P-GRB}\label{short_marco_sec:pgrb}

As mentioned in Sec.~\ref{short_marco_sec:fireshell}, the arrival time separation between the P-GRB and the peak of the extended afterglow is a function of $E_{e^+e^-}^{tot}$ and $B$ and depends on the detailed profile of the CBM density.
For $B \sim 4\times10^{-5}$ (see Fig.~\ref{short_marco_fig:2c}) the time separation is $\sim 10^{-3}$--$10^{-2}$ s in the source cosmological rest frame.
In this light, there is an interface between reaching transparency in the P-GRB and the early part of the extended afterglow. 
This connection has already been introduced in the literature, see e.g. Refs.~\refcite{2012MNRAS.420..468P,2012A&A...543A..10I,2012A&A...538A..58P}. 

\begin{table}
\tbl{The density mask of GRB 090227B: in the first column we list the number of CBM clouds, in the second one their distance away from the BH, and in the third one the number density with the associated error box.}
{\begin{tabular}{ccc}
\hline\hline
\textbf{Cloud}  &  \textbf{Distance [cm]}  &  \textbf{$n_{CBM}$ [$\#$/cm$^3$]}      \\
\hline
$1^{th}$        &  $1.76\times10^{15}$     &  $(1.9\pm0.2)\times10^{-5}$            \\
$2^{th}$        &  $1.20\times10^{17}$     &  $(3.5\pm0.6)\times10^{-6}$            \\
$3^{th}$        &  $1.65\times10^{17}$     &  $(9.5\pm0.5)\times10^{-6}$            \\
$4^{th}$        &  $1.80\times10^{17}$     &  $(5.0\pm0.5)\times10^{-6}$            \\
$5^{th}$        &  $2.38\times10^{17}$     &  $(2.6\pm0.2)\times10^{-5}$            \\
$6^{th}$        &  $2.45\times10^{17}$     &  $(1.0\pm0.5)\times10^{-7}$            \\
$7^{th}$        &  $4.04\times10^{17}$     &  $(6.0\pm1.0)\times10^{-5}$            \\
\hline
\end{tabular}
\label{short_marco_npar}}
\end{table}

\begin{figure}
\centering
\includegraphics[width=0.7\hsize,clip]{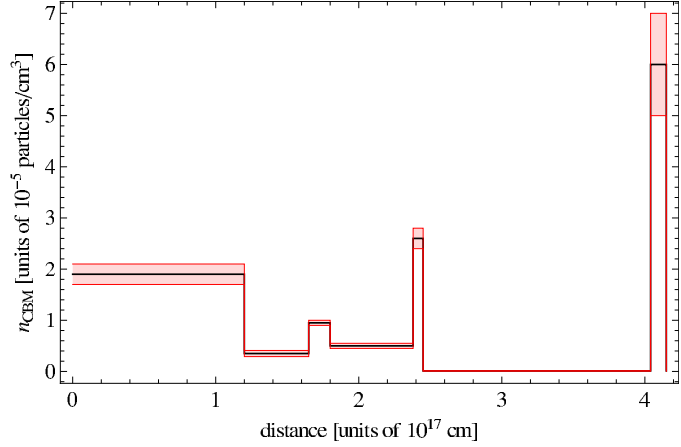} 
\caption{The radial CBM density distribution of GRB 090227B (black line) and its range of validity (red shaded region).}
\label{short_marco_fig:2e}
\end{figure}

\begin{figure}
\centering
\includegraphics[height=0.7\hsize,angle=-90,clip]{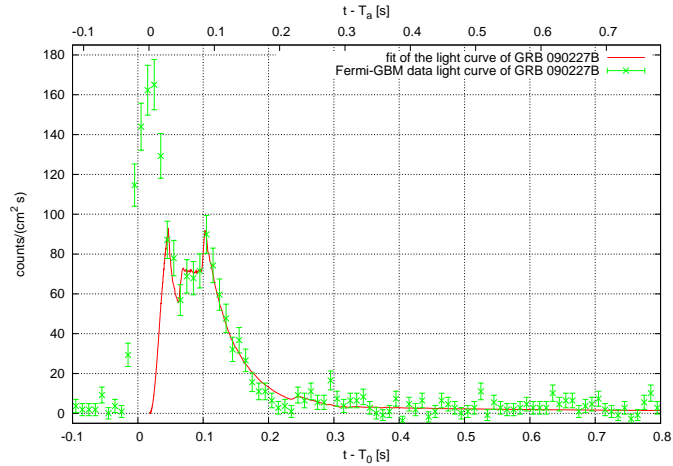}
\caption{The NaI-n2 simulated light curve of the extended-afterglow of GRB 090227B; each spike corresponds to the CBM density profile described in Table~\ref{short_marco_npar} and Fig.~\ref{short_marco_fig:2e}. The zero of the lower $x$-axis corresponds to the trigtime $T_0$; the zero of the upper $x$-axis is the time from which we have started the simulation of the extended afterglow, $T_a$, namely $0.017$ s after $T_0$.}
\label{short_marco_fig:2d1}
\end{figure} 

\begin{figure}
\centering
\includegraphics[height=0.49\hsize,angle=-90,clip]{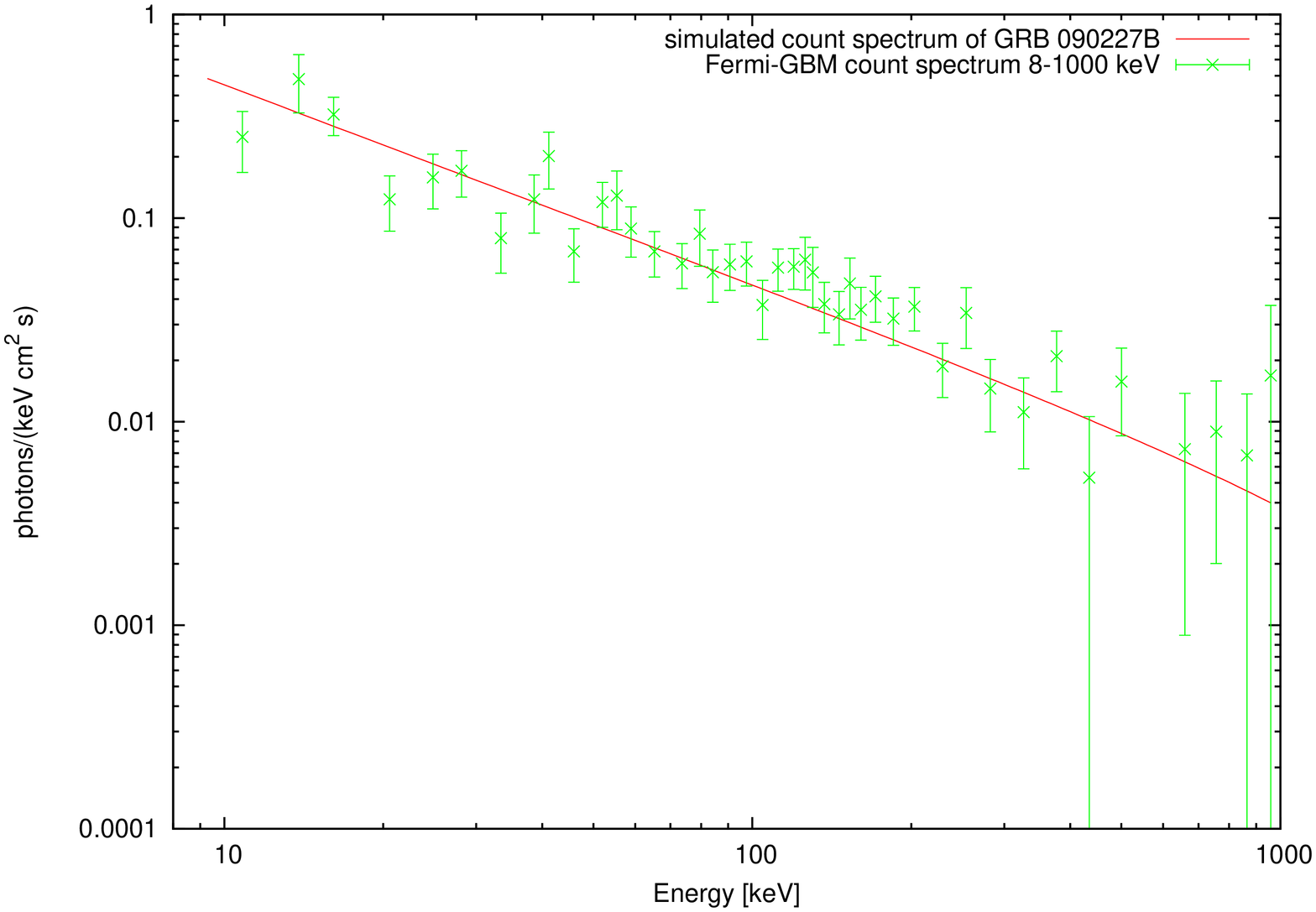}
\includegraphics[height=0.49\hsize,angle=-90,clip]{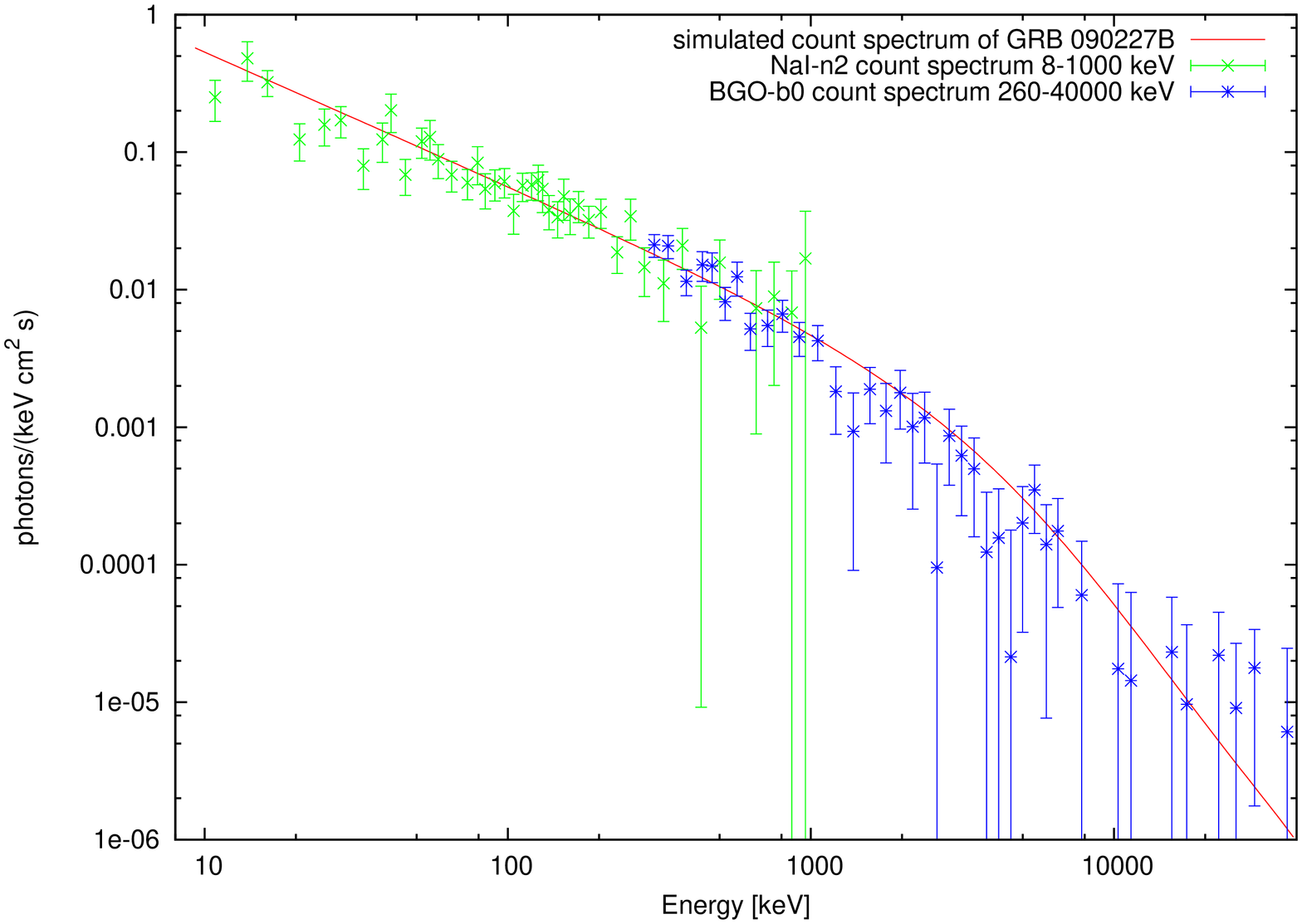}
\caption{Left panel: the simulated photon number spectrum of the extended-afterglow of GRB 090227B (from $T_0+0.015$ s to $T_0+0.385$ s) in the energy band $8$--$1000$ keV, compared to the NaI-n2 data in the same time interval. Right panel: the same simulated spectrum, with the same parameters, extended up to $40$ MeV and compared to the NaI-n2 and the BGO-b0 data in the same time interval.}
\label{short_marco_fig:2d2}
\end{figure} 

From the determination of the initial values of the energy $E^{tot}_{e^+e^-}=2.83\times10^{53}$ ergs, the baryon load $B=4.13\times10^{-5}$, and the Lorentz factor $\Gamma_{tr}=1.44\times10^4$, we have simulated the light curve of the extended afterglow by deriving the radial distribution of the CBM clouds around the burst site (see Table~\ref{short_marco_npar} and Fig.~\ref{short_marco_fig:2e}). 
In particular, each spike in Fig.~\ref{short_marco_fig:2e} corresponds to a CBM cloud.
The error boxes on the number density on each cloud is defined as the maximum possible tolerance to ensure agreement between the simulated light curve and the observed data.
The average value of the CBM density is $\langle n \rangle = (1.90\pm0.20)\times10^{-5}$ particles/cm$^3$ with an average density contrast $\langle \delta n/n \rangle = 0.82\pm0.11$ (see also Table~\ref{short_marco_par}).
These values are typical of the galactic halo environment.  
The filling factor varies in the range $9.1\times10^{-12} \leq \mathcal{R} \leq 1.5\times10^{-11}$, up to $2.38\times10^{17}$ cm away from the burst site, and then drops to the value $\mathcal{R} = 1.0\times10^{-15}$.
The value of the $\alpha$ parameter has been found to be $-1.99$ along the entire duration of the GRB.
In Fig.~\ref{short_marco_fig:2d1} we show the NaI-n2 simulated light curve ($8$--$1000$ keV) of GRB 090227B and in Fig.~\ref{short_marco_fig:2d2} (left panel) the corresponding spectrum in the early $\sim 0.4$ s of the emission, using the spectral model described by Eq.~\ref{090618_eq:no2}.
The simulation of the extended afterglow starts $T_a-T_0\sim0.017$ s after the trigtime $T_0$.
At the 13$^{th}$ Marcel Grossmann Meeting in 2012, G.~Vianello suggested extending our simulations from $1$ MeV all the way to $40$ MeV, since significant data are available from the BGO detector.
Without changing the parameters used in the theoretical simulation of the NaI-n2 data, we have extended the simulation up to $40$ MeV and have compared the results with the BGO-b0 data (see Fig.~\ref{short_marco_fig:2d2}, right panel).
The theoretical simulation we performed, optimized on the NaI-n2 data alone, is perfectly consistent with the observed data all over the \emph{entire} range of energies covered by the Fermi-GBM detector, both NaI and BGO.

\begin{figure}
\centering
\includegraphics[width=0.49\hsize,clip]{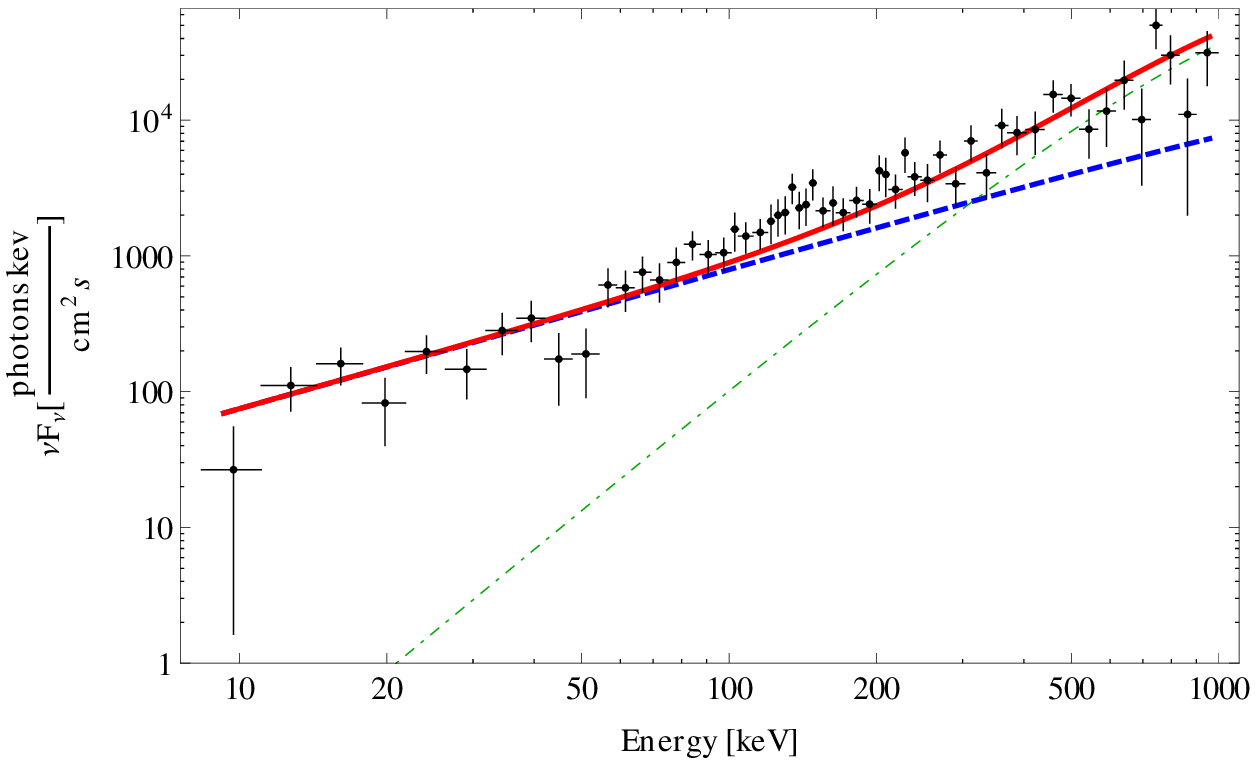}
\includegraphics[width=0.49\hsize,clip]{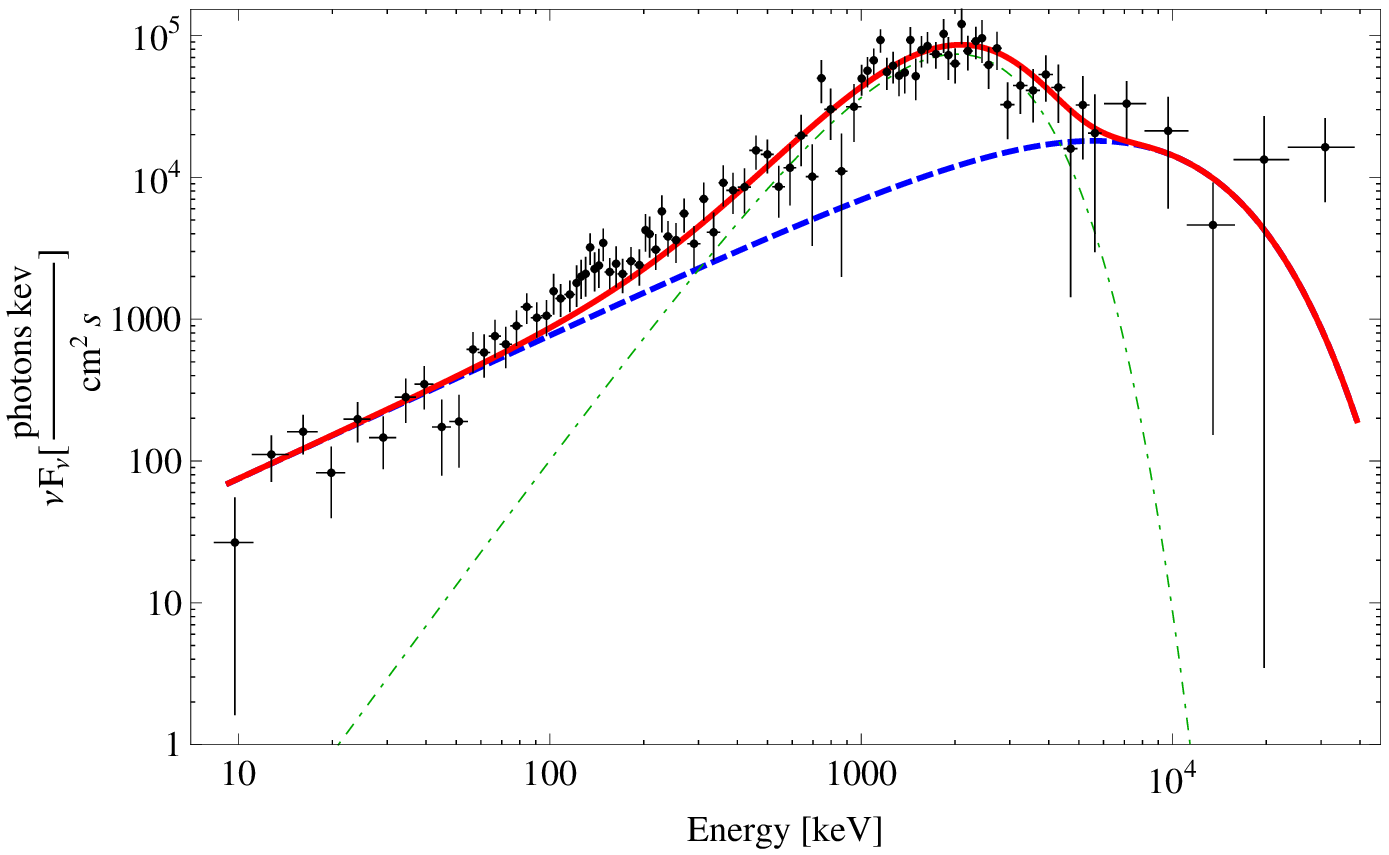}
\caption{Left panel: the time-integrated (from $T_0+0.015$ s to $T_0+0.080$ s) fireshell simulation in the energy band $8$--$1000$ keV, dashed blue line, and the BB emission, dashed-dotted green line; the sum of the two components, the solid red line, is compared to the observed P-GRB emission. Right panel: the same considerations including the BGO data up to $40$ MeV.}
\label{short_marco_fig:2f}
\end{figure}

We turn now to the emission of the early $96$ ms.
We have studied the interface between the P-GRB emission and the on-set of the extended afterglow emission.
In Fig.~\ref{short_marco_fig:2f} we have plotted the thermal spectrum of the P-GRB and the fireshell simulation (from $T_0+0.015$ s to $T_0+0.080$ s) of the early interaction of the extended afterglow.
The sum of these two components is compared with the observed spectrum from the NaI-n2 detector in the energy range $8$--$1000$ keV (see Fig.~\ref{short_marco_fig:2f}, left panel).
Then again, from the theoretical simulation in the energy range of the NaI-n2 data, we have verified the consistency of the simulation extended up to $40$ MeV with the observed data all over the range of energies covered by the Fermi-GBM detector, both NaI and BGO.
The result is shown in Fig.~\ref{short_marco_fig:2f} (right panel).

\subsection{Conclusions}\label{short_marco_sec:5}

The comprehension of this short GRB has been improved by analyzing the different spectra in the $T_{90}$, $T_{spike}$ and $T_{tail}$ time intervals. 
We have shown that within the $T_{90}$ and the $T_{spike}$ all the considered models (BB+Band, Band+PL, Compt+PL) are viable, while in the $T_{tail}$ time interval, the presence of a thermal component is ruled out.
The result of the analysis in the $T_{tail}$ time interval gives an additional correspondence between the fireshell model (see Sec.~\ref{exaft}) and the observational data. 
According to this picture, the emission within the $T_{spike}$ time interval is related to the P-GRB and it is expected to have a thermal spectrum with in addition an extra NT component due to an early onset of the extended afterglow. 
In this time interval a BB with an additional band component has been observed and we have shown that it is statistically equivalent to the Compt+PL and the Band+PL models.
Our theoretical interpretation is consistent with the observational data and statistical analysis.
From an astrophysical point of view the BB+Band model is preferred over the Compt+PL and the Band+PL models, being described by a consistent theoretical model.

GRB 090227B is the missing link between the genuine short GRBs, with the baryon load $B \lesssim 5\times10^{-5}$ and theoretically predicted by the fireshell model \cite{2001ApJ...555L.117R,2001ApJ...555L.113R,2001ApJ...555L.107R}, and the long bursts.

From the observations, GRB 090227B has an overall emission lasting $\sim 0.9$ s with a fluence of $3.79 \times 10^{-5}$ erg/cm$^2$ in the energy range $8$ keV -- $40$ MeV.
In absence of an optical identification, no determination of its cosmological redshift and of its energetics was possible.

Thanks to the excellent data available from Fermi-GBM \cite{2009ApJ...702..791M}, it has been possible to probe the comparison between the observations and the theoretical model.
In this sense, we have then performed a more detailed spectral analysis on the time scale as short as $16$ ms of the time interval $T_{spike}$.
As a result we have found in the early $96$ ms a thermal emission which we have identified with the theoretically expected P-GRB component.
The subsequent emission of the time interval $T_{spike}$ has been interpreted as part the extended afterglow.
Consequently, we have determined the cosmological redshift $z = 1.61\pm0.14$, as well as the baryon load $B = (4.13\pm0.05)\times10^{-5}$, its energetics,$E^{tot}_{e^+e^-} = (2.83\pm0.15)\times10^{53}$ ergs, and the extremely high Lorentz $\Gamma$ factor at the transparency $\Gamma_{tr} = (1.44\pm0.01)\times10^4$.

We are led to the conclusion \cite{2012arXiv1205.6915R} that the progenitor of this GRB is a binary neutron star, which for simplicity we assume to have the same mass, by the following considerations:
\begin{enumerate}
\item the very low average number density of the CBM, $\langle n_{CBM}\rangle \sim 10^{-5}$ particles/cm$^3$; this fact points to two compact objects in a binary system that have spiraled out in the halo of their host galaxy \cite{2007A&A...474L..13B,2008AIPC..966....7B,2008AIPC..966...12B,2009A&A...498..501C,2010A&A...521A..80C,2011A&A...529A.130D};
\item the large total energy, $E^{tot}_{e^+e^-} = 2.83\times10^{53}$ ergs, which we can indeed infer in view of the absence of beaming, and the very short time scale of the emission also point to two neutron stars. We are led to a binary neutron star with total mass $m_1+m_2$ larger than the neutron star critical mass, $M_{cr}$. In light of the recent neutron star theory in which all the fundamental interactions are taken into account, see Ref.~\refcite{2012NuPhA.883....1B}, we obtain for simplicity in the case of equal neutron star masses, $m_1 = m_2 = 1.34M_\odot$, radii $R_1 = R_2 = 12.24$ km, where we have used the NL3 nuclear model parameters for which $M_{cr}=2.67M_\odot$;
\item the very small value of the baryon load $B = 4.13\times10^{-5}$ is consistent with the above two neutron stars which have crusts $\sim 0.47$ km thick. The new theory of the neutron stars developed in Ref.~\refcite{2012NuPhA.883....1B} leads to the prediction of GRBs with still smaller baryon load and consequently shorter periods. We indeed infer an absolute upper limit on the energy emitted via gravitational waves of $\sim 9.6\times10^{52}$ ergs \cite{2012arXiv1205.6915R}.
\end{enumerate}

\begin{figure}
\centering
\includegraphics[width=0.7\hsize,clip]{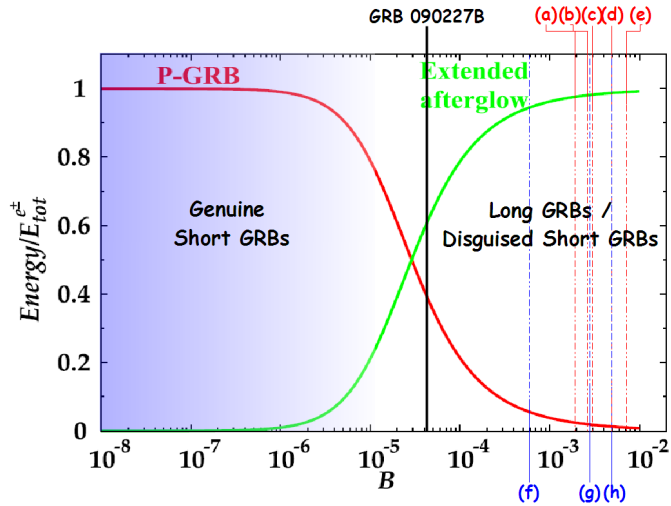}
\caption{The energy emitted in the extended afterglow (green curve) and in the P-GRB (red curve) in units of the total energy $E_{e^+e^-}^{tot} = 1.77 \times 10^{53}$ erg are plotted as functions of the parameter $B$. In the figure are also shown some values of the baryon load: in black GRB 090227B and in red and blue some values corresponding to, respectively, some long and some disguised short GRBs that we analyzed.}
\label{short_marco_fig:s}
\end{figure}

\begin{table}
\tbl{List of the long and disguised short GRBs labeled in Fig.~\ref{short_marco_fig:s} with in addition GRB 090227B. For each burst the total energy of the plasma, the baryon load, and the average CBM density are indicated.}{\tiny\begin{tabular}{ccccc}
\hline\hline
\textbf{label} & \textbf{GRB} & $E^{tot}_{e^+e^-}$ [erg] & $B$ & $\langle n_{CBM}\rangle$ [\#/cm$^3$] \\
\hline
(a) & 090618  & $2.49\times10^{53}$ & $1.98\times10^{-3}$ & $1.0$ \\
(b) & 080319B & $1.32\times10^{54}$ & $2.50\times10^{-3}$ & $6.0$ \\
(c) & 991216  & $4.83\times10^{53}$ & $3.00\times10^{-3}$ & $1.0$ \\
(d) & 030329  & $2.12\times10^{52}$ & $4.80\times10^{-3}$ & $2.0$ \\
(e) & 031203  & $1.85\times10^{50}$ & $7.40\times10^{-3}$ & $0.3$ \\
\hline
(f) & 050509B & $5.52\times10^{48}$ & $6.00\times10^{-4}$ & $1.0\times10^{-3}$ \\
(g) & 060614  & $2.94\times10^{51}$ & $2.80\times10^{-3}$ & $1.0\times10^{-3}$ \\
(h) & 970228  & $1.45\times10^{54}$ & $5.00\times10^{-3}$ & $9.5\times10^{-4}$ \\
\hline
    & 090227B & $2.83\times10^{53}$ & $4.13\times10^{-5}$ & $1.9\times10^{-5}$ \\
\hline
\end{tabular}
\label{short_marco_pars}}
\end{table}

We can then generally conclude the existence of three different possible structures for the canonical GRBs (see Fig.~\ref{short_marco_fig:s} and Table~\ref{short_marco_pars}):
\begin{enumerate}
\item[a.] long GRBs with baryon load $3.0\times10^{-4} \lesssim B \leq 10^{-2}$, exploding in a CBM with average density of $\langle n_{CBM} \rangle \approx 1$ particle/cm$^3$, typical of the inner galactic regions;
\item[b.] disguised short GRBs with the same baryon load as the previous class, but occurring in a CBM with $\langle n_{CBM} \rangle \approx 10^{-3}$ particle/cm$^3$, typical of galactic halos \cite{2007A&A...474L..13B,2008AIPC..966....7B,2008AIPC..966...12B,2009A&A...498..501C,2010A&A...521A..80C,2011A&A...529A.130D};
\item[c.] genuine short GRBs which occur for $B \lesssim 10^{-5}$ with the P-GRB predominant with respect to the extended afterglow and exploding in a CBM with $\langle n_{CBM} \rangle \approx 10^{-5}$ particle/cm$^3$, typical again of galactic halos,  GRB 090227B being the first example.
\end{enumerate}

Finally, if we turn to the theoretical model within a general relativistic description of the gravitational collapse to a $10 M_\odot$ BH, see e.g. Refs.~\refcite{2003PhLB..573...33R,2005IJMPD..14..131R} and Fig.~2 in Ref.~\refcite{2006NCimB.121.1477F}, we find it necessary to use time resolutions on the order of a fraction of a ms, possibly down to $\mu$s, in order to follow such a process. 
One would need new space missions with larger collecting area to prove with great accuracy the identification of a thermal component.
It is likely that an improved data acquisition with high signal to noise ratio on a shorter time scale would show more clearly the thermal component as well as distinguish more effectively different fitting procedures.

\section{Unveiling the GRB-SN Connection}

\subsection{Introduction}

Until recently, all the X- and $\gamma$-ray activities of a signal sufficiently short in time, less than $10^2$--$10^3$ s, and of extragalactic origin have been called a GRB. 
A new situation has occurred with the case of GRB 090618 \cite{TEXAS} in which the multi-component nature of GRBs has been illustrated. 
This GRB is a member of a special class of bursts associated with a SN. 
It is now clear from the detailed analysis that there are at least three different components in the nature of this GRB: episode 1 which corresponds to the early emission of the SN event with Lorentz factor $\Gamma \sim 1$; episode 2 which corresponds to the GRB with Lorentz factor $10^2\lesssim\Gamma\lesssim10^4$; and episode 3 which appears to be related to the activities of the newly born NS.
I will describe a few key moments in the recent evolution of our understanding of this system which is very unique within physics and astrophysics.

\subsection{The case of GRB 090618}

GRB 090618 represents the prototype of a class of energetic ($E_{iso} \geq 10^{52}$ erg) GRBs, characterized by the presence of a supernova observed  10 (1+z) days after the trigger time, and the observation of two distinct emission episodes in their hard X-ray light curve (see details in Ref.~\refcite{2012A&A...543A..10I}).

It was discovered by the Swift satellite \cite{2009GCN..9512....1S}. The BAT light curve shows a multi-peak structure, whose total estimated duration is $\sim$ 320 s and whose T$_{90}$ duration in the (15--350) keV range was 113 s \cite{2009GCN..9530....1B}. 
The first 50 s of the light curve shows a smooth decay trend followed by a spiky emission, with three prominent peaks at 62, 80, and 112 s after the trigger time, respectively, and each have the typical appearance of a FRED pulse \cite{1994ApJS...92..229F}, see Fig.~\ref{090618_fig:1}. 
The time-integrated spectrum, (t$_0$ - 4.4, t$_0$ + 213.6) s in the (15--150)keV range, was found to agree with a power-law spectral model with an exponential cut-off, whose photon index is $\gamma$ = 1.42 $\pm$ 0.08 and a cut-off energy $E_{peak}$ = 134 $\pm$ 19 keV \cite{2009GCN..9534....1S}.  
The XRT observations started 125 s after the BAT trigger time and lasted $\sim$ 25.6 ks \cite{2009GCN..9528....1B} and reported an initially bright uncatalogued source, identified as the afterglow of GRB 090618.
Its early decay is very steep, ending at 310 s after the trigger time, when it starts a shallower phase, the plateau.
Then the light curve breaks into a steeper late phase.

GRB 090618 was observed also by the Gamma-ray Burst Monitor (GBM) on board the Fermi satellite \cite{2009ApJ...702..791M}.
From a first analysis, the time-integrated spectrum, ($t_0$, $t_0$ + 140) s in the (8--1000)keV range, was fit by a band \cite{1993ApJ...413..281B} spectral model, with a peak energy $E_{peak}$ = 155.5  keV, $\alpha$ = $-1.26$ and $\beta$ = $-2.50$ \cite{2009GCN..9535....1M}, but with strong spectral variations within the considered time interval.

The redshift of the source is $z =  0.54$ and it was determined thanks to the identification of the MgII, Mg I, and FeII absorption lines using the KAST spectrograph mounted at the 3 m Shane telescope at the Lick observatory \cite{2009GCN..9518....1C}.
Given the redshift and the distance of the source, we computed the emitted isotropic energy in the 8 -- 10000 keV energy range, with the Schaefer formula \cite{2007ApJ...660...16S}: using the fluence in the (8--1000 keV) as observed by Fermi-GBM, S$_{obs}$ = 2.7 $\times$ 10$^{-4}$ \cite{2009GCN..9535....1M}, and the $\Lambda$ Cold Dark Matter (CDM) cosmological standard model $H_0$ = 70 km/s/Mpc, $\Omega_m$ = 0.27, $\Omega_{\Lambda}$ = 0.73, we obtain for the emitted isotropic energy the value of E$_{iso}$ = 2.90 $\times$ 10$^{53}$ erg.

This GRB was observed also by Konus-WIND \cite{2009GCN..9553....1G}, Suzaku-WAM \cite{2009GCN..9568....1K}, and by the AGILE satellite \cite{2009GCN..9524....1L}, which detected emission in the (18--60) keV and in the MCAL instrument, operating at energies greater than 350 keV, but it did not observe high-energy photons above 30 MeV.
GRB 090618 was the first GRB observed by the Indian payloads RT-2 on board the Russian satellite CORONAS-PHOTON \cite{2008cosp...37.1596K,2009arXiv0912.4126N,2011ApJ...728...42R}.

Thanks to the complete data coverage of the optical afterglow of GRB 090618, the presence of a supernova underlying the emission of its optical afterglow was reported \cite{2011MNRAS.413..669C}.
The evidence of a supernova emission came from the presence of several bumps in the light curve and by the change in $R_c$ - $i$ color index over time: in the early phases, the blue color is dominant, typical of the GRB afterglow, but then the color index increases, suggesting a core-collapse SN. 
At late times, the contribution from the host galaxy was dominant.

\begin{figure}
\centering
\includegraphics[width=0.7\hsize,angle=270]{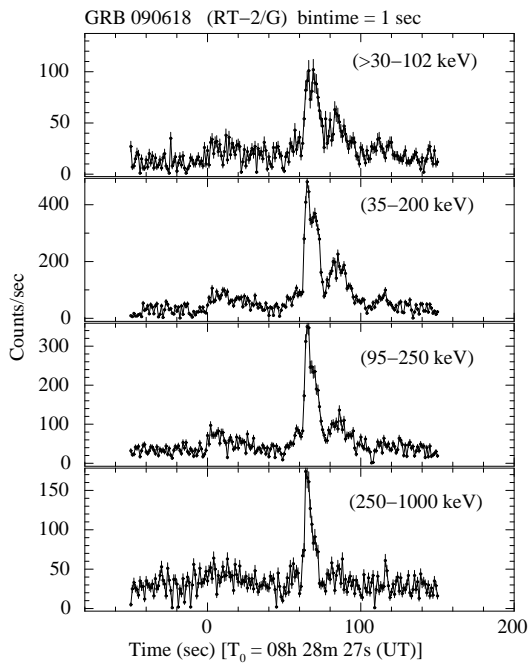}
\caption{RT2 light curves of GRB 090618. }
\label{090618_fig:Chak}
\end{figure}

\subsubsection{Data analysis}\label{090618_sec:2.2}

We have analyzed GRB 090618, considering the BAT and XRT data of the Swift satellite together with the Fermi-GBM and RT2 data of the Coronas-PHOTON satellite (see Fig.~\ref{090618_fig:Chak}).
The data reduction was made with the Heasoft v6.10 packages\footnote{http://heasarc.gsfc.nasa.gov/lheasoft/} for BAT and XRT, and the Fermi-Science tools for GBM. The details of the data reduction and analysis are given in Ref.~\refcite{2012A&A...543A..10I}.

In Table \ref{090618_tab:no1} we give the results of our spectral analysis. 
The time reported in the first column corresponds to the time after the GBM trigger time t$_{trig}$ = 267006508 s, where the $\beta$ parameter was not constrained, we used its averaged value, $\beta$ = -2.3 $\pm$ 0.10, as delineated in Ref.~\refcite{2011A&A...525A..53G}. We considered the chi-square statistic for testing our data fitting procedure. The reduced chi-square $\tilde{\chi}^2 = \chi^2/N$, where $N$ is the number of degrees of freedom (dof), which is $N = 82$ for the NaI dataset and $N = 121$ for that of the BGO.

\begin{figure}
\centering
\begin{tabular}{|c|}
\hline
\includegraphics[width=0.48\hsize,clip]{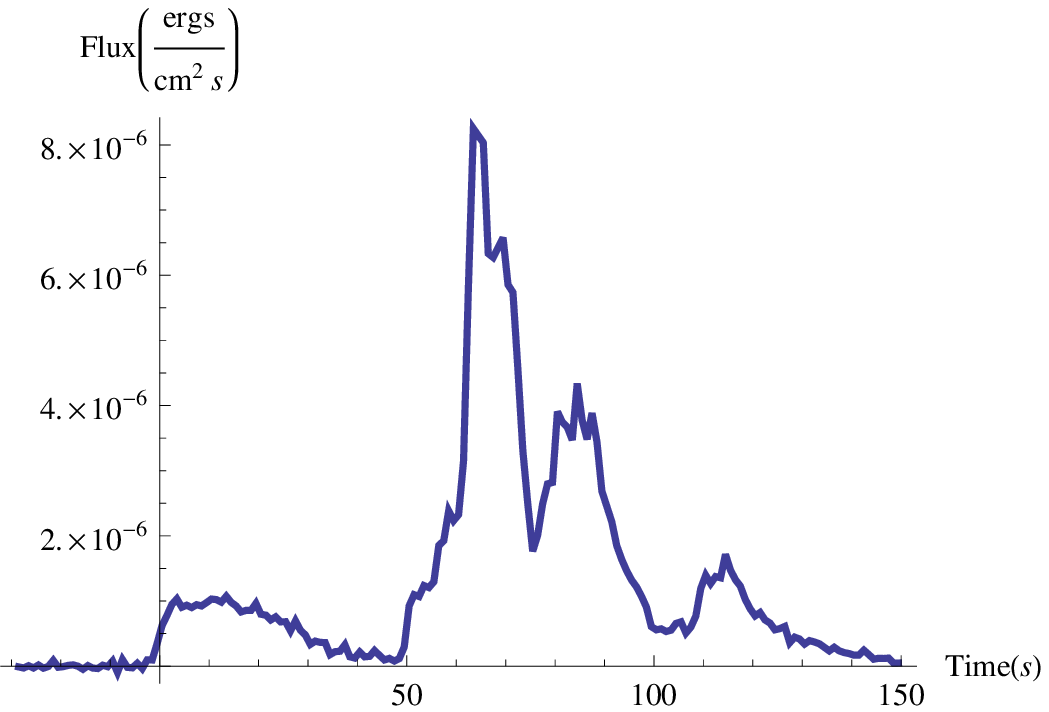}\vline
\includegraphics[width=0.48\hsize,clip]{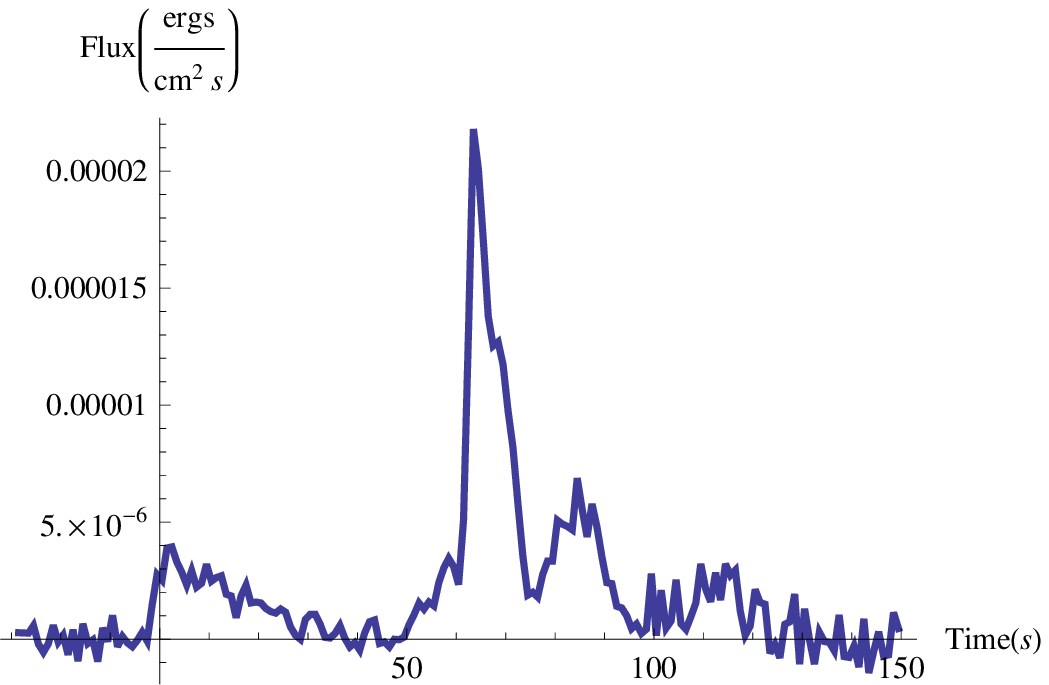}\\
\hline
\end{tabular}\label{090618_fig:1}
\caption{Fermi-GBM flux light curve of GRB 090618 referring to the NaI (8--440 keV, \emph{left panel}) and BGO (260 keV -- 40 MeV, \emph{right panel}) detectors.}
\end{figure}

For the last pulse of the second episode, the band model is not very precise ($\tilde{\chi}^2$ = 2.24), but a slightly better approximation is given by a power-law with an exponential cut-off, whose fit results are shown for the same intervals in the last two columns.
From these values, we built the flux light curves for both detectors, which are shown in Fig.~\ref{090618_fig:1}.

\begin{table}
\tbl{Time-resolved spectral analysis of GRB 090618. We considered six time intervals, each one corresponding to a particular emission feature in the light curve. We fit the GBM (8 keV -- 10 MeV) observed emission with a band model \protect\cite{1993ApJ...413..281B} and a power-law function with an exponential cut-off. In columns 2--4 we list the band model low-energy index $\alpha$, the high-energy $\beta$ and the break energy $E_0^{BAND}$, with the reduced chi-square value in the 6$^{th}$ column. The last three columns list the power-law index $\gamma$, the cut-off energy $E_0^{cut}$ and the reduced chi-square value respectively, as obtained from the spectral fit with the cut-off power-law spectral function.} 
{\tiny \begin{tabular}{l c c c c c c c}
\hline\hline
Time Interval  & $\alpha$ & $\beta$ & $E_0^{BAND}$ (keV) &  $\tilde{\chi}^2_{BAND}$ & $\gamma$ & $E_0^{cut}$ (keV) &  $\tilde{\chi}^2_{cut}$\\ 
\hline 
0 - 50 & -0.77$^{+0.38}_{-0.28}$ & -2.33$^{+0.33}_{-0.28}$ & 128.12$^{+109.4}_{-56.2}$  & 1.11 & 0.91$^{+0.18}_{-0.21}$ & 180.9$^{+93.1}_{-54.2}$ & 1.13\\
50 - 57 & -0.93$^{+0.48}_{-0.37}$ & -2.30 $\pm$ 0.10 & 104.98$^{+142.3}_{-51.7}$ & 1.22 & 1.11$^{+0.25}_{-0.30}$ & 168.3$^{+158.6}_{-70.2}$ & 1.22 \\
57 - 68 & -0.93$^{+0.09}_{-0.08}$ & -2.43$^{+0.21}_{-0.67}$ & 264.0$^{+75.8}_{-54.4}$ & 1.85 & 1.01$^{+0.06}_{-0.06}$ & 340.5$^{+56.0}_{-45.4}$ & 1.93 \\
68 - 76 & -1.05$^{+0.08}_{-0.07}$ & -2.49$^{+0.21}_{-0.49}$ & 243.9$^{+57.1}_{-53.0}$  & 1.88  & 1.12$^{+0.04}_{-0.04}$ & 311.0$^{+38.6}_{-32.9}$ & 1.90 \\
76 - 103 & -1.06$^{+0.08}_{-0.08}$ & -2.65$^{+0.19}_{-0.34}$ & 125.7$^{+23.27}_{-19.26}$ & 1.23 & 1.15$^{+0.06}_{-0.06}$ & 157.7$^{+22.2}_{-18.6}$ & 1.39 \\          
103 - 150 & -1.50$^{+0.20}_{-0.18}$ & -2.30 $\pm$ 0.10 & 101.1$^{+58.3}_{-30.5}$ & 1.07  & 1.50$^{+0.18}_{-0.20}$ & 102.8$^{+56.8}_{-30.4}$ & 1.06\\
\hline
\end{tabular}
\label{090618_tab:no1}}
\end{table}

\subsubsection{Spectral analysis of GRB 090618}\label{090618_sec:4}

We proceed now to the detailed spectral analysis of GRB 090618. 
We divide the emission into six time intervals, shown in Table~\ref{090618_tab:no1}, each one identifying a significant feature in the emission process.
We then fit for each time interval the spectra by a band model and a blackbody with an extra power-law component, following Ref.~\refcite{2004ApJ...614..827R}. 
In particular, we are interested in estimating the temperature $kT$ and the observed energy flux $\phi_{obs}$ of the blackbody component.
The specific intensity of emission of a thermal spectrum at energy $E$ in energy range $d E$ into solid angle $\Delta \Omega$ is 
\begin{equation}\label{090618_eq:no2a}
I(E)dE = \frac{2}{h^3 c^2} \frac{E^3}{\exp(E/kT) - 1} \Delta \Omega dE .
\end{equation}
The source of radius $R$ is seen within a solid angle $\Delta \Omega = \pi R^2/D^2$, and its full luminosity is $L = 4 \pi R^2 \sigma T^4$. What we are fitting, however, is the background-subtracted photon spectra $A(E)$, which is obtained by dividing the specific intensity $I(E)$ by the energy $E$:
\begin{eqnarray}
A(E)dE \equiv \frac{I(E)}{E}dE 
&=& \frac{k^4 L}{2\sigma (kT)^4 D^2 h^3 c^2}\frac{E^2 d E}{\exp(E/kT)-1} 
\nonumber\\
&=& \frac{15 \phi_{obs}}{\pi^4 (kT)^4}\frac{E^2 d E}{\exp(E/kT)-1} ,
\label{090618_eq:no2b}
\end{eqnarray}
where $h$, $k$ and $\sigma$ are the Planck, Boltzmann, and Stefan-Boltzmann constants respectively, $c$ is the speed of light and $\phi_{obs} = L/(4 \pi D^2)$ is the observed energy flux of the blackbody emitter.
The great advantage of Eq.~(\ref{090618_eq:no2b}) is that it is written in terms of the observables $\phi_{obs}$ and $T$, so from a spectral fitting procedure we can obtain the values of these quantities for each time interval considered.
To determine these parameters, we must perform an integration of the actual photon spectrum $A(E)$ over the instrumental response $R(i,E)$ of the detector that observes the source, where $i$ denotes the different instrument energy channels. The result is a predicted count spectrum 
\begin{equation}
C_{p}(i) = \int_{E_{min}(i)}^{E_{max}(i)} A(E) R(i,E) dE,
\label{090618_eq_int}
\end{equation}
where $E_{min}(i)$ and $E_{max}(i)$ are the boundaries of the $i$-th energy channel of the instrument. Eq.~(\ref{090618_eq_int}) must be compared with the observed data by a fit statistic.

\begin{figure}
\centering
\includegraphics[width=0.7\hsize,clip]{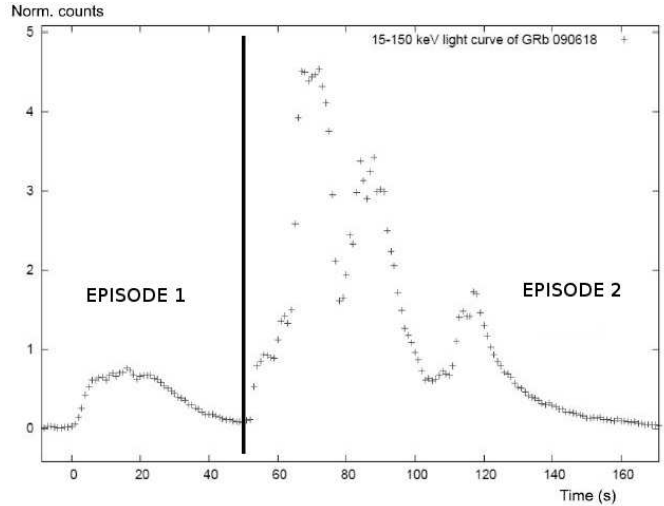}
\caption{Two episode nature of GRB 090618.}\label{090618_fig:cospar}
\end{figure}

\begin{figure}
\centering
\includegraphics[height=0.37\hsize,clip]{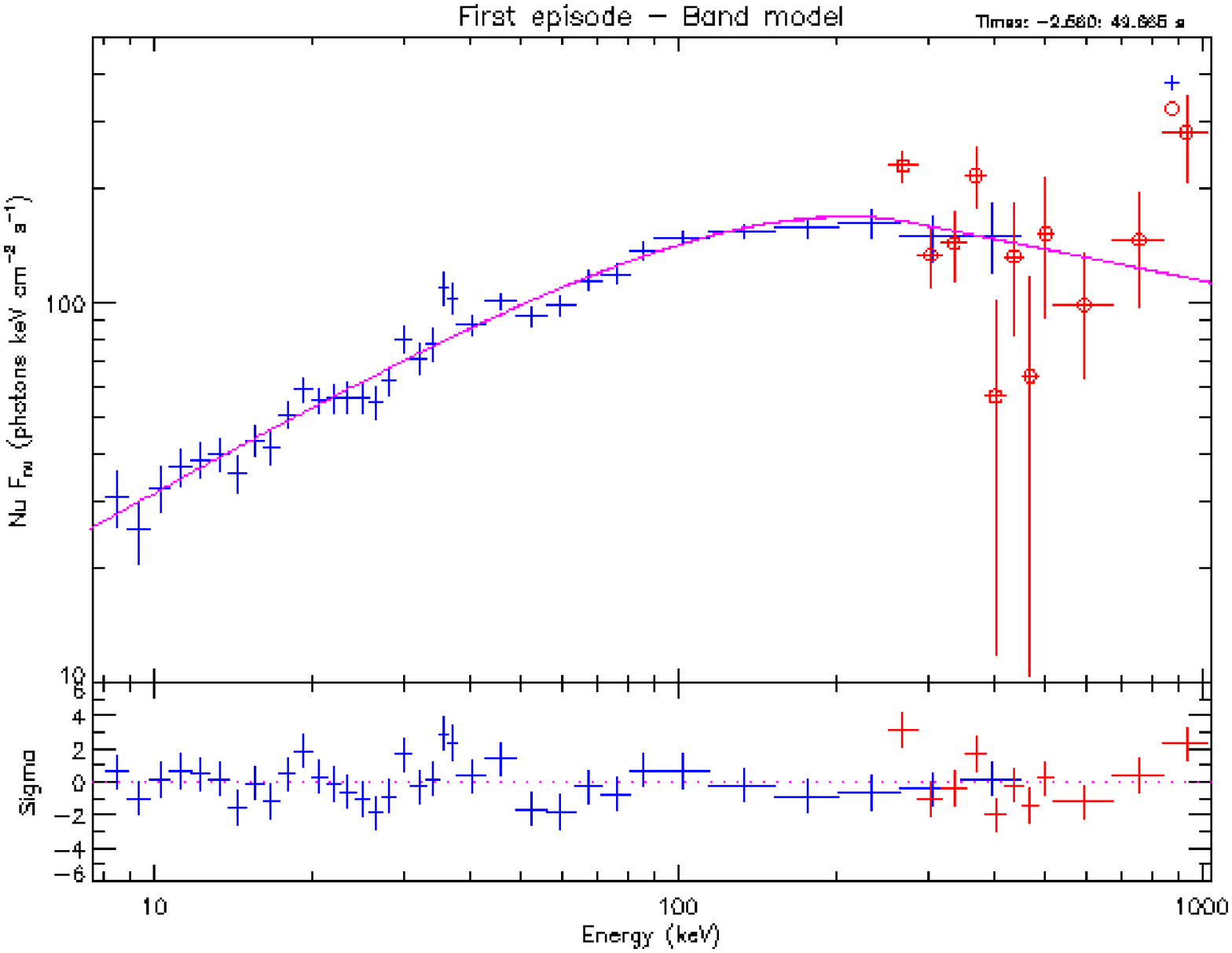}
\includegraphics[height=0.37\hsize,clip]{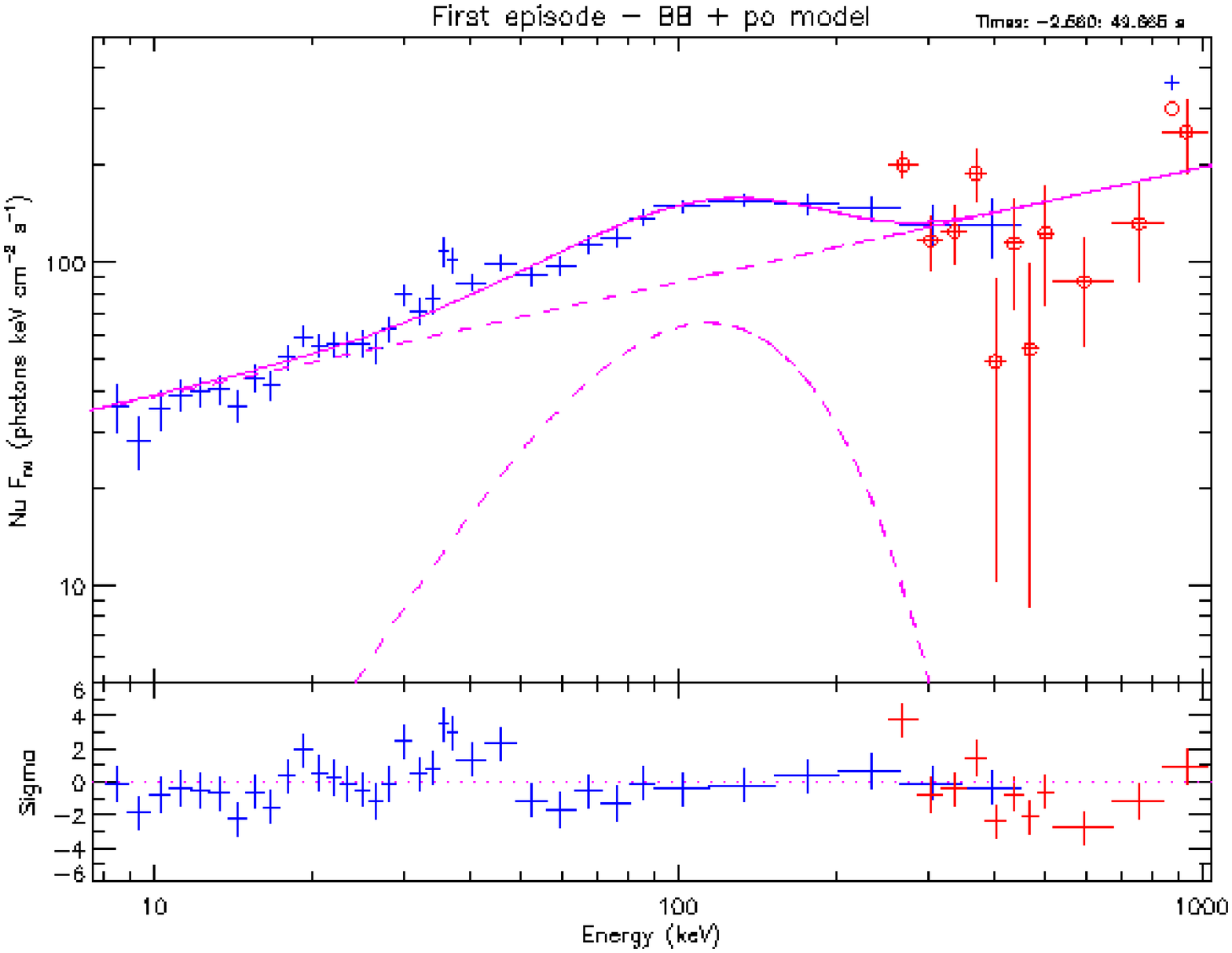}
\caption{Time-integrated spectra for the first episode (from 0 to 50 s) of GRB 090618 fit with the band, $\tilde{\chi}^2$ = 1.12 (left) and blackbody + power-law (right) models, $\tilde{\chi}^2$ = 1.28. In the following we will consider the case of a blackbody + power-law model and infer some physical consequences. The corresponding considerations for the band model are in progress and will be published elsewhere.}
\label{090618_fig:big1}
\end{figure}

\begin{figure}
\centering
\includegraphics[height=0.37\hsize,clip]{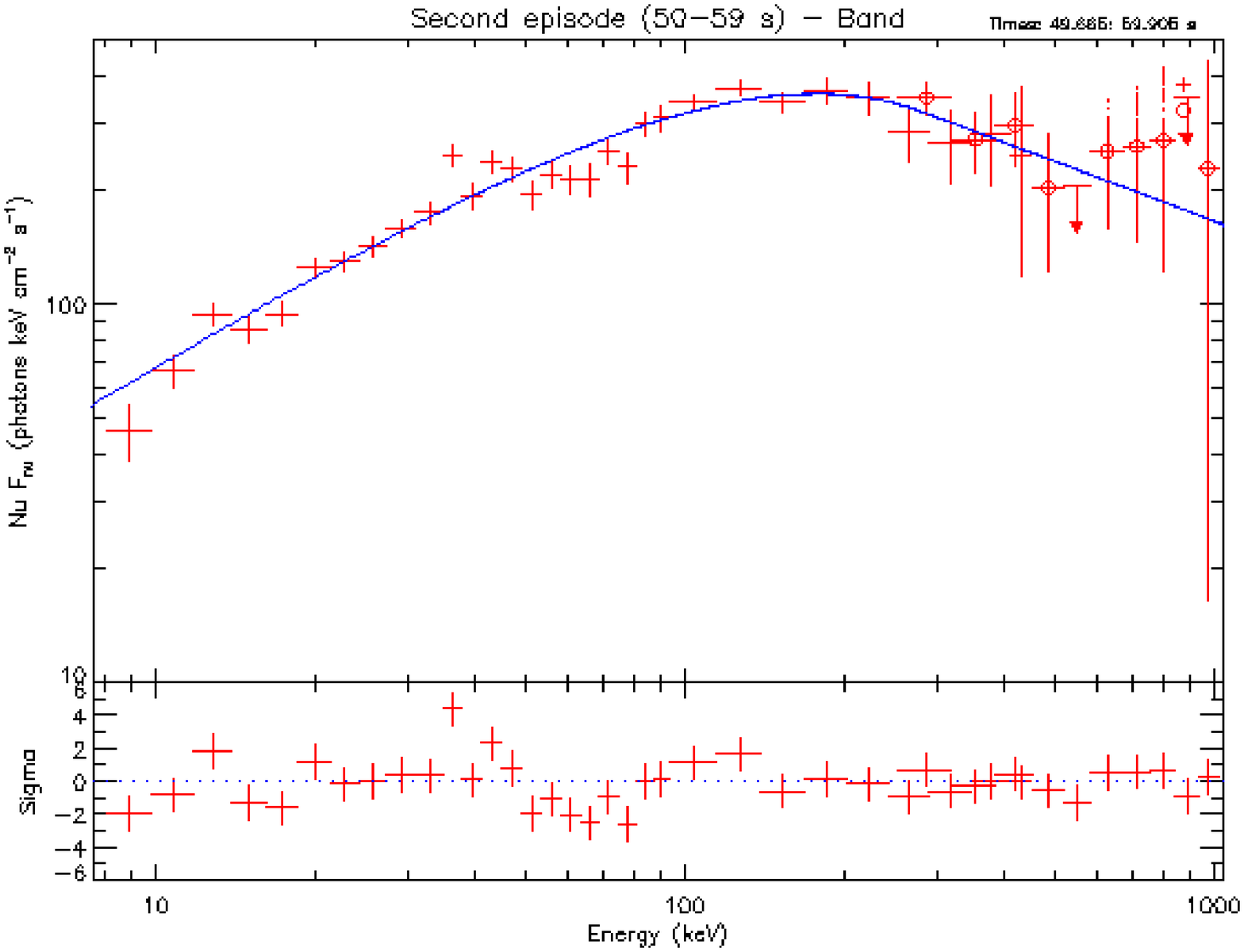}
\includegraphics[height=0.37\hsize,clip]{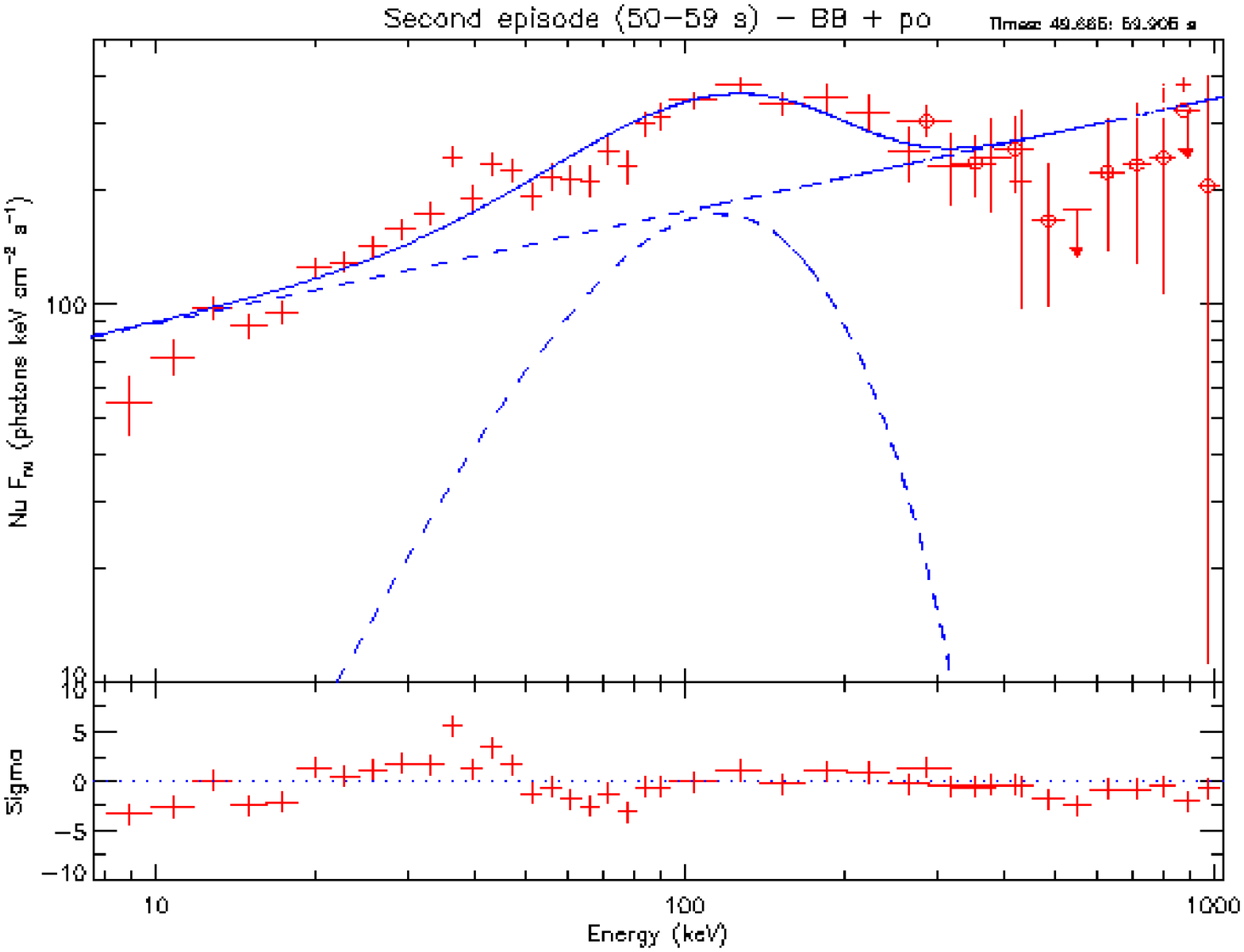}
\caption{Time-integrated spectra for the first 9 s of the second episode (from 50 to 59 s after the trigger time) of GRB 090618 fit with the band, $\tilde{\chi}^2$ = 1.23 (left) and blackbody + power-law (right) models, $\tilde{\chi}^2$ = 1.52. }\label{090618_fig:big2}
\end{figure}

The main parameters obtained from the fitting procedure are shown in Table~\ref{090618_tab:no2b}.
We divide the entire GRB into two main episodes, as proposed in Ref.~\refcite{TEXAS}: one lasting the first 50 s and the other from 50 to 151 s after the GRB trigger time, see Fig.~\ref{090618_fig:cospar}. 
Clearly, the first 50 s of emission, corresponding to the first episode, are well-fit by a band model as well as a blackbody with an extra power-law model, Fig.~\ref{090618_fig:big1}. 
The same happens for the first 9 s of the second episode (from 50 to 59 s after the trigger time), Fig.~\ref{090618_fig:big2}. 
For the subsequent three intervals corresponding to the main peaks in the light curve, the blackbody plus a power-law model does not provide a satisfactory fit. 
Only the band model fits the spectrum with good accuracy, with the exception of the first main spike (compare the values of $\chi^2$ in the table). 
We find also that the last peak can be fit by a simple power-law model with a photon index $\gamma$ = 2.20 $\pm$ 0.03, better than by a band model.

The result of this analysis points to a different emission mechanism in the first 50 s of GRB 090618 and in the next 9 s. A sequence of very strong pulses follows, whose spectral energy distribution is not attributable either to a blackbody or a blackbody and an extra power-law component.
Good evidence for the transition is shown by the test of the data fitting, whose indicator is given by the changing of $\tilde{\chi^2}$ ($N_{dof} = 169$) for the blackbody plus a power-law model for the different time intervals, see Table~\ref{090618_tab:no2b}.
Although the band spectral model is an empirical model without a clear physical origin, we checked its validity in all time-detailed spectra with the sole exception of the first main pulse of the second episode.
The $\chi^2$ corresponding to the band model for this main pulse, although better than that corresponding to the blackbody and power-law case, is unsatisfactory. 
We now directly apply the fireshell model to make the above conclusions more stringent and reach a better understanding of the source. 

\begin{table}
\centering
\tbl{Time-resolved spectral analysis (8 keV -- 10 MeV) of the second episode in GRB 090618.}
{\tiny
\begin{tabular}{l l c c c c c c c}
\hline\hline
 & Time Interval (s) & $\alpha$ & $\beta$ & $E_0(keV)$ & $\tilde{\chi}^2_{BAND}$ & $kT(keV)$ & $\gamma$ & $\tilde{\chi}^2_{BB+po} $\\ 
\hline
A & 0 - 50 & -0.74 $\pm$ 0.10 & -2.32 $\pm$ 0.16 & 118.99 $\pm$ 21.71  & 1.12 & 32.07 $\pm$ 1.85  & 1.75 $\pm$ 0.04 & 1.21 \\
B & 50 - 59 & -1.07 $\pm$ 0.06 & -3.18 $\pm$ 0.97 & 195.01 $\pm$ 30.94  & 1.23 & 31.22 $\pm$ 1.49  & 1.78 $\pm$ 0.03 & 1.52 \\
C & 59 - 69 & -0.99 $\pm$ 0.02 & -2.60 $\pm$ 0.09 & 321.74 $\pm$ 14.60 & 2.09 & 47.29 $\pm$ 0.68  & 1.67 $\pm$ 0.08   & 7.05\\
D & 69 - 78 & -1.04 $\pm$ 0.03 & -2.42 $\pm$ 0.06 & 161.53 $\pm$ 11.64    &  1.55 & 29.29 $\pm$ 0.57   & 1.78 $\pm$ 0.01 & 3.05\\
E & 78 - 105 &  -1.06 $\pm$ 0.03 & -2.62 $\pm$ 0.09 & 124.51 $\pm$ 7.93    &  1.20 & 24.42 $\pm$ 0.43    &  1.86 $\pm$ 0.01  & 2.28\\     
F & 105 - 151 & -2.63 $\pm$ -1    & -2.06 $\pm$ 0.02 & unconstrained    &   1.74 & 16.24 $\pm$ 0.84  & 2.23 $\pm$ 0.05   & 1.15\\
\hline
\end{tabular}
\label{090618_tab:no2b}}
\end{table}

\subsection{Analysis of GRB 090618 in the fireshell scenario: from a single GRB to a multi-component GRB}\label{090618_sec:5}

\subsubsection{Attempt for a single GRB scenario: the role of the first episode}

We first approach the analysis of GRB 090618 by assuming that we observe a single GRB and attempt identification of the P-GRB emission of a canonical GRB within the fireshell scenario (see panel A in Fig.~\ref{090618_fig:big2} and Table~\ref{090618_tab:no2b}).
This has been shown to be inconsistent (see details in Ref.~\refcite{2012A&A...543A..10I}).
We then turn to a multicomponent emission.

\subsubsection{The multi-component scenario: the second episode as an independent GRB}

\paragraph{The identification of the P-GRB of the second episode.}

\begin{figure}
\centering
\includegraphics[height=0.37\hsize,clip]{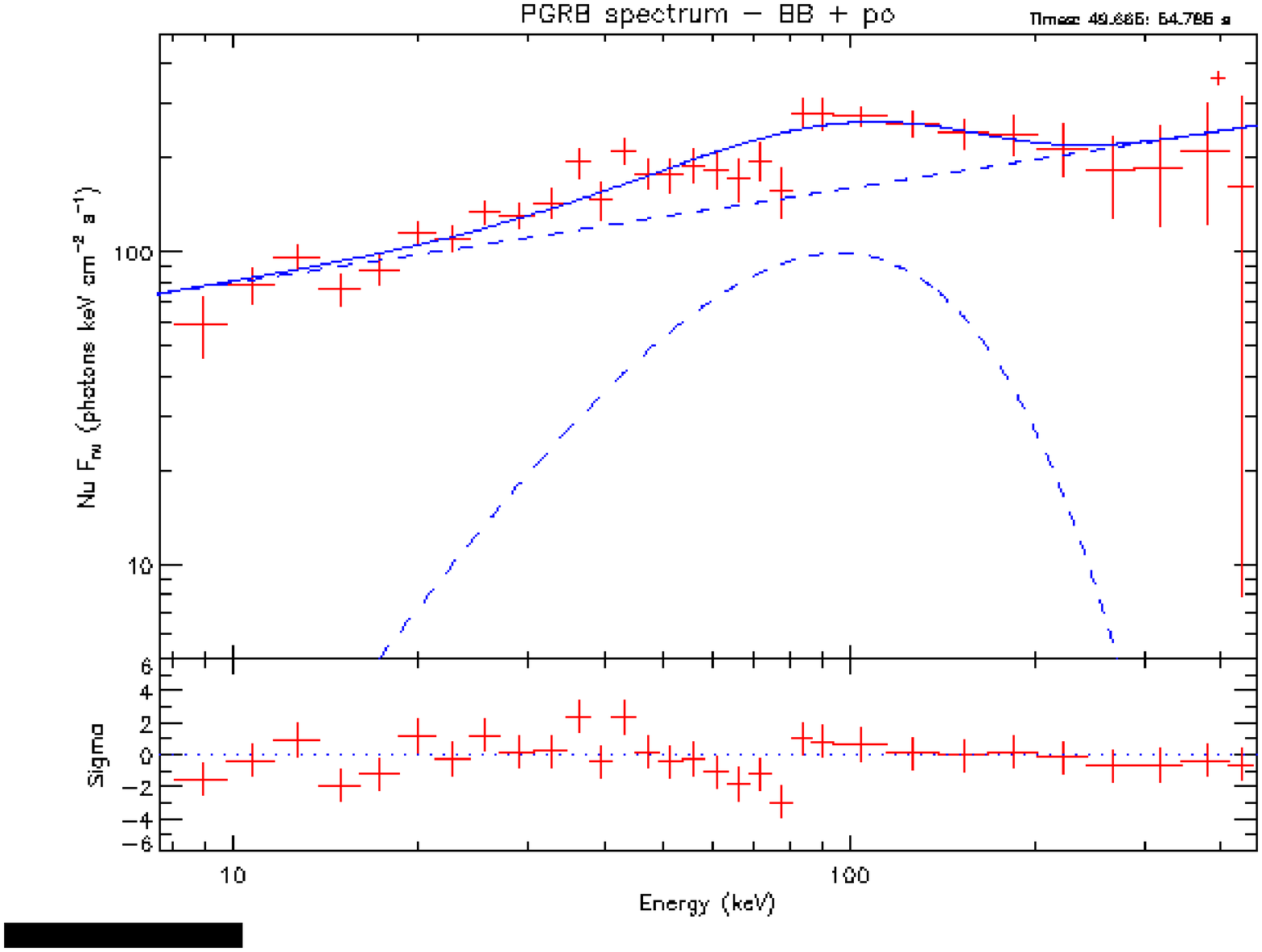}
\includegraphics[height=0.37\hsize,clip]{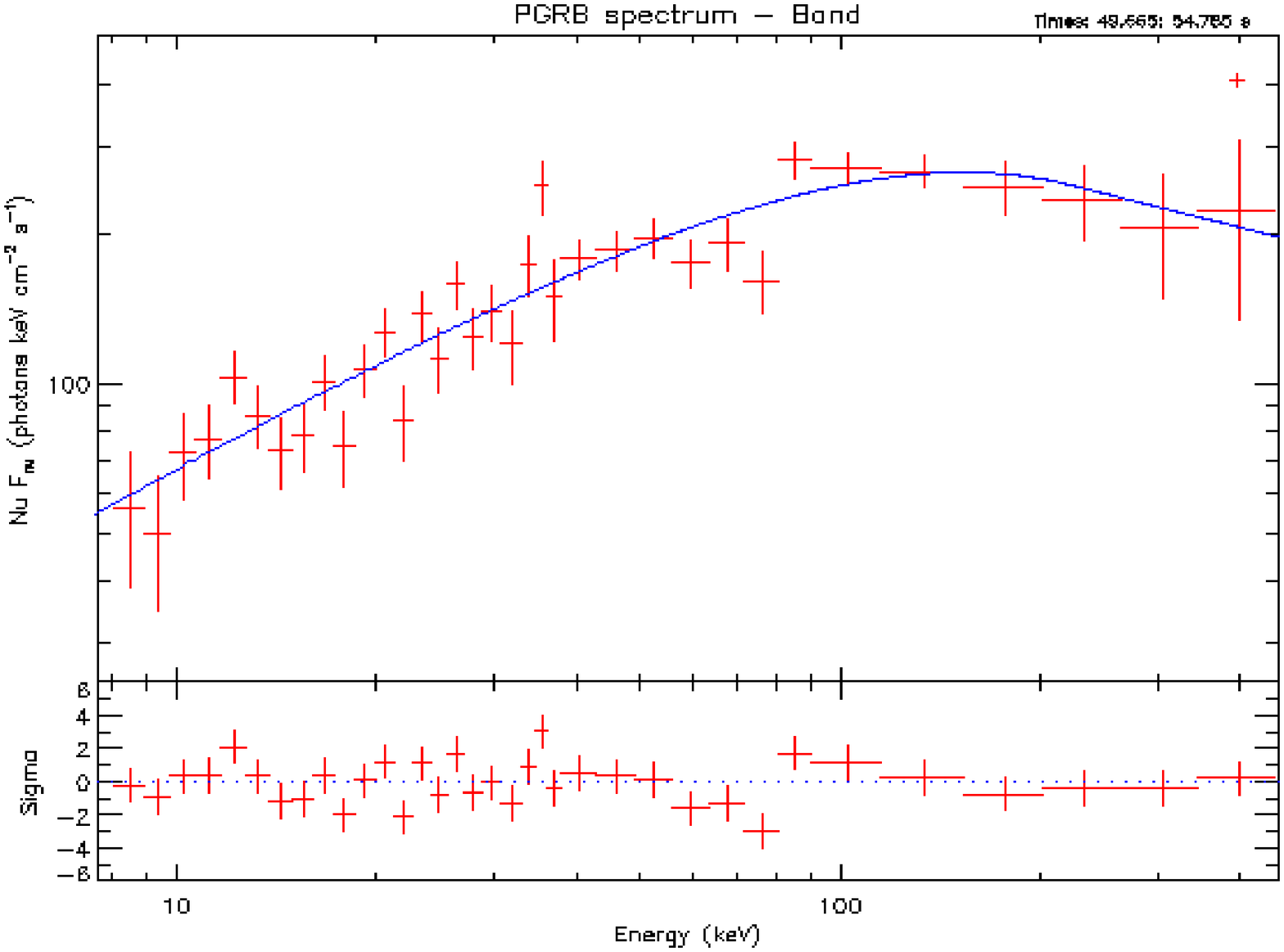}
\caption{Left panel, the time-integrated spectrum (8--440 keV) for the P-GRB emission episode (from 50 to 54 s after the trigger time) of GRB 090618 fit with the blackbody + power-law models, $\tilde{\chi}^2$ = 1.15, while the right panel shows the fit with a band model, $\tilde{\chi}^2$ = 1.25.}
\label{090618_fig:pgrb}
\end{figure}

We now proceed to the analysis of the data between 50 and 150 s after the trigger time as a canonical GRB in the fireshell scenario, namely the second episode \cite{TEXAS}, see Fig.~\ref{090618_fig:cospar}.
We proceed to identify the P-GRB within the emission between 50 and 59 s, since we find a blackbody signature in this early second-episode emission.
Considerations based on the time variability of the thermal component bring us to conclude that the first 4 s of this time interval to due to the P-GRB emission. 
The corresponding spectrum (8--440 keV) is well fit ($\tilde{\chi}^2 = 1.15$) with a blackbody of a temperature $kT = 29.22 \pm 2.21$ keV (norm = 3.51 $\pm$ 0.49), and an extra power-law component with photon index $\gamma$ = 1.85 $\pm$ 0.06, (norm = 46.25 $\pm$ 10.21), see Fig.~\ref{090618_fig:pgrb}. 
The fit with the band model is also acceptable ($\tilde{\chi}^2 = 1.25$), which gives a low-energy power-law index $\alpha=-1.22 \pm 0.08$, a high-energy index $\beta=-2.32 \pm 0.21$ and a break energy $E_0 = 193.2 \pm 50.8$, see Fig.~\ref{090618_fig:pgrb}.
In view of the theoretical understanding of the thermal component in the P-GRB (see Section 3.2), we focus below on the blackbody + power-law spectral model.

The isotropic energy of the second episode is $E_{iso}$ = (2.49 $\pm$ 0.02) $\times$ 10$^{53}$ ergs.
The simulation within the fireshell scenario is made assuming $E_{tot}^{e^+e^-} \equiv E_{iso}$. 
From the observed temperature, we can then derive the corresponding value of the baryon load.
The observed temperature of the blackbody component is $kT = 29.22 \pm 2.21$, so that we can determine a value of the baryon load of $B = 1.98 \pm 0.15 \times$ 10$^{-3}$, and deduce the energy of the P-GRB as a fraction of the total $E_{tot}^{e^+e^-}$. We therefore obtain a value of the P-GRB energy of 4.33$^{+0.25}_{-0.28}$ $\times$ 10$^{51}$ erg.

Now we can derive the radius of the transparency condition, to occur at $r_{tr}$ = 1.46 $\times$ 10$^{14}$ cm.
From the third panel we derive the bulk Lorentz factor of $\Gamma_{th}$ = 495.
We compare this value with the energy measured only in the blackbody component of $E_{BB}$ = 9.24$^{+0.50}_{-0.58}$ $\times$ 10$^{50}$ erg, and with the energy in the blackbody plus the power-law component of $E_{BB+po}$ = 5.43$^{+0.07}_{-0.11}$ $\times$ 10$^{51}$ erg, and verify that the theoretical value is in between these observed energies.
We have found this result to be quite satisfactory: it represents the first attempt to relate the GRB properties to the details of the BH responsible for the overall GRB energetics. The above theoretical estimates were based on a nonrotating BH of 10 M$_{\odot}$, a total energy of $E_{tot}^{e^+e^-}$ = 2.49 $\times$ 10$^{53}$ erg and a mean temperature of the initial $e^+ e^-$ plasma of 2.4 MeV, derived from the expression for the dyadosphere radius, Eq.~\ref{090618_eq:rh}. Any refinement of the direct comparison between theory and observations will have to address a variety of fundamental problems such as 
1) the possible effect of rotation of the BH, leading to a more complex dyadotorus structure, 
2) a more detailed analysis of the transparency condition of the $e^+e^-$ plasma, simply derived from the condition $\tau$ = $\int_R dr (n_{e^{\pm}} + n_{e^-}^b) \sigma_T = 0.67$ \cite{1999A&AS..138..513R}, and 
3) an analysis of the  general relativistic, electrodynamical, strong interaction descriptions of the gravitational core collapse leading to BH formation \cite{2009PhRvD..79l4002C,2003AIPC..668...16R,1999A&AS..138..513R}.

\begin{figure}
\centering
\includegraphics[height=0.5\hsize,clip]{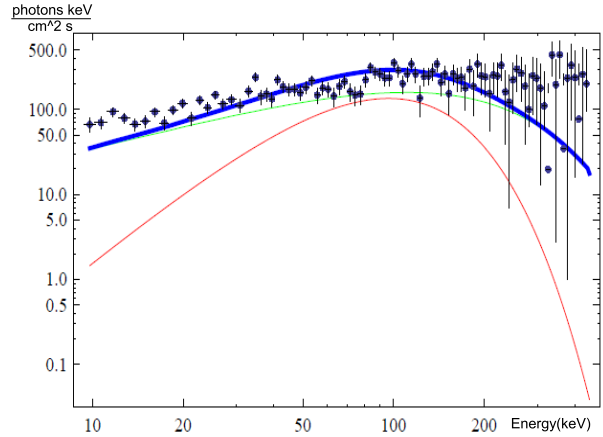}
\caption{Fireshell simulation, green line, and the sole blackbody emission, red line, of the time-integrated (t0+50, t0+54 s) spectrum of the P-GRB emission. The sum of the two components, the blue line, is the total simulated emission in the first 4 s of the second episode.}
\label{090618_fig:pgrb2}
\end{figure}

\paragraph{The analysis of the extended afterglow of the second episode.}

\begin{figure}
\centering
\includegraphics[height=0.5\hsize,clip]{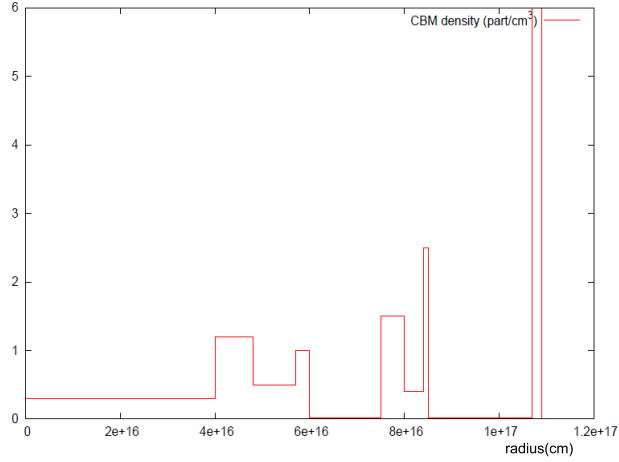}
\caption{Radial CBM density distribution for GRB 090618. The characteristic masses of each cloud are on the order of $\sim$ 10$^{22-24}$ g and 10$^{16}$ cm in radii.}
\label{090618_fig:rad}
\end{figure}

The extended afterglow starts at the above given radius of the transparency, with an initial value of the Lorentz $\Gamma$ factor of $\Gamma_0$ = 495.
To simulate the extended-afterglow emission, we need to determine the radial distribution of the CBM around the burst site, which we assume for simplicity to be spherically symmetric, from which we infer a characteristic size of $\Delta R = 10^{15--16}$ cm.
We already described above how the simulation of the spectra and of the observed multi-band light curves have to be performed together and need to be jointly optimized, leading to the determination of the fundamental parameters characterizing the CBM medium \cite{2007ESASP.622..561R}.
This radial distribution is shown in Fig.~\ref{090618_fig:rad} and is characterized by a mean value of $\langle n\rangle$ = 0.6 part/cm$^3$ and an average density contrast with a $\langle \delta n/n \rangle$ $\approx$ 2, see Fig.~\ref{090618_fig:rad} and Table~\ref{090618_tab:dens}.
The data up to 8.5 $\times$ 10$^{16}$ cm are simulated with a value for the filling factor $\mathcal{R} = 3 \times 10^{-9}$, while the data from this value on with $\mathcal{R} = 9 \times 10^{-9}$.
From the radial distribution of the CBM density, and considering the $1/\Gamma$ effect on the fireshell visible area, we found that the CBM clumps causing the spikes in the extended-afterglow emission have masses on the order of $10^{22--24}$ g.
The value of the $\alpha$ parameter was found to be $-1.8$ along the total duration of the GRB. 

\begin{figure}
\centering
\includegraphics[width=0.34\hsize,angle=270]{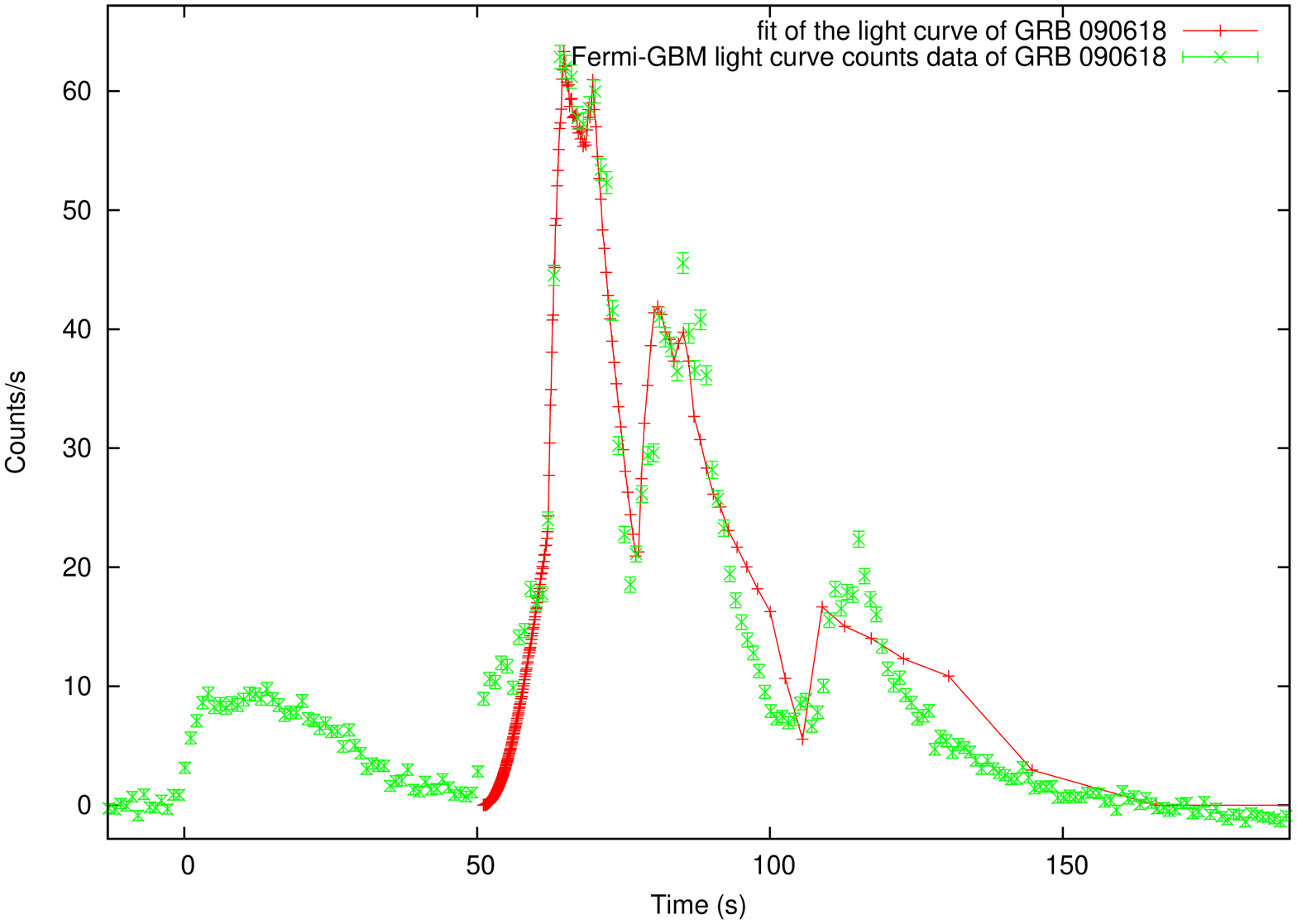}
\includegraphics[width=0.35\hsize,angle=270]{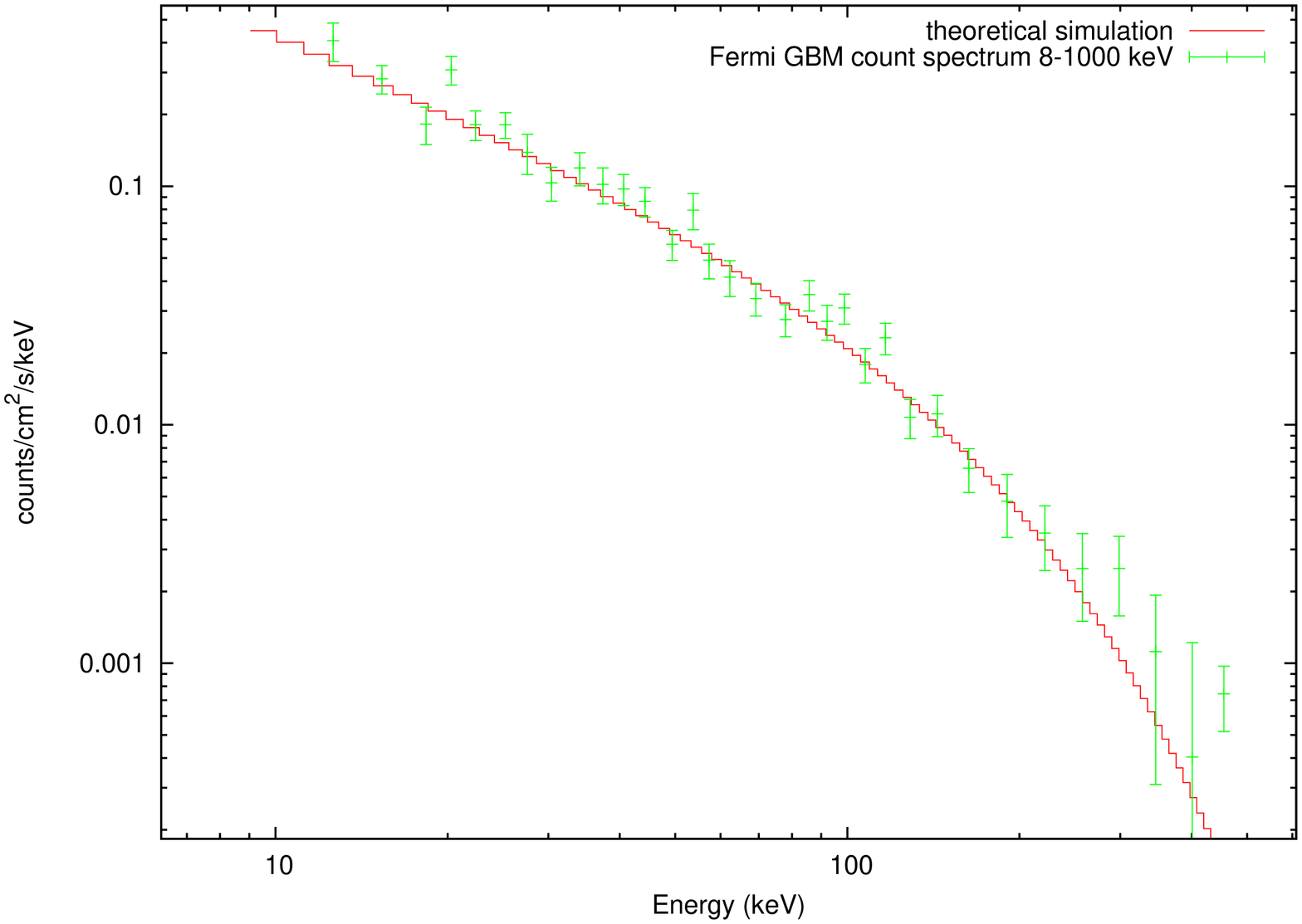}
\caption{Simulated light curve and time integrated (t0+58, t0+150 s) spectrum (8--440 keV) of the extended-afterglow of GRB 090618.}
\label{090618_fig:firesh}
\end{figure}

In Fig.~\ref{090618_fig:firesh} we show the simulated light curve (8--1000 keV) of the GRB and the corresponding spectrum, using the spectral model described in Refs.~\refcite{2004ApJ...605L...1B,2011IJMPD..20.1983P}.
 
\begin{figure}
\centering
\includegraphics[width=0.619\hsize,clip]{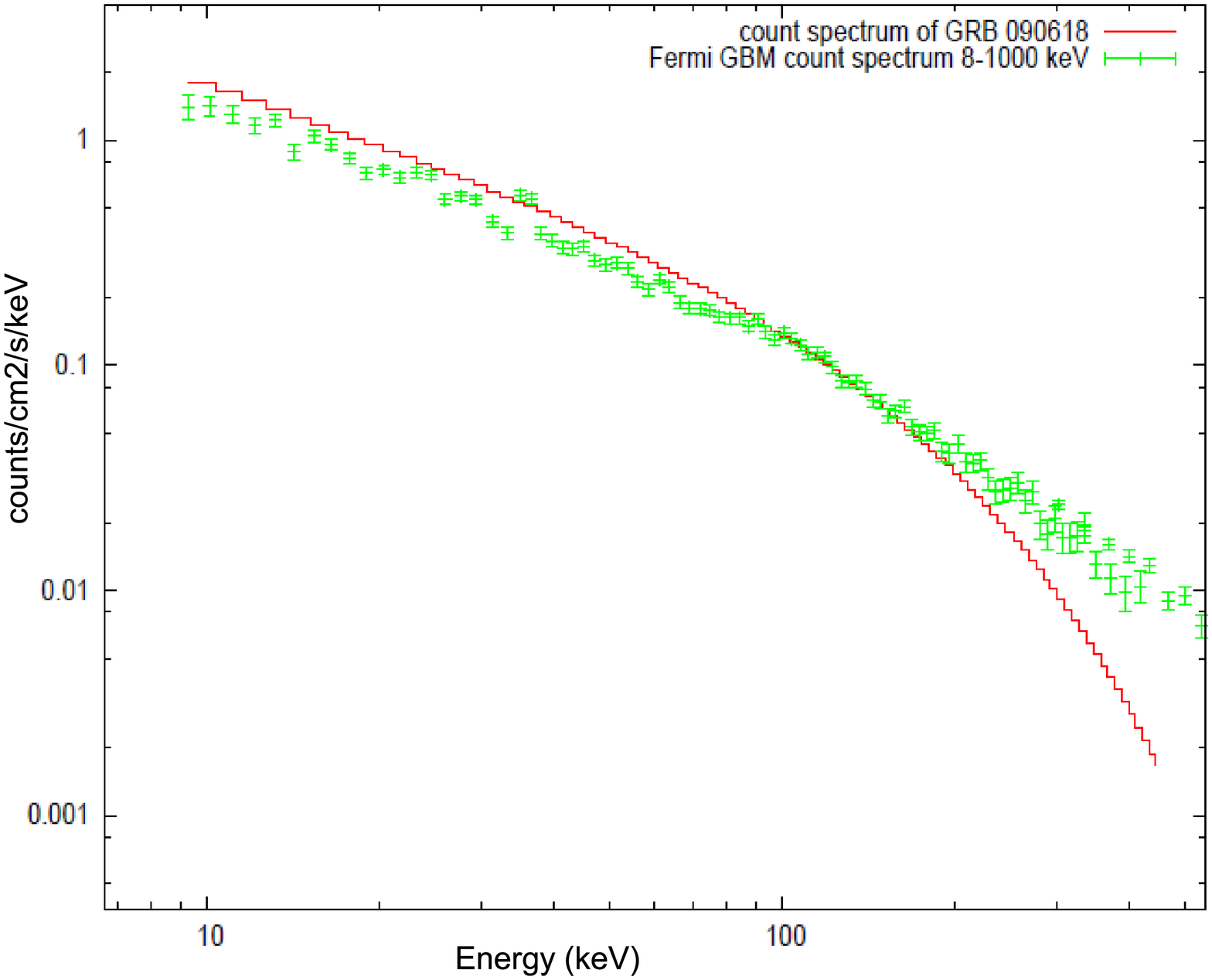}\\
\includegraphics[width=0.619\hsize,clip]{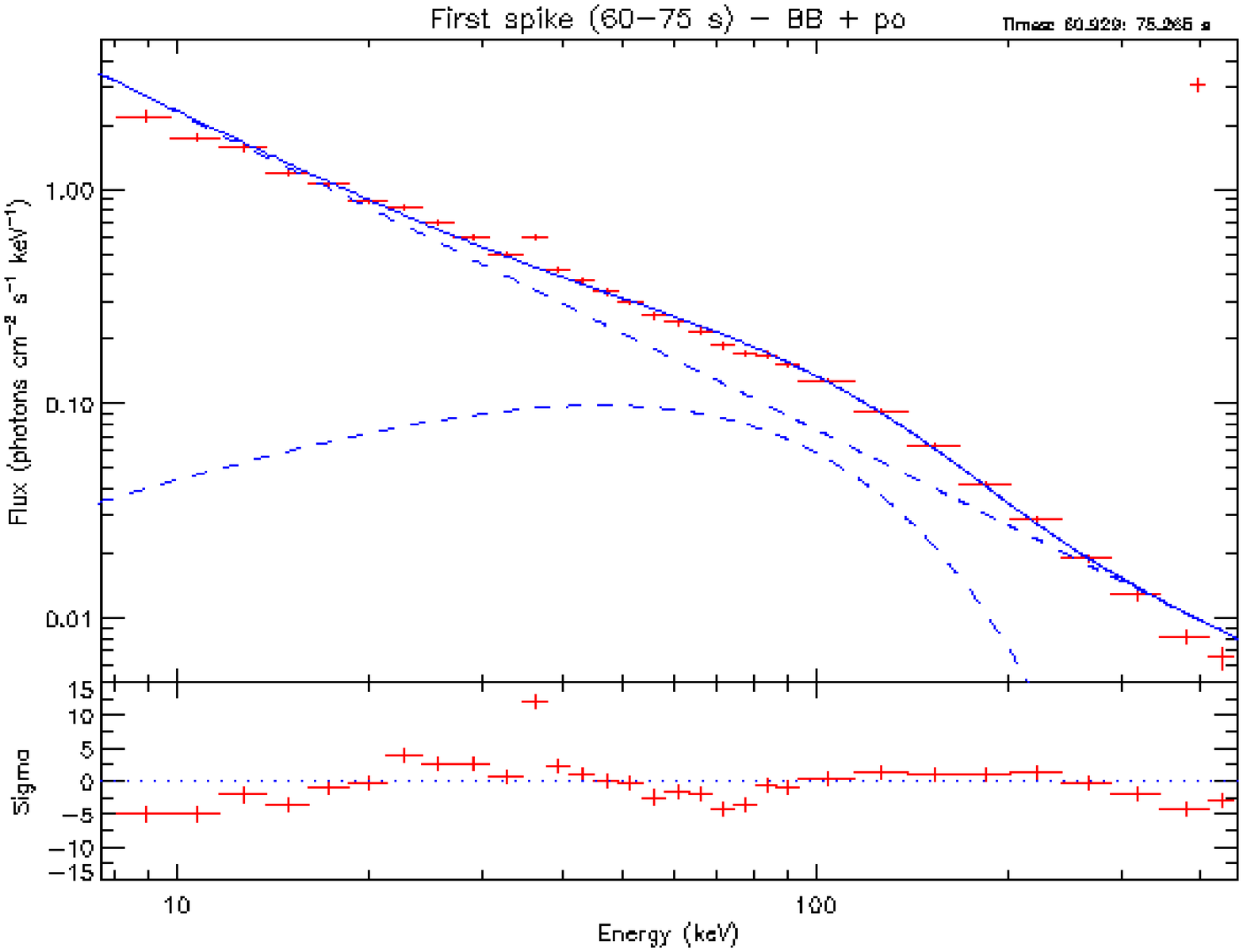}\\
\includegraphics[width=0.619\hsize,clip]{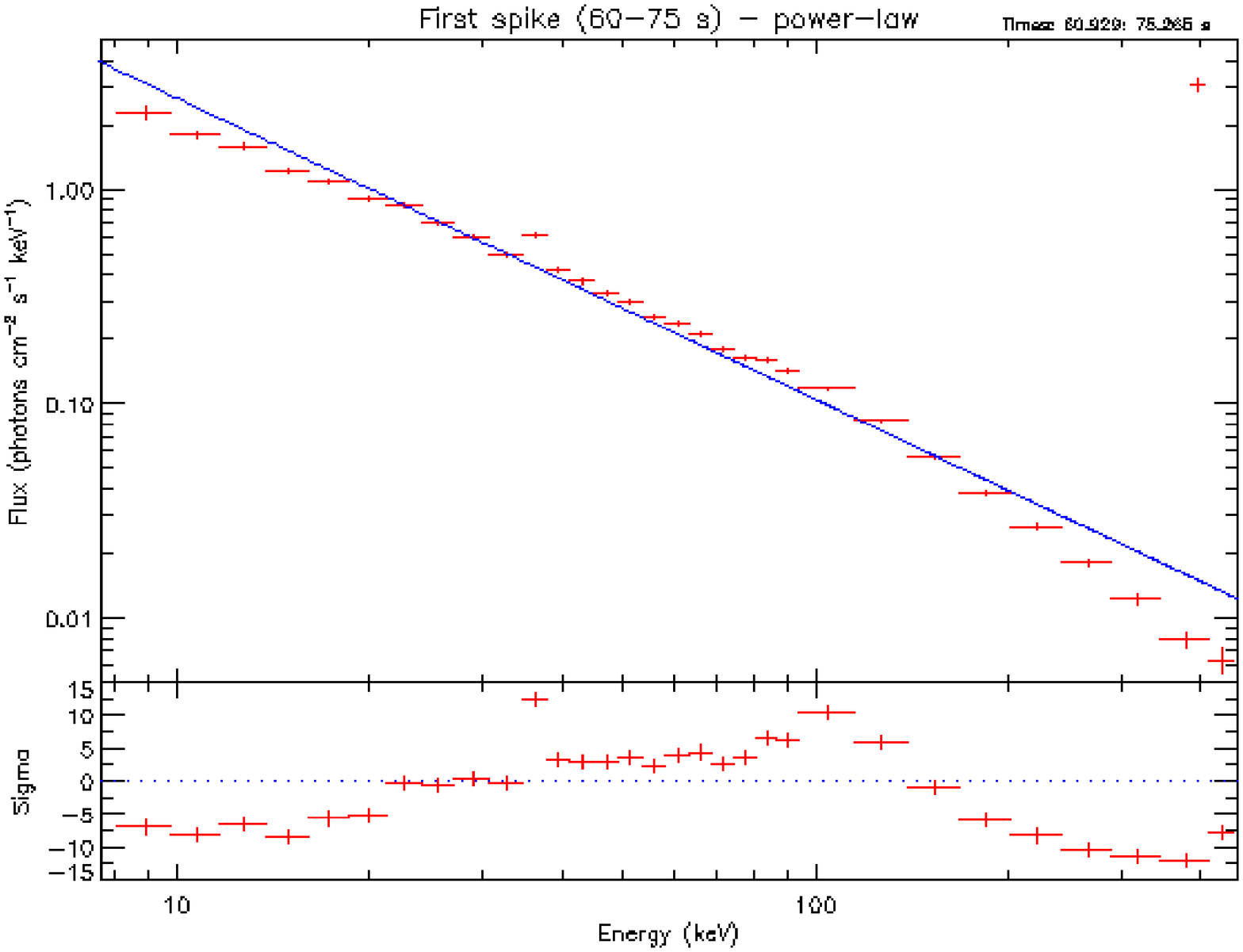}
\caption{Simulated time-integrated (t0+58, t0+66 s) count spectrum (8-440 keV) of the extended afterglow of GRB 090618 (upper panel), count spectrum (8 keV -- 10 MeV) of the main pulse emission (t0+58, t0+66), and best fit with a blackbody + power-law model (middle panel) and a simple power-law model (lower panel).\label{090618_fig:comp}}
\end{figure}

We focus our attention on the structure of the first spikes. 
The comparison between the spectra of the first main spike (t$_0$+59, t$_0$+66 s) of the extended afterglow of GRB 090618 obtained with three different assumptions is shown in Fig.~\ref{090618_fig:comp}: in the upper panel we show the fireshell simulation of the integrated spectrum (t0+59, t0+66 s) of the first main spike, in the middle panel we show the best fit with a blackbody and a power-law component model and in the lower panel the best fit using a simple power-law spectral model.

We can see that the fit with the last two models is not satisfactory: the corresponding $\tilde{\chi}^2$ is 7 for the blackbody + power-law and $\sim$ 15 for the simple power-law.
We cannot give the $\tilde{\chi}^2$ of the fireshell simulation, since it is not represented by an explicit analytic fitting function, but it originates in a sequence of complicated highly nonlinear procedures.
It is clear from a direct scrutiny that it correctly reproduces the low-energy emission, thanks in particular to the role of the $\alpha$ parameter, which was described previously.
At higher energies, the theoretically predicted spectrum is affected by the cut-off induced by the thermal spectrum.
The temporal variability of the first two spikes is well simulated. 

We are not able to accurately reproduce the last spikes of the light curve, since the equations of motion of the accelerated baryons become very complicated after the first interactions of the fireshell with the CBM \cite{2007ESASP.622..561R}.
This happens for various reasons.
First, a possible fragmentation of the fireshell can occur \cite{2007ESASP.622..561R}.
Moreover, at larger distances from the progenitor, the fireshell visible area becomes larger than the transverse dimension of a typical blob of matter, consequently a modification of the code for a three-dimensional description of the interstellar medium will be needed.
This is unlike the early phases in the prompt emission, which is the main topic we address at the moment, where a spherically symmetric approximation applies. 
The fireshell visible area is smaller than the typical size of the CBM clouds in the early phases of the prompt radiation \cite{2010JKPS...57..551I}.

The second episode, lasting from 50 to 151 s, agrees with a canonical GRB in the fireshell scenario.
Particularly relevant is the problematic presented by the P-GRB.
It interfaces with the fundamental physics problems, related to the physics of the gravitational collapse and the BH formation.
There is an interface between reaching transparency of the P-GRB and the early part of the extended afterglow.
This connection has already been introduced in the literature, see e.g. Ref.~\refcite{2012MNRAS.420..468P}.
We studied this interface in the fireshell by analyzing the thermal emission at the transparency with the early interaction of the baryons with the CBM matter, see Fig.~\ref{090618_fig:pgrb2}.

We now aim to reach a better understanding of the meaning of the first episode, between 0 and 50 s of the GRB emission.
To this end we examine the two episodes with respect to 1) the Amati relation, 2) the hardness variation, and 3) the observed time lag.

\begin{table}
\centering
\tbl{Final results of the simulation of GRB 090618 in the fireshell scenario.}
{\begin{tabular}{l c}
\hline\hline
Parameter & Value \\ 
\hline 
$E_{tot}^{e^+e^-}$ &  2.49 $\pm$ 0.02 $\times$ 10$^{53}$ ergs\\
$B$ &  1.98 $\pm$ 0.15 $\times$ 10$^{-3}$\\
$\Gamma_0$ & 495 $\pm$ 40\\
$kT_{th}$ & 29.22 $\pm$ 2.21 keV\\
$E_{P\mbox{-}GRB,th}$ & 4.33 $\pm$ 0.28 $\times$ 10$^{51}$ ergs\\     
$<n>$ & $0.6 \,  part/cm^3$\\
$<\delta n/n>$ & $2 \, part/cm^3$\\
\hline
\end{tabular}
\label{090618_tab:ris}}
\end{table}

\begin{table}
\centering
\tbl{Physical properties of the three clouds surrounding the burst site: the distance from the burst site (column 2), the radius $r$ of the cloud (column 3), the particle density $\rho$ (column 4), and the mass $M$ (the last column).}
{\begin{tabular}{l c c c c}
\hline\hline
Cloud & Distance (cm) & r (cm) & $\rho$ (\#/cm$^3$) & M (g)\\ 
\hline 
First & 4.0 $\times$ 10$^{16}$ & 1 $\times$ 10$^{16}$ & 1 & 2.5 $\times$ 10$^{24}$ \\
Second & 7.4 $\times$ 10$^{16}$ & 5 $\times$ 10$^{15}$ & 1 & 3.1 $\times$ 10$^{23}$ \\
Third & 1.1 $\times$ 10$^{17}$ & 2 $\times$ 10$^{15}$ & 4 &  2.0 $\times$ 10$^{22}$ \\
\hline
\end{tabular}
\label{090618_tab:dens}}
\end{table}

\subsubsection{A different emission process in the first episode}\label{090618_sec:7}

\paragraph{The time-resolved spectra and temperature variation.}

One of the most significant outcomes of the multi-year work of Felix Ryde and his collaborators Ref.~\refcite{2010ApJ...709L.172R} has been the identification and the detailed analysis of the thermal plus power-law features observed in time-limited intervals in selected BATSE GRBs.
Similar features have also been observed in the data acquired by the Fermi satellite \cite{2010ApJ...709L.172R,2011ApJ...727L..33G}.
We propose to divide these observations into two broad families.
The first family presents a thermal plus power-law(s) feature, with a temperature changing in time following a precise power-law behavior.
The second family is also characterized by a thermal plus power-law component, but with the blackbody emission generally varying without a specific power-law behavior and on shorter time scales.
It is our goal to study these features within the fireshell scenario to possibly identify the underlying physical processes.
We have already shown in Sec.~\ref{sec:pgrb} that the emission of the thermal plus power-law component characterizes the P-GRB emission.
We have also emphasized that the P-GRB emission is the most relativistic regime occurring in GRBs, uniquely linked to the process of BH formation, see Sec.~\ref{sec:pgrb}.
This process appears to belong to the second family considered above. 
Our aim here is to see if the first episode of GRB 090618 can lead to the identification of the first family of events: those whose temperature changes with time following a power-law behavior on time scales from 1 to 50 s.
We have already pointed out in the previous section that the hardness-ratio evolution and the long time lag observed for the first episode \cite{2011ApJ...728...42R} points to a distinct origin for the first 50 s of emission, corresponding to the first episode.

We made a detailed time-resolved analysis of the first episode, considering different time bin durations to obtain good statistics in the spectra and to take into account the sub-structures in the light curve.
We then used two different spectral models to fit the observed data, a classical band spectrum \cite{1993ApJ...413..281B}, and a blackbody with a power-law component.

\begin{figure}
\centering
\includegraphics[width=0.7\hsize,clip]{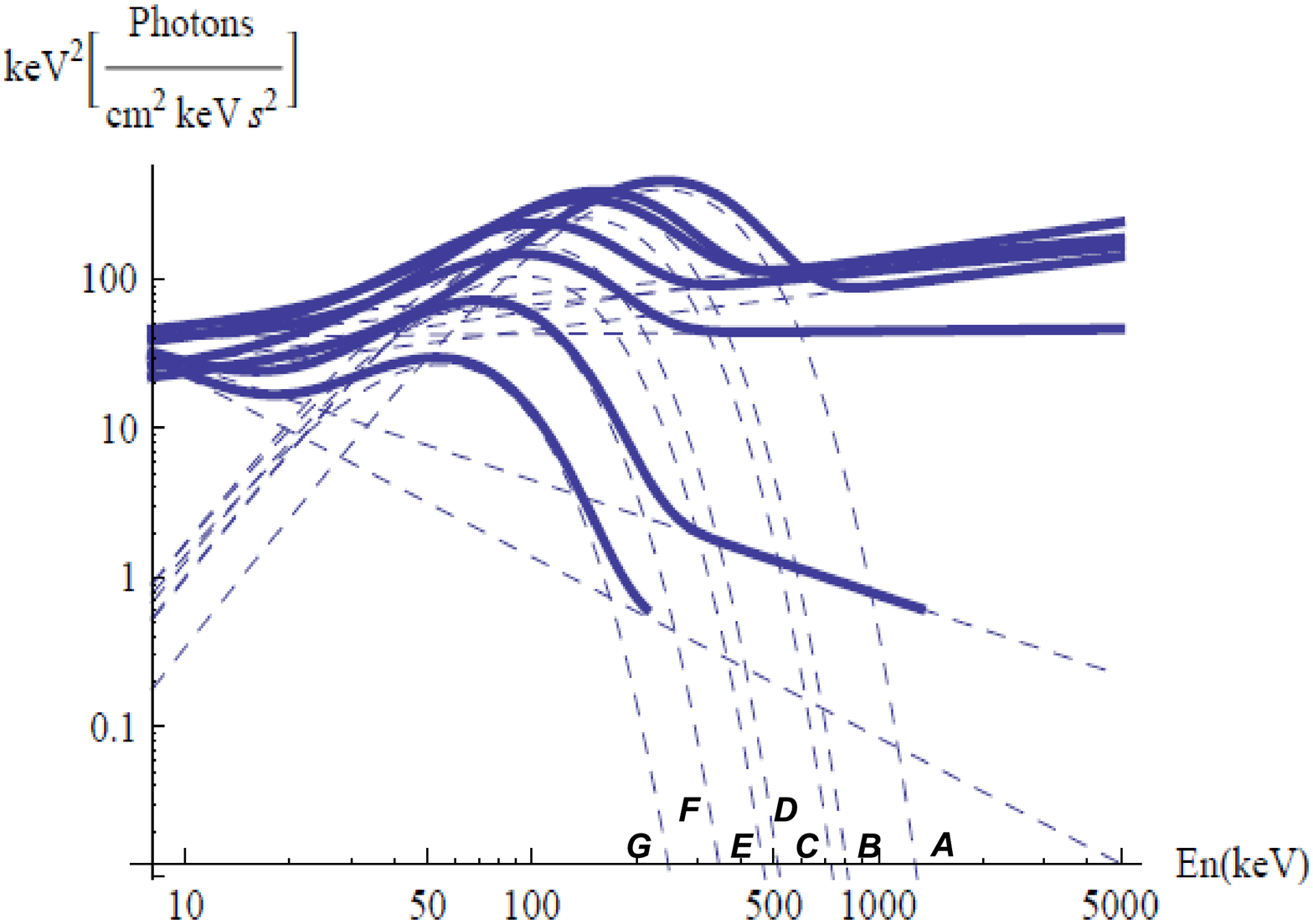}
\caption{Evolution of the BB+powerlaw spectral model in the $\nu\,F(\nu)$ spectrum of the first emission of GRB 090618. It shows the cooling with time of the blackbody and  associated nonthermal components. We only plot the fitting functions for clarity.}
\label{090618_fig:no6}
\end{figure}

\begin{figure}
\centering
\begin{tabular}{|c|}
\hline
\includegraphics[width=0.48\hsize,clip]{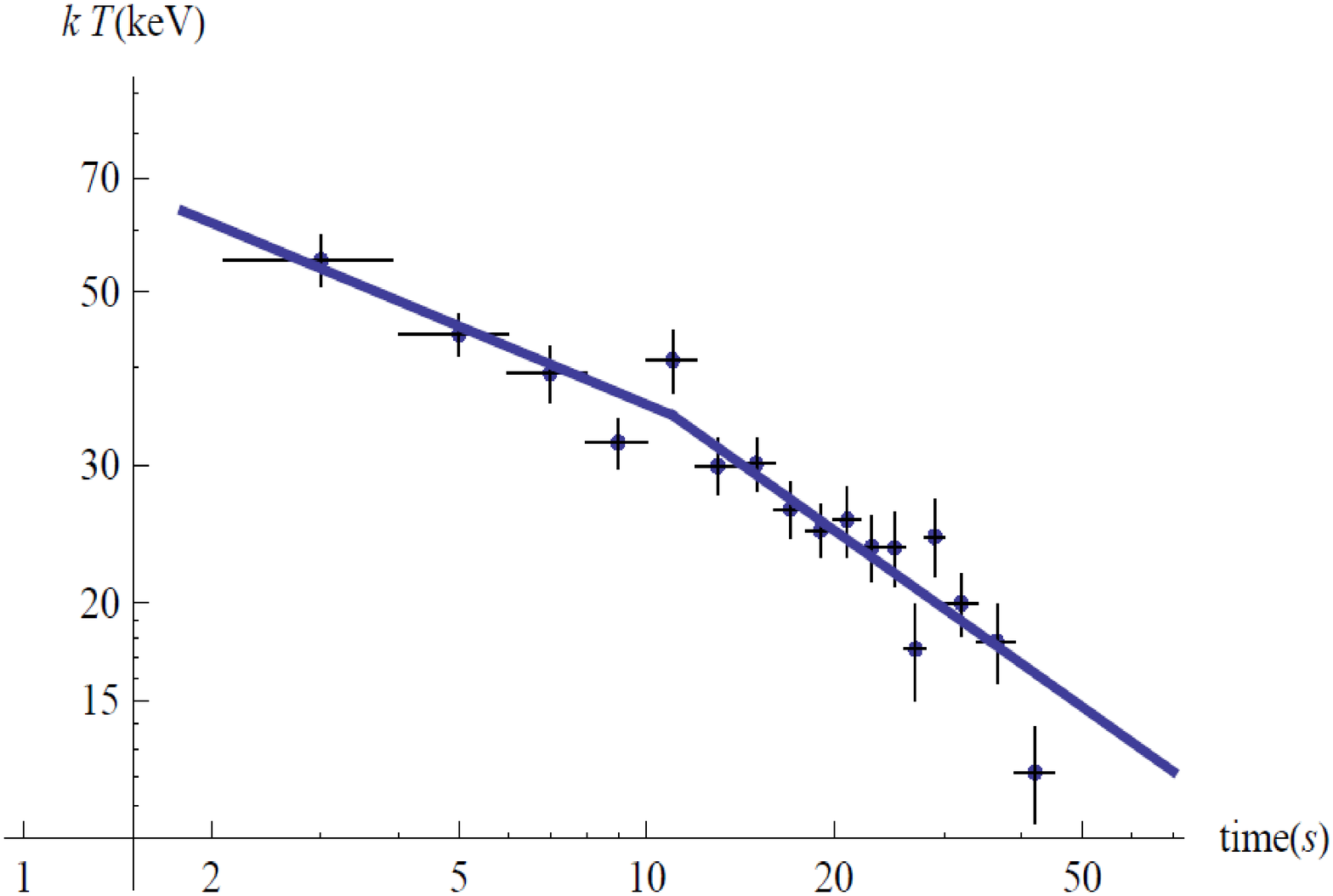}\vline
\includegraphics[width=0.48\hsize,clip]{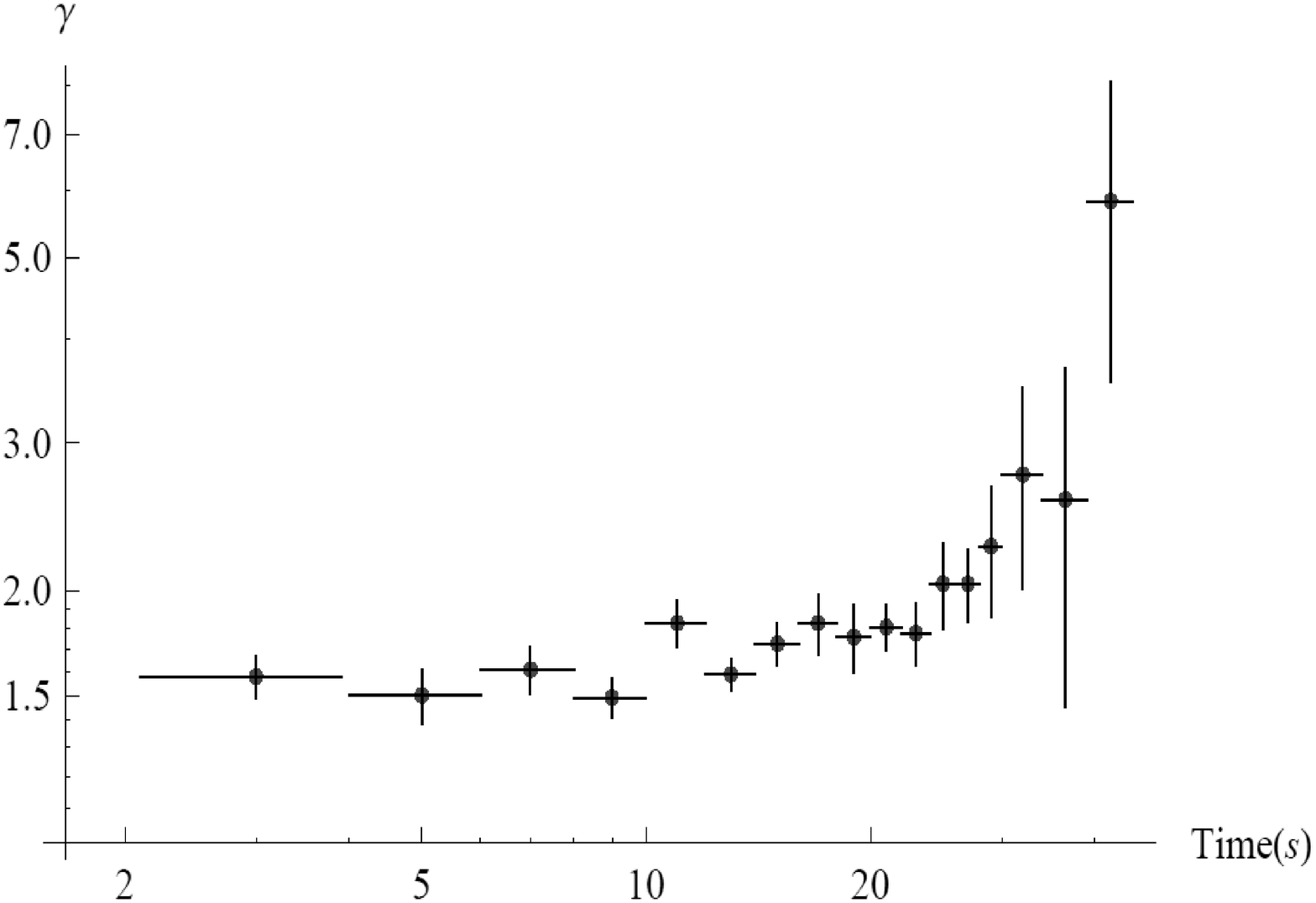}\\
\hline
\end{tabular}
\caption{Evolution of theobserved temperature $kT$  of the blackbody component and the corresponding evolution of the power-law photon index. The blue line in the upper panel corresponds to the fit of the time evolution of the temperature with a broken power-law function. It shows a break time $t_b$ around 11 s after the trigger time, as obtained from the fitting procedure.}
\label{090618_fig:no17}
\end{figure}

To obtain more accurate constraints on the spectral parameters, we made a joint fit considering the observations from both the n4 NaI and the b0 BGO detectors, covering a wider energy range in this way, from 8 keV to 40 MeV.
To avoid some bias from low-photon statistics, we considered an energy upper limit of the value of 10 MeV.
In the last three columns of Table~\ref{090618_tab:no} we report the spectral analysis performed in the energy range of the BATSE LAD instrument ($20$--$1900$ keV), as analyzed in Ref.~\refcite{2009ApJ...702.1211R} as a comparison tool with the results described in that paper.
Our analysis is summarized in Figs.~\ref{090618_fig:no6} and \ref{090618_fig:no17}, and in Table~\ref{090618_tab:no}, where we report the residual ratio diagram and the reduced-$\chi^2$ values for the spectral models. 

\begin{table}
\tbl{Time-resolved spectral analysis of the first episode in GRB 090618. We considered seven time intervals and used two spectral models, whose best-fit parameters are shown here. }
{\tiny
\begin{tabular}{l c c c c c c c c c c}
\hline\hline
Time  & $\alpha$ & $\beta$ & $E_0$ (keV) & $\tilde{\chi}^2_{BAND}$ & $kT$ (keV) & $\gamma$ & $\tilde{\chi}^2_{BB+po}$ \\ 
\hline 
A:0 - 5 & -0.45 $\pm$ 0.11 & -2.89 $\pm$ 0.78 & 208.9 $\pm$ 36.13 & 0.93 & 59.86 $\pm$ 2.72 & 1.62 $\pm$ 0.07 & 1.07 \\
B:5 - 10 & -0.16 $\pm$ 0.17 & -2.34 $\pm$ 0.18 & 89.84 $\pm$ 17.69 & 1.14 & 37.57 $\pm$ 1.76 & 1.56 $\pm$ 0.05 & 1.36 \\
C:10 - 17 & -0.74 $\pm$ 0.08 & -3.36 $\pm$ 1.34 & 149.7 $\pm$ 21.1 & 0.98 & 34.90 $\pm$ 1.63 & 1.72 $\pm$ 0.05 & 1.20 \\
D:17 - 23 & -0.51 $\pm$ 0.17 & -2.56 $\pm$ 0.26 & 75.57 $\pm$ 16.35 & 1.11 & 25.47 $\pm$ 1.38 & 1.75 $\pm$ 0.06 & 1.19 \\
E:23 - 31 & -0.93 $\pm$ 0.13 & unconstr. & 104.7 $\pm$ 21.29 & 1.08 & 23.75 $\pm$ 1.68 & 1.93 $\pm$ 0.10 & 1.13 \\           
F:31 - 39 & -1.27 $\pm$ 0.28 & -3.20 $\pm$ 1.00 & 113.28 $\pm$ 64.7 & 1.17 & 18.44 $\pm$ 1.46 & 2.77 $\pm$ 0.83 & 1.10 \\
G:39 - 49 & -3.62 $\pm$ 1.00 & -2.19 $\pm$ 0.17 & 57.48 $\pm$ 50.0 & 1.15 & 14.03 $\pm$ 2.35 & 3.20 $\pm$ 1.38 & 1.10 \\
\hline
\end{tabular}
\label{090618_tab:no}}
\end{table}

We conclude that both the band and the proposed BB+PL spectral models fit the observed data very well.
Particularly interesting is the clear evolution in the time-resolved spectra, which corresponds to the blackbody and power-law component, see Fig.~\ref{090618_fig:no6}.
 In particular the $kT$ parameter of the blackbody  shows a strong decay, with a temporal behavior well-described by a double broken power-law function, see the upper panel in Fig.~\ref{090618_fig:no17}.
From a fitting procedure we find that the best fit (R$^2$-statistic = 0.992) for the two decay indexes for the temperature variation are $a_{kT}$ = -0.33 $\pm$ 0.07 and $b_{kT}$ = -0.57 $\pm$ 0.11.
In Ref.~\refcite{2009ApJ...702.1211R} an average value for these parameters on a set of 49 GRBs is given: $\left\langle a_{kT} \right\rangle$ = -0.07 $\pm$ 0.19 and $\left\langle b_{kT} \right\rangle$ = -0.68 $\pm$ 0.24.

The results presented in Figs.~\ref{090618_fig:no6} and \ref{090618_fig:no17}, and Table~\ref{090618_tab:no} point to a rapid cooling of the thermal emission with time of the first episode.
The evolution of the corresponding power-law spectral component also appears to be strictly related to the change of the temperature $kT$.
The power-law $\gamma$ index falls, or softens, with temperature, see Fig.~\ref{090618_fig:no6}.
An interesting feature appears to occur at the transition of the two power-laws describing the observed decrease of the temperature. 
The long time lag observed in the first episode has a clear explanation in the power-law behavior of the temperature and corresponding evolution of the photon index $\gamma$ (see Figs.~\ref{090618_fig:no6} and \ref{090618_fig:no17}).

\paragraph{The radius of the emitting region.}

We turn now to estimate an additional crucial parameter for identifying the nature of the blackbody component: the radius of the emitter $r_{em}$.
We proved that the first episode is not an independent GRB and not part of a GRB.
We can therefore provide the estimate of the emitter radius from nonrelativistic considerations, just corrected for the cosmological redshift $z$.
In fact we find that the temperature of the emitter $T_{em} = T_{obs} (1+z)$, and that the luminosity of the emitter, due to the blackbody emission, is
\begin{equation}\label{090618_eq:8.1}
L = 4 \pi r_{em}^2 \sigma T_{em}^4 = 4 \pi r_{em}^2 \sigma T_{obs}^4 (1+z)^4 ,
\end{equation}
where $r_{em}$ is the emitter radius and $\sigma$ is the Stefan-Boltzmann constant. From the luminosity distance definition, we also have that the observed flux $\phi_{obs}$ is given by
\begin{equation}\label{090618_eq:8.2}
\phi_{obs} = \frac{L}{4 \pi D^2} = \frac{r_{em}^2 \sigma T_{obs}^4 (1+z)^4}{D^2}.
\end{equation}
We then obtain
\begin{equation}\label{090618_eq:radius}
r_{em} = \left(\frac{\phi_{obs}}{\sigma T_{ob}^4}\right)^{1/2} \frac{D}{(1+z)^2}.
\end{equation}

The above radius differs from the radius $r_{ph}$ given in Eq.~(1) of Ref.~\refcite{2009ApJ...702.1211R}, which was also clearly obtained by interpreting the early evolution of GRB 970828 as belonging to the photospheric emission of a GRB and assuming a relativistic expansion with a Lorentz gamma factor $\Gamma$
\begin{equation}
r_{ph} = \hat{\mathcal{R}} D \left(\frac{\Gamma}{(1.06) (1+z)^2}\right),
\end{equation}
where $\hat{\mathcal{R}} = \left(\phi_{obs}/(\sigma T_{ob}^4)\right)^{1/2}$ and the prefactor 1.06 arises from the dependence of $r_{ph}$ on the angle of the line of sight \cite{2008ApJ...682..463P}.
Typical values of $r_{ph}$ are at least two orders of magnitude higher than our radius $r_{em}$.

\begin{figure}
\centering
\includegraphics[width=0.7\hsize,clip]{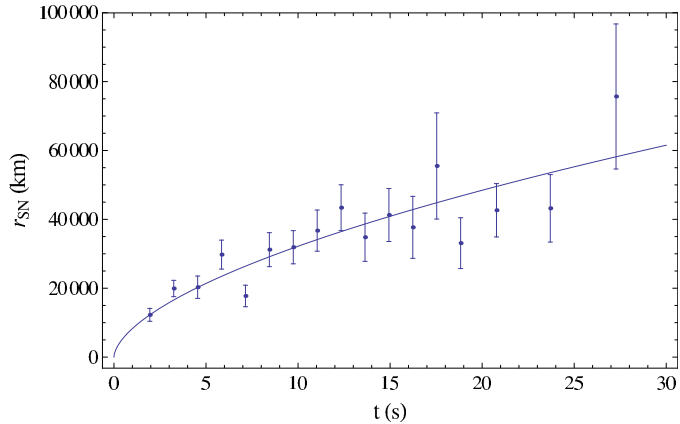}
\caption{Evolution of the first episode emitter radius given by Eq.~(\ref{090618_eq:radius}).}
\label{090618_fig:no18}
\end{figure}

Assuming a standard cosmological model ($H_0 = 70$ km/s/Mpc, $\Omega_m = 0.27$ and $\Omega_{\Lambda} = 0.73$) for estimating the
luminosity distance $D$, and using the values for the observed flux $\phi_{obs}$ and the temperature $kT_{obs}$, we give in Fig.~\ref{090618_fig:no18} the evolution of the surface radius that emits the blackbody $r_{em}$ as a function of time.

Assuming an exponential evolution with time $t^{\delta}$ of the radius in the comoving frame, we obtain the value $\delta = 0.59 \pm 0.11$ from a fitting procedure, which is well compatible with $\delta =0.5$.
We also notice a steeper behavior for the variation of the radius with time corresponding to the first 10 s, which corresponds to the emission before the break of the double power-law behavior of the temperature.
We estimate an average velocity of $\bar{v} = 4067 \pm 918$ km/s, R$^2$ = 0.91 in these first 10 s of emission.
In episode 1 the observations lead to a core of an initial radius of $\sim$ 12000 km expanding in the early phase with a higher initial velocity of $\sim$ 4000 km/s.
The effective Lorentz $\Gamma$ factor is very low, $\Gamma - 1 \sim$ 10$^{-5}$.

I propose to identify this first episode as the early phases of the SN explosion in the IGC scenario which I discuss in the next paragraph.

\section{The GRB-SN in the IGC Scenario}

\subsection{Induced gravitational collapse of a NS to a BH by a type Ib/c SN}

The systematic and spectroscopic analysis of GRB-SN events, following the pioneering discovery of the temporal coincidence of GRB 980425\cite{2000ApJ...536..778P} and SN 1998bw\cite{1998Natur.395..670G}, has revealed evidence for the association of other nearby GRBs with Type Ib/c SNe (see Ref.~\refcite{2011arXiv1104.2274H} for a recent review of all the GRB-SN systems). It has also been clearly understood that SN Ib/c lack Hydrogen (H) and Helium (He) in their spectra, and the most likely explanation is that the SN progenitor star is in a binary system with a compact companion, a neutron star (see e.g. Refs.~\refcite{1988PhR...163...13N,nomoto1994,1994ApJ...437L.115I}, for details). 

In the current literature there has been an attempt to explain both the SN and the GRB as two aspects of the same astrophysical phenomenon. Hence, GRBs have been assumed to originate from a specially violent SN process, a hypernova or a collapsar (see e.g. Ref.~\refcite{2006ARA&A..44..507W} and references therein). Both of these possibilities imply a very dense and strong wind-like CBM structure. Such a dense medium appears to be in contrast with the CBM density found in most GRBs (see e.g. Fig.~10 in Ref.~\refcite{2012A&A...548L...5I}). In fact, the average CBM density, inferred from the analysis of the afterglow, has been shown to be in most of the cases of the order of 1 particle cm$^{-3}$ (see e.g. Ref.~\refcite{2011IJMPD..20.1797R}). The only significant contribution to the baryonic matter component in the GRB process is the one represented by the baryon load \cite{2000A&A...359..855R}. In a GRB, the electron-positron plasma, loaded with a certain amount of baryonic matter, is expected to expand at ultra-relativistic velocities with Lorentz factors $\Gamma\gtrsim 100$\cite{1990ApJ...365L..55S,1993MNRAS.263..861P,1993ApJ...415..181M}. Such an ultra-relativistic expansion can actually occur if the amount of baryonic matter, quantifiable through the baryon load parameter, does not exceed the critical value $B \sim 10^{-2}$ (see Ref.~\refcite{2000A&A...359..855R}, for details). 

In our approach we have  consistently assumed that the GRB has to originate from the gravitational collapse to a BH. The SN follows instead the complicated pattern of the final evolution of a massive star, possibly leading to a NS or to a complete explosion but never to a BH. There is a further general argument in favor of our explanation, namely the extremely different energetics of SNe and GRBs. While the SN energy range is $10^{49}$--$10^{51}$ erg, the GRBs are in a larger and wider range of energies $10^{49}$--$10^{54}$ erg. It is clear that in no way a GRB, being energetically dominant, can originate from the SN. We explain the temporal coincidence of the two phenomena, the SN explosion and the GRB, within the concept of \emph{induced gravitational collapse} \cite{2001ApJ...555L.117R,2008mgm..conf..368R}.

In recent years we have outlined two different possible scenarios for the GRB-SN connection. In the first version \cite{2001ApJ...555L.117R}, we have considered the possibility that GRBs may have caused the trigger of the SN event. For  this scenario to occur, the companion star has to be in a very special phase of its thermonuclear evolution (see Ref.~\refcite{2001ApJ...555L.117R} for details).

\begin{figure}
\centering
\includegraphics[width=0.7\hsize,clip]{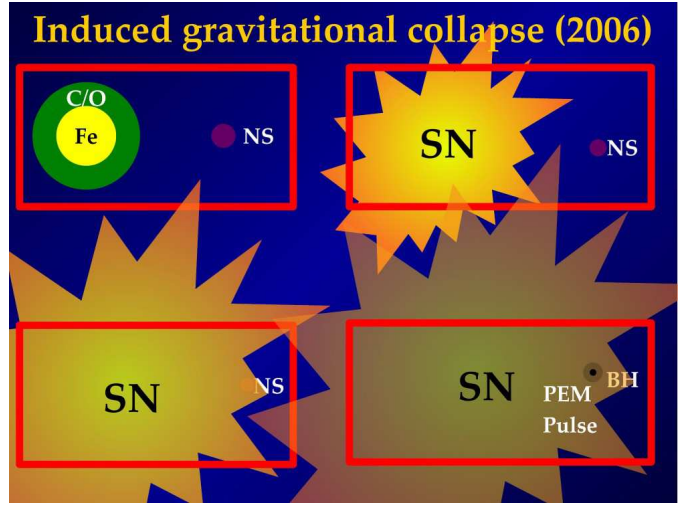}
\caption{Process of gravitational collapse to a BH induced by the type Ib/c SN on a companion NS in a close binary system. Figure reproduced from Ref.~\protect\refcite{2008mgm..conf..368R}.}\label{fig:inducedcollapse}
\end{figure}

\begin{figure}
\centering
\includegraphics[width=0.7\hsize,clip]{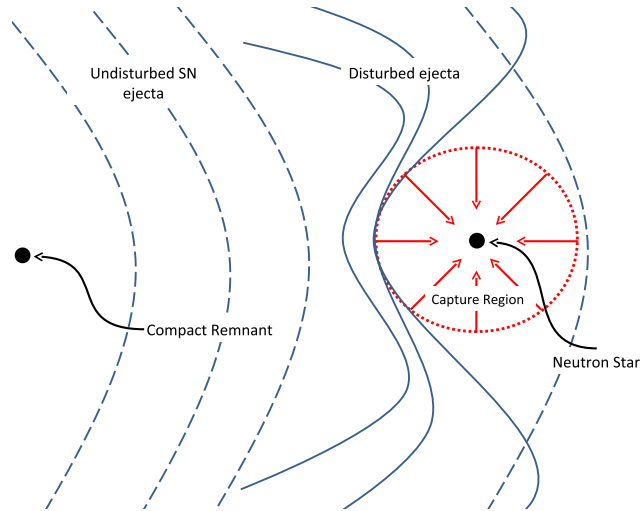}
\caption{Sketch of the binary scenario for GRB 090618: core collapse of an evolved star in close binary with a NS. A rapid accretion rate of the ejected material onto the NS is established reaching in a few seconds the critical mass and undergoes gravitational collapse to a BH, emitting the GRB.}\label{fig:scenario}
\end{figure}

More recently, I have proposed in Ref.~\refcite{2008mgm..conf..368R} a different possibility occurring at the final stages of the evolution of a close binary system: the explosion in such a system of a Ib/c SN leads to an accretion process onto the NS companion. The NS will reach the critical mass value, undergoing gravitational collapse to a BH. The process of gravitational collapse to a BH leads to the emission of the GRB (see Figs.~\ref{fig:inducedcollapse} and \ref{fig:scenario}). Here we evaluate the accretion rate onto the NS and give the explicit expression of the accreted mass as a function of the nature of the components and the binary parameters following Ref.~\refcite{2012ApJ...758L...7R}.

\begin{figure}
\centering
\includegraphics[width=0.7\hsize,clip]{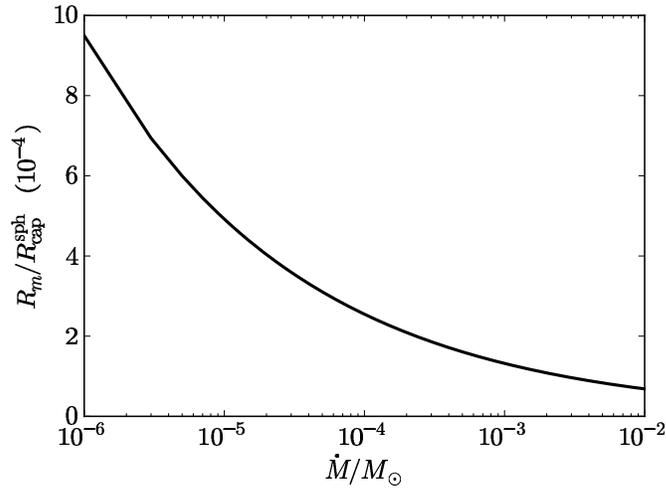}
\caption{Ratio between the magnetospheric radius and the gravitational capture radius of a NS of $B=10^{12}$ Gauss, $M_{\rm NS}=1.4 M_\odot$, $R=10^6$ cm, in the spherically symmetric case. The flow velocity has been assumed to be $v=10^9$ cm s$^{-1}$.}\label{fig:RmRcap}
\end{figure}

We turn now to the details of the accretion process of the SN material onto the NS. In a spherically symmetric accretion process, the magnetospheric radius is \cite{2012MNRAS.420..810T}
\begin{equation}\label{eq:Rm}
R_m = \left( \frac{B^2 R^6}{\dot{M} \sqrt{2 G M_{\rm NS}}}\right)^{2/7}\, ,
\end{equation}
where $B$, $M_{\rm NS}$, $R$ are the NS magnetic field, mass, radius, and $\dot{M}\equiv dM/dt$ is the mass-accretion rate onto the NS. We now estimate the relative importance of the NS magnetic field for the accretion process. At the beginning of a SN explosion, the ejecta moves at high velocities $v\sim 10^9$ cm s$^{-1}$ and the NS will capture matter at a radius approximately given by $R^{\rm sph}_{\rm cap} \sim 2 G M/v^2$. For $R_m << R^{\rm sph}_{\rm cap}$, we can neglect the effects of the magnetic field. It is already clear from Eq.~(\ref{eq:Rm}) that a high accretion rate might reduce the magnetospheric radius drastically. In Fig.~\ref{fig:RmRcap} we plot the ratio between the magnetospheric radius and the gravitational capture radius as a function of the mass accretion rate onto a NS of $B=10^{12}$ Gauss, $M_{\rm NS}=1.4 M_\odot$, $R=10^6$ cm, and for a flow with velocity $v=10^9$ cm s$^{-1}$. It can be seen that for high accretion rates the influence of the magnetosphere will be negligible.

We therefore assume for simplicity hereafter that the NS is nonrotating and neglect the effects of the magnetosphere. The NS captures the material ejected from the core collapse of the companion star in a region delimited by the radius $R_{\rm cap}$ from the NS center
\begin{equation}\label{eq:Rcap}
R_{\rm cap} = \frac{2 G M_{\rm NS}}{v^2_{\rm rel,ej}}\, ,
\end{equation}
where $M_{\rm NS}$ is the initial NS mass and $v_{\rm rel,ej}$ is the velocity of the ejecta relative to the orbital motion of the NS around the supernova progenitor star
\begin{equation}\label{eq:vrel}
v_{\rm rel,ej}=\sqrt{v_{\rm orb}^2+v^2_{\rm ej}}\, ,
\end{equation}
with $v_{\rm ej}$ the ejecta velocity in the frame of the supernova progenitor star with mass $M_{\rm SN-prog}$ and $v_{\rm orb}$ is the orbital velocity of the NS, given by
\begin{equation}
v_{\rm orb}= \sqrt{\frac{G (M_{\rm SN-prog}+M_{\rm NS})}{a}}\, ,
\end{equation}
where $a$ is the binary separation, and thus the orbital period of the binary system is
\begin{equation}\label{eq:period}
P=\sqrt{\frac{4\pi^2 a^3}{G (M_{\rm SN-prog}+M_{\rm NS})}}\, .
\end{equation}

The NS accretes the material that enters into its capture region defined by Eq.~(\ref{eq:Rcap}). The mass-accretion rate is given by \cite{1944MNRAS.104..273B}
\begin{equation}\label{eq:Mdot}
\dot{M}= \xi \pi \rho_{\rm ej} R^2_{\rm cap} v_{ej} = \xi \pi \rho_{ej} \frac{(2 G M_{\rm NS})^2}{(v_{\rm orb}^2+v^2_{ej})^{3/2}}\, , 
\end{equation}
where the parameter $\xi$ is lies in the range $1/2\leq\xi\leq 1$, $\rho_{\rm ej}$ is the density of the accreted material, and in the last equality we have used Eqs.~(\ref{eq:Rcap}) and (\ref{eq:vrel}). The upper value $\xi=1$ corresponds to the Hoyle-Lyttleton accretion rate \cite{1939PCPS...35..405H}. The actual value of $\xi$ depends on the properties of the medium in which the accretion process occurs, e.g. vacuum or wind. The velocity of the SN ejecta $v_{\rm ej}$ will be much larger than the sound speed $c_s$ of the already existing material between the C+O star and the NS due to the prior mass transfer, namely the Mach number of the SN ejecta will certainly satisfy ${\cal M}=v_{\rm ej}/c_s>>$ 1. Thus in practical calculations we can assume the value $\xi=1$ in Eq.~(\ref{eq:Mdot}) and the relative velocity $v_{\rm rel,ej}$ of the SN ejecta with respect to the NS companion is given only by the NS orbital velocity and the ejecta velocity as given by Eq.~(\ref{eq:vrel}). In Fig.~\ref{fig:scenario} we have sketched the accreting process of the supernova ejected material onto the NS. 

The density of the ejected material can be assumed to decrease in time following the simple power-law\cite{1989ApJ...346..847C}
\begin{equation}\label{eq:rhoej}
\rho_{\rm ej}=\frac{3 M_{\rm ej}}{4\pi r^3}=\frac{3 M_{\rm ej}}{4\pi \sigma^3 t^{3 n}}\, ,
\end{equation}
where without loss of generality we have assumed that the radius of the SN ejecta expands as $r_{\rm ej}=\sigma t^n$, with $\sigma$ and $n$ constants. Therefore the velocity of the ejecta obeys $v_{\rm ej}=n r_{\rm ej}/t$.

One can integrate Eq.~(\ref{eq:Mdot}) to obtain the accreted mass in a given time interval
\begin{eqnarray}\label{eq:deltaMarr}
\Delta M (t) &=& \int \dot{M} dt= \pi (2 G M_{\rm NS})^2\frac{3 M_{\rm ej}}{4\pi n^3 \sigma^6} {\cal F} + {\rm constant}\, ,
\end{eqnarray}
where 
\begin{eqnarray}
{\cal F} &=& t^{-3 (n+1)} [-4 n (2 n-1) t^{4 n} \sqrt{k t^{2-2 n}+1} \, _2F_1\left(1/2,1/(n-1);n/(n-1);-k
   t^{2-2 n}\right)\nonumber\\
&-&k^2 \left(n^2-1\right) t^4+2 k (n-1) (2 n-1) t^{2 n+2}+4 n (2 n-1) t^{4 n}]\times \nonumber\\
&& [k^3 (n-1) (n+1) (3 n-1)\sqrt{k+t^{2 n-2}}]^{-1}\, ,
\end{eqnarray}
with $k=v^2_{\rm orb}/(n\,\sigma)^2$ and $_{2}F_{1}(a,b;c;z)$ is the hypergeometric function. The integration constant is computed with the condition $\Delta M (t)=0$ for $t \leq t^{\rm acc}_0$, where $t^{\rm acc}_0$ is the time at which the accretion process starts, namely the time at which the SN ejecta reaches the NS capture region (see Fig.~\ref{fig:scenario}). 

We discuss now the problem of the maximum stable mass of a NS. Nonrotating NS equilibrium configurations have been recently constructed taking into proper account the strong, weak, electromagnetic, and gravitational interactions within general relativity. The equilibrium equations are given by the general relativistic Thomas-Fermi equations coupled with the Einstein-Maxwell equations to form the Einstein-Maxwell-Thomas-Fermi system of equations, which must be solved under the condition of global charge neutrality \cite{2012NuPhA.883....1B}. These equations supersede the traditional Tolman-Oppenheimer-Volkoff ones that impose the condition of local charge neutrality throughout the configuration. The maximum stable mass $M_{\rm crit}=2.67 M_\odot$ of nonrotating NSs has been obtained in Ref.~\refcite{2012NuPhA.883....1B}.

The high and rapid accretion rate of the SN material can lead the NS mass to reach the critical value $M_{\rm crit}=2.67 M_\odot$. This system will undergo gravitational collapse to a BH, producing a GRB. The initial NS mass is likely to be rather high due to the highly nonconservative mass transfer during the previous history of the evolution of the binary system (see e.g.~Refs.~\refcite{1988PhR...163...13N,nomoto1994,1994ApJ...437L.115I}, for details). Thus the NS could reach the critical mass in just a few seconds. Indeed we can see from Eq.~(\ref{eq:Mdot}) that for an ejecta density $10^6$ g cm$^{-3}$ and velocity $10^9$ cm s$^{-1}$, the accretion rate might be as large as $\dot{M} \sim 0.1 M_\odot s^{-1}$.

The occurrence of a GRB-SN event in the scenario depends on some specific conditions satisfied by the binary progenitor system, such as a short binary separation and an orbital period $<1$ h. This is indeed the case with GRB 090618 and 110709B that we have already analyzed within the context of this scenario in Refs.~\refcite{2012A&A...548L...5I,2013A&A...551A.133P}, respectively (see below in the next subsections). In addition to offering an explanation for the GRB-SN temporal coincidence, the considerations presented here lead to an astrophysical implementation of the concept of proto-BH, generically introduced in our previous works on GRBs 090618, 970828, and 101023 (see Refs.~\refcite{2012A&A...548L...5I,2012arXiv1205.6651I,2012A&A...538A..58P}). The proto-BH represents the first stage $20 \lesssim t \lesssim 200$ s of the SN evolution.


It is appropriate now to discuss the possible progenitors of such binary systems. A viable progenitor is represented by X-ray binaries such as Cen X-3 and Her X-1\cite{1972ApJ...172L..79S,1972ApJ...174L..27W,1972ApJ...174L.143T,1973ApJ...180L..15L,1973ApJ...179..585D,1975ASSL...48.....G,2011ApJ...730...25R}. The binary system is expected to follow an evolutionary track\cite{1988PhR...163...13N,nomoto1994,1994ApJ...437L.115I}: the initial binary system is composed of main-sequence stars 1 and 2 with a mass ratio $M_2/M_1 \gtrsim 0.4$. The initial mass of the star 1 is likely $M_1 \gtrsim 11 M_\odot$, leaving a NS through a core-collapse event. The star 2, now with $M_2 \gtrsim 11 M_\odot$ after some almost conservative mass transfer, evolves filling its Roche lobe. It then starts a spiralling in of the NS into the envelope of the star 2. If the binary system does not merge, it will be composed of a helium star and a NS in close orbit. The helium star expands filling its Roche lobe and a nonconservative mass transfer to the NS takes place. This scenario naturally leads to a binary system composed of a C+O star and a massive NS, as the one considered here.

We point out that the systems showing a temporal GRB-SN coincidence  form a special class of GRBs: 

(1) There exist type Ib/c SNe without an associated GRB, see e.g. the observations of the type Ib/c SN 1994I\cite{2002ApJ...573L..27I} and SN 2002ap\cite{2004A&A...413..107S}. Also this class of apparently isolated SNe may be in a binary system with a NS companion at a large binary separation $a$ and long orbital period $P$ (\ref{eq:period}) and therefore the accretion as given by Eqs.~(\ref{eq:Mdot}) and (\ref{eq:deltaMarr}) is not sufficiently high to trigger the gravitational collapse of the NS. 

(2) There are GRBs that do not show the presence of an associated SN. This is certainly the case of GRBs at large cosmological distances $z\gtrsim 0.6$ when the SN is not detectable even by the current high power optical telescopes. This is likely the case of GRB 101023 \cite{2012A&A...538A..58P}. 

(3) There is the most interesting case of GRBs that do not show a SN, although it would be detectable. This is the case of GRB 060614 \cite{2009A&A...498..501C} in which a possible progenitor has been indicated in a binary system formed of a white dwarf and a NS, which clearly departs from the considered binary class. Finally there are systems giving rise to genuinely short GRBs which have been proved to have their progenitors in binary NSs, and clearly do not have an associated SN, e.g.~GRB 090227B\cite{2013ApJ...763..125M,2012arXiv1205.6915R}.

It is clear that after the occurrence of the SN and the GRB emission, the outcome is represented, respectively, by a NS and a BH. A possible strong evidence of the NS formation is represented by the observation of a characteristic late ($t=10^8$--$10^9$ s) X-ray emission (called URCA sources, see Ref.~\refcite{2005tmgm.meet..369R}) that has been interpreted as originating from the young ($t \sim$ 1 minute--$(10$--$100)$ years), hot ($T \sim 10^7$--$10^8$ K) NS, which we have called neo-NS (see Ref.~\refcite{2012A&A...540A..12N}, for details). This has been indeed observed in GRB 090618 \cite{2012A&A...543A..10I} and also in GRB 101023 \cite{2012A&A...538A..58P}. If the NS and the BH are gravitationally bound they give rise to a new kind of binary system, which can lead itself to the merging of the NS and the BH and consequently to a new process of gravitational collapse of the NS into the BH. In this case the system could originate yet another process of GRB emission and possibly a predominant emission in gravitational waves.

\subsection{The application to GRB 090618}

We apply the previous considerations of Ref.~\refcite{2012ApJ...758L...7R} to the specific case of GRB 090618 and its associated SN (see Ref.~\refcite{2012A&A...548L...5I}, for details). We have shown that GRB 090618 \cite{2012A&A...548L...5I} is composed of two sharply different emission episodes. A time-resolved spectral analysis showed that the first episode, which lasts $\sim 32$ s in the rest frame, is characterized by a black-body emission that evolves due to a temperature decreasing  with time (see Fig.~17 in Ref.~\refcite{2012A&A...543A..10I}). Associated to the decreasing black-body temperature, the radius of the emitter has been found to increase with time (see Fig.~18 in Ref.~\refcite{2012A&A...543A..10I}). From the evolution of the radius of the black-body emitter, we find that it expands at nonrelativistic velocities (see Eq.~(\ref{eq:rph}), below). Consequently, the first episode cannot be associated to a GRB. Because it happens prior to the GRB and therefore to the BH formation, this first episode emission has been temporally called a proto-BH, from the ancient Greek $\pi \rho \tilde{\omega} \tau o \varsigma$, meaning before in space and time.

We here identify the proto-BH of the first episode as the first stages of the SN expansion. The black-body-emitting surface in the first episode evolves during the first $\sim 32$ s, as observed in the rest frame, following a power-law behavior
\begin{equation}\label{eq:rph}
r_{\rm SN} = \sigma t^n\, ,\qquad v_{\rm SN} = n \frac{r_{\rm SN}}{t} = n \sigma t^{n-1}\, ,
\end{equation}
where $\sigma=8.048\times 10^8$ cm s$^{-n}$, $n\approx 3/5$ as shown in Fig.~\ref{090618_fig:no18}, and $v_{\rm SN}=dr_{\rm SN}/dt$ is the corresponding early SN velocity of the SN, so $\sim 4\times 10^8$ cm s$^{-1}$ at the beginning of the expansion.

When the mass accreted onto the NS triggers the gravitational collapse of the NS into a BH, the authentic GRB emission is observed in the subsequent episode at $t-t_0 \gtrsim 50$ s (observer frame). The characteristics of GRB 090618 are shown in Table 3 of Ref.~\refcite{2012A&A...543A..10I} and we refer to that reference for more details on the GRB light curve and spectrum simulation. 

We now turn to the details of the accretion process of the SN material onto the NS. The NS of initial mass $M_{NS}$ accretes mass from the SN ejecta at a rate given by \cite{2012ApJ...758L...7R}
\begin{eqnarray}\label{eq:deltaM}
\dot{M}_{acc}(t)= \pi \rho_{ej}(t) \frac{(2 G M_{NS})^2}{v^3_{\rm rel,ej}}\, ,\qquad \rho_{ej}(t)=\frac{3 M_{\rm ej}(t)}{4 \pi r^3_{\rm SN}(t)}\, ,
\end{eqnarray}
where $r^3_{\rm SN}(t)$ given by Eq.~(\ref{eq:rph}) and $M_{\rm ej}(t)=M_{\rm ej,0}-M_{acc}(t)$ is the available mass to be accreted by the NS as a function of time, with $M_{\rm ej,0}$ the mass ejected in the SN. $v_{\rm rel,ej}=\sqrt{v^2_{orb}+v^2_{\rm SN}}$ is the velocity of the ejecta relative to the NS, where $v_{\rm SN}$ is the SN ejecta velocity given by Eq.~(\ref{eq:rph}) and $v_{\rm orb} = \sqrt{G (M_{\rm core}+M_{NS})/a}$ is the orbital velocity of the NS. Here $M_{\rm core}$ is the mass of the SN core progenitor and $a$ the binary separation. Hereafter we assume $a=9\times 10^9$ cm, a value higher than the maximum distance traveled by the SN material during the total time interval of Episode 1, $\Delta t\simeq 32$ s, $\Delta r \sim 7\times 10^9$ cm (see Fig.~\ref{090618_fig:no18}). 

If the accreted mass onto the NS is much smaller than the initial mass of the ejecta, i.e., $M_{acc}/M_{\rm ej,0}<<1$, the total accreted mass can be obtained from the formula  given by Eq.~(8) of Ref.~\refcite{2012ApJ...758L...7R}, which for GRB 090618 leads to
\begin{eqnarray}\label{eq:deltaMapp}
M_{acc}(t) &=& \left.\int_{t^{\rm acc}_0}^t \dot{M}_{acc}(t) dt\approx (2 G M_{\rm NS})^2\frac{15 M_{\rm ej,0} t^{2/5}}{8 n^3 \sigma^6 \sqrt{1+k t^{4/5}}}\right|_{t^{\rm acc}_0}^t\, ,
\end{eqnarray}
where $k=v^2_{orb}/(n\sigma)^2$ and $t^{\rm acc}_0$ is the time at which the accretion process starts, namely the time at which the SN ejecta reaches the NS capture region, $R_{cap}=2 G M_{NS}/v^2_{\rm rel,ej}$, so for $t \leq t^{\rm acc}_0$ we have $M_{acc} (t)=0$. The accretion process leads to the gravitational collapse of the NS onto a BH when it reaches the critical mass value. Here we adopt the critical mass $M_{\rm crit}=2.67 M_\odot$ computed recently in Ref.~\refcite{2012NuPhA.883....1B}. Eq.~(\ref{eq:deltaMapp}) is more accurate for massive NSs since the amount of mass needed to reach the critical mass by accretion is much smaller than $M_{\rm ej,0}$. In general, the total accreted mass must be computed from the numerical integration of Eq.~(\ref{eq:deltaM}), which we present below for GRB 090618.

\begin{figure}
\centering
\includegraphics[width=0.7\hsize,clip]{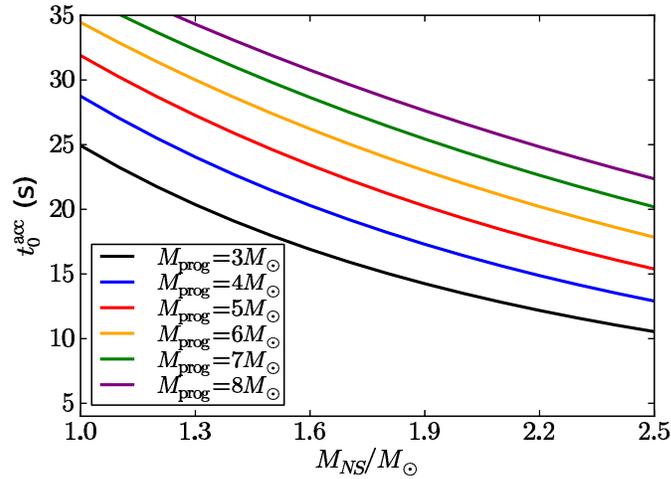}
\caption{Time $t^{\rm acc}_0$ since the SN explosion when the accretion process onto the NS starts as a function of the initial mass of the NS $M_{NS}$ and for selected values of the initial ejected mass $M_{\rm ej,0}$, for GRB 090618.}\label{fig:t0acc}
\end{figure}

The occurrence of a GRB-SN event in the accretion induced collapse scenario is subject to some specific conditions of the binary progenitor system such as a short binary separation and orbital period. The orbital period in the present case is
\begin{equation}\label{eq:period2}
P=\sqrt{\frac{4\pi^2 a^3}{G (M_{\rm core}+M_{NS})}}=9.1 \left(\frac{M_{\rm core}+M_{NS}}{M_\odot}\right)^{-1/2}\,\,{\rm min}\, .
\end{equation}

\begin{figure}
\centering
\includegraphics[width=0.7\hsize,clip]{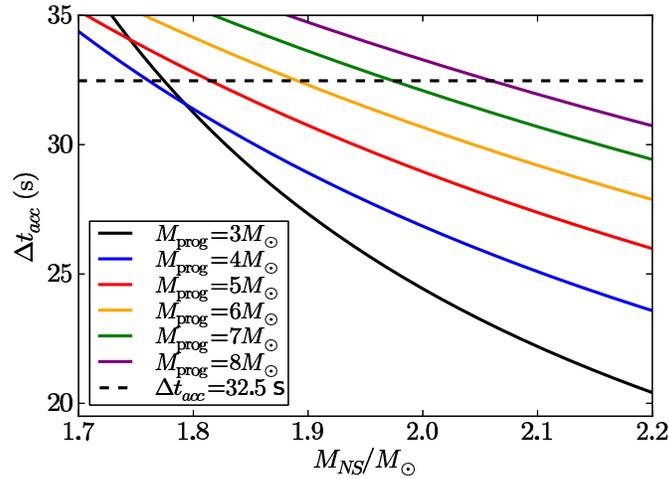}
\caption{Time interval $\Delta t_{acc}$ of the accretion process onto the NS as a function of initial NS mass $M_{NS}$ for selected values of the SN core progenitor mass $M_{\rm core}$. The horizontal dashed line is the duration $\Delta t=32.5$ s of the first episode of GRB 090618, which constrains the duration of the time needed by the NS to reach the critical mass. The crossing points between the dashed horizontal line and the solid curves give the NSs with $M_{NS}$ that reach the critical mass in the time $\Delta t$.}\label{fig:deltaT}
\end{figure}

We denote by $\Delta t_{\rm acc}$ the total time interval since the beginning of the SN ejecta expansion all the way up to the instant where the NS reaches the critical mass. In Fig.~\ref{fig:deltaT} we plot $\Delta t_{\rm acc}$ as a function of the initial NS mass and for different masses of the SN core progenitor mass. The mass of the SN ejecta is assumed to be $M_{\rm ej,0}=M_{\rm core}-M_{\rm rem}$, where $M_{\rm rem}$ is the mass of the central compact remnant (NS) left by the SN explosion. Here we assumed $M_{\rm core}=(3$--$8) M_\odot$ at the epoch of the SN explosion, and $M_{\rm rem}=1.3 M_\odot$, following some of the type Ic SN progenitors studied in Refs.~\refcite{1988PhR...163...13N,nomoto1994,1994ApJ...437L.115I}.

We can see from Fig.~\ref{fig:deltaT} that, for GRB 090618, the mass of the NS companion that collapses onto a BH should be in the range $1.8\lesssim M_{NS}/M_\odot \lesssim 2.1$ corresponding to the SN Ic progenitors $3\leq M_{\rm core}/M_\odot \leq 8$. The massive NS companion of the evolved star is in line with the binary scenario proposed in Ref.~\refcite{2008mgm..conf..368R}. These results also agree with the well-understood Ib/c nature of the SN associated with GRBs. The most likely explanation for SN Ib/c, which lack H and He in their spectra, is that the SN progenitor star is in a binary system with an NS; see also Refs.~\refcite{1988PhR...163...13N,nomoto1994,1994ApJ...437L.115I} and also \refcite{2007ARep...51..291T,2012ApJ...752L...2C}.


It is also interesting to compare the results on the IGC of an NS to a BH by a type Ib/c SN \cite{2012ApJ...758L...7R} with the results of Chevalier \cite{1989ApJ...346..847C} on the accretion of a supernova material by the central NS generated by the supernova. A total accreted mass of up to $0.1 M_\odot$ in a time of a few hours was obtained there for a normal type II SN. Thus a similar amount of mass can be accreted in the two cases, but in the latter the accretion occurs over a longer time. To reach a high accretion rate of the inner SN material onto the central NS, a mechanism is needed that helps to increase the density of the NS surrounding layers, which is decreasing due to the expansion after being unbound by the SN explosion. Ref.~\refcite{1989ApJ...346..847C} analyzed the possibility of having a reverse shock wave as this mechanism while it moves back through the SN core. The reverse shock is formed in the interaction of the mantle gas with the low-density envelope. The time scale of the accretion process is thus determined by the time it takes the reverse shock to reach the vicinity of the central newly born NS, which is a few hours in the case of SN II progenitors. 
However, the existence of a low-density outer envelope, e.g. H and He outer layers, is essential for the strength of the reverse shock. Fall-back accretion onto the central NS is expected to be relevant only in SN II but not in SN Ic like those associated to GRBs, where H and He are absent. 
 
The argument presented in \refcite{2012ApJ...758L...7R} naturally explains the sequence of events: SN explosion -- IGC-BH formation -- GRB emission.
Correspondingly, the accretion of the material ejected by the SN into the nearby NS of the IGC model presented here occurs almost instantaneously. Indeed for the SN expansion parameters obtained from the observations of episode 1 in GRB 090618 (see Eq.~(\ref{eq:rph}), the accretion of the SN material onto the nearby NS occurs in a few seconds (see Figs.~\ref{fig:t0acc} and \ref{fig:deltaT}). The binary parameters are such that the ejecta density does not decrease too much (from $10^6$ to $\sim 10^4$ g cm$^{-3}$) before reaching the capture region of the NS, leading to a high accretion rate. 
As pointed out in Ref.~\refcite{1989ApJ...346..847C}, radiative diffusion will lower the accretion rate up to the Eddington limit (and then to even lower rates) when the trapping radius of the radiation in the flow $r_{tr}=\kappa \dot{M}_{acc}/(4\pi c)$ \cite{1989ApJ...346..847C}, where $\kappa$ is the opacity, is equal to the Bondi radius $r_{B}=G M_{NS}/v^2_{\rm rel,ej}$, the gravitational capture radius. The radius $r_{tr}$ is located where the outward diffusion luminosity is equal to the inward convective luminosity. It can be checked that for the parameters of our system given by Eqs.~(\ref{eq:rph})--(\ref{eq:deltaMapp}), the equality $r_{tr}=r_B$ occurs in a characteristic time $\sim 200$ days, where we used $\kappa = 0.2$ cm$^2$ g$^{-1}$. Thus, this regime is not reached in the present case since the NS is brought to its critical mass just in a few seconds. In the case analyzed by Ref.~\refcite{1989ApJ...346..847C}, it happens in a time $\sim 8$ days.

In conclusion, the IGC binary scenario applied here to the specific case of GRB 090618 naturally leads to understanding the energetics and the temporal coincidence of SN and GRBs, as well as their astrophysical scenario and their origins.
It also provides new predictions of the final outcome, originating from a binary system composed of an evolved core and an NS. It is clear, however, that these GRBs and their associated SNe form a special class of long GRBs and of SNe Ib/c. There are in fact SNe Ib/c that are not associated to a GRB, e.g. SN 1994I\cite{2002ApJ...573L..27I} and SN 2002ap\cite{2004A&A...413..107S}. Their observations refer to late phases of the SN evolution typically $\sim 15$--$20$ days after the original collapse process. The existing descriptions of these late phases after 15--20 days from the original explosion make use of a Sedov-type behavior $r\propto t^{2/5}$, see Refs.~\refcite{sedov46,1959sdmm.book.....S}. In the present case of the IGC we present here for the first time, the first $\sim 30$ s of the very early evolution of an SN Ib/c associated to a GRB (see Eq.~(\ref{eq:rph}). The energetic of this SN Ib/c, as shown from episode 1, appears to be much higher than the ones of the usual SNe Ib/c not associated to GRBs, $E_{iso,Epi1} \propto 10^{52}$ erg\cite{2012A&A...543A..10I}. The reason for this marked difference is certainly due to the accretion process during an SN explosion into the companion NS and consequent gravitational collapse of the NS onto a BH. The description of this challenging process, although clear from a general energetic point of view, has still to be explored in detail theoretically and certainly does not show any relation to the Sedov-type solution.

\section{On a Possible Distance Indicator from GRB-SN-IGC}

It is appropriate to remember an important selection effect  occurring in the study of the IGC scenario. Only for systems with cosmological redshift $z \lesssim 1$ does the current optical instrumentation allow the observation of the related SN Ib/c. A particularly challenging analysis is that of the system GRB 101023 \cite{2012A&A...538A..58P} in which the SN is not detectable but the IGC nature of the source is clearly recognized by the two different episodes in the GRB sources and the spectral features of the first episode. Following the case of GRB 101023, we have found and analyzed the X-ray emission of a sample of 8 GRBs having $E_{iso} \geq 10^{52}$ erg and satisfying at least one of the following three requirements:
 \begin{itemize}
 \item the detection of a SN after about 10 days in the rest frame from the GRB trigger,
 \item the presence of a double emission episode in the prompt emission: episode 1, with a decaying thermal feature, and episode 2, a canonical GRB, as in GRB 090618 \cite{2012A&A...548L...5I} and GRB 101023 \cite{2012A&A...538A..58P}, and
\item the presence of a shallow phase followed by a final steeper decay, namely episode 3.
 \end{itemize} 

\begin{table}
\tbl{The GRB sample considered in this work. The redshifts of GRB 101023 and GRB 110709B, which are marked by an asterisk, were deduced theoretically by using the method outlined in  Ref.~\protect\refcite{2012A&A...538A..58P}
and the corresponding isotropic energy computed by assuming these redshifts.}
{\begin{tabular}{l c c}
\hline\hline
GRB  & $z$ & $E_{iso} (erg)$ \\
\hline 
GRB 060729 & $0.54$ & $1.6 \times 10^{52}$ \\
GRB 061007 & $1.261$ & $1.2 \times 10^{54}$ \\
GRB 080319B & $0.937$ & $1.4 \times 10^{54}$ \\
GRB 090618 & $0.54$ & $2.7 \times 10^{53}$ \\
GRB 091127 & $0.49$ & $1.4 \times 10^{52}$ \\
GRB 111228 & $0.713$ & $2.3 \times 10^{52}$ \\
\hline
GRB 101023 & $0.9^*$ & $1.3 \times 10^{53}$ \\
GRB 110709B & $0.75^*$ & $2.72 \times 10^{53}$ \\
\hline\\
\end{tabular}\label{table1}}
\end{table}

\begin{figure}
\centering
\includegraphics[width=0.7\hsize,clip]{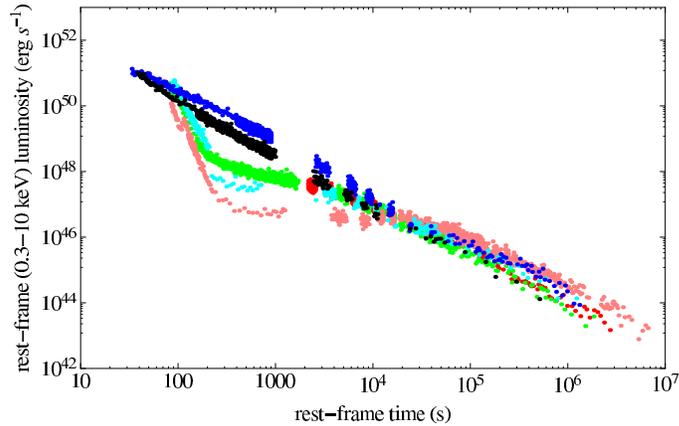}
\caption{The X-ray luminosity light curves of the six GRBs with measured redshift in the $0.3\,$--$\,10$ keV rest frame energy range: in pink GRB 060729, $z=0.54$; in black GRB 061007, $z=1.261$; in blue GRB 080319B, $z=0.937$; in green GRB 090618, $z=0.54$, in red GRB 091127, $z=0.49$, in cyan GRB 111228, $z=0.713$.}\label{fig:sample}
\end{figure}

The characteristics of the 8 GRBs are the following:

\textit{GRB 060729}. In this source a SN bump was observed in the optical GRB afterglow \cite{Cano2011}. It is at the same redshift $z=0.54$ of GRB 090618 and shows a small precursor plus a main event in the prompt light curve and a peculiar prolonged duration for the X-ray afterglow \cite{Grupe2007b}. The isotropic energy emitted in this burst is $E_{iso} = 1.6 \times 10^{52}$ erg.

\textit{GRB 061007}. This GRB has no associated SN  but is characterized by the presence of an almost long precursor where a clear evolving thermal emission was reported \cite{Larsson2011}. With an energetic of $E_{iso} = 1.2 \times 10^{54}$ erg at $z=1.261$, it is the farthest GRB in our sample. The large distance directly implies difficulties in the detection of a SN from this GRB.

\textit{GRB 080319B}. A debatable SN was reported also for GRB 080319B, well known as the naked-eye GRB, whose prompt emission shows also a possible double emission episode \cite{Kann2008}. Its measured redshift is $z=0.937$. This is one of the most energetic GRBs with $E_{iso} = 1.4 \times 10^{54}$ and its X-ray light curve is well described by a simple decaying power-law.

\textit{GRB 090618}. This GRB is the prototype of the IGC GRB-SN subclass. Its prompt emission shows a clear episode 1 plus episode 2 structure in light curve and spectrum. The measured redshift is $z=0.54$ and the isotropic energy emitted by the burst is $E_{iso} = 2.7 \times 10^{53}$ erg. There is a clear identification in the afterglow light curve of GRB 090618 of a late $\sim10$ day optical bump associated to the SN emission \cite{Cano2011}. The characteristic parameters of this GRB, including baryon load ($B=1.98 \times 10^{-3}$), the Lorentz gamma factor at trasparency ($\Gamma_{tr}=495$) and the nature of the CBM ($\langle n_{CBM} \rangle=0.6 \, part/cm^3$) have been estimated \cite{2012A&A...548L...5I}.

\textit{GRB 091127}. GRB 091127 is associated with SN 2009nz at a distance of $z=0.49$ \cite{Cobb2010}. The isotropic energy emitted in this burst is $E_{iso} = 1.4 \times 10^{52}$ erg \cite{Stamatikos2009}.

\textit{GRB 111228}. A SN feature is reported in the literature also for GRB 111228 \cite{DAvanzo2012}, which shows a multiply peaked prompt light curve in the Fermi-GBM data. The measured redshift of this GRB is $z=0.713$, its isotropic energy is $E_{iso} = 2.3 \times 10^{52}$ erg and a dedicated analysis of this GRB will be presented elsewhere. The detection of a SN in GRB 111228 is debatable, since the eventual optical bump has the same flux than the host galaxy of the source, but SN features were observed in the differential photometry between the last epochs of observations, where a transient component was detected unrelated to the afterglow and consequently associated to the SN. 

\textit{GRB 101023}. This GRB shows clear episode 1 plus episode 2 emission in the prompt light curve and spectrum, but there is no detection of a SN and no measured redshift because of the lack of optical observations at late times. We have estimated the redshift of this source as $z=0.9$ in analogy with the late X-ray afterglow decay observed in the 6 GRBs with a measured redshift. This leads to the estimation of an isotropic energy of $E_{iso} = 1.3 \times 10^{53}$ erg, a baryon load of $B=3.8 \times 10^{-3}$, a Lorentz gamma factor at transparency of $\Gamma_{tr}=260$, and an average density for the CBM of ($\langle n_{CBM} \rangle \approx 16 \, part/cm^3$ \cite{2012A&A...538A..58P}.

\textit{GRB 110709B}. Like GRB 101023, this GRB shows a clear episode 1 plus episode 2 emission in the prompt light curve and spectrum, but there is no detection of a SN. This can be explained by the fact that it is a dark GRB, so its emission is strongly influenced by absorption. Particularly interesting is the detection of a clear radio emission from GRB 110709B  \cite{Zauderer2012}. There is no measure for the redshift but, as for the case of GRB 101023, we have estimated it as $z=0.75$ in analogy with the late X-ray afterglow decay observed in the 6 GRBs with measured redshifts. This leads to the estimation of an isotropic energy of $E_{iso} = 2.43 \times 10^{52}$ erg, a baryon load of $B=5.7 \times 10^{-3}$, a Lorentz gamma factor at transparency of $\Gamma_{tr}=174$ and an average density of the CBM of $\langle n_{CBM} \rangle \approx 76 part/cm^3$ \cite{2013A&A...551A.133P}.

We have focused our attention on the analysis of all the available XRT data of these sources \cite{2013A&A...552L...5P}. Characteristically, XRT follow-up starts only about 100 seconds after the BAT trigger (typical repointing time of Swift after the BAT trigger). Since the behavior was similar in all the sources, we have performed an analysis to compare the XRT luminosity light curve $L_{rf}$ for the six GRBs with measured redshift $z$ in the common rest frame energy range $0.3\,$--$\,10$ keV. To perform this computation, the first step is to convert the observed XRT flux $f_{obs}$ to the one in the $0.3\,$--$\,10$ keV rest frame energy range. In the detector frame, the $0.3\,$--$\,10$ keV rest frame energy range becomes $[0.3/(1+z)]\,$--$\,[10/(1+z)]$ keV where $z$ is the redshift of the GRB. We assume a simple power-law function as the best-fit for the spectral energy distribution of the XRT data\footnote{http://www.swift.ac.uk/}:
\begin{equation}
\frac{dN}{dA\,dt\,dE} \propto E^{-\gamma} \,.
\label{spettro_pl}
\end{equation}
We can then write the flux light curve $f_{rf}$ in the $0.3\,$--$\,10$ keV rest frame energy range as:
\begin{equation}
f_{rf} = f_{obs} \frac{\int_{\frac{0.3\,keV}{1+z}}^{\frac{10\,keV}{1+z}}E^{-\gamma}dE}{\int_{0.3\,keV}^{10\,keV}E^{-\gamma}dE} = f_{obs} (1+z)^{\gamma-1} \,.
\label{flusso_1}
\end{equation}
Then, we have to multiply $f_{rf}$ by the luminosity distance to get $L_{rf}$:
\begin{equation}
L_{rf} = 4 \, \pi \, d_l^2(z) f_{rf} \,,
\label{luminosity}
\end{equation}
where we assume a standard cosmological model $\Lambda$CDM with $\Omega_m = 0.27$ and $\Omega_{\Lambda}=0.73$. Clearly, this luminosity must be plotted as a function of the rest frame time $t_{rf}$, namely:
\begin{equation}
\label{time_correction}
t_{rf} = \frac{t_{obs}}{1+z}\,.
\end{equation}

The X-ray luminosity light curves of the six GRBs with measured redshift in the $0.3$--$10$ keV rest frame energy band are plotted together in Fig.~\ref{fig:sample}. What is most striking is that these six GRBs, with redshift in the range $0.49\,$--$\,1.261$, show a remarkably common behavior of the late X-ray afterglow luminosity light curves (episode 3) despite that their prompt emissions (episode 1 and 2) are very different and that their energetics spans more than two orders of magnitude. Such a common behavior starts between $10^4\,$--$\,10^5$ s after the trigger and continues up to when the emission falls below the XRT threshold. This standard behavior of episode 3 represents strong evidence of very low or even the absence of beaming in this particular phase of the X-ray afterglow emission process. We have proposed that this late time X-ray emission in episode 3 is related to the process of the SN explosion within the IGC scenario, possibly emitted by the newly born NS, and not by the GRB itself \cite{2012A&A...540A..12N}. This scaling law, when confirmed in sources presenting the episode 1 plus the episode 2 emissions, offers a powerful tool to estimate the redshift of GRBs belonging to this subclass of events. 

\begin{figure}
\includegraphics[width=0.49\hsize,clip]{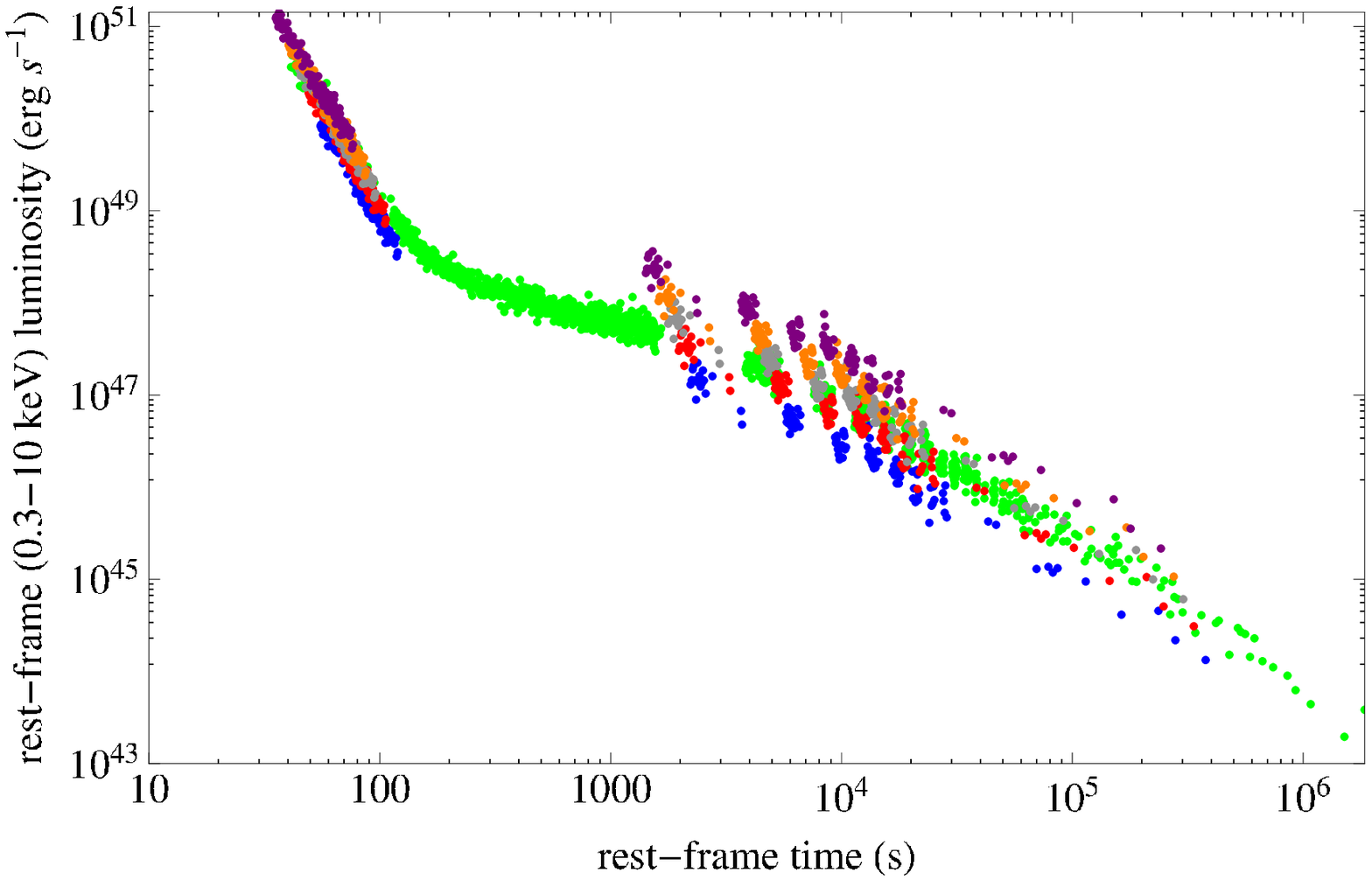}
\includegraphics[width=0.49\hsize,clip]{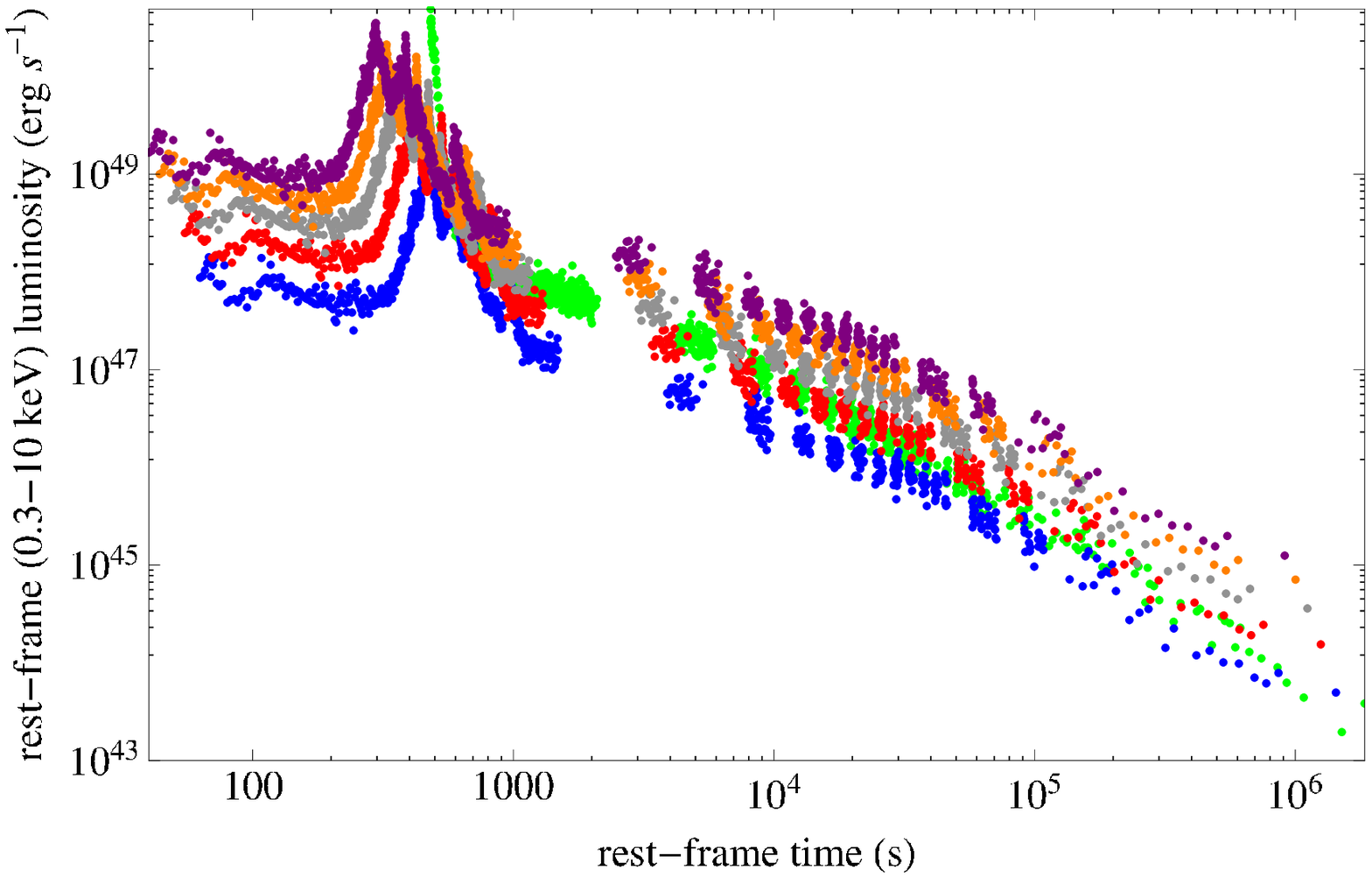}
\caption{In green we show the rest frame X-ray luminosity light curve of GRB 090618 in the $0.3$--$10$ keV energy range in comparison with the one of GRB 101023 (left) and GRB 110709B (right), computed for different hypothetical redshifts: respectively, from blue to purple: $z=0.6, 0.8, 1.0, 1.2, 1.5$ (left) and $z=0.4, 0.6, 0.8, 1.0, 1.2$ (right). The overlapping at late time of the two X-ray luminosity light curves is obtained for a redshift of $z=0.9$ (left) and $z=0.75$ (right). For further details see Ref.~\protect\refcite{2012A&A...538A..58P,2013A&A...551A.133P}.
\label{fig:101023}}
\end{figure}

As an example, we present in Fig.~\ref{fig:101023} the rest frame X-ray luminosity (0.3 -- 10 keV) light curve of GRB 090618 (considered as a prototype for the common behavior shown in Fig.~\ref{fig:sample}) with the rest frame X-ray luminosity light curves of GRB 110709B estimated for selected values of its redshifts, $z=0.4, 0.6, 0.8, 1.0, 1.2$, and similarly the correspondent analysis for GRB 101023 for selected values of the redshift, $z=0.6, 0.8, 1.0, 1.2, 1.5$. We then find that GRB 101023 should have been located at $z \sim 0.9$ and GRB 110709B at $z \sim 0.75$. These redshift estimations are within the range expected using the Amati relation as shown in Ref.~\refcite{2012A&A...538A..58P,2013A&A...551A.133P}. This is an important independent confirmation of validity for this new redshift estimator we propose for the family of IGC GRB-SN systems. It should be stressed,  however, that the determination of the redshift is done assuming the validity of the standard $\Lambda$CDM cosmological model for sources with redshift in the range $z=0.49\,$--$\,1.216$. We are currently testing the validity of this assumption for sources at larger cosmological redshifts.

Concerning the nature of the late X-ray emission discussed in \refcite{2013A&A...552L...5P}, I am currently exploring the possibility that the emission process is linked to the decay of transuranic elements produced by the interaction of the GRB with the SNe through the $r$-process \cite{Burbridge1957} and accreted onto the newly-formed NS.

\subsection{Conclusions}

The nature of GRBs is presenting itself as one of the richest diagnostics ever encountered within physics and astrophysics. 
It is clear that phenomena never before explored in this domain can now be submitted to theoretical and observational scrutiny.
In the GRB-SN connection we have introduced, in analogy with the S-matrix of particle physics, a cosmic matrix (C-matrix) in which the in-states are a NS and an evolved core undergoing a SN explosion in a binary system, and the out-states are a BH and a newly-born NS.
With the same spirit, the C-matrix of a genuine short GRB has as in-states two NSs and as out-states GW emission and the formation of a BH.


\begin{thebibliography}{100}

\bibitem{1975ASSL...48.....G}
H.~{Gursky} and R.~{Ruffini} (eds.), {\em Neutron stars, black holes and binary
  X-ray sources; PROCEEDINGS of the Annual Meeting, San Francisco, Calif.,
  February 28, 1974}, Astrophysics and Space Science Library Vol.~48, 1975.

\bibitem{2000ApJ...536..778P}
E.~{Pian}, L.~{Amati}, L.~A. {Antonelli}, R.~C. {Butler}, E.~{Costa},
  G.~{Cusumano}, J.~{Danziger}, M.~{Feroci}, F.~{Fiore}, F.~{Frontera},
  P.~{Giommi}, N.~{Masetti}, J.~M. {Muller}, L.~{Nicastro}, T.~{Oosterbroek},
  M.~{Orlandini}, A.~{Owens}, E.~{Palazzi}, A.~{Parmar}, L.~{Piro}, J.~J.~M.
  {in't Zand}, A.~{Castro-Tirado}, A.~{Coletta}, D.~{Dal Fiume}, S.~{Del
  Sordo}, J.~{Heise}, P.~{Soffitta} and V.~{Torroni}, {\em ApJ} {\bf 536},
  778 (June 2000).

\bibitem{1998Natur.395..670G}
T.~J. {Galama}, P.~M. {Vreeswijk}, J.~{van Paradijs}, C.~{Kouveliotou},
  T.~{Augusteijn}, H.~{B{\"o}hnhardt}, J.~P. {Brewer}, V.~{Doublier}, J.-F.
  {Gonzalez}, B.~{Leibundgut}, C.~{Lidman}, O.~R. {Hainaut}, F.~{Patat},
  J.~{Heise}, J.~{in't Zand}, K.~{Hurley}, P.~J. {Groot}, R.~G. {Strom}, P.~A.
  {Mazzali}, K.~{Iwamoto}, K.~{Nomoto}, H.~{Umeda}, T.~{Nakamura}, T.~R.
  {Young}, T.~{Suzuki}, T.~{Shigeyama}, T.~{Koshut}, M.~{Kippen},
  C.~{Robinson}, P.~{de Wildt}, R.~A.~M.~J. {Wijers}, N.~{Tanvir},
  J.~{Greiner}, E.~{Pian}, E.~{Palazzi}, F.~{Frontera}, N.~{Masetti},
  L.~{Nicastro}, M.~{Feroci}, E.~{Costa}, L.~{Piro}, B.~A. {Peterson},
  C.~{Tinney}, B.~{Boyle}, R.~{Cannon}, R.~{Stathakis}, E.~{Sadler}, M.~C.
  {Begam} and P.~{Ianna}, {\em Nature} {\bf 395}, 670 (October 1998).

\bibitem{2011Sci...331..736T}
M.~{Tavani}, A.~{Bulgarelli}, V.~{Vittorini}, A.~{Pellizzoni}, E.~{Striani},
  P.~{Caraveo}, M.~C. {Weisskopf}, A.~{Tennant}, G.~{Pucella}, A.~{Trois},
  E.~{Costa}, Y.~{Evangelista}, C.~{Pittori}, F.~{Verrecchia}, E.~{Del Monte},
  R.~{Campana}, M.~{Pilia}, A.~{De Luca}, I.~{Donnarumma}, D.~{Horns},
  C.~{Ferrigno}, C.~O. {Heinke}, M.~{Trifoglio}, F.~{Gianotti},
  S.~{Vercellone}, A.~{Argan}, G.~{Barbiellini}, P.~W. {Cattaneo}, A.~W.
  {Chen}, T.~{Contessi}, F.~{D'Ammando}, G.~{DeParis}, G.~{Di Cocco}, G.~{Di
  Persio}, M.~{Feroci}, A.~{Ferrari}, M.~{Galli}, A.~{Giuliani}, M.~{Giusti},
  C.~{Labanti}, I.~{Lapshov}, F.~{Lazzarotto}, P.~{Lipari}, F.~{Longo},
  F.~{Fuschino}, M.~{Marisaldi}, S.~{Mereghetti}, E.~{Morelli}, E.~{Moretti},
  A.~{Morselli}, L.~{Pacciani}, F.~{Perotti}, G.~{Piano}, P.~{Picozza},
  M.~{Prest}, M.~{Rapisarda}, A.~{Rappoldi}, A.~{Rubini}, S.~{Sabatini},
  P.~{Soffitta}, E.~{Vallazza}, A.~{Zambra}, D.~{Zanello}, F.~{Lucarelli},
  P.~{Santolamazza}, P.~{Giommi}, L.~{Salotti} and G.~F. {Bignami}, {\em
  Science} {\bf 331}, 736 (February 2011).

\bibitem{1968Natur.217..709H}
A.~{Hewish}, S.~J. {Bell}, J.~D.~H. {Pilkington}, P.~F. {Scott} and R.~A.
  {Collins}, {\em Nature} {\bf 217}, 709 (February 1968).

\bibitem{1968ApJ...153..865F}
A.~{Finzi} and R.~A. {Wolf}, {\em ApJ} {\bf 153}, p.~865 (September 1968).

\bibitem{LesHouches}
W.~{Arnett}, in {\em From Nuclei to White Dwarfs and Neutron Stars\/},  eds.
  A.~{Mezzacappa} and R.~{Ruffini}  (Singapore: World Scientific, in press).

\bibitem{1974bhgw.BOOK.....R}
M.~{Rees}, R.~{Ruffini} and J.~A. {Wheeler}, {\em Black holes, gravitational
  waves and cosmology: an introduction to current research}, Topics in
  Astrophysics and Space Physics, Vol.~10  (New York: Gordon and Breach, Science
  Publishers, Inc., 1974).

\bibitem{1971PhRvD...4.3552C}
D.~{Christodoulou} and R.~{Ruffini}, {\em Phys. Rev. D} {\bf 4}, 3552  (1971).

\bibitem{RRKerr}
R.~{Ruffini}, The ergosphere and dyadosphere of black holes, in {\em The Kerr
  Spacetime\/},  eds. D.~L. {Wiltshire}, M.~{Visser} and S.~{Scott}  (Cambridge
  University Press, 2009).

\bibitem{RRKl}
R.~{Ruffini}, Analogies, new paradigms and observational data as growing
  factors of relativistic astrophysics, in {\em Fluctuating Paths and
  Fields\/},  eds. W.~{Janke}, A.~{Pelster}, H.~J. {Schmidt} and M.~{Bachmann}
   (Singapore: World Scientific, 2001).

\bibitem{1975PhRvL..35..463D}
T.~{Damour} and R.~{Ruffini}, {\em Physical Review Letters} {\bf 35},
  463 (August 1975).

\bibitem{1999A&A...350..334R}
R.~{Ruffini}, J.~D. {Salmonson}, J.~R. {Wilson} and S.-S. {Xue}, {\em A\&A}
  {\bf 350}, 334 (October 1999).

\bibitem{2000A&A...359..855R}
R.~{Ruffini}, J.~D. {Salmonson}, J.~R. {Wilson} and S.-S. {Xue}, {\em A\&A}
  {\bf 359}, 855 (July 2000).

\bibitem{1999ApJS..122..465P}
W.~S. {Paciesas}, C.~A. {Meegan}, G.~N. {Pendleton}, M.~S. {Briggs},
  C.~{Kouveliotou}, T.~M. {Koshut}, J.~P. {Lestrade}, M.~L. {McCollough}, J.~J.
  {Brainerd}, J.~{Hakkila}, W.~{Henze}, R.~D. {Preece}, V.~{Connaughton}, R.~M.
  {Kippen}, R.~S. {Mallozzi}, G.~J. {Fishman}, G.~A. {Richardson} and
  M.~{Sahi}, {\em ApJSS} {\bf 122}, 465 (June 1999).

\bibitem{2013ApJ...763..125M}
M.~{Muccino}, R.~{Ruffini}, C.~L. {Bianco}, L.~{Izzo} and A.~V. {Penacchioni},
  {\em ApJ} {\bf 763}, p.~125 (February 2013).

\bibitem{2012A&A...543A..10I}
L.~{Izzo}, R.~{Ruffini}, A.~V. {Penacchioni}, C.~L. {Bianco}, L.~{Caito}, S.~K.
  {Chakrabarti}, J.~A. {Rueda}, A.~{Nandi} and B.~{Patricelli}, {\em A\&A} {\bf
  543}, p.~A10 (July 2012).

\bibitem{2012A&A...548L...5I}
L.~{Izzo}, J.~A. {Rueda} and R.~{Ruffini}, {\em A\&A} {\bf 548}, p.~L5 (December
  2012).

\bibitem{2013A&A...552L...5P}
G.~B. {Pisani}, L.~{Izzo}, R.~{Ruffini}, C.~L. {Bianco}, M.~{Muccino}, A.~V.
  {Penacchioni}, J.~A. {Rueda} and Y.~{Wang}, {\em A\&A} {\bf 552}, p.~L5 (April
  2013).

\bibitem{1968PhRv..174.1559C}
B.~{Carter}, {\em Physical Review} {\bf 174}, 1559 (October 1968).

\bibitem{2009PhRvD..79l4002C}
C.~{Cherubini}, A.~{Geralico}, H.~{J.~A.~Rueda} and R.~{Ruffini}, {\em Phys.
  Rev. D} {\bf 79}, p.~124002 (June 2009).

\bibitem{1978MNRAS.183..359C}
G.~{Cavallo} and M.~J. {Rees}, {\em MNRAS} {\bf 183}, 359 (May 1978).

\bibitem{2007PhRvL..99l5003A}
A.~{Aksenov}, R.~{Ruffini} and G.~{Vereshchagin}, {\em Phys. Rev. Lett.} {\bf
  99}, p.~125003 (September 2007).

\bibitem{2001ApJ...555L.113R}
R.~{Ruffini}, C.~L. {Bianco}, P.~{Chardonnet}, F.~{Fraschetti} and S.-S. {Xue},
  {\em ApJ} {\bf 555}, L113 (July 2001).

\bibitem{2008pint.conf..207R}
R.~{Ruffini}, {The Role of Thomas-Fermi Approach in Neutron Star Matter}, in
  {\em Path Integrals --- New Trends and Perspectives\/},  eds. W.~{Janke} and
  A.~{Pelster}November 2008.

\bibitem{2012NuPhA.883....1B}
R.~{Belvedere}, D.~{Pugliese}, J.~A. {Rueda}, R.~{Ruffini} and S.-S. {Xue},
  {\em Nuclear Physics A} {\bf 883}, 1 (June 2012).

\bibitem{1939PhRv...55..364T}
R.~C. {Tolman}, {\em Physical Review} {\bf 55}, 364 (February 1939).

\bibitem{1939PhRv...55..374O}
J.~R. {Oppenheimer} and G.~M. {Volkoff}, {\em Physical Review} {\bf 55},
  374 (February 1939).

\bibitem{1966ARA&A...4..393W}
J.~A. {Wheeler}, {\em ARAA} {\bf 4}, p.~393  (1966).

\bibitem{1949RvMP...21..531K}
O.~{Klein}, {\em Reviews of Modern Physics} {\bf 21}, 531 (July 1949).

\bibitem{2011NuPhA.872..286R}
J.~A. {Rueda}, R.~{Ruffini} and S.-S. {Xue}, {\em Nuclear Physics A} {\bf 872},
  286 (December 2011).

\bibitem{2009ApJ...702..791M}
C.~{Meegan}, G.~{Lichti}, P.~N. {Bhat}, E.~{Bissaldi}, M.~S. {Briggs},
  V.~{Connaughton}, R.~{Diehl}, G.~{Fishman}, J.~{Greiner}, A.~S. {Hoover},
  A.~J. {van der Horst}, A.~{von Kienlin}, R.~M. {Kippen}, C.~{Kouveliotou},
  S.~{McBreen}, W.~S. {Paciesas}, R.~{Preece}, H.~{Steinle}, M.~S. {Wallace},
  R.~B. {Wilson} and C.~{Wilson-Hodge}, {\em ApJ} {\bf 702}, 791 (September
  2009).

\bibitem{2002ApJ...581L..19R}
R.~{Ruffini}, C.~L. {Bianco}, P.~{Chardonnet}, F.~{Fraschetti} and S.-S. {Xue},
  {\em ApJ} {\bf 581}, L19 (December 2002).

\bibitem{2012arXiv1205.6915R}
J.~A. {Rueda} and R.~{Ruffini}, {\em arXiv:1205.6915}  (May 2012).

\bibitem{2004RvMP...76.1143P}
T.~{Piran}, {\em Reviews of Modern Physics} {\bf 76}, 1143 (October 2004).

\bibitem{1986ApJ...308L..47G}
J.~{Goodman}, {\em ApJ} {\bf 308}, L47 (September 1986).

\bibitem{1986ApJ...308L..43P}
B.~{Paczynski}, {\em ApJ} {\bf 308}, L43 (September 1986).

\bibitem{PhysRep}
R.~{Ruffini}, G.~{Vereshchagin} and S.-S. {Xue}, {\em Phys. Rep.} {\bf 487},
  p.~1 (February 2010).

\bibitem{2011IJMPD..20.1797R}
R.~{Ruffini}, {\em International Journal of Modern Physics D} {\bf 20}, 1797
   (2011).

\bibitem{2000ApJ...539..712R}
E.~{Ramirez-Ruiz} and E.~E. {Fenimore}, {\em ApJ} {\bf 539}, 712 (August 2000).

\bibitem{1994ApJ...430L..93R}
M.~J. {Rees} and P.~{Meszaros}, {\em ApJ} {\bf 430}, L93 (August 1994).

\bibitem{1992MNRAS.258P..41R}
M.~J. {Rees} and P.~{Meszaros}, {\em MNRAS} {\bf 258}, 41P (September 1992).

\bibitem{1996ApJ...466..768T}
M.~{Tavani}, {\em ApJ} {\bf 466}, p.~768 (August 1996).

\bibitem{2000ApJS..127...59F}
F.~{Frontera}, L.~{Amati}, E.~{Costa}, J.~M. {Muller}, E.~{Pian}, L.~{Piro},
  P.~{Soffitta}, M.~{Tavani}, A.~{Castro-Tirado}, D.~{Dal Fiume}, M.~{Feroci},
  J.~{Heise}, N.~{Masetti}, L.~{Nicastro}, M.~{Orlandini}, E.~{Palazzi} and
  R.~{Sari}, {\em ApJSS} {\bf 127}, 59 (March 2000).

\bibitem{1997ApJ...479L..39C}
A.~{Crider}, E.~P. {Liang}, I.~A. {Smith}, R.~D. {Preece}, M.~S. {Briggs},
  G.~N. {Pendleton}, W.~S. {Paciesas}, D.~L. {Band} and J.~L. {Matteson}, {\em
  ApJ} {\bf 479}, p.~L39 (April 1997).

\bibitem{2002ApJ...581.1248P}
R.~D. {Preece}, M.~S. {Briggs}, T.~W. {Giblin}, R.~S. {Mallozzi}, G.~N.
  {Pendleton}, W.~S. {Paciesas} and D.~L. {Band}, {\em ApJ} {\bf 581},
  1248 (December 2002).

\bibitem{2002A&A...393..409G}
G.~{Ghirlanda}, A.~{Celotti} and G.~{Ghisellini}, {\em A\&A} {\bf 393},
  409 (October 2002).

\bibitem{2003A&A...406..879G}
G.~{Ghirlanda}, A.~{Celotti} and G.~{Ghisellini}, {\em A\&A} {\bf 406},
  879 (August 2003).

\bibitem{2008MNRAS.384...33K}
P.~{Kumar} and E.~{McMahon}, {\em MNRAS} {\bf 384}, 33 (February 2008).

\bibitem{2009MNRAS.393.1107P}
T.~{Piran}, R.~{Sari} and Y.~{Zou}, {\em MNRAS} {\bf 393}, 1107 (March 2009).

\bibitem{1976PhFl...19.1130B}
R.~D. {Blandford} and C.~F. {McKee}, {\em Physics of Fluids} {\bf 19},
  1130 (August 1976).

\bibitem{2007A&A...469L..13M}
E.~{Molinari}, S.~D. {Vergani}, D.~{Malesani}, S.~{Covino}, P.~{D'Avanzo},
  G.~{Chincarini}, F.~M. {Zerbi}, L.~A. {Antonelli}, P.~{Conconi}, V.~{Testa},
  G.~{Tosti}, F.~{Vitali}, F.~{D'Alessio}, G.~{Malaspina}, L.~{Nicastro},
  E.~{Palazzi}, D.~{Guetta}, S.~{Campana}, P.~{Goldoni}, N.~{Masetti}, E.~J.~A.
  {Meurs}, A.~{Monfardini}, L.~{Norci}, E.~{Pian}, S.~{Piranomonte},
  D.~{Rizzuto}, M.~{Stefanon}, L.~{Stella}, G.~{Tagliaferri}, P.~A. {Ward},
  G.~{Ihle}, L.~{Gonzalez}, A.~{Pizarro}, P.~{Sinclaire} and J.~{Valenzuela},
  {\em A\&A} {\bf 469}, L13 (July 2007).

\bibitem{2009ApJ...702..489R}
E.~S. {Rykoff}, F.~{Aharonian}, C.~W. {Akerlof}, M.~C.~B. {Ashley}, S.~D.
  {Barthelmy}, H.~A. {Flewelling}, N.~{Gehrels}, E.~{G{\"o}{\v g}{\"u}{\c s}},
  T.~{G{\"u}ver}, {\"U}.~{Kizilo{\v g}lu}, H.~A. {Krimm}, T.~A. {McKay},
  M.~{{\"O}zel}, A.~{Phillips}, R.~M. {Quimby}, G.~{Rowell}, W.~{Rujopakarn},
  B.~E. {Schaefer}, D.~A. {Smith}, W.~T. {Vestrand}, J.~C. {Wheeler},
  J.~{Wren}, F.~{Yuan} and S.~A. {Yost}, {\em ApJ} {\bf 702}, 489 (September
  2009).

\bibitem{1999ApJ...520..641S}
R.~{Sari} and T.~{Piran}, {\em ApJ} {\bf 520}, 641 (August 1999).

\bibitem{1999A&AS..138..527G}
G.~{Ghisellini} and A.~{Celotti}, {\em A\&AS} {\bf 138}, 527 (September 1999).

\bibitem{1991ApJ...366..343Z}
A.~A. {Zdziarski}, R.~{Svensson} and B.~{Paczynski}, {\em ApJ} {\bf 366},
  343 (January 1991).

\bibitem{1994MNRAS.269.1112S}
A.~{Shemi}, {\em MNRAS} {\bf 269}, p.~1112 (August 1994).

\bibitem{2006ApJ...653..454P}
A.~{Pe'er} and B.~{Zhang}, {\em ApJ} {\bf 653}, 454 (December 2006).

\bibitem{2000ApJ...540..704M}
M.~V. {Medvedev}, {\em ApJ} {\bf 540}, 704 (September 2000).

\bibitem{2000ApJ...544L..17P}
A.~{Panaitescu} and P.~{M{\'e}sz{\'a}ros}, {\em ApJ} {\bf 544}, L17 (November
  2000).

\bibitem{2004MNRAS.352L..35S}
B.~E. {Stern} and J.~{Poutanen}, {\em MNRAS} {\bf 352}, L35 (August 2004).

\bibitem{2000ApJ...529..146E}
D.~{Eichler} and A.~{Levinson}, {\em ApJ} {\bf 529}, 146 (January 2000).

\bibitem{2000ApJ...530..292M}
P.~{M{\'e}sz{\'a}ros} and M.~J. {Rees}, {\em ApJ} {\bf 530}, 292 (February
  2000).

\bibitem{2002ARA&A..40..137M}
P.~{M{\'e}sz{\'a}ros}, {\em ARAA} {\bf 40}, 137  (2002).

\bibitem{2002MNRAS.336.1271D}
F.~{Daigne} and R.~{Mochkovitch}, {\em MNRAS} {\bf 336}, 1271 (November 2002).

\bibitem{2006A&A...457..763G}
D.~{Giannios}, {\em A\&A} {\bf 457}, 763 (October 2006).

\bibitem{2009ApJ...702.1211R}
F.~{Ryde} and A.~{Pe'er}, {\em ApJ} {\bf 702}, 1211 (September 2009).

\bibitem{2010ApJ...725.1137L}
D.~{Lazzati} and M.~C. {Begelman}, {\em ApJ} {\bf 725}, 1137 (December 2010).

\bibitem{1999A&AS..138..513R}
R.~{Ruffini}, {\em A\&AS} {\bf 138}, 513 (September 1999).

\bibitem{2006RPPh...69.2259M}
P.~{Meszaros}, {\em Reports of Progress in Physics} {\bf 69}, 2259  (2006).

\bibitem{2004ApJ...605L...1B}
C.~L. {Bianco} and R.~{Ruffini}, {\em ApJ} {\bf 605}, L1 (April 2004).

\bibitem{2005ApJ...620L..23B}
C.~L. {Bianco} and R.~{Ruffini}, {\em ApJ} {\bf 620}, L23 (February 2005).

\bibitem{1993ApJ...415..181M}
P.~{Meszaros}, P.~{Laguna} and M.~J. {Rees}, {\em ApJ} {\bf 415}, 181 (September
  1993).

\bibitem{1997ApJ...489L..37S}
R.~{Sari}, {\em ApJ} {\bf 489}, p.~L37 (November 1997).

\bibitem{1998ApJ...494L..49S}
R.~{Sari}, {\em ApJ} {\bf 494}, p.~L49 (February 1998).

\bibitem{1997ApJ...491L..19W}
E.~{Waxman}, {\em ApJ} {\bf 491}, p.~L19 (December 1997).

\bibitem{1998ApJ...496L...1R}
M.~J. {Rees} and P.~{Meszaros}, {\em ApJ} {\bf 496}, p.~L1 (March 1998).

\bibitem{1999ApJ...513..679G}
J.~{Granot}, T.~{Piran} and R.~{Sari}, {\em ApJ} {\bf 513}, 679 (March 1999).

\bibitem{1998ApJ...493L..31P}
A.~{Panaitescu} and P.~{Meszaros}, {\em ApJ} {\bf 493}, p.~L31 (January 1998).

\bibitem{1999ApJ...511..852G}
A.~{Gruzinov} and E.~{Waxman}, {\em ApJ} {\bf 511}, 852 (February 1999).

\bibitem{2000ARA&A..38..379V}
J.~{van Paradijs}, C.~{Kouveliotou} and R.~A.~M.~J. {Wijers}, {\em ARAA} {\bf
  38}, 379  (2000).

\bibitem{2005ApJ...633L..13B}
C.~L. {Bianco} and R.~{Ruffini}, {\em ApJ} {\bf 633}, L13 (November 2005).

\bibitem{2001ApJ...555L.107R}
R.~{Ruffini}, C.~L. {Bianco}, P.~{Chardonnet}, F.~{Fraschetti} and S.-S. {Xue},
  {\em ApJ} {\bf 555}, L107 (July 2001).

\bibitem{2011IJMPD..20.1983P}
B.~{Patricelli}, M.~G. {Bernardini}, C.~L. {Bianco}, L.~{Caito}, L.~{Izzo},
  R.~{Ruffini} and G.~{Vereshchagin}, {\em IJMPD} {\bf 20}, 1983  (2011).

\bibitem{1999ApJ...526..697M}
M.~V. {Medvedev} and A.~{Loeb}, {\em ApJ} {\bf 526}, 697 (December 1999).

\bibitem{2008ApJ...673L..39S}
A.~{Spitkovsky}, {\em ApJ} {\bf 673}, L39 (January 2008).

\bibitem{2009ApJ...700..956M}
M.~V. {Medvedev} and A.~{Spitkovsky}, {\em ApJ} {\bf 700}, 956 (August 2009).

\bibitem{2005AIPC..782...42R}
R.~{Ruffini}, M.~G. {Bernardini}, C.~L. {Bianco}, P.~{Chardonnet},
  F.~{Fraschetti}, V.~{Gurzadyan}, L.~{Vitagliano} and S.-S. {Xue}, The
  blackholic energy: long and short gamma-ray bursts  (new perspectives in
  physics and astrophysics from the theoretical understanding of gamma-ray
  bursts, ii), in {\em XI Brazilian School of Cosmology and Gravitation\/},
  eds. M.~{Novello} and S.~E. {Perez Bergliaffa}, American Institute of Physics
  Conference Series, Vol.~782August 2005.

\bibitem{2007A&A...474L..13B}
M.~G. {Bernardini}, C.~L. {Bianco}, L.~{Caito}, M.~G. {Dainotti}, R.~{Guida}
  and R.~{Ruffini}, {\em A\&A} {\bf 474}, L13 (October 2007).

\bibitem{2009A&A...498..501C}
L.~{Caito}, M.~G. {Bernardini}, C.~L. {Bianco}, M.~G. {Dainotti}, R.~{Guida}
  and R.~{Ruffini}, {\em A\&A} {\bf 498}, 501 (May 2009).

\bibitem{2010A&A...521A..80C}
L.~{Caito}, L.~{Amati}, M.~G. {Bernardini}, C.~L. {Bianco}, G.~{de Barros},
  L.~{Izzo}, B.~{Patricelli} and R.~{Ruffini}, {\em A\&A} {\bf 521}, p.
  A80 (October 2010).

\bibitem{2011A&A...529A.130D}
G.~{de Barros}, L.~{Amati}, M.~G. {Bernardini}, C.~L. {Bianco}, L.~{Caito},
  L.~{Izzo}, B.~{Patricelli} and R.~{Ruffini}, {\em A\&A} {\bf 529}, p.
  A130 (May 2011).

\bibitem{2011AN....332...92P}
Z.~Y. {Peng}, Y.~{Yin}, X.~W. {Bi}, Y.~Y. {Bao} and L.~{Ma}, {\em Astronomische
  Nachrichten} {\bf 332}, p.~92 (January 2011).

\bibitem{2002A&A...396..705Q}
Y.-P. {Qin}, {\em A\&A} {\bf 396}, 705 (December 2002).

\bibitem{1949PhRv...75.1169F}
E.~{Fermi}, {\em Physical Review} {\bf 75}, 1169 (April 1949).

\bibitem{1954ApJ...119....1F}
E.~{Fermi}, {\em ApJ} {\bf 119}, p.~1 (January 1954).

\bibitem{1997AJ....114..258S}
M.~M. {Shara}, D.~R. {Zurek}, R.~E. {Williams}, D.~{Prialnik}, R.~{Gilmozzi}
  and A.~F.~J. {Moffat}, {\em AJ} {\bf 114}, p.~258 (July 1997).

\bibitem{2009MNRAS.398.2152D}
L.~{Ducci}, L.~{Sidoli}, S.~{Mereghetti}, A.~{Paizis} and P.~{Romano}, {\em
  MNRAS} {\bf 398}, 2152 (October 2009).

\bibitem{2005Natur.437..851G}
N.~{Gehrels}, C.~L. {Sarazin}, P.~T. {O'Brien}, B.~{Zhang}, L.~{Barbier}, S.~D.
  {Barthelmy}, A.~{Blustin}, D.~N. {Burrows}, J.~{Cannizzo}, J.~R. {Cummings},
  M.~{Goad}, S.~T. {Holland}, C.~P. {Hurkett}, J.~A. {Kennea}, A.~{Levan},
  C.~B. {Markwardt}, K.~O. {Mason}, P.~{Meszaros}, M.~{Page}, D.~M. {Palmer},
  E.~{Rol}, T.~{Sakamoto}, R.~{Willingale}, L.~{Angelini}, A.~{Beardmore},
  P.~T. {Boyd}, A.~{Breeveld}, S.~{Campana}, M.~M. {Chester}, G.~{Chincarini},
  L.~R. {Cominsky}, G.~{Cusumano}, M.~{de Pasquale}, E.~E. {Fenimore},
  P.~{Giommi}, C.~{Gronwall}, D.~{Grupe}, J.~E. {Hill}, D.~{Hinshaw},
  J.~{Hjorth}, D.~{Hullinger}, K.~C. {Hurley}, S.~{Klose}, S.~{Kobayashi},
  C.~{Kouveliotou}, H.~A. {Krimm}, V.~{Mangano}, F.~E. {Marshall},
  K.~{McGowan}, A.~{Moretti}, R.~F. {Mushotzky}, K.~{Nakazawa}, J.~P. {Norris},
  J.~A. {Nousek}, J.~P. {Osborne}, K.~{Page}, A.~M. {Parsons}, S.~{Patel},
  M.~{Perri}, T.~{Poole}, P.~{Romano}, P.~W.~A. {Roming}, S.~{Rosen},
  G.~{Sato}, P.~{Schady}, A.~P. {Smale}, J.~{Sollerman}, R.~{Starling},
  M.~{Still}, M.~{Suzuki}, G.~{Tagliaferri}, T.~{Takahashi}, M.~{Tashiro},
  J.~{Tueller}, A.~A. {Wells}, N.~E. {White} and R.~A.~M.~J. {Wijers}, {\em
  Nature} {\bf 437}, 851 (October 2005).

\bibitem{2008AIPC..966....7B}
M.~G. {Bernardini}, C.~L. {Bianco}, L.~{Caito}, M.~G. {Dainotti}, R.~{Guida}
  and R.~{Ruffini}, Grb970228 and the class of grbs with an initial spikelike
  emission: do they follow the amati relation?, in {\em Relativistic
  Astrophysics\/},  eds. C.~L. {Bianco} and S.~S. {Xue}, American Institute of
  Physics Conference Series, Vol.~966January 2008.

\bibitem{1997Natur.387R.476S}
K.~C. {Sahu}, M.~{Livio}, L.~{Petro}, F.~D. {Macchetto}, J.~{van Paradijs},
  C.~{Kouveliotou}, G.~J. {Fishman}, C.~A. {Meegan}, P.~J. {Groot} and
  T.~{Galama}, {\em Nature} {\bf 387}, 476 (May 1997).

\bibitem{1997Natur.386..686V}
J.~{van Paradijs}, P.~J. {Groot}, T.~{Galama}, C.~{Kouveliotou}, R.~G. {Strom},
  J.~{Telting}, R.~G.~M. {Rutten}, G.~J. {Fishman}, C.~A. {Meegan},
  M.~{Pettini}, N.~{Tanvir}, J.~{Bloom}, H.~{Pedersen}, H.~U.
  {N{\o}rdgaard-Nielsen}, M.~{Linden-V{\o}rnle}, J.~{Melnick}, G.~{van der
  Steene}, M.~{Bremer}, R.~{Naber}, J.~{Heise}, J.~{in't Zand}, E.~{Costa},
  M.~{Feroci}, L.~{Piro}, F.~{Frontera}, G.~{Zavattini}, L.~{Nicastro},
  E.~{Palazzi}, K.~{Bennet}, L.~{Hanlon} and A.~{Parmar}, {\em Nature} {\bf
  386}, 686 (April 1997).

\bibitem{2006ApJ...638..354B}
J.~S. {Bloom}, J.~X. {Prochaska}, D.~{Pooley}, C.~H. {Blake}, R.~J. {Foley},
  S.~{Jha}, E.~{Ramirez-Ruiz}, J.~{Granot}, A.~V. {Filippenko},
  S.~{Sigurdsson}, A.~J. {Barth}, H.~{Chen}, M.~C. {Cooper}, E.~E. {Falco},
  R.~R. {Gal}, B.~F. {Gerke}, M.~D. {Gladders}, J.~E. {Greene}, J.~{Hennanwi},
  L.~C. {Ho}, K.~{Hurley}, B.~P. {Koester}, W.~{Li}, L.~{Lubin}, J.~{Newman},
  D.~A. {Perley}, G.~K. {Squires} and W.~M. {Wood-Vasey}, {\em ApJ} {\bf 638},
  354 (February 2006).

\bibitem{2008MNRAS.385L..10T}
E.~{Troja}, A.~R. {King}, P.~T. {O'Brien}, N.~{Lyons} and G.~{Cusumano}, {\em
  MNRAS} {\bf 385}, L10 (March 2008).

\bibitem{2010ApJ...708....9F}
W.~{Fong}, E.~{Berger} and D.~B. {Fox}, {\em ApJ} {\bf 708}, 9 (January 2010).

\bibitem{2011NewAR..55....1B}
E.~{Berger}, {\em New. Astron. Rev.} {\bf 55}, 1 (January 2011).

\bibitem{2012arXiv1203.1864K}
D.~{Kopa{\v c}}, P.~{D'Avanzo}, A.~{Melandri}, S.~{Campana}, A.~{Gomboc},
  J.~{Japelj}, M.~G. {Bernardini}, S.~{Covino}, S.~D. {Vergani},
  R.~{Salvaterra} and G.~{Tagliaferri}, {\em MNRAS} {\bf 424}, 2392 (August
  2012).

\bibitem{2009GCN..9568....1K}
K.~{Kono}, A.~{Daikyuji}, E.~{Sonoda}, N.~{Ohmori}, H.~{Hayashi}, K.~{Noda},
  Y.~{Nishioka}, M.~{Yamauchi}, M.~{Ohno}, M.~{Suzuki}, M.~{Kokubun},
  T.~{Takahashi}, K.~{Yamaoka}, S.~{Sugita}, Y.~E. {Nakagawa}, T.~{Tamagawa},
  S.~{Hong}, N.~{Vasquez}, Y.~{Hanabata}, T.~{Uehara}, Y.~{Fukazawa},
  W.~{Iwakiri}, M.~{Tashiro}, Y.~{Terada}, A.~{Endo}, K.~{Onda},
  T.~{Sugasahara}, Y.~{Urata}, T.~{Enoto}, K.~{Nakazawa} and K.~{Makishima},
  {\em GCN Circ.} {\bf 9568}, p.~1  (2009).

\bibitem{2009GCN..8925....1G}
S.~{Golenetskii}, R.~{Aptekar}, E.~{Mazets}, V.~{Pal'Shin}, D.~{Frederiks},
  P.~{Oleynik}, M.~{Ulanov}, D.~{Svinkin}, T.~{Cline}, K.~{Yamaoka}, M.~{Ohno},
  Y.~{Fukazawa}, T.~{Takahashi}, M.~{Tashiro}, Y.~{Terada}, T.~{Murakami},
  K.~{Makishima}, Y.~{Hanabata}, V.~{Connaughton}, M.~{Briggs}, A.~{von
  Kienlin}, G.~{Lichti}, A.~{Rau} and K.~{Hurley}, {\em GCN Circ.} {\bf 8925},
  p.~1  (2009).

\bibitem{2009GCN..8926....1G}
S.~{Golenetskii}, R.~{Aptekar}, E.~{Mazets}, V.~{Pal'Shin}, D.~{Frederiks},
  P.~{Oleynik}, M.~{Ulanov}, D.~{Svinkin} and T.~{Cline}, {\em GCN Circ.} {\bf
  8926}, p.~1  (2009).

\bibitem{1993ApJ...413..281B}
D.~{Band}, J.~{Matteson}, L.~{Ford}, B.~{Schaefer}, D.~{Palmer},
  B.~{Teegarden}, T.~{Cline}, M.~{Briggs}, W.~{Paciesas}, G.~{Pendleton},
  G.~{Fishman}, C.~{Kouveliotou}, C.~{Meegan}, R.~{Wilson} and P.~{Lestrade},
  {\em ApJ} {\bf 413}, 281 (August 1993).

\bibitem{2010ApJ...725..225G}
S.~{Guiriec}, M.~S. {Briggs}, V.~{Connaugthon}, E.~{Kara}, F.~{Daigne},
  C.~{Kouveliotou}, A.~J. {van der Horst}, W.~{Paciesas}, C.~A. {Meegan}, P.~N.
  {Bhat}, S.~{Foley}, E.~{Bissaldi}, M.~{Burgess}, V.~{Chaplin}, R.~{Diehl},
  G.~{Fishman}, M.~{Gibby}, M.~M. {Giles}, A.~{Goldstein}, J.~{Greiner},
  D.~{Gruber}, A.~{von Kienlin}, M.~{Kippen}, S.~{McBreen}, R.~{Preece},
  A.~{Rau}, D.~{Tierney} and C.~{Wilson-Hodge}, {\em ApJ} {\bf 725},
  225 (December 2010).

\bibitem{2012MNRAS.420..468P}
A.~{Pe'Er}, B.-B. {Zhang}, F.~{Ryde}, S.~{McGlynn}, B.~{Zhang}, R.~D. {Preece}
  and C.~{Kouveliotou}, {\em MNRAS} {\bf 420}, 468 (February 2012).

\bibitem{2012A&A...538A..58P}
A.~V. {Penacchioni}, R.~{Ruffini}, L.~{Izzo}, M.~{Muccino}, C.~L. {Bianco},
  L.~{Caito}, B.~{Patricelli} and L.~{Amati}, {\em A\&A} {\bf 538}, p.
  A58 (February 2012).

\bibitem{2001ApJ...555L.117R}
R.~{Ruffini}, C.~L. {Bianco}, P.~{Chardonnet}, F.~{Fraschetti} and S.-S. {Xue},
  {\em ApJ} {\bf 555}, L117 (July 2001).

\bibitem{2008AIPC..966...12B}
C.~L. {Bianco}, M.~G. {Bernardini}, L.~{Caito}, M.~G. {Dainotti}, R.~{Guida}
  and R.~{Ruffini}, The ``fireshell'' model and the ``canonical'' grb
  scenario., in {\em Relativistic Astrophysics\/},  eds. C.~L. {Bianco} and
  S.~S. {Xue}, American Institute of Physics Conference Series, Vol.~966January
  2008.

\bibitem{2003PhLB..573...33R}
R.~{Ruffini}, L.~{Vitagliano} and S.-S. {Xue}, {\em Physics Letters B} {\bf
  573}, 33 (October 2003).

\bibitem{2005IJMPD..14..131R}
R.~{Ruffini}, F.~{Fraschetti}, L.~{Vitagliano} and S.-S. {Xue}, {\em IJMPD}
  {\bf 14}, 131  (2005).

\bibitem{2006NCimB.121.1477F}
F.~{Fraschetti}, R.~{Ruffini}, L.~{Vitagliano} and S.~S. {Xue}, {\em Nuovo
  Cimento B} {\bf 121}, 1477 (December 2006).

\bibitem{TEXAS}
R.~{Ruffini}, L.~{Izzo}, A.~V. {Penacchioni}, C.~L. {Bianco}, L.~{Caito}, S.~K.
  {Chakrabarti} and A.~{Nandi}, {\em PoS (Texas2010)} , p.~101  (2011).

\bibitem{2009GCN..9512....1S}
P.~{Schady}, W.~H. {Baumgartner}, A.~P. {Beardmore}, S.~{Campana}, P.~A.
  {Curran}, C.~{Guidorzi}, J.~A. {Kennea}, J.~{Mao}, R.~{Margutti}, J.~P.
  {Osborne}, K.~L. {Page}, P.~{Romano}, M.~H. {Siegel}, G.~{Stratta} and T.~N.
  {Ukwatta}, {\em GCN Circ.} {\bf 9512}  (2009).

\bibitem{2009GCN..9530....1B}
W.~H. {Baumgartner}, S.~D. {Barthelmy}, J.~R. {Cummings}, E.~E. {Fenimore},
  N.~{Gehrels}, H.~A. {Krimm}, C.~B. {Markwardt}, D.~M. {Palmer},
  T.~{Sakamoto}, G.~{Sato}, P.~{Schady}, M.~{Stamatikos}, J.~{Tueller} and
  T.~N. {Ukwatta}, {\em GCN Circ.} {\bf 9530}, p.~1  (2009).

\bibitem{1994ApJS...92..229F}
G.~J. {Fishman}, C.~A. {Meegan}, R.~B. {Wilson}, M.~N. {Brock}, J.~M. {Horack},
  C.~{Kouveliotou}, S.~{Howard}, W.~S. {Paciesas}, M.~S. {Briggs}, G.~N.
  {Pendleton}, T.~M. {Koshut}, R.~S. {Mallozzi}, M.~{Stollberg} and J.~P.
  {Lestrade}, {\em ApJSS} {\bf 92}, 229 (May 1994).

\bibitem{2009GCN..9534....1S}
T.~{Sakamoto}, T.~N. {Ukwatta} and S.~D. {Barthelmy}, {\em GCN Circ.} {\bf
  9534}, p.~1  (2009).

\bibitem{2009GCN..9528....1B}
A.~P. {Beardmore} and P.~{Schady}, {\em GCN Circ.} {\bf 9528}, p.~1  (2009).

\bibitem{2009GCN..9535....1M}
S.~{McBreen}, {\em GCN Circ.} {\bf 9535}  (2009).

\bibitem{2009GCN..9518....1C}
S.~B. {Cenko}, D.~A. {Perley}, V.~{Junkkarinen}, M.~{Burbidge}, U.~S. {Diego}
  and K.~{Miller}, {\em GCN Circ.} {\bf 9518}  (2009).

\bibitem{2007ApJ...660...16S}
B.~E. {Schaefer}, {\em ApJ} {\bf 660}, 16 (May 2007).

\bibitem{2009GCN..9553....1G}
S.~{Golenetskii}, R.~{Aptekar}, E.~{Mazets}, V.~{Pal'Shin}, D.~{Frederiks},
  P.~{Oleynik}, M.~{Ulanov}, D.~{Svinkin} and T.~{Cline}, {\em GCN Circ.} {\bf
  9553}, p.~1  (2009).

\bibitem{2009GCN..9524....1L}
F.~{Longo}, E.~{Moretti}, G.~{Barbiellini}, E.~{Vallazza}, M.~{Trifoglio},
  A.~{Bulgarelli}, F.~{Gianotti}, F.~{Fuschino}, M.~{Marisaldi}, C.~{Labanti},
  M.~{Galli}, G.~{Di Cocco}, S.~{Cutini}, C.~{Pittori}, M.~{Tavani},
  E.~{Striani}, G.~{Pucella}, F.~{D'Ammando}, V.~{Vittorini}, A.~{Argan},
  A.~{Trois}, G.~{Piano}, S.~{Sabatini}, E.~M. {Del}, M.~{Feroci},
  Y.~{Evangelista}, I.~{Donnarumma}, L.~{Pacciani}, P.~{Soffitta}, E.~{Costa},
  F.~{Lazzarotto}, I.~{Lapshov}, M.~{Rapisarda}, A.~{Giuliani}, A.~{Chen},
  S.~{Mereghetti}, F.~{Perotti}, P.~{Caraveo}, A.~{Pellizzoni}, M.~{Pilia},
  S.~{Vercellone}, P.~{Picozza}, A.~{Morselli}, M.~{Prest}, P.~{Lipari},
  D.~{Zanello}, A.~{Rappoldi}, P.~{Cattaneo}, P.~{Giommi}, P.~{Santolamazza},
  F.~{Verrecchia} and L.~{Salotti}, {\em GCN Circ.} {\bf 9524}  (2009).

\bibitem{2008cosp...37.1596K}
Y.~{Kotov}, A.~{Kochemasov}, S.~{Kuzin}, V.~{Kuznetsov}, J.~{Sylwester} and
  V.~{Yurov}, {Set of instruments for solar EUV and soft X-ray monitoring
  onboard satellite Coronas-Photon}, in {\em 37th COSPAR Scientific
  Assembly\/}, , COSPAR Meeting Vol.~372008.

\bibitem{2009arXiv0912.4126N}
A.~{Nandi}, A.~R. {Rao}, S.~K. {Chakrabarti}, J.~P. {Malkar}, S.~{Sreekumar},
  D.~{Debnath}, M.~K. {Hingar}, T.~{Kotoch}, Y.~{Kotovk} and
  A.~{Arkhangelskiy}, {\em ArXiv:0912.4126}  (December 2009).

\bibitem{2011ApJ...728...42R}
A.~R. {Rao}, J.~P. {Malkar}, M.~K. {Hingar}, V.~K. {Agrawal}, S.~K.
  {Chakrabarti}, A.~{Nandi}, D.~{Debnath}, T.~B. {Kotoch}, R.~{Sarkar}, T.~R.
  {Chidambaram}, P.~{Vinod}, S.~{Sreekumar}, Y.~D. {Kotov}, A.~S. {Buslov},
  V.~N. {Yurov}, V.~G. {Tyshkevich}, A.~I. {Arkhangelskij}, R.~A. {Zyatkov} and
  S.~{Naik}, {\em ApJ} {\bf 728}, p.~42 (February 2011).

\bibitem{2011MNRAS.413..669C}
Z.~{Cano}, D.~{Bersier}, C.~{Guidorzi}, R.~{Margutti}, K.~M. {Svensson},
  S.~{Kobayashi}, A.~{Melandri}, K.~{Wiersema}, A.~{Pozanenko}, A.~J. {van der
  Horst}, G.~G. {Pooley}, A.~{Fernandez-Soto}, A.~J. {Castro-Tirado}, A.~D.~U.
  {Postigo}, M.~{Im}, A.~P. {Kamble}, D.~{Sahu}, J.~{Alonso-Lorite},
  G.~{Anupama}, J.~L. {Bibby}, M.~J. {Burgdorf}, N.~{Clay}, P.~A. {Curran},
  T.~A. {Fatkhullin}, A.~S. {Fruchter}, P.~{Garnavich}, A.~{Gomboc},
  J.~{Gorosabel}, J.~F. {Graham}, U.~{Gurugubelli}, J.~{Haislip}, K.~{Huang},
  A.~{Huxor}, M.~{Ibrahimov}, Y.~{Jeon}, Y.-B. {Jeon}, K.~{Ivarsen},
  D.~{Kasen}, E.~{Klunko}, C.~{Kouveliotou}, A.~{Lacluyze}, A.~J. {Levan},
  V.~{Loznikov}, P.~A. {Mazzali}, A.~S. {Moskvitin}, C.~{Mottram}, C.~G.
  {Mundell}, P.~E. {Nugent}, M.~{Nysewander}, P.~T. {O'Brien}, W.-K. {Park},
  V.~{Peris}, E.~{Pian}, D.~{Reichart}, J.~E. {Rhoads}, E.~{Rol},
  V.~{Rumyantsev}, V.~{Scowcroft}, D.~{Shakhovskoy}, E.~{Small}, R.~J. {Smith},
  V.~V. {Sokolov}, R.~L.~C. {Starling}, I.~{Steele}, R.~G. {Strom}, N.~R.
  {Tanvir}, Y.~{Tsapras}, Y.~{Urata}, O.~{Vaduvescu}, A.~{Volnova},
  A.~{Volvach}, R.~A.~M.~J. {Wijers}, S.~E. {Woosley} and D.~R. {Young}, {\em
  MNRAS} {\bf 413}, 669 (May 2011).

\bibitem{2011A&A...525A..53G}
D.~{Guetta}, E.~{Pian} and E.~{Waxman}, {\em A\&A} {\bf 525}, p.~A53 (January
  2011).

\bibitem{2004ApJ...614..827R}
F.~{Ryde}, {\em ApJ} {\bf 614}, 827 (October 2004).

\bibitem{2003AIPC..668...16R}
R.~{Ruffini}, C.~L. {Bianco}, P.~{Chardonnet}, F.~{Fraschetti}, L.~{Vitagliano}
  and S.-S. {Xue}, New perspectives in physics and astrophysics from the
  theoretical understanding of gamma-ray bursts, in {\em Cosmology and
  Gravitation\/},  eds. M.~{Novello} and S.~E. {Perez Bergliaffa}, American
  Institute of Physics Conference Series, Vol.~668June 2003.

\bibitem{2007ESASP.622..561R}
R.~{Ruffini}, M.~G. {Bernardini}, C.~L. {Bianco}, L.~{Caito}, P.~{Chardonnet},
  M.~G. {Dainotti}, F.~{Fraschetti}, R.~{Guida}, G.~{Vereshchagin} and S.-S.
  {Xue}, The role of grb 031203 in clarifying the astrophysical grb scenario,
  in {\em The $6^{th}$ Integral Workshop - The Obscured Universe\/},  eds.
  S.~{Grebenev}, R.~{Sunyaev}, C.~{Winkler}, A.~{Parmar} and L.~{Ouwehand}, ESA
  Special Publication, Vol.~SP-6222007.

\bibitem{2010JKPS...57..551I}
L.~{Izzo}, M.~G. {Bernardini}, C.~L. {Bianco}, L.~{Caito}, B.~{Patricelli} and
  R.~{Ruffini}, {\em Journal of Korean Physical Society} {\bf 57}, p.
  551 (September 2010).

\bibitem{2010ApJ...709L.172R}
F.~{Ryde}, M.~{Axelsson}, B.~B. {Zhang}, S.~{McGlynn}, A.~{Pe'er},
  C.~{Lundman}, S.~{Larsson}, M.~{Battelino}, B.~{Zhang}, E.~{Bissaldi},
  J.~{Bregeon}, M.~S. {Briggs}, J.~{Chiang}, F.~{de Palma}, S.~{Guiriec},
  J.~{Larsson}, F.~{Longo}, S.~{McBreen}, N.~{Omodei}, V.~{Petrosian},
  R.~{Preece} and A.~J. {van der Horst}, {\em ApJ} {\bf 709}, L172 (February
  2010).

\bibitem{2011ApJ...727L..33G}
S.~{Guiriec}, V.~{Connaughton}, M.~S. {Briggs}, M.~{Burgess}, F.~{Ryde},
  F.~{Daigne}, P.~{M{\'e}sz{\'a}ros}, A.~{Goldstein}, J.~{McEnery},
  N.~{Omodei}, P.~N. {Bhat}, E.~{Bissaldi}, A.~{Camero-Arranz}, V.~{Chaplin},
  R.~{Diehl}, G.~{Fishman}, S.~{Foley}, M.~{Gibby}, M.~M. {Giles},
  J.~{Greiner}, D.~{Gruber}, A.~{von Kienlin}, M.~{Kippen}, C.~{Kouveliotou},
  S.~{McBreen}, C.~A. {Meegan}, W.~{Paciesas}, R.~{Preece}, A.~{Rau},
  D.~{Tierney}, A.~J. {van der Horst} and C.~{Wilson-Hodge}, {\em ApJ} {\bf
  727}, p.~L33 (February 2011).

\bibitem{2008ApJ...682..463P}
A.~{Pe'er}, {\em ApJ} {\bf 682}, 463 (July 2008).

\bibitem{2011arXiv1104.2274H}
J.~{Hjorth} and J.~S. {Bloom}, {\em ArXiv e-prints}  (April 2011).

\bibitem{1988PhR...163...13N}
K.~{Nomoto} and M.~{Hashimoto}, {\em Phys. Rep.} {\bf 163}, 13  (1988).

\bibitem{nomoto1994}
K.~{Nomoto}, H.~{Yamaoka}, O.~R. {Pols}, E.~P.~J. {van den Heuvel},
  K.~{Iwamoto}, S.~{Kumagaiparallel} and T.~{Shigeyama}, {\em Nature} {\bf
  371}, 227  (1994).

\bibitem{1994ApJ...437L.115I}
K.~{Iwamoto}, K.~{Nomoto}, P.~{Hoflich}, H.~{Yamaoka}, S.~{Kumagai} and
  T.~{Shigeyama}, {\em ApJ} {\bf 437}, L115 (December 1994).

\bibitem{2006ARA&A..44..507W}
S.~E. {Woosley} and J.~S. {Bloom}, {\em ARAA} {\bf 44}, 507 (September 2006).

\bibitem{1990ApJ...365L..55S}
A.~{Shemi} and T.~{Piran}, {\em ApJ} {\bf 365}, L55 (December 1990).

\bibitem{1993MNRAS.263..861P}
T.~{Piran}, A.~{Shemi} and R.~{Narayan}, {\em MNRAS} {\bf 263}, p.~861 (August
  1993).

\bibitem{2008mgm..conf..368R}
R.~{Ruffini}, M.~G. {Bernardini} and C.~L. {Bianco et al.}, {On Gamma-Ray
  Bursts}, in {\em The Eleventh Marcel Grossmann Meeting\/},  eds.
  H.~{Kleinert}, R.~T. {Jantzen} and R.~{Ruffini}  (September 2008).

\bibitem{2012ApJ...758L...7R}
J.~A. {Rueda} and R.~{Ruffini}, {\em ApJ} {\bf 758}, p.~L7 (October 2012).

\bibitem{2012MNRAS.420..810T}
O.~D. {Toropina}, M.~M. {Romanova} and R.~V.~E. {Lovelace}, {\em MNRAS} {\bf
  420}, 810 (February 2012).

\bibitem{1944MNRAS.104..273B}
H.~{Bondi} and F.~{Hoyle}, {\em MNRAS} {\bf 104}, p.~273  (1944).

\bibitem{1939PCPS...35..405H}
F.~{Hoyle} and R.~A. {Lyttleton}, {\em Proceedings of the Cambridge
  Philosophical Society} {\bf 35}, p.~405  (1939).

\bibitem{1989ApJ...346..847C}
R.~A. {Chevalier}, {\em ApJ} {\bf 346}, 847 (November 1989).

\bibitem{2013A&A...551A.133P}
A.~V. {Penacchioni}, R.~{Ruffini}, C.~L. {Bianco}, L.~{Izzo}, M.~{Muccino},
  G.~B. {Pisani} and J.~A. {Rueda}, {\em A\&A} {\bf 551}, p.~A133 (March 2013).

\bibitem{2012arXiv1205.6651I}
L.~{Izzo}, R.~{Ruffini} and C.~L. {Bianco et al.}, {\em A\&A, submitted; arXiv:
  1205.6651}  (May 2012).

\bibitem{1972ApJ...172L..79S}
E.~{Schreier}, R.~{Levinson}, H.~{Gursky}, E.~{Kellogg}, H.~{Tananbaum} and
  R.~{Giacconi}, {\em ApJ} {\bf 172}, p.~L79 (March 1972).

\bibitem{1972ApJ...174L..27W}
R.~E. {Wilson}, {\em ApJ} {\bf 174}, p.~L27 (May 1972).

\bibitem{1972ApJ...174L.143T}
H.~{Tananbaum}, H.~{Gursky}, E.~M. {Kellogg}, R.~{Levinson}, E.~{Schreier} and
  R.~{Giacconi}, {\em ApJ} {\bf 174}, p.~L143 (June 1972).

\bibitem{1973ApJ...180L..15L}
R.~W. {Leach} and R.~{Ruffini}, {\em ApJ} {\bf 180}, p.~L15 (February 1973).

\bibitem{1973ApJ...179..585D}
K.~{Davidson} and J.~P. {Ostriker}, {\em ApJ} {\bf 179}, 585 (January 1973).

\bibitem{2011ApJ...730...25R}
M.~L. {Rawls}, J.~A. {Orosz}, J.~E. {McClintock}, M.~A.~P. {Torres}, C.~D.
  {Bailyn} and M.~M. {Buxton}, {\em ApJ} {\bf 730}, p.~25 (March 2011).

\bibitem{2002ApJ...573L..27I}
S.~{Immler}, A.~S. {Wilson} and Y.~{Terashima}, {\em ApJ} {\bf 573}, L27 (July
  2002).

\bibitem{2004A&A...413..107S}
R.~{Soria}, E.~{Pian} and P.~A. {Mazzali}, {\em A\&A} {\bf 413}, 107 (January
  2004).

\bibitem{2005tmgm.meet..369R}
R.~{Ruffini}, M.~G. {Bernardini}, C.~L. {Bianco}, L.~{Vitagliano}, S.-S. {Xue},
  P.~{Chardonnet}, F.~{Fraschetti} and V.~{Gurzadyan}, Black hole physics and
  astrophysics: The grb-supernova connection and urca-1 - urca-2, in {\em The
  Tenth Marcel Grossmann Meeting. On recent developments in theoretical and
  experimental general relativity, gravitation and relativistic field
  theories\/},  eds. M.~{Novello}, S.~{Perez Bergliaffa} and R.~{Ruffini}
   (Singapore: World Scientific, January 2005).

\bibitem{2012A&A...540A..12N}
R.~{Negreiros}, R.~{Ruffini}, C.~L. {Bianco} and J.~A. {Rueda}, {\em A\&A} {\bf
  540}, p.~A12 (April 2012).

\bibitem{2007ARep...51..291T}
A.~V. {Tutukov} and A.~V. {Fedorova}, {\em Astronomy Reports} {\bf 51},
  291 (April 2007).

\bibitem{2012ApJ...752L...2C}
R.~A. {Chevalier}, {\em ApJ} {\bf 752}, p.~L2 (June 2012).

\bibitem{sedov46}
L.~I. {Sedov}, {\em Compt. Rend.  (Doklady) Acad. Sci. URSS} {\bf 52}, 17
   (1946).

\bibitem{1959sdmm.book.....S}
L.~I. {Sedov}, {\em {Similarity and Dimensional Methods in Mechanics}} 1959.

\bibitem{Cano2011}
Z.~{Cano}, D.~{Bersier}, C.~{Guidorzi}, R.~{Margutti}, K.~M. {Svensson},
  S.~{Kobayashi}, A.~{Melandri}, K.~{Wiersema}, A.~{Pozanenko}, A.~J. {van der
  Horst}, G.~G. {Pooley}, A.~{Fernandez-Soto}, A.~J. {Castro-Tirado}, A.~D.~U.
  {Postigo}, M.~{Im}, A.~P. {Kamble}, D.~{Sahu}, J.~{Alonso-Lorite},
  G.~{Anupama}, J.~L. {Bibby}, M.~J. {Burgdorf}, N.~{Clay}, P.~A. {Curran},
  T.~A. {Fatkhullin}, A.~S. {Fruchter}, P.~{Garnavich}, A.~{Gomboc},
  J.~{Gorosabel}, J.~F. {Graham}, U.~{Gurugubelli}, J.~{Haislip}, K.~{Huang},
  A.~{Huxor}, M.~{Ibrahimov}, Y.~{Jeon}, Y.-B. {Jeon}, K.~{Ivarsen},
  D.~{Kasen}, E.~{Klunko}, C.~{Kouveliotou}, A.~{Lacluyze}, A.~J. {Levan},
  V.~{Loznikov}, P.~A. {Mazzali}, A.~S. {Moskvitin}, C.~{Mottram}, C.~G.
  {Mundell}, P.~E. {Nugent}, M.~{Nysewander}, P.~T. {O'Brien}, W.-K. {Park},
  V.~{Peris}, E.~{Pian}, D.~{Reichart}, J.~E. {Rhoads}, E.~{Rol},
  V.~{Rumyantsev}, V.~{Scowcroft}, D.~{Shakhovskoy}, E.~{Small}, R.~J. {Smith},
  V.~V. {Sokolov}, R.~L.~C. {Starling}, I.~{Steele}, R.~G. {Strom}, N.~R.
  {Tanvir}, Y.~{Tsapras}, Y.~{Urata}, O.~{Vaduvescu}, A.~{Volnova},
  A.~{Volvach}, R.~A.~M.~J. {Wijers}, S.~E. {Woosley} and D.~R. {Young}, {\em
  MNRAS} {\bf 413}, 669 (May 2011).

\bibitem{Grupe2007b}
D.~{Grupe}, C.~{Gronwall}, X.-Y. {Wang}, P.~W.~A. {Roming}, J.~{Cummings},
  B.~{Zhang}, P.~{M{\'e}sz{\'a}ros}, M.~D. {Trigo}, P.~T. {O'Brien}, K.~L.
  {Page}, A.~{Beardmore}, O.~{Godet}, D.~E. {vanden Berk}, P.~J. {Brown},
  S.~{Koch}, D.~{Morris}, M.~{Stroh}, D.~N. {Burrows}, J.~A. {Nousek},
  M.~{McMath Chester}, S.~{Immler}, V.~{Mangano}, P.~{Romano}, G.~{Chincarini},
  J.~{Osborne}, T.~{Sakamoto} and N.~{Gehrels}, {\em ApJ} {\bf 662}, 443 (June
  2007).

\bibitem{Larsson2011}
J.~{Larsson}, F.~{Ryde}, C.~{Lundman}, S.~{McGlynn}, S.~{Larsson}, M.~{Ohno}
  and K.~{Yamaoka}, {\em MNRAS} {\bf 414}, 2642 (July 2011).

\bibitem{Kann2008}
D.~A. {Kann}, S.~{Schulze} and A.~C. {Updike}, {\em GRB Coordinates Network}
  {\bf 7627}, p.~1  (2008).

\bibitem{Cobb2010}
B.~E. {Cobb}, J.~S. {Bloom}, D.~A. {Perley}, A.~N. {Morgan}, S.~B. {Cenko} and
  A.~V. {Filippenko}, {\em ApJ} {\bf 718}, L150 (August 2010).

\bibitem{Stamatikos2009}
M.~{Stamatikos}, S.~D. {Barthelmy}, W.~H. {Baumgartner}, J.~R. {Cummings},
  E.~E. {Fenimore}, N.~{Gehrels}, H.~A. {Krimm}, C.~B. {Markwardt}, D.~M.
  {Palmer}, T.~{Sakamoto}, E.~{Troja}, J.~{Tueller} and T.~N. {Ukwatta}, {\em
  GRB Coordinates Network} {\bf 10197}, p.~1  (2009).

\bibitem{DAvanzo2012}
P.~{D'Avanzo}, A.~{Melandri}, E.~{Palazzi}, S.~{Campana}, M.~{Della Valle},
  E.~{Pian}, R.~{Salvaterra} and G.~{Tagliaferri}, {\em GRB Coordinates
  Network} {\bf 13069}, p.~1  (2012).

\bibitem{Zauderer2012}
B.~A. {Zauderer}, E.~{Berger}, R.~{Margutti}, A.~J. {Levan}, F.~{Olivares},
  D.~A. {Perley}, W.~{Fong}, A.~{Horesh}, A.~C. {Updike}, J.~{Greiner}, N.~R.
  {Tanvir}, T.~{Laskar}, R.~{Chornock}, A.~M. {Soderberg}, K.~M. {Menten},
  E.~{Nakar}, J.~{Carpenter} and P.~{Chandra}, {\em ArXiv e-prints}  (September
  2012).

\bibitem{Burbridge1957}
E.M.~Burbidge, G.R.~Burbidge, W.A.~Fowler, and F.~Hoyle, 
{\em Rev.\ Mod.\ Phys.} {\bf 29}, p.~547   (1957).

\end{thebibliography}
\end{document}